\documentclass[oneside]{book}

\usepackage{geometry}
\newgeometry{vmargin={25mm}, hmargin={30mm}} 
\title{Learning the physics of open quantum systems from experiments}
\author{Alexandra Ramôa\\
\small Supervised by: Ernesto Galvão and Luís Soares Barbosa\\\\
\small Master Thesis in Engineering Physics\\
\small University of Minho\\\\
\textit{November 2021}}

\date{}
\usepackage{hyperref} 
\usepackage{epigraph} 
\usepackage[nomain, acronym, toc]{glossaries-extra}
\makeatletter
\newlength\interepigraphskip
\renewcommand\epigraph[3][\interepigraphskip]{\vspace{\beforeepigraphskip}
  {\epigraphsize\begin{\epigraphflush}\begin{minipage}{\epigraphwidth}
    \@epitext{#2}\\[#1] \@episource{#3}
    \end{minipage}\end{\epigraphflush}
    \vspace{\afterepigraphskip}}}
\makeatother
\usepackage[table]{xcolor} 
\usepackage{mathtools}
\usepackage{amssymb}
\usepackage{amsmath}
\usepackage{subcaption}
\usepackage{siunitx}
\usepackage{multicol} 
\usepackage{scalerel}
\usepackage{stackengine}
\usepackage{fbox}
\usepackage{dashbox}
\usepackage{bm}
\usepackage{enumitem}
\setlist{topsep=4pt,itemsep=0pt,parsep=2pt}

\newcommand{\op}[1]{\mathbf{\hat{#1}}}
\newcommand{\opsub}[2]{\mathbf{\hat{#1}_#2}}
\newcommand{\norm}[1]{\left\lVert#1\right\rVert}
\newcommand{\bra}[1]{\left\langle\ #1\ \right\rvert}
\newcommand{\ket}[1]{\left\lvert \ #1 \ \right\rangle}
\newcommand{\braket}[2]{\left\langle\ #1 \mid #2 \ \right\rangle}
\newcommand{\ketbra}[2]{\left\lvert\ #1\ \right\rangle \!
\left\langle\ #2\ \right\rvert}
\newcommand{\pbra}[2]{\prescript{}{#2}{\bra{#1}}}
\newcommand\equalhat{\mathrel{\stackon[1.5pt]{=}{\stretchto{%
    \scalerel*[\widthof{=}]{\wedge}{\rule{1ex}{3ex}}}{0.5ex}}}}
\newcommand\dboxed[1]{\dbox{\ensuremath{#1}}}
\newcommand{\dd}[0]{\mathrm{d}}
\newcommand{\abs}[1]{\left\lvert#1\right\rvert}
\mathchardef\hyph="2D
\newcolumntype{g}{>{\columncolor{gray!15}}c}

\usepackage{float}
\floatstyle{boxed}
\newfloat{algorithm}{h}{loa}
\floatname{algorithm}{Algorithm}
\newacronym{bPM}{bPM}{block pseudo-marginal}
\newacronym{CLT}{CLT}{central limit theorem}
\newacronym{ECS}{ECS}{energy conserving subsampling}
\newacronym{ESS}{ESS}{effective sample size}
\newacronym{GRF}{GRF}{Gaussian rejection filtering}
\newacronym{HL}{HL}{Heisenberg limit}
\newacronym{HMC}{HMC}{Hamiltonian Monte Carlo}
\newacronym{IPE}{IPE}{iterative phase estimation}
\newacronym{IQFT}{IQFT}{inverse quantum Fourier transform}
\newacronym{KDE}{KDE}{kernel density estimate}
\newacronym{LWF}{LWF}{Liu-West filtering} 
\newacronym{MC}{MC}{Monte Carlo}
\newacronym{MCMC}{MCMC}{Markov Chain Monte Carlo}
\newacronym{MH}{MH}{Metropolis-Hastings}
\newacronym{NISQ}{NISQ}{noisy intermediate scale quantum}
\newacronym{NUTS}{NUTS}{no U-turn sampler}
\newacronym{OSR}{OSR}{operator-sum representation}
\newacronym{pCN}{pCN}{preconditioned Crank-Nicolson}
\newacronym{PM}{PM}{pseudo-marginal}
\newacronym{QAOA}{QAOA}{quantum approximate optimization algorithm}
\newacronym{QFT}{QFT}{quantum Fourier transform}
\newacronym{RWM}{RWM}{random walk Metropolis}
\newacronym{SG}{SG}{stochastic gradient}
\newacronym{SHMC}{SHMC}{sequential Hamiltonian Monte Carlo}
\newacronym{SIR}{SIR}{sequential importance resampling}
\newacronym{SIS}{SIS}{sequential importance sampling}
\newacronym{SISR}{SIS/R}{sequential importance sampling and resampling}
\newacronym{SMC}{SMC}{sequential Monte Carlo}
\newacronym{SQL}{SQL}{standard quantum limit}
\newacronym{TLE}{TLE}{tempered likelihood estimation}
\begin{document}

\maketitle

\vspace*{\fill} 
    \begin{flushright}
    \textit{``[...] all knowledge degenerates into probability ; and this probability is greater or less, according to our experience of the veracity or deceitfulness of our understanding, and according to the simplicity or intricacy of the question.``}\\
    \vspace{1em}
    --- David Hume
    \end{flushright}
\vspace*{\fill}
\vspace*{\fill}

\chapter*{Abstract}

The ability to efficiently determine the dynamics to which a quantum device conforms is vital for its reliable operation. Thus, as quantum machines evolve, the means for their characterization must evolve alongside them, especially as they reach the limits of classical tractability. Bayesian inference has been proposed as a solution, as it offers a flexible way of using experimental data to learn the dynamical parameters - or even models - governing the evolution of quantum systems. It gives rise to noise-resilient protocols, which are capable of quantifying their own uncertainty, of learning from scarce information, and of real-time estimation. Importantly, and apart from the obvious applications in sensing devices, online processing enables adaptivity, a stepping stone for achieving fundamental limits of metrology via what is called \textit{quantum-enhanced} estimation. Like so, the exploitation of quantum control as a resource within the Bayesian paradigm opens the door to uncertainties scaling at the Heisenberg limit.

Most work on the subject has relied on simple methods for portraying the Bayesian posterior, but they quickly become a bottleneck in realistic scenarios. The difficulty of the task grows with the complexity of the devices to be verified, which adds to the fact that they are ultimately open systems and as such undergo uncontrollable interactions with their surroundings. While capturing unwanted processes like decoherence is by itself a valuable endeavor, this poses extra challenges, and requires careful computational treatment. In general, when scaling up, the biggest difficulty of Bayesian learning is numerical integration. In this context, pairing it with advanced Monte Carlo methods makes for remarkably robust algorithms, which can succeed under complex features and control a multitude of trade-offs. 

\vspace*{\fill}


\tableofcontents


\chapter{Introduction}

Speaking about quantum mechanics, Albert Einstein famously expressed distaste at the idea that God would play dice with the universe. Yet, as of today, this renders our best means of describing reality. What then is the object of science if not the movement of these dice?

If measurements are dice play, metrology is a game of chance; its ultimate goal is to put the odds in our favor. It should then come as no surprise that methods borrowed from statistics may be especially handy when dealing with quantum systems. In particular, statistical inference is well tailored to the characterization of quantum systems - the pillar of tasks as vital as certification and simulation.

Inference paradigms sketch a path between experimental data and system properties, these data consisting of measurements performed on the system. This work focuses on \textit{Bayesian} inference, a method that relies on the Bayes' theorem to update one's \textit{confidence} in a hypothesis given observations - which we call the \textit{probability} of this hypothesis. This stands in contrast with frequentist statistics, where probabilities are relative frequencies. A downside of the Bayesian method is requiring that we consider multiple hypotheses instead of a single one; a benefit is \textit{allowing us} to consider multiple hypotheses instead of a single one. 

In other words, as much as this statistical paradigm may bring added complexity, it is not in vain that it does so. Improved descriptions can be had by averaging over hypothesis, be they different parameters or even different models. What this lacks in intuitiveness, it makes up for in predictive power; and science is ultimately concerned with reality, not some immaterial definition of truth. Projecting \textit{truth} into a stiff mold is no more natural - as in frequentist statistics, where doubt is artificially sifted out of the main theory and left up to meta-statistics.

Furthermore, acknowledged doubt paves the way for shaping strong convictions. In practice, systems do not always behave as our idealized version of themselves. This is doubly true of quantum systems, where noise is joined by fundamental randomness that amplifies our uncertainty. By embracing that uncertainty and welcoming it into our framework, we can reduce it from within - which makes for remarkably reliable and informative methods. Reliable, because they can recover from errors and estimate their own uncertainty; informative, because they can pick out the questions whose answers are the most meaningful. 

Another point in favour of the Bayesian approach is the fact that it is quite general, forgoing the need for a new solution in the face of a new problem. What is more, allows for incorporating previous knowledge into the estimation process, something which is often available in physically interesting scenarios. Lastly, it naturally allows for leveraging quantum resources, and for exploiting them beyond what is requisite. The characterization of a quantum nature calls for quantum resources, much like its simulation.

\section{Characterization of quantum systems}

If building a quantum device is a challenging task, so too is certifying its behaviour. A closed quantum system, such as an analogue quantum simulator, abides by a Hamiltonian. Being given the intended Hamiltonian, how can one check that it does? The knee-jerk solution is to compare its behaviour with classical predictions. However, this defeats the purpose of a quantum computer, which is of little interest if it can so easily be replicated classically. 

As quantum devices approach the frontier of classical tractability, more apt verification methods are in order \cite{Granade_2012}. Bringing together quantum resources and Bayesian statistics creates a powerful characterization tool for working in the quantum regime, which is crucial for approaching fundamental estimation limits \cite{Higgins_2007,Ferrie_2011}.

Bayesian inference systematizes the use of quantum resources for quantum characterization tasks; this could mean state or process tomography, parameter learning, establishing noise models, or others. It relies on comparing the expected behaviour of a comprehensive set of quantum simulators with the behaviour of the quantum system of interest, to obtain the configuration most similar to it. Owing to the broad scope of this analysis, the final product is a probability distribution over a continuum of hypotheses, called the \textit{posterior} distribution. As compared to a single point estimate, this provides an extensive insight rather than a lone guess.

Apart from being a useful assessment of the confidence the results should inspire, this insight arms us with the tools to perform \textit{quantum-enhanced} estimation. Although a probabilistic system will never unequivocally behave like itself, it will under some circumstances more than in others. We can choose those circumstances mindfully if we construct a holistic understanding of it, and allow it to interact with the way we observe it - a technique termed Bayesian experimental design. It leverages the Bayesian insight to forecast the outcomes resulting from the possible choices, aiming to choose the experimental controls that are expected to bring the best results in light of current knowledge. This offers the means to make the most out of each experiment, bringing the greatest possible returns out of the fewest possible measurements.

\section{Research questions, literature overview and objectives}

The main open problems pertaining to Bayesian Hamiltonian learning arise in two different fronts. One goal is to balance the quality of the experiments with the computational cost of finding them; and another is to choose the aforementioned \textit{comprehensive set} of quantum simulators, in order to capture relevant information while economizing simulation runs. 

The first of these pillars is intrinsic to the inference process, whereas the second one concerns its computational treatment. Together they dictate the usefulness of the inference-produced distribution, being responsible for shaping it and extracting information from it respectively. In short, the two main focuses are:
\begin{itemize}
    \item \textbf{Inference}: choose the experiment controls to maximize the utility of the experiments. Another relevant topic in this category is the choice of models, but this is more context-dependent. Should exact inference be within reach, these and kindred concerns would be the only ones, as they alone shape the posterior distribution. 
    \item \textbf{Representation}: discretize the probability distributions so as to best capture the inference results. There is generally a trade-off between resources and quality of the representation, where the resources are simulation runs. They provide the Bayesian \textit{likelihoods}, which tell how likely the simulator would have been to replicate the observed behaviour. 
\end{itemize}

Both problems become harder when scaling up complexity. In particular, when dealing with open quantum systems - which realistically are all quantum devices -, a greater number of parameters is expected to be at play. Their descriptions are given by more intricate formalisms, and learning all of their underlying variables commands both informative data and robust processing. Higher dimensional models are much less forgiving of careless measurements and brute-force algorithms. Further, this difference is not merely quantitative, as they tend to give rise to more complex features - such as multi-modality, the embodiment of redundancy in the probabilistic domain.

These problems require provisions of their own, namely more refined representation methods. Representing the posterior distribution appropriately is key for the correction of the relayed system behaviour, given in terms of expectation integrals over the posterior. \gls{MC} methods propose to estimate these expectations numerically, by probabilistically choosing samples to be used in numerical integration. Interestingly, they just like inference itself exploit the relation between \textit{simulation} and \textit{characterization}, but now in a statistical (rather than quantum-mechanical) sense. They offer a vast collection of solutions, which can differ greatly in how they produce samples and thus in how efficient they are. Unlike deterministic approaches such as grid-based quadrature, their required resources don't necessarily scale exponentially with the dimension, which is a first step toward escaping what is dramatically called \textit{the curse of dimensionality}. 

We now consider the main sources for this project, most of which concern Bayesian learning applied to quantum systems, and their associated discretization strategies. Arguably the most popular representation is via \gls{SIR} - a type of \gls{SMC}/particle filter - with \gls{LWF}, adopted in \cite{Granade_2012,Chase_2009,Wiebe_2014a,Wiebe_2014b,Wang_2017,Granade_2016,Hincks_2018} as well as by the available software \cite{qinfer}. This is a quite simple and lightweight strategy, especially due to \gls{LWF}. Its main advantage is keeping quantum simulation to a minimum; the drawback is that it preserves only the two first statistical moments of the distribution. In other words, normal distributions are kept invariant, but others are only assured to maintain their mean and variance. In the latter case, a sensitive trade-off between structure preservation and efficiency is imposed, endangering the viability of the method. This is a glaring defect; not only does it impose severe restrictions on accuracy, it may even lead to absolute failure. In particular, this happens in the case of multimodality, be it native or due to redundancy in the early stages of the inference process (when the information isn't yet decisive). Still less structure preserving approaches are the rejection filtering of \cite{Wiebe_2016} and Gaussian conjugate posteriors of \cite{Evans_2019}, which work under the assumption of absolute normality.

In \cite{Granade_2017}, the \gls{SIR}-\gls{LWF} protocol is adapted to multimodal distributions by means of clustering, in addition to other sophistications. This provides a working solution, but solves only the most blatant symptom and not the root cause - inaccuracy may still be an issue. Some physically motivated problems are concerned only with approximations and their uncertainty, rejecting the need for more complex or accurate expectations. In this case, \gls{SIR} and \gls{LWF} may suffice, and they are quite enticing due to their sparing use of quantum resources. Even still, simplistic and/or mode-seeking methods are likely to fail in more complex applications (namely in high dimensional parameter spaces). Further, this fix employs costly classical processing, which could be scraped entirely simply by replacing either \gls{LWF} or \gls{SIR} altogether with a more conservative approach. \gls{MCMC} can stand in for either, and is the most obvious solution. The downside is the increase in the computational cost, and if dispensing with \gls{SMC} a likely decrease in the overall convergence speed. 

Simple \gls{MCMC} methods have more recently been applied to Bayesian learning for quantum problems. Naive \gls{MH} is applied to quantum tomography in \cite{Mai_2017}; \cite{Williams_2017} points out the drawbacks of that type of strategy and uses slice sampling instead, but acknowledges that e.g. multimodal distributions require more advanced methods, mentioning \gls{SMC} as a possible solution. It is remarked that \gls{SMC} may incur inaccuracies, but this strongly depends on the implementation -  while it may be true of \gls{SIR} with \gls{LWF}, this is not necessarily so for more robust variants. Another source using slice sampling is \cite{Lu_2019}.

Still more recently, \gls{pCN} has been applied to pseudo-Bayesian quantum tomography \cite{Lukens_2020,Lukens_2020b,Mai_2021}, improving upon vanilla \gls{MH}. This type of \gls{MCMC} is rather recent, having been introduced in \cite{Cotter_2013}. By adapting the stepsize, it offers better acceptance probabilities for high dimensions as compared to standard \gls{MH} algorithms. However, it is not informed by the geometry of the target distribution, and so is still expected to produce diffusive transitions. Another more popular and widely studied variation that scales favourably with the dimension is \gls{HMC}\footnote{Note that the ideas behind \gls{pCN} and \gls{HMC} are not mutually exclusive, and \gls{pCN}-like improvements can be adopted for \gls{HMC} too. References \cite{Cotter_2013,Beskos_2017,Alenlov_2019} elaborate on the relation between these methods.} \cite{Duane_1987,Neal_2011,Betancourt_2018}, which brings two added benefits: good mixing, and robustness under complex distribution features \cite{Beskos_2017}. This is one of the methods tested here. Also, none of the above consider embedding \gls{MCMC} in an \gls{SMC} scheme, which may be very advantageous. \gls{SIR} is more reliable than isolated \gls{MCMC} and suitable for online estimation, enabling adaptivity. If post-processing, a still more reliable option is \gls{TLE} \cite{Neal2001}, a well-established method for complex statistical simulation problems.

Other publications on this subject typically consider less general approaches, or focus on other points instead. In \cite{Sergeevich_2011}, an exact representation (a Fourier series) makes a simple single parameter problem analytically tractable, but this method doesn't generalize well to more complex inference problems. Similarly, rough approximations such as the Gaussian mixtures of \cite{Craigie_2021} can be quite agile for few parameters, but are generally not scalable. Some papers focus mostly on the experimental design and barely \cite{Huszar_2012,Fiderer_2021} or not at all \cite{Ferrie_2011} in the discretization; \cite{Ferrie_2012} tackles the same problem but 
makes a normality assumption to make it at once general and analytically tractable. Likewise, many attempt to improve yet other aspects, while briefly mentioning \gls{SMC} strategies and referring to the sources above. Some notable examples are \cite{Ferrie_2014}, which relates the effect of relaxing the simulation requirements, \cite{Ferrie_2014b}, which considers embodying model uncertainty to improve robustness, and \cite{Gentile_2021}, which presents a more general algorithm that loosens the parametrization assumptions.

This dissertation intends to overview the main foundational topics of Bayesian learning, and to combine them with more advanced numerical methods that enjoy widespread acceptance by the statistical community. The final objective is to test the performance of these algorithms with an emphasis on quantum applications, while also implementing some of the standard strategies for comparison. The central statistical methods to be approached are all Monte Carlo algorithms, and fall into three categories:
\begin{itemize}
    \item \gls{MCMC} (\gls{MH}): \gls{HMC} and \gls{RWM}. This includes enhancements of these methods, such as subsampling and extensions of \gls{HMC} (namely the \gls{NUTS}).
    \item Improvements to the \gls{SIR}-\gls{LWF} algorithm. More specifically, replacements for \gls{LWF} with stronger assurances are considered. They precisely match the \gls{MCMC} methods mentioned above.
        \item \gls{TLE}, an \gls{SMC} algorithm distinct from \gls{SIR}.
\end{itemize}

\gls{HMC}/\gls{NUTS} and \gls{SMC} with \gls{MCMC} kernels (namely \gls{TLE}) are techniques used in leading software, such as state-of-the-art probabilistic packages and programming languages for Bayesian inference and statistical simulation: PyMC3 \cite{Salvatier_2016} supports both, and Stan \cite{Gelman_2015,Stan} intends to despite currently leaving \gls{SMC} out. Another technique they both use is variational inference, perhaps the most interesting lower precision and lower cost alternative to Monte Carlo algorithms - but this dissertation will focus on the latter.

Overall, the aims of this dissertation fall into four main topics, each corresponding to a chapter. First, to overview Bayesian inference, with an emphasis on the problem of learning the dynamical parameters of quantum systems. This includes experimental design, namely adaptive experimental design. Second, to explore numerical methods well-tailored to this paradigm and evaluate their relative efficiency. These are mostly Monte Carlo methods. Third, to apply the considered strategies to representative test cases in order to assess their merits. And lastly, to use this knowledge to characterize quantum devices, considering in particular quantum channels capturing decoherence processes or other open system effects. For this end, the theoretical foundations necessary to realize simple open system descriptions are considered as well.

\section{Related topics}

We will now allude to some well-known subjects that are connected with the ones to be explored in this dissertation, and the differences that separate them.

Quantum state tomography is the most related concept; however, it targets an instantaneous quantum state and not the entirety of the systems' dynamics. Quantum process tomography, on the other hand, means to characterize processes in the sense of operations or maps, and not continuous time evolution. By contrast, here we consider the dynamics undergone by a determined initial quantum state; furthermore, this is realized under some practical assumptions, resulting a full description of the system's time evolution. This can be given by operator(s), or a function; either way, the entity to be characterized is time-dependent, and so too are the measurements. As an example, this can mean determining the evolution of the excited state population as opposed to an instantaneous density matrix. More generally, any dynamical parameters could be targeted in the learning process. And finally, unlike in tomography, all the experiments here use measurements on a single basis - though this is a choice and not the method's own limitation.

Another related topic is quantum machine learning. While Bayesian learning can broadly speaking be considered machine learning, quantum machine learning in its usual sense is outside of the scope of this work. It generally refers to quantum-\textit{enhanced} machine learning, i.e. machine learning algorithms which seek to extract an advantage from quantum resources. In this case, the learning techniques are entirely classical, only they're \textit{applied} to quantum systems. In a few words, in our case quantum systems are \textit{a subject} and not a tool.

Likewise, quantum Monte Carlo - which deals directly with quantum problems - is quite distant from the Monte Carlo techniques approached here, inasmuch as the latter are not quantum by construction. Rather, they are structurally application-blind, their reliance on quantum simulation notwithstanding. They simply require resources that match the problem; should the problem be quantum, so too must they. 

Finally, quantum Bayesianism is associated with both quantum systems and Bayesian statistics, but on an epistemic level. It gives rise to a multitude of interpretations of quantum mechanics, the most notable being QBism. This dissertation focuses merely on the mathematical aspects of Bayesian statistics, and how it can serve quantum-mechanical systems (\textit{quantitatively}).

\section{Document structure}

\begin{figure}[!ht]
    \centering
    \includegraphics[width=\textwidth]{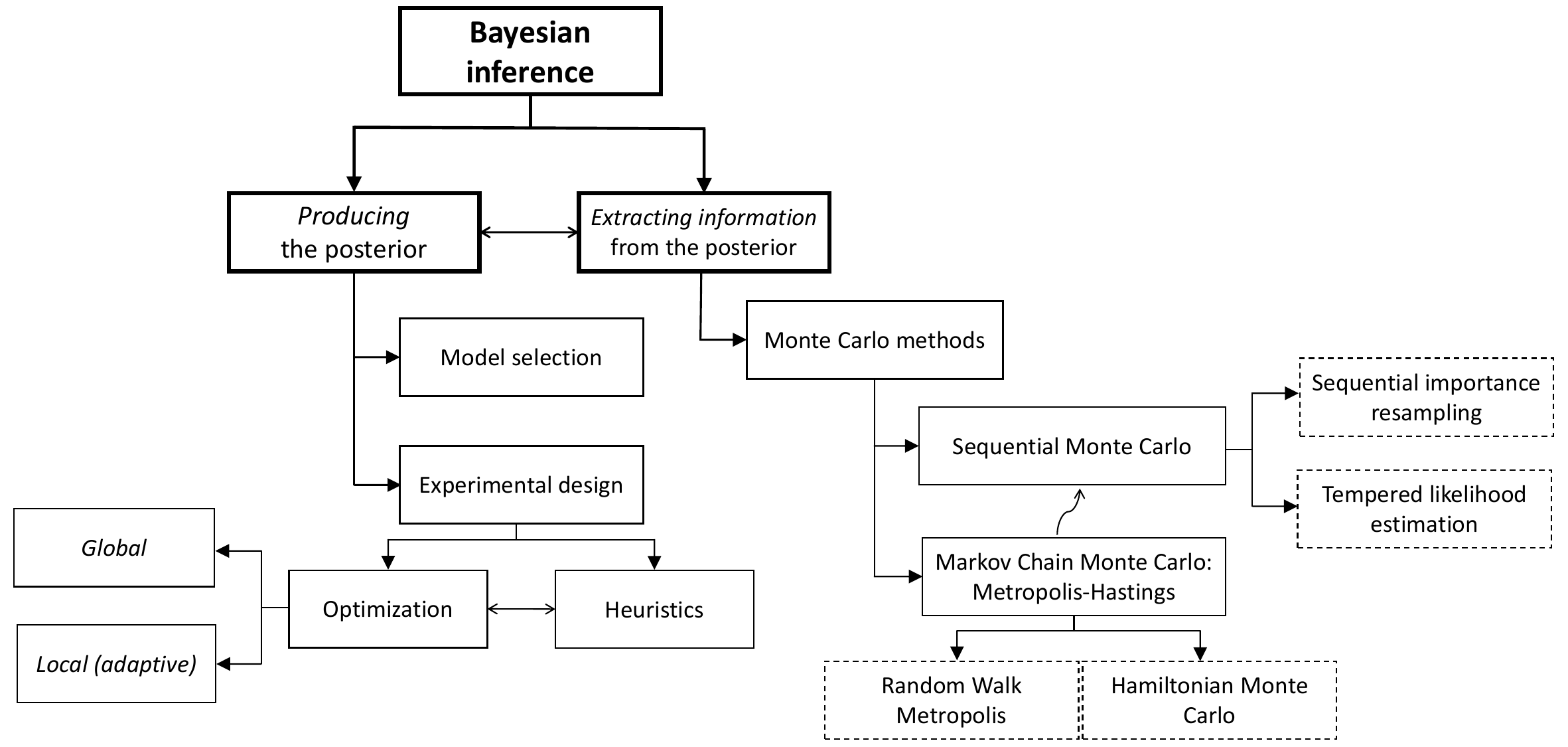}
    \caption{Diagram of the topics to be covered in this dissertation. Dashed boxes denote specific algorithms.}
    \label{fig:general_diagram}
\end{figure}

For a clearer picture, figure \ref{fig:general_diagram} schematically represents the topics to be covered throughout the document. The left half of the tree corresponds to chapter \ref{cha:quantum_parameter_estimation}, while the right half respects chapter \ref{cha:monte_carlo_posterior_sampling}. Chapters \ref{cha:applied_examples} and \ref{cha:experiments_quantum_hardware} showcase the results of applying this collection of methods to a selection of test cases; they consider statistical simulation and characterization of quantum devices from IBMQ respectively.

More specifically, the dissertation is structured as follows. Chapter \ref{cha:quantum_parameter_estimation} begins with the basics of Bayesian statistics and inference (sections \ref{sec:bayes_rule}
and \ref{sec:bayesian_inference}), including basic methods for exact and approximate inference (subsection \ref{sub:estimation_in_practice}). After that, section \ref{sec:application_to_quantum} discusses how this applies to quantum systems and the resources involved (\ref{sub:likelihood_evaluations_quantum}), and provides illustrative examples (\ref{sub:quantum_characterization_examples}). It also touches upon model selection (\ref{sub:model_selection}), which opens the door to more general algorithms. Finally, section \ref{sec:bayesian_experimental_design} respects Bayesian experimental design strategies. It covers both global optimization strategies and local adaptive ones (\ref{sub:adaptivity}), as well as how they relate to the fundamental limits on measurement precision (\ref{sub:beating_shot_noise}). This section finishes with some cost-effective heuristics for a typical quantum example (\ref{sub:precession_heuristics}).

After this overview of Bayesian inference, chapter \ref{cha:monte_carlo_posterior_sampling} focuses on complementary numerical methods whose purpose is to characterize the resulting distribution. It starts with the motivation and idea behind \gls{MC} methods in section \ref{sec:monte_carlo}, and follows with
\gls{SMC} in \ref{sec:sequential_monte_carlo}. Contained therein is \gls{LWF} (\ref{sub:liu_west}), as well as \gls{SIR} (\ref{sub:sir}) and \gls{TLE} (\ref{sub:tle}). Then, section \ref{sec:mcmc} introduces \gls{MCMC}, exemplifying the algorithm with the elementary \gls{RWM} variant (\ref{sub:rwm}) before discussing \gls{HMC} (\ref{sub:hmc}) and the usefulness of \gls{MCMC} for \gls{SMC} (and vice versa; subsection \ref{sub:mcmc_smc}). Finally, section \ref{sec:hmc_implementations} delves deeper into \gls{HMC}, namely its applicability (\ref{sub:applicability_hmc}), progressive biased variants (\ref{sub:trajectory_generation}), and dynamic implementations such as the \gls{NUTS} (\ref{sub:nuts}). Section \ref{sec:subsampling} finishes with an overview of subsampling strategies, with an emphasis on \gls{HMC}: control variate-enhanced estimators (\ref{sub:control_variates}), stochastic gradients (\ref{sub:sg_hmc}) and energy conserving subsampling (\ref{sub:shmc_ecs}).

Chapter \ref{cha:applied_examples} considers a few applications of statistical simulation. Three simple sampling problems are considered first in section \ref{sec:sampling_problems} (6-dimensional Gaussian in subsection \ref{sub:6d_gaussian}, Rosenbrock function in \ref{sub:rosenbrock}, a smiley face in \ref{sub:smiley_kde}), followed by two inference problems using generated data in section \ref{sec:inference_problems} (phase estimation in subsection \ref{sub:phase_estimation}, multivariate sum of squared sinusoids in \ref{sub:multi_cos}). 

As for chapter \ref{cha:experiments_quantum_hardware}, it starts with some rudiments on open quantum systems (mixed states in the Bloch picture in subsection \ref{sub:bloch}, and the Kraus \gls{OSR} in \ref{sub:kraus}) and the grounds for the experiments to be performed in section \ref{sec:models_open_quantum_systems} (subsections \ref{sub:amplitude_damping} and \ref{sub:phase_damping} present simple noise channels depicting amplitude and phase damping respectively, and \ref{sub:spin_precession} discusses Larmor/Rabi/Ramsey oscillations). The results of these experiments follow on section \ref{sec:characterization_quantum}. Relevant considerations, including an overview of the methods, are laid out in subsection \ref{sub:exp_setup}, while the following sections present and compare the results of using measurements to characterize qubit dynamics for a few physical processes: Hahn echo sequences (subsection \ref{sub:exp_t2}), energy relaxation (\ref{sub:exp_t1}), Ramsey sequences (\ref{sub:exp_ramsey_2d}), and \textit{Hahn-Ramsey} sequences (\ref{sub:exp_ramsey_1d}).

Lastly, chapter \ref{cha:conclusions} briefly discusses the results and recapitulates the main takeaways. 

The code used for this project can be found online at the GitHub repository \cite{repository}.

\chapter{Quantum Parameter Estimation}
\label{cha:quantum_parameter_estimation}

The attribution of limited predictive power to reality itself has been a source of controversy in quantum mechanics, and one whose implications reach well beyond conceptual perplexity. Ultimately, to reject predictability is to resign to ambiguity; that truth leaves the future to chance precludes this future from conveying its past.

Less abstractly, the fact that quantum measurements are inherently probabilistic bars any single one from disclosing the complete state of the system. This indetermination may be overcome by repeating different measurements multiple times on an ensemble of states to be characterized. But evidently, this adds a layer of complexity to metrology, effectively placing data collection on the same level as the usual experimental limitations (such as the precision of the measuring apparatus).

In any case, should we want to describe the physics of quantum systems, we must honor their non-deterministic essence. This can only be achieved by adopting characterization techniques appropriate to phenomena of probabilistic nature. Hence, methods borrowed from statistical inference may be of use when one means to work with technology in the quantum regime.

One such method is Bayesian inference. Interestingly, Bayesianism is another theory which has been subjected to criticism due to holding probabilities a little too close to heart. However, it is also one more which came to enjoy widespread acceptance on account of its practical successes, once they were made possible by the available technological means.

This chapter will present basic concepts of Bayesian statistics and inference, and how they can be of use to quantum science. We will use a few basic examples to illustrate these techniques and contrast them with their more traditional counterparts. The formal justification for using such techniques will also be briefly touched on, as well as some refinements that the framework enables.

\section{Bayes' rule}
\label{sec:bayes_rule}

If we repeatedly toss a coin and it always lands tails up, we might strongly suspect that it is biased. But what if the trend is more subtle than that, or multi-factorial? The outcome distribution may depend on its temperature, or the throwing angle. That doesn't make it any less of a tendency - simply one we struggle more to understand.

From determining whether a coin is fair to getting a thorough description of its behaviour, the leap is an monotonous one, and relies mostly on uninspired processing. Of course, \textit{uninspired} doesn't necessarily mean easy. While the human mind may excel at many things, analysing large amounts of data is not one of them. 

Fortunately, there exist machines to do that for us; we have but to instruct them on how to. This brings the additional complication of doing so - but ever since the advent of computers, it has often been rewarding to translate intuition into reason, and reason into mechanical work. 

Bayes' rule renders the former transition. It arises from what is essentially a \textit{common sense} observation, and translates empirical intuition into a powerful mathematical expression. 

In order to present it, it is only natural that we start with its constituents. Firstly, we have an entity $A$, whose \textit{probability} we write as:
\begin{equation}
    \mathbf{P}(A) \nonumber
\end{equation}

As basic as this expression may seem, its interpretation captures the essence of Bayesian thinking. In frequentist statistics, the probability of an event $\mathbf{P}(A)$ is simply a frequency: it asymptotically matches the relative frequency of occurrence of $A$ (in the scope of an experiment). In Bayesian statistics, on the other hand, $\mathbf{P}(A)$ is viewed as a \textit{degree of belief} in $A$. In this interpretation, a probability describes the knowledge held by some \textit{agent} (such as ourselves), rather than a property of the phenomenon.

For instance, $\mathbf{P}(A)$ could represent how likely we initially think it is that the aforementioned coin is fair (i.e. that it yields equal probability of flipping heads or tails).  Let's say somebody told us that it was, but we only moderately trust them. Then we might attribute the fairness a confidence of 50\%:
\begin{equation}
    \mathbf{P}(\text{fair coin}) = 0.5 
\end{equation}

This notion of \textit{confidence} or \textit{degree of belief} may not seem very \textit{mathematical}, insofar as it appears to hold some subjectivity - but it lies at the heart of Bayesian thinking. For all the things that set it apart from a frequentist approach, it is this this shift from \textit{probabilities as frequencies} to \textit{probabilities as quantifications of confidence} what best illustrates their fundamental divergence. 

One may argue that this stands in the way of impartiality, but such is not necessarily the case. We may choose to rely essentially on experimental data alone to establish these \textit{beliefs}, thereby removing the practitioner's own partisanship from the process.

A more concrete take on what exactly this uncontentious choice entails is left for section \ref{sec:bayesian_inference}, but the key point is that we leave it up to the data to shape our convictions, as well as the confidence we have in them. Because these data are produced precisely by the reality we mean to study, one would be hard pressed to find a more unbiased source of insight. 

Next up are \textit{conditional} probabilities, which play a central role in Bayesian statistics. We write the probability of $A$, conditional on $B$, as:
\begin{equation}
    \mathbf{P}(A \mid B) \nonumber
\end{equation}

This can be read as the probability of $A$, given $B$. Note that this \textit{probability} continues to be interpreted in the Bayesian sense. For example, it could indicate how strongly we are convinced that the coin is fair, given that we have tossed it three times and always got a tails outcome:
\begin{equation}
    \label{eq:likelihood_tails}
    \mathbf{P}(\text{fair coin} \mid \{\text{tails},\text{tails},\text{tails}\})
\end{equation}

We may be skeptical that such is the case, since a fair coin would have been unlikely to only produce tails outcomes - but it would not be impossible, so there's still a chance. Of course, how plausible it is depends on the total amount of coin tosses. In this case, they were indeed not many.

At this point, we may start wondering whether these \textit{beliefs} actually belong to the realm of science. \textit{How exactly} are we supposed to quantify how much we trust that the coin is fair based on some outcomes?

The answer is, \textit{we're not} - that's precisely where Bayes' rule steps in.

We can straightforwardly derive it by remarking on the fact that the joint probability of two events, namely $A$ and $B$, cannot depend on the arbitrary order according to which we choose to write them.
\begin{equation}
\label{eq:joint_probability}
\mathbf{P}(A \cap B) = \mathbf{P}(B \cap A)
\end{equation}

Then we move on to \textit{conditional} probabilities. Naturally, the joint probability of $A$ and $B$, $\mathbf{P}(A \cap B)$, is the conjunction of $B$ occurring with $A$ occurring given that $B$ occurs. Since the conjunction of two events translates into the product of their probabilities, this means that we write equation \ref{eq:joint_probability} as:
\begin{gather}
    \mathbf{P}(A \mid B)\mathbf{P}(B) = \mathbf{P}(B \mid A)\mathbf{P}(A) \nonumber\\ 
    \leftrightarrow \mathbf{P}(A \mid B) = \frac{\mathbf{P}(B \mid A)\mathbf{P}(A)}{\mathbf{P}(B)} \label{eq:bayesI}
\end{gather}

, where we have assumed $\mathbf{P}(B)$ to be non-zero (with little to no loss of generality, as it's rather unlikely we would need to compute the aftermath of an impossible event).

The last form of equation \ref{eq:bayesI} is the statement of Bayes' theorem. It provides a quantitative measure of how we should change our degree of belief in $A$, should $B$ happen.

Note that this brings somewhat of an asymmetry between $A$ and $B$, as is evident in the rearrangement of equation \ref{eq:bayesI}. And as a matter of fact, these variables assume quite different roles. $A$ is an underlying factor that influences an observation; $B$ is the object of that observation. $A$ is or is not true; $B$ does or does not happen. $A$ is a hypothesis (e.g. "$\mathbf{P}(\text{tails})=0.5$"); $B$ is an event (e.g. "tails landed up"). 

As such, with the purpose of making notation clearer, $A$ is often replaced by $\theta$ or $x$ (denoting a possible value of an unknown parameter), and $B$ by $D$ (denoting data, or a single datum). With this choice of notation, equation \ref{eq:bayesI} becomes:
\begin{gather}
    \label{eq:bayesII}
    \mathbf{P}(\theta \mid D) = \frac{\mathbf{P}(D \mid \theta)\mathbf{P}(\theta)}{\mathbf{P}(D)}
\end{gather}

Relatedly, there is a more straightforward manner of computing $\mathbf{P}(D \mid \theta)$ than there is its antipode $\mathbf{P}(\theta \mid D)$. Reverting equation \ref{eq:likelihood_tails}, we get:
\begin{equation}
    \mathbf{P}\left(\{\text{tails},\text{tails},\text{tails}\} \mid \text{fair coin}\right)
\end{equation}

Here we simply wish to know how likely these outcomes are, given that the coin is fair. This can be readily assessed, without much in the way of a philosophical discussion:
\begin{equation}
    \mathbf{P}\left(\{\text{tails},\text{tails},\text{tails}\} \mid \text{fair coin}\right) = \prod_{i=1}^{3} \mathbf{P}(\text{tails} \mid \text{fair coin}) = \left(\frac{1}{2}\right)^3 = \frac{1}{8} 
\end{equation}

Note the strategic avoidance of the word \textit{probability}. Our previous interpretation of probability as strength of belief is somewhat uncalled for here: having flipped tails or not is hardly a matter of conviction. Are the relative frequencies of random events demoted from their status as probabilities in Bayesian statistics? As one might expect - or hope -, they are not.

The Bayesian concept of probability encompasses those too, and their probabilities may still be read as frequencies; the Bayesian perspective is simply more flexible than the frequentist one. That flexibility is made apparent by the fact that Bayesian strategies can be (and are) gainfully employed in otherwise frequentist paradigms. We will return to this fact later.

Even still, we typically distinguish the roles of the parameter $\theta$ and the event $D$, and call  $\mathbf{P}(D \mid \theta)$ the \textit{likelihood} of $\theta$, given $D$. Again, it still identifies a probability; it \textit{is} a probability by definition.
\begin{equation}
    \label{eq:likelihood}
    \mathbf{L}(D \mid \theta) \equiv \mathbf{P}(D \mid \theta)
\end{equation}

Regardless, the likelihood formulation is not only correct, but also more descriptive. Considering how likely the parameter was to have generated the data, rather than how probable the data was given the parameter, emphasizes the position of each element. The order reversal of the arguments in \ref{eq:likelihood} suits this purpose as well: it evidences the central standing of $\theta$, while placing $D$ in a supporting role. Conveniently, we also avoid overburdening the term \textit{probability}, which streamlines any discussions relating \ref{eq:bayesII}.

Plugging this into equation \ref{eq:bayesII}, we get our final statement of the Bayes' rule:
\begin{equation}
    \label{eq:bayesIII}
    \boxed{\mathbf{P}(\theta \mid D) = \frac{\mathbf{L}(\theta \mid D)\mathbf{P}(\theta)}{\mathbf{P}(D)}}
\end{equation}

Now that we have our final formula, let's try to make it a bit clearer by getting back to our coin example. In that case, $\theta$ would be the probability of tails (or equivalently, of heads):
\begin{equation}
    \theta \equiv p_\text{tails}
\end{equation}

In more complex cases where we have not a binomial distribution, which is described by a single variable - the probability of success -, but one which requires multiple parameters, $\theta$ will be a vector as long as the number of parameters. For instance, if its components are real numbers:
\begin{equation}
    \vec{\theta} \in \mathbb{R}^d
\end{equation}

\noindent, where $d$ denotes the dimension of the parameter space where $\vec{\theta}$ lives.

On the other hand, $D$ would be an outcome:
\begin{equation}
    \label{eq:outcome}
    D \equiv \text{outcome} \in \{0,1\}
\end{equation}

\noindent, where for the sake of conciseness we have mapped to 0 and 1 the two possible outcomes (heads and tails respectively).
\begin{equation}
    \label{eq:binary_outcome}
    \{\text{heads},\text{tails}\} \equiv \{0,1\}
\end{equation}

Naturally, other contexts could give rise to different sets of outcomes. In our case of a binomial distribution, only two exist; the likelihood is then given by:
\begin{equation}
    \label{eq:binomial_likelihood}
    \mathbf{L}(\theta \mid D) = \theta^{D}  (1-\theta)^{1-{D}}
\end{equation}

\noindent, as it represents the probability that the specific parameter value $\theta$ would have produced outcome $D$.

Returning to the general case given by equation \ref{eq:bayesIII}, we can identify as the central element of the right-hand side $\mathbf{P}(\theta \mid D)$, the \textit{likelihood} that the parameter(s) $\theta$ would result in $D$ happening. On the other hand, $\mathbf{P}(\theta)$ represents our belief in $\theta$. Simply put, it stands for the probability of its being true, to the best of our knowledge - which doesn't have to be much, as we shall get to shortly.

Regardless, once we get experimental data, such as heads/tails outcomes, this belief if obsolete. If we thought that the probability of tails was 100\%, but flip the coin and get heads, then that conviction no longer stands. Likewise, any outcome should slightly tweak our knowledge about $\theta$, hopefully getting it a bit closer to the truth.

This fine tuning process is achieved by applying equation \ref{eq:bayesIII}. Schematically:
\begin{equation}
    \label{eq:prior_to_posterior}
    \mathbf{P}(\theta) \xrightarrow[]{D} \mathbf{P}(\theta \mid D)
\end{equation}

, where the arrow with superscript $D$ denotes evolution contingent on the occurrence of $D$ (i.e. on having observed the datum $D$).

The pivotal idea is that we get a fresh stance, based on data collection. This stance reflects both our \textit{prior} beliefs, as evinced by the term $\mathbf{P}(\theta)$, and the experimental results we observed, as evinced by the term $\mathbf{L}(\theta \mid D)$. This produces our \textit{posterior} beliefs, which have been revised in light of the new evidence:
\begin{equation}
    \label{eq:bayes_prop}
    \mathbf{P}(\theta \mid D) \propto \mathbf{L}(\theta \mid D)\mathbf{P}(\theta)
\end{equation}

If we gather a lot of experimental data that our $\theta$ hypothesis would have been very likely to generate, we have more of a reason to believe that it \textit{did} generate these data. That is portrayed in the factor $\mathbf{L}(\theta \mid D)$. If on the other hand our hypothesis parameter $\theta$ would have been unlikely to replicate the events that took place, the very same factor will penalize its probability. 

Lastly, $\mathbf{P}(D)$ is the prior probability of $D$, also called the marginal probability, marginal likelihood, or the model \textit{evidence} (this refers to the statistical model, which provides the likelihood and the prior probability). It represents the \textit{a priori} probability of $D$ happening for any possible $\theta$ (i.e. integrated or marginalized over the prior for $\theta$). It has been left out of equation \ref{eq:bayes_prop} because it doesn't take as much of a conceptual significance as the other elements. In fact, for most of our purposes it works as a normalization factor, assuring that $\mathbf{P}(\theta \mid D)$ integrates to one as any probability should. It is given by:
\begin{equation}
    \label{eq:marginal}
    \mathbf{P}(D) = \int \mathbf{P}(D \mid \theta)\mathbf{P}(\theta)\mathrm{d}\theta
\end{equation}

By this we don't mean to disregard its importance; it may be of relevance when considering other applications, such as Bayesian model comparison (subsection \ref{sub:model_selection}). This divisor will again be briefly touched upon in section \ref{sec:bayesian_inference}, as will the practical details regarding its computation (or that of an equivalent counterpart within the scope of our framework).

With all the elements at hand, one key aspect is perhaps worth stressing. As conveyed by the representation in \ref{eq:bayes_prop}, there is an implicit \textit{evolution} from $\mathbf{P}(\theta)$ to $\mathbf{P}(\theta \mid D)$. This isn't just a separate calculation which depends on $\mathbf{P}(\theta)$ - rather, it provides an improved version of it. Hence, there's a hierarchical relation of \textit{substitution} between these mathematical entities.

So if we register a datum $D_1$, $\mathbf{P}(\theta \mid D_1)$ is our \textit{updated} $\mathbf{P}(\theta)$. The "$\mid D_1$``  annotation is purely denoting that this calculation takes $D_1$ into consideration, whereas in $\mathbf{P}(\theta)$ no such explicit indications were present. If we subsequently want to add yet another datum, namely $D_2$, to the estimate, then $\mathbf{P}(\theta \mid D)$ takes the place of $\mathbf{P}(\theta)$ in equation \ref{eq:bayesIII}, and so on for any additional data. 

This is in accordance with the fact that adding the data one by one or all at once should have no bearing in our findings. In equation \ref{eq:outcome}, we defined $D$ as an outcome. Alternatively, we can define it as a \textit{list} of $N$ outcomes:
\begin{equation}
    \vec{D} \equiv \langle \text{outcome} \rangle \in \{0,1\}^N
\end{equation}

But this is in truth no more general, because the probability given a vector of independent data is simply the product of the probabilities of its individual constituents.
\begin{equation}
    \mathbf{P}(\theta \mid \vec{D}) = \mathbf{P} \left( \theta \mid \bigwedge_{j=1}^N D_j \right) = \prod_{j=1}^N \mathbf{P}(\theta \mid D_j)
\end{equation}

This is true if both the likelihood and the marginal probability are themselves products over all the data (which they clearly are, by virtue of their respective definitions as joint probabilities), and $\mathbf{P}(\theta)$ is replaced consecutively by 
\begin{gather}
    \mathbf{P'}(\theta) \equiv \mathbf{P}(\theta \mid D_1) \nonumber\\
    \mathbf{P''}(\theta) \equiv \mathbf{P}\left((\theta \mid D_1) \mid D_2\right) \equiv \mathbf{P}(\theta \mid (D_1 \land D_2)) \equiv \mathbf{P}(\theta \mid D_1, D_2) \nonumber\\
    \dots\nonumber
\end{gather}

\noindent in expression \ref{eq:bayesIII} (when adding data one by one). So whether we consider multiple data or a data vector is of no consequence, as should be. Should we wish to do so, we can consider a series of outcomes at once when updating our distribution.
\begin{equation}
    \label{eq:bayes_datavector}
    \mathbf{P}(\theta \mid \vec{D})
    = \frac{\mathbf{P}(\vec{D} \mid \theta)\mathbf{P}(\theta)}{\mathbf{P}(\vec{D})}
    = \frac{\mathbf{L}(\theta \mid \vec{D})\mathbf{P}(\theta)}{\mathbf{P}(\vec{D})}
\end{equation}

In short, we have constructed a framework that allows us to adjust our knowledge so that it better matches our observations - which ultimately means that it better matches reality. This provides a way of correcting our expectations about probabilistic phenomena,  such as coin flipping, by proceeding in accordance with the results they produce. In a nutshell, we characterize a random phenomenon by sampling from it.

This approach is in order when a phenomenon we are interested in is governed by an unknown - some hidden variable(s) that dictate its (probabilistic) behaviour, such as the probability of flipping tails. Our expectations then take the form of degrees of belief, which we assign to specific instances of this unknown. But what good is that by itself? And what if we don't have any knowledge to begin with? We do not merely want to evaluate how good an \textit{isolated} hypothesis is - we want to secure a reliable one.

Clearly, a broader approach is due. We must build on this framework and devise a systematic procedure - one which will allow us to retrieve more serviceable information. In other words, we want a protocol for inference.

\section{Bayesian inference}
\label{sec:bayesian_inference}

In frequentist statistics, the very definition of probabilities equipped us with a direct way of computing their estimates; unfortunately, the same cannot be said for its Bayesian analogue. 

In the former case, we could effortlessly derive an estimate for a coin's probability of tails by flipping it several times. That's because we would identify the true probability of the \textit{tails up} event with the limiting value of its relative frequency, i.e. with its propensity of occurrence as the number of trials (flips) approaches infinity:
\begin{equation}
    \label{eq:frequentist_probability}
    \theta \equiv \lim_{\text{coin flips} \rightarrow \infty} \left( \frac{\text{number of tails up outcomes}}{\text{coin flips}} \right) \equiv
    \lim_{N \rightarrow \infty} \frac{\sum_{j=1}^N D_j}{N}
\end{equation}

\noindent, where the quotient has been written more compactly by defining $N$ as the total number of coin flips, recalling convention \ref{eq:binary_outcome}, and representing with $D_j$ the \textit{i}th datum/outcome (so $j \in \{1,\dots,N\}$). 

Of course, this is unachievable, so we would truncate the expression after some finite number of experiments.
\begin{equation}
    \tilde{\theta} \approx \frac{\text{number of tails up outcomes}}{\text{coin flips}}
\end{equation}

By contrast, Bayes' theorem gives us solely $\mathbf{P}(\theta)$, our degree of trust in a given hypothesis $\theta$ (conditioned on some information). In the previous case, at no point did we even attribute probabilities to different values of $\theta$, or contemplate these values at all - but now this is all we have. As for the expression being general with respect to $\theta$, that can be dealt with by sweeping over the possible $\theta$s. But how can we procure our $\tilde{\theta}$ while relying on the same information as before? And how will the result differ?

Let us start by focusing on equation \ref{eq:bayes_prop}. The likelihood $\mathbf{L}(\theta \mid D)$ can be obtained for any $\theta$, using only the vector of outcomes. We will use $\vec{D}$ to emphasize that this vector encompasses the results $D_j$ from multiple trials (coin flips). Then:
\begin{equation}
    \label{eq:binomial_ordered}
    \mathbf{L}(\theta \mid \vec{D}) = \prod_{j=1}^N \theta^{D_j}  (1-\theta)^{1-{D_j}}
\end{equation}

\noindent, where we have applied \ref{eq:binomial_likelihood} and used the same conventions as in equation \ref{eq:frequentist_probability}. This highlights their shared dependence on $\vec{D}$, the list of outcomes, and nothing else ($N$ is simply the cardinality of $\vec{D}$).

Here we have chosen to consider an ordered vector, in anticipation for other applications where our data won't all be equivalent (due to additional flexibility in the measurements). If we disregard the order and consider a set of outcomes instead, we get the more familiar form of a binomial distribution,
\begin{equation}
    \label{eq:binomial}
    \mathbf{L}(\theta \mid \vec{D}) = C(N,s) \cdot \theta^{s} (1-\theta)^{N- s}
\end{equation}

\newcommand*\colvec[3][]{
    \begin{pmatrix}\ifx\relax#1\relax\else#1\\\fi#2\\#3\end{pmatrix}
}

\noindent, where $C(N,s) = \big(\begin{smallmatrix}N\\s\end{smallmatrix} \big)$ is the binomial coefficient and $s \equiv \sum_{j=1}^N D_j$ is the number of successes ($1$ outcomes) among the $N$ trials.

At this point, we already have something interesting: a general formula for the likelihood of specific instances of the variable $\theta$, which is provided by the data alone (along with the description, or model, of the phenomenon - in this case, that's just a binomial distribution). Now all we're missing in proportionality \ref{eq:bayes_prop} is $\mathbf{P}(\theta)$.

As said before, this $\mathbf{P}(\theta)$ consists of the initial probabilities we assign to possible values of $\theta$ - that is, of the probability distribution choose to we start with. If we were inclined to believe it is nearly fair, we could start out with a Gaussian with some spread centered at $\theta=1/2$. Of course, we should be cautious in making our choice. We don't want to bias our estimate or slow down convergence by making unwarranted assumptions. Yet, we might not know much about our coin, nor about its fairness, and as such not be confident enough to include any such inclination in our reasoning. In that scenario, the probability of tails could be anywhere between 0 and 1, and we wouldn't have any sound reasons to favor some values over others. How then could we define this prior distribution?

We already have. We have established that it:
\begin{itemize}
    \item Must treat all possible $\theta$\textit{s} equally, i.e. cannot depend on $\theta$;
    \item Must integrate to one, because it is a probability distribution;
    \item Must be 0 for  $\theta \not\in [0,1]$, because $\theta$ itself is a probability.
\end{itemize}

These three characteristics uniquely define a probability distribution. It consists of a flat distribution covering all possible values:
\begin{equation}
\label{eq:flat_prior}
\mathbf{P}(\theta) = 
    \begin{cases}
    1, & \text{if } 0 \leq \theta \leq 1\\
    0, & \text{otherwise}
\end{cases}
\end{equation}

While in more complex cases plenty can be said about how a \textit{representation of ignorance} can be best achieved, for many applications of interest a so-called \textit{objective} prior distribution can be found - that is, one which conveys little to no information, and so barely affects the inference results. This is the case of our example above, where the prior simply expressed problem constraints and otherwise followed the \textit{principle of indifference} by assuming a uniform distribution over the plausible region (the one delimited by these constraints).

What is more, often the prior beliefs barely affect the posterior ones at all, so long as enough data is considered (although this defense has in some cases come to be criticized \cite{Edwards_1993}, it stands for many others). As put by \cite{Edwards_1963}, even if two individuals initially disagree, the inference process is expected to reconcile their positions. That means that by collecting enough evidence, we could expect to bring their viewpoints together up to unbounded like-mindedness (assuming they're not Bayesian skeptics). For this reason, it may be unnecessary to overly stress the choice of a prior; resources could be more gainfully employed in gathering the data that should cause its relevance to subside.

This is more formally expressed by the Bernstein–von Mises theorem \cite[chapter~10]{Vaart_1998}, which proves under some fairly inclusive assumptions that the posterior distribution can be made not to depend on the prior by means of data collection. These assumptions often hold for physically motivated estimation problems, under the typical circumstances in which these problems occur (namely the class of priors that they prompt).

It may be worth mentioning that this matter brings forth the major divergence within Bayesianism, which is split into what are called objective and subjective interpretations. The former is more temperate in how it adopts the idea of probabilities as beliefs, trying to stay by \textit{rational} beliefs; the latter interprets it quite literally (as in personal beliefs). In practice, this distinction mostly concerns the prior.

\begin{figure}[!ht]
    \centering
    \includegraphics[width=0.7\textwidth]{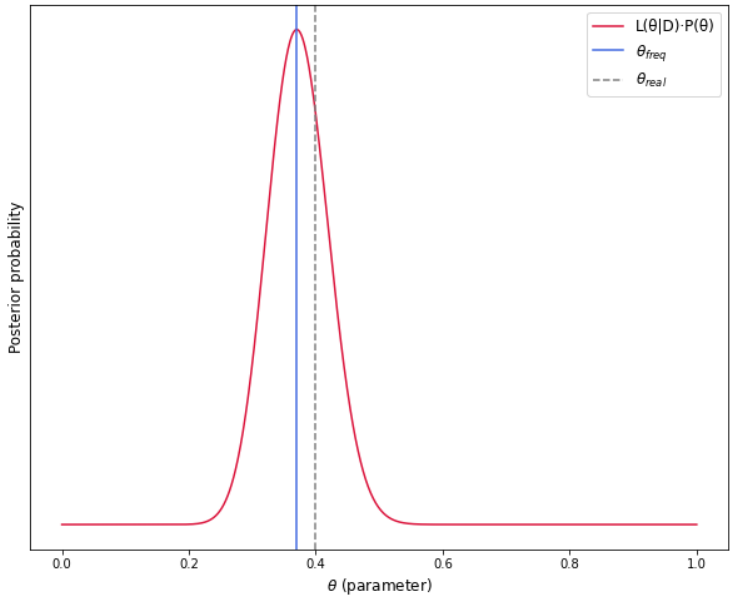}
    \caption{Bayesian posterior probability density for a binomial likelihood. The dashed and full vertical lines mark the real parameter used to generate the data and its frequentist estimate respectively. The same data vector contemplating 100 trials was used for the Bayesian and frequentist calculations. }
    \label{fig:coin_bayes_vs_freq}
\end{figure}

Finally, having both a prior and a likelihood, we can compute the right-hand side of proportionality \ref{eq:bayes_prop} in full. By evaluating it on a grid and interpolating a continuous curve, we can visualize the result. Figure \ref{fig:coin_bayes_vs_freq} represents the posterior probability density - which is proportional to $\mathbf{P}(\theta \mid D)$ - that it yields for a randomly generated data vector. 

The omission of the y-axis scale on figure \ref{fig:coin_bayes_vs_freq} is analogous to the omission of the divisor in proportionality \ref{eq:bayes_prop}. Clearly, these pieces of information are not strictly necessary for all purposes: even visually, we can approximate the maximum just by looking at the probability density graph. For many applications, relative probabilities are sufficient.

But even if not, we don't need a \textit{separate} or very precise computation of the marginal $\mathbf{P}(D)$. What we need is given by our distribution already, because it is a normalization factor. In effect, that is \textit{the best possible} estimate of $\mathbf{P}(D)$, because it is the only one which guarantees that our probability distribution behaves like one.

This discussion may be somewhat unclear without discussing the practical representation of the distributions, or how the integrals are to be evaluated. Figure \ref{fig:coin_bayes_vs_freq} shows a smooth curve achieved through interpolation, but this does not have much bearing in practice because in general a continuous representation is not viable. Further clarification is left for subsection \ref{sub:estimation_in_practice}, which will cover the most conceptually simple implementations.  But before that, we will look into Bayesian inference a little bit more.

\subsection{Estimators and expressions of uncertainty}
\label{sub:estimators_and_expressions_of_uncertainty}

The furthest we got so far was a probability density function given by equation \ref{eq:bayes_prop} and represented in figure \ref{fig:coin_bayes_vs_freq}. This function quantifies how likely each of the values that the parameter $\theta$ could take would have been to produce the outcomes we observed \footnote{Recalling that a flat prior was used, we know that the posterior is maximized precisely at the likelihood maximum. The prior probabilities $\mathbf{P}(\theta)$ - and of course $\mathbf{P}(D)$, the marginal - exactly match for all points.}. From there, we must still extract an estimate $\tilde{\theta}$.

An estimator, usually denoted by a circumflex - as in $\hat{\theta}$ -, is a construction which maps the posterior (or equivalently, the conjunction of the prior and the data) to an estimate. Looking at the graph from figure  \ref{fig:coin_bayes_vs_freq}, the answer seems obvious: we should just take the value of $\theta$ corresponding to the posterior peak. In this case, the choice is made painless by the fact that the values are approximately normally distributed: for normal distributions, the mean, mode and median exactly match. We can pick any - all - of them as our estimator. 

By contrast, we will generally have to pick a single one when looking at less regular posteriors. Which to select is then an implementation choice. When discussing this selection, we generally speak in terms of \textit{loss functions} (or \textit{utility functions}, a symmetric construction). They typically penalize inexactness (i.e. deviations of the estimators with respect to the real values) in different ways:
\begin{equation}
    \label{eq:loss}
    \mathcal{L}(\theta,\hat{\theta}(\vec{D})) = f\left(|\theta-\hat{\theta}(\vec{D})|\right)
\end{equation}

Note the calligraphic $\mathcal{L}$, which means to avoid confusion with the $L$ symbol we use for the likelihood.

We will mention some popular choices of loss functions shortly, as well as the estimators they prompt. But the process of determining these estimators requires further consideration: as is, equation \ref{eq:loss} is unusable, by virtue of its dependence on $\vec{D}$ and $\theta$. We would like to somehow keep the loss to a minimum; this expression doesn't allow us to evaluate it at all, still less minimize it.

Of course, the actual loss incurred by an estimator will depend on the collected data; this is made clear by the fact that we don't get our estimate until the inference process has terminated. We can however evaluate the \textit{a priori} merit of an estimator by considering the expected loss over all possible outcomes. This expectation over the data yields what is called the \textit{risk} of the estimator, $R(\theta,\hat{\theta})$:
\begin{equation}
    \label{eq:risk}
    R(\theta,\hat{\theta}) = 
    \sum_{\vec{D}} \mathbf{P}(\vec{D} \mid \theta) \mathcal{L}(\theta,\hat{\theta}(\vec{D})) \equiv
    \mathbb{E}_{\mathbf{P}(\vec{D} \mid \theta)} \left[ \mathcal{L}(\theta,\hat{\theta}(\vec{D})) \right]
\end{equation}

\noindent, where we have introduced the $\mathbb{E}_{\mathbf{P}(-)}$ notation for the expected value (average) over a distribution $\mathbf{P}(-)$. We will discuss expectations in more detail in the next subsection, but for now the notation will mostly be used to symbolize ideas that are described in the text.

We have removed the dependence on the data by using the likelihood to transfer it to $\theta$, the real parameter. This $\theta$ is also an argument to the loss function itself. We don't have a clear-cut estimate of this parameter in advance, so we can't exactly pinpoint the values of these quantities (if we did, we would choose the estimator $\hat{\theta}=\theta$, making the problem trivial and removing the need for this discussion entirely). But we \textit{do} have our best-effort estimate, embodied in our prior. So to remove this unwanted variable, we just integrate it out, getting the expectation of \ref{eq:risk} over $\theta$:
\begin{gather}
    r\left(\mathbf{P}(\theta),\hat{\theta}\right) = 
    \mathbb{E}_{\mathbf{P}(\theta)}
    \left[ R(\theta,\hat{\theta}) \right]
    = \mathbb{E}_{\mathbf{P}(\theta),\mathbf{P}(\vec{D} \mid \theta)}
    \left[ 
    \mathcal{L}(\theta,\hat{\theta}(\vec{D})) 
    \right] 
    \\
    = \int\sum_{\vec{D}} \mathbf{P}(\vec{D} \mid \theta) \mathcal{L}(\theta,\hat{\theta}(\vec{D}))\mathbf{P}(\theta)\mathrm{d}\theta
    \label{eq:bayes_risk}
\end{gather}

\noindent, where the interiormost distribution in the subscript is the distribution over which to take the expectation first.

We will henceforth omit the dependence of this quantity on the prior and simply write $r(\hat{\theta})$, to avoid overcrowding equations; the use of a flat prior distribution within some bounds (which dictate the integration domain) can be assumed, as this strategy will be adopted throughout the document.

The removal of our last unwanted variable $\theta$ in equation \ref{eq:bayes_risk} by averaging over the prior then yields $r(\hat{\theta})$, which is called the \textit{Bayes risk}. It depends on how the estimator function $\hat{\theta}$ maps data to estimates. Once a loss function is chosen, the \textit{Bayes estimator} is the one which minimizes the the Bayes risk, or equivalently the posterior expected loss; this expectation is supplied by the prior (along with the likelihood function).

To say that the risk represents the posterior expected loss may be somewhat confusing. It is intuitive that it should be so, since the posterior effectively encodes our results, and it is these results' expected performance we are concerned with. However, at no point did the posterior come up during the discussion. As a matter of fact, we took the expectation of the risk over the prior. So where exactly does the posterior come in, and what do we really mean by \textit{posterior expected loss}?

The key point resides in the step represented by equation \ref{eq:risk}, which takes place before the integration of \ref{eq:bayes_risk}. After taking the expectation of the loss over the data, when performing the integration the distribution already resembles our posterior more than our prior. This is made clear by some algebraic manipulation:
\begin{gather}
    \label{eq:bayes_riskII}
    r(\hat{\theta}) = 
    \int \sum_{\vec{D}} \mathbf{P}(\vec{D} \mid \theta) \mathcal{L}(\theta,\hat{\theta}(\vec{D}))\mathbf{P}(\theta)\mathrm{d}\theta 
    = \sum_{\vec{D}} \int \mathcal{L}(\theta,\hat{\theta}(\vec{D})) 
    \cdot  \mathbf{P}(\vec{D}\mid \theta) \mathbf{P}(\theta)\mathrm{d}\theta
    \\
    =  \sum_{\vec{D}} \int \mathcal{L}(\theta,\hat{\theta}(\vec{D})) 
    \cdot \mathbf{P}(\vec{D}) \left( \frac{\mathbf{P}(\vec{D} \mid \theta) \mathbf{P}(\theta)}{\mathbf{P}(\vec{D})} \right) \mathrm{d}\theta
    \\
    = \sum_{\vec{D}}  \mathbf{P}(\vec{D}) 
    \int \mathcal{L}(\theta,\hat{\theta}(\vec{D})) 
    \cdot \mathbf{P}(\theta \mid \vec{D}) \mathrm{d}\theta
\end{gather}

\noindent, where we have recognized equation \ref{eq:bayes_datavector}.

More compactly:
\begin{equation}
    \label{eq:bayes_riskIII}
    r(\hat{\theta}) = \sum_{\vec{D}}  \mathbf{P}(\vec{D}) 
    \mathbb{E}_{\mathbf{P}(\theta \mid \vec{D})}
    \left[ \mathcal{L}(\theta,\hat{\theta}(\vec{D}))  \right] 
    =
    \mathbb{E}_{\mathbf{P}(\vec{D}),\mathbf{P}(\theta \mid \vec{D})}
    \left[ \mathcal{L}(\theta,\hat{\theta}(\vec{D}))  
    \right]
\end{equation}

\noindent, where again the expectations are taken in right-to-left order to remove all dependence in both $\theta$ and $\vec{D}$.

That is, the Bayes' risk of an estimator is sum of its posterior expected loss conditional on every possible dataset, weighted by the respective datasets' probabilities. We can obtain the posterior expectation as long as we condition it on a specific list of outcomes: we proceed just as we would if we had actually gathered these hypothetical data. As for the probabilities of the data, they are the likelihoods averaged over the prior; $\mathbf{P}(\vec{D})$ is by definition of the marginal distribution (equation \ref{eq:marginal}).

In hindsight, we can see that we did in equation \ref{eq:bayes_risk}, when we took the conditional (on the parameter) expectation over the data and then averaged over the prior, was equivalent to taking the conditional (on the data) posterior average and then the expectation over the data. 

Comparing the subscripts for the expectations in \ref{eq:bayes_risk} and \ref{eq:bayes_riskIII}, they look like they make up a Bayes' theorem of their own; and the proof of their likeness (\ref{eq:bayes_riskII}) was indeed achieved by means of that theorem. This ties into the relation between the the probabilities at hand (prior, posterior, likelihood, and marginal). The equivalence can be seen by taking 
\begin{equation}
   \mathbf{P}(\theta)\mathbf{P}(\vec{D} \mid \theta) =
   \mathbf{P}(\vec{D})\mathbf{P}(\theta \mid \vec{D} )
\end{equation}

\noindent, multiplying by the loss, integrating over $\theta$, and summing over $\vec{D}$. By linearity of integration, the order of the two last operations is irrelevant.

Having clarified our objective and terms, we can finally discuss specific estimators. A popular choice is the posterior mean. It minimizes a \textit{quadratic} loss function - that is, large errors are more penalized than small ones, which tends to be appropriate in scientific applications.
\begin{equation}
    \label{eq:quadratic_loss}
    \mathcal{L}^{\text{(quadratic)}}(\theta,\hat{\theta}(\vec{D})) = \left(\theta-\hat{\theta}(\vec{D})\right)^2
\end{equation}

In the one-dimensional (single parameter) case, the quadratic loss function reduces to the mean squared error. Clearly, for each possible dataset, the expectation value, or mean, minimizes the expected mean squared difference (this can readily be checked by differentiation), which is then the posterior variance (or more generally, the covariance). 
\begin{equation}
    \label{eq:expected_quadratic_loss}
        \mathbb{E}_{_{\mathbf{P}(\theta \mid \vec{D})}}
    \left[ \left( \theta-\bar{\theta}\right)^2\right]
    = \mathbb{V}\left[\mathbf{P}(\theta \mid \vec{D}) \right]
    \equalhat \mathcal{L}^{\text{(quadratic)}}(\theta,\hat{\theta}(\vec{D}))
\end{equation}

\noindent, where $A \equalhat B$ denotes \textit{$A$ estimates $B$}. Note the fixed $\vec{D}$, which is assuming an after data collection point of view. The fact that the Bayes' estimator is the mean is unchanged after marginalizing over \textit{a priori} possible data records, because for all of them the mean minimizes the variance.

In general, these are the estimators we will use (the mean estimates the real parameter, and the variance its mean squared error); these tend to be popular choices in research.

A different possibility would be the median. It minimizes a \textit{linear} loss function. 
\begin{equation}
    \label{eq:linear_loss}
    \mathcal{L}^{\text{(linear)}}(\theta,\hat{\theta}(\vec{D})) = \left|\theta-\hat{\theta}(\vec{D})\right|
\end{equation}

The median divides the distribution into two even parts containing equal probability mass, i.e. it falls in the middle of the distribution. That's where the weighted average of the absolute distances to all other parameter locations (which is in this case the expected loss) is smallest.

Finally, yet another option would be the mode, which minimizes a \textit{binary} (0/1) loss function.
\begin{equation}
    \label{eq:binary_loss}
    \mathcal{L}^{\text{(binary)}}(\theta,\hat{\theta}(\vec{D})) = 
    \begin{cases}
    0, & \text{if } \hat{\theta}(\vec{D})=\theta\\
    1, & \text{otherwise}
\end{cases}
\end{equation}

It attempts to maximize the chance of correctness by taking the posterior maximum, i.e. the highest probability point. The expected loss will simply consist of the probability mass assigned to the region where the loss is non-zero, $\hat{\theta}(\vec{D})\neq \theta$. For continuous distributions one can relax the equality condition by defining a tolerance (otherwise the contribution of a point estimate to an integral is always null, as the corresponding volume is infinitesimal), or use an alternative definition involving the Dirac delta construction.

When the mode is chosen as estimator, the estimation method is generally termed \textit{maximum a posteriori probability estimation} (or \textit{MAP} for short). If additionally a flat prior is used, the designation further simplifies to \textit{maximum likelihood estimation} (\textit{MLE}), for plain reasons. Clearly, these techniques are less general than their common framework, because they automatically produce a single point estimate rather than a whole distribution to pick from (and from where to retrieve additional information).

Of these three estimators, the mode is perhaps the one which most easily escapes the criticism directed at Bayesian inference, or at least much of it. This is owning to the fact that it converges to correct values even in cases where other estimates fail due to (for some choices of a prior) not meeting the premises for asymptotic correctness \cite{Freedman_1963,Freedman_1965}. However, it may also display inherently less robust behaviour \cite{Sivaganesan_1991}, namely in the presence of noise. Finding the mode is finding the posterior maximizer, and optimization based strategies are prone to overfitting. 

Since irregularities invalidating asymptotic correctness won't be of concern in our case studies, we will generally - as said before - use the posterior mean. In any case, this choice is not very critical if the posterior is uni-modal and relatively regular, since the three estimators we mentioned will fall within close proximity of each other. 

Something that stands out is the fact that when we took a frequentist stance, \textit{we never had this discussion}. We didn't need an estimator, because we automatically produced an estimate, $\tilde{\theta}$. Initially, it seemed like somewhat of an unnecessary complication not to directly get one. Arguably, it is a nuisance to have to extract an estimate, and that too after having meticulously built up a thorough state of knowledge. But also, \textit{we meticulously built up a thorough state of knowledge}.

This contrast is effectively illustrated by figure \ref{fig:coin_bayes_vs_freq}, where a single line (identified by a a number) characterizes the frequentist's knowledge, but a full distribution is required to describe the Bayesian viewpoint. This exhaustive characterization may seem more fastidious, but it also provides a more holistic overview.

\begin{figure}[!ht]
    \centering
    \includegraphics[width=0.7\textwidth]{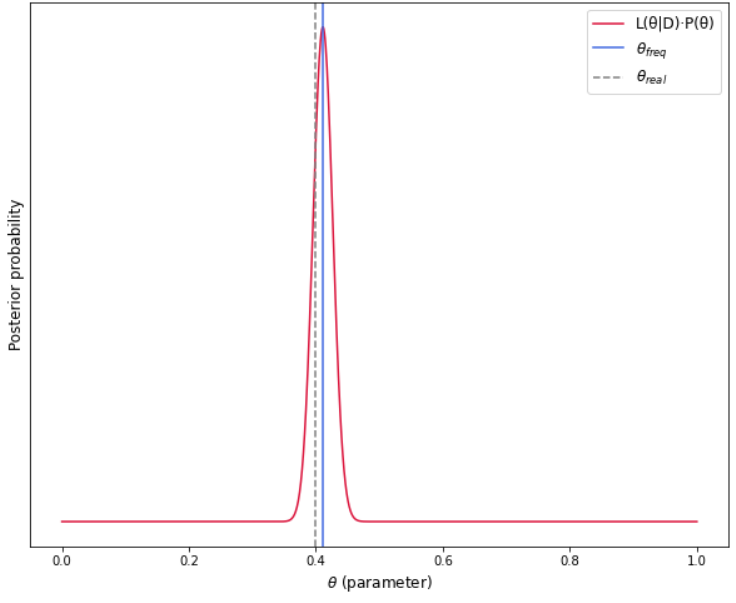}
    \caption{Bayesian posterior probability density for a binomial likelihood. The dashed and full vertical lines mark the real parameter used to generate the data and its frequentist estimate respectively. The same data vector contemplating 1000 trials was used for the Bayesian and frequentist calculations. }
    \label{fig:coin_bayes_vs_freq_sharper}
\end{figure}

For example, let's see what happens if we add data to the distribution in figure \ref{fig:coin_bayes_vs_freq}. The aftermath is represented in figure \ref{fig:coin_bayes_vs_freq_sharper}, along with a line marking the frequentist guess based on the same data.

Comparing figures \ref{fig:coin_bayes_vs_freq} and \ref{fig:coin_bayes_vs_freq_sharper}, we can see that the posterior has gotten \textit{sharper} with the inclusion of extra trials. We can tell that our knowledge has been refined since the previous checkpoint, and that would not change even if we were unaware of how many trials each of the posteriors actually contemplated. Contrastingly, the immediate frequentist result continues to be reported by the same unsavoury single line; its position has solely been readjusted.

Put differently, the Bayesian framework inherently provides not only an estimate of the real parameter, but also one of our trust in it. Furthermore, the Bernstein–von Mises theorem (which we already referred to in connection with the prior) proves that both these results are asymptotically correct from a frequentist point of view, under some conditions. This is convenient inasmuch as it allows us to perform estimation without subscribing to an entire new philosophical doctrine.

More specifically, the aforementioned theorem states that the posterior converges to a normal distribution centered at an efficient estimator\footnote{This efficiency is, as before, relative to a loss function.}. We can notice that in the simple case of figures \ref{fig:coin_bayes_vs_freq} and \ref{fig:coin_bayes_vs_freq_sharper} the posterior is approximately a Gaussian curve whose center coincides with the frequentist estimate.

Furthermore, this asymptotic form's variance is known, and it matches its frequentist analogue \cite{Bontemps_2011}. This holds provided that the prior fulfils some criteria; note that this statement is otherwise independent of the prior, in agreement with our previous claims.

This effectively harmonizes Bayesian and frequentist statistics in many contexts. In particular, it enables the approximation of frequentist quantifications of uncertainty by means of the Bayesian posterior, via the legitimization of its frequentist interpretation. It does so by proving that the two frameworks' measurements of uncertainty - expressed in terms of intervals or, more generally, regions -, mathematically match in the asymptotic limit (in spite of their conceptual disagreement). Summarily, it sustains the use of Bayesian means for non-Bayesian purposes. 

Credible regions delimit a section of the parameter space where we believe the real parameter is contained to some level of certainty. This \textit{belief} denotes Bayesian probability. So more specifically, a 95\% credible interval would be one where our $\theta$ is 95\% likely to fall (given our state of knowledge).

In figure \ref{fig:coin_bayes_vs_freq_sharper}, this interval would be one over which the posterior probability (which, contrary to the probability density, is normalized by definition) integrates to 0.95\footnote{We can observe that this condition does not uniquely define an interval. Several methods for picking one can be selected; this selection is akin to the designation of an estimator.}. Put another way, the confidence level is the fraction of probability mass bounded within the confidence interval.

As is evident, this definition is implausible from a frequentist perspective. The \textit{distribution} over possible real values could be but a pulse on a single one, regardless of whether we know it or not. To speak of the \textit{probability} that the true parameter lies within a designated region is nonsense: as a static entity, it either does or it doesn't. We can at most be more or less certain that one of these cases is true. 

Hence in this paradigm the \textit{credibility} of the parameters is replaced by the \textit{confidence} in the estimate. The (so to say) \textit{active} role shifts from the true variable to the interval. Instead of speaking of the probability that \textit{the parameter is contained} in the interval, we speak of the probability that \textit{the interval would have contained} the parameter. This \textit{confidence} concerns the intervals we might have obtained \textit{in general}, whereas the Bayesian credible region pertains to the particular one we got. 

This key divergence is brought on by the fact that in the latter case we regard the interval as probabilistic, whereas in the former it is the parameter itself which we consider to be so. The origin of the observed data's randomness is ascribed to either the nature of the measurement operation isolatedly, or also that of the parameter that generated it. 

Such a parting is attributable to the fact that unlike frequentism, what Bayesianism seeks to describe is a subjective reality. This \textit{subjectivity} - or lack thereof - dictates whether uncertainty is assimilated into the theory, or is simply an immaterial assessment pertaining to the reliability of its findings. Accordingly, in frequentist statistics our 95\% confidence interval would be defined as one that would have contained the parameter with 95\% probability. Now this is a frequentist probability: it is the limit of a relative frequency.

More specifically, by proceeding the way we proceeded to obtain our 95\% confidence interval estimate, we could have arrived at a \textit{correct} interval (i.e. one holding the true parameter) or not. It follows from  definition of confidence intervals that the former scenario would occur 95\% of the times if we continued to repeat the attempts eternally. In this case, the intervals are determined under a normality assumption (for the sample means). The desired confidence level then determines - along with some summary statistics - the width of the region, which is centered at the normal curve's peak.

For instance, a 95\% confidence interval would cover 1.96 standard deviations to each side of the Gaussian's maximum. This is a property of the Gaussian, and can be calculated or looked up in a table. The only extra details needed to identify the interval are the the population variance $\sigma^2$ (or an estimator) and sample size $N$.

We should bear in mind that these statistics respect the sample means, \textit{not} a single sample. This is in line with our discussion about how confidence regards all of the possible end products our estimation could have yielded. That's why the sample size is required: the standard deviation of the sample means is approximated by $\sigma/\sqrt{N}$ (in subsection \ref{sub:beating_shot_noise} we will come across this again, when discussing shot noise). We need this factor to convert the interval width from units of standard deviations to our actual units. 

These techniques occur within the scope of \textit{interval estimation}, which improves upon the single point frequentist estimates discussed above to provide a more descriptive conclusion. One may then question whether any benefit is to be had from the Bayesian strategy. Wouldn't the discrepancy in thoroughness be redeemed by this stepping up of frequentist statistics?

That is not quite the case. First of all, we can observe that while it does allow for uncertainty estimation, it is not entirely as general as the Bayesian approach. A posterior distribution allows for computing the expectation value of nearly \textit{any} function of the parameters.

In practice we can do this simply by sampling from the posterior\footnote{An overview of what this \textit{simply} understates is left for chapter \ref{cha:monte_carlo_posterior_sampling}.}. Statistical functions, such as the mean or variance, are just particular cases of expectations. While distributions are completely described by these parameters when they are normal, in general they are not. 

Secondly, the method itself is more universal: it requires only a parametric model and the ability to replicate it (for the likelihoods; this requirement will be presented more clearly in subsection \ref{sub:likelihood_evaluations_quantum}). Moreover, its degrees of freedom are built into the algorithm by default, independently of the model's complexity. Contrarily, the frequentist process demands more structure. These characteristics again betray an overarching context of unmatched flexibilities.

As a consequence, it is often more viable to demarcate Bayesian credible regions than it is frequentist confidence regions \cite{Bontemps_2011}. It should then be clear why we would be interested in a theorem which substantiates their interchangeability. By stating  that the credible intervals will converge to their frequentist counterpart, the Bernstein-von Mises theorem produces extra paths toward the computation of the latter. 

\subsection{Estimation in practice: approximate inference and conjugate priors}
\label{sub:estimation_in_practice}

Just before subsection \ref{sub:estimators_and_expressions_of_uncertainty}, we downplayed the role of the marginal probability, on the grounds of its being an unimportant normalization factor as far as parameter estimation is concerned. We will now return to pick up on that statement.

Let's begin by considering the computation of this constant, $\mathbf{P}(D)$. Recall that we defined it as:
\begin{equation} \tag{\ref{eq:marginal}}
    \mathbf{P}(D) = \int \mathbf{P}(D \mid \theta)\mathbf{P}(\theta)\mathrm{d}\theta
\end{equation}

\noindent, and said that it represented the probability of the data $D$ occurring independent of $\theta$.

In fact, it is the integral of the denominator in equation \ref{eq:bayesII} over the parameter space. As the term marginal probability suggests, the variable $\theta$ is integrated out (by marginalizing the conditional probability over the prior). For a flat prior (re-scaled to a normalized domain $\theta \in [0,1]$), it is also called the marginal likelihood. 

More often than not, the most feasible way of computing integrals like \ref{eq:marginal} is through numerical integration. That is, we replace the infinite summation over infinitesimal volume units with a finite one, by selecting some set of points at which to evaluate the integral.

These points can be chosen deterministically (e.g. by arranging them on a dense enough grid) or randomly (as in Monte Carlo integration). Then we simply take the average:
\begin{equation}
    \label{eq:numerical_marginal}
    \mathbf{P}(D) \approx \frac{V}{M}\sum_{i=1}^{M} \mathbf{P}(D \mid \theta_i)\mathbf{P}(\theta_i)
\end{equation}

\noindent, where we have considered $M$ evaluation sites, and $V$ is the volume covered by the original integral. For instance, in our coin case we have  a single-variable integral, so $V$ is a length. Since $\theta \in [0,1]$, we have that $V=1$. Clearly, $V/M$ is the finite unit of volume that we attribute to each point, which comes to replace the infinitesimal one $\mathrm{d}\theta$.

We should note approximations such as \ref{eq:numerical_marginal} can be improved upon. This discussion is left for chapter \ref{cha:monte_carlo_posterior_sampling}, as it is extraneous to our current purposes; the analysis that follows holds regardless.

All that's left is to choose the $M$ points ($\theta$s) at which to evaluate the function. This is one of the aspects that strongly affect the cost-to-performance ratio, but without much of a thought, we can simply choose them to be evenly spaced and as many as we're willing to spend computational resources with.

\begin{figure}[!ht]
    \centering
    \includegraphics[width=0.7\textwidth]{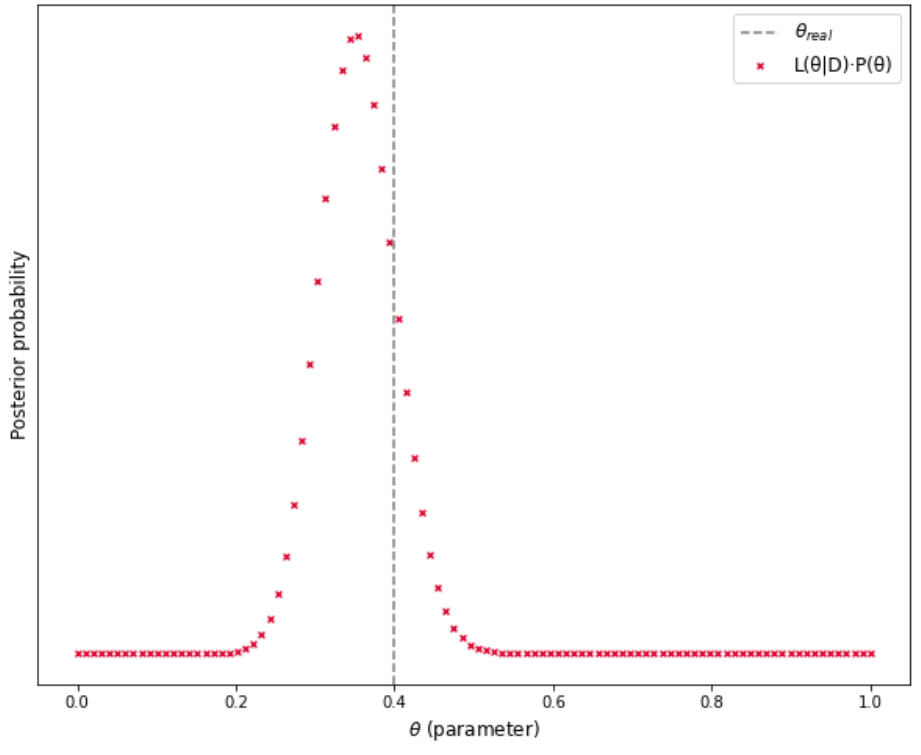}
    \caption{Bayesian posterior probability evaluations for a binomial likelihood. The dashed line marks the real parameter used to generate the data. 100 trials were included in the data vector. }
    \label{fig:coin_discrete}
\end{figure}

Actually, we already did something similar back in section \ref{sec:bayesian_inference}, where we evaluated Bayes' rule on a grid to get the smooth curve of figure \ref{fig:coin_bayes_vs_freq}. In reality, our immediate result would look more like figure \ref{fig:coin_discrete}, since there's no way to plot a continuous function directly (as much as we can bring neighbouring points close up to any desired precision).

\begin{figure}[!ht]
    \centering
    \includegraphics[width=0.7\textwidth]{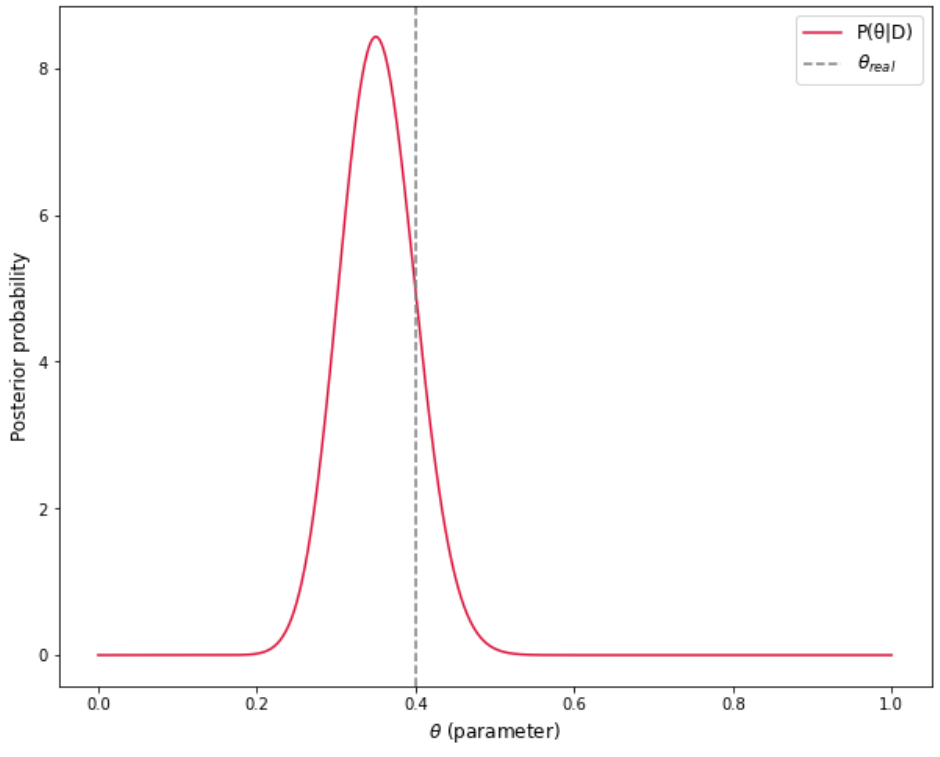}
    \caption{Bayesian posterior probability distribution for a binomial likelihood. The dashed line marks the real parameter used to generate the data. 100 trials were included in the data vector. }
    \label{fig:coin_normalized}
\end{figure}

Let's again opt for that strategy, and evaluate approximation \ref{eq:numerical_marginal} using those same sites. With that, all details are fixed: we can finally normalize our distribution. That can be done by dividing the $\mathbf{L}(\theta \mid D)\mathbf{P}(\theta)$ calculations by the approximation of $\mathbf{P}(D)$ given by \ref{eq:numerical_marginal} to get the complete form of the Bayes' rule as per equation \ref{eq:bayesIII}. Figure \ref{fig:coin_normalized} shows the result, again represented with a smoothed line.

So finally, we have a normalized probability distribution, which integrates to one (asymptotically). Which seems like a good time to ask ourselves, \textit{what exactly do we want that for?}

The short answer is, because that's how probability distributions work - they \textit{have} to be normalized. But it is also not a very satisfactory one, and somewhat of a lapalissade - why should \textit{we} particularly care that they are?

In the beginning of this section, we remarked that we could point out an estimate just from looking at the unnormalized probability density, and further noted that this construct sufficed for some applications. Where is it that this falls apart? We can do better by looking deeper into our motives.

In section \ref{sub:estimators_and_expressions_of_uncertainty}, we talked about estimators of the real value and of the uncertainties, and suggested we'd adopt the mean and the variance respectively. These are two particular cases of expectation values that we can extract from a posterior distribution, much like we extracted $\mathbf{P}(D)$ from its unnormalized version. In general, we can calculate the expected value (or equivalently, the mean) of any function of the parameters by integrating over the posterior distribution:
\begin{equation} \label{eq:expectation}
    \mathbb{E}_{\mathbf{P}(\theta \mid D)}\left[f(\theta)\right] = \int f(\theta) \mathbf{P}(\theta \mid D)\mathrm{d}\theta
\end{equation}

Of course, this holds for \textit{any} probability distribution, not just the posterior. We could replace the occurrences of $\mathbf{P}(\theta \mid D)$ in equation \ref{eq:expectation} by some other distribution, such as our prior, to get the corresponding expectation values.

Ultimately, \textit{this} is what characterizes a probability distribution: the behaviour that it induces, described in terms of expectations.  For the mean $\bar{\theta}$ we simply have $f(\theta) = \theta$ (hence the interchangeability of the terms \textit{expectation} and \textit{mean}: this definition can be generalized to functions other than the identity), and for the variance $f(\theta) =(\theta-\bar{\theta})^2$. The fact that the two metrics that are most commonly used to describe a distribution consist of a particular case of expression \ref{eq:expectation} only comes to underline its centrality\footnote{Of course, if one seeks to find the posterior maximizer, single point evaluations are to be employed in numerical optimization rather than in computing expectations. This is one of the cases where normalization is not required. However, averages enjoy a theoretical soundness that optimization does not. This ties into the discussion of chapter \ref{cha:monte_carlo_posterior_sampling}.}.

Thus, the foremost goal of the inference process is to enable the computation of integral \ref{eq:expectation}. This unlocks all the information we may want, which takes the form of estimates. And interestingly, \textit{we already know} what that computation entails in practice, because all that was said in the beginning of this subsection about approximating the marginal probability \ref{eq:marginal} holds here as well. So as before, we approximate the integral in \ref{eq:expectation} as:
\begin{equation}
    \label{eq:numerical_expectation}
    \mathbb{E}_{\mathbf{P}( \theta \mid D)}\left[f(\theta)\right] \approx \frac{V}{M}\sum_{i=1}^{M} f(\theta_i) \cdot \mathbf{P}( \theta_i \mid D)
\end{equation}

Once more, we have replaced the distribution by single point evaluations; these can be the same grid points as before. Which raises the question, what do we need \textit{the rest} of the distribution for?

If our goal is to compute equation \ref{eq:expectation} and we do so through equation \ref{eq:numerical_expectation}, we don't need it at all. For this reason, the distribution can be more conveniently expressed discretely, as a \textit{probability mass function} with support over a finite set of points.

In this case, normalization demands that the points \textit{sum} to one, as opposed to \textit{integrating} to one. So instead of using the exact equation \ref{eq:bayesIII}, we will apply a discretized version of it to each of our $M$ grid points $\theta_i$ $(i \in \{1..M\})$.

First we sweep of all these points and attribute them unnormalized \textit{probabilities} (this is clearly an abuse of terminology, because there is no such thing). These tell of their \textit{relative} likelihoods given the data. This simply amounts to repeatedly applying the formula:
\begin{equation}
    \label{eq:bayes_prop_discrete}
     W_i(D) \equiv \mathbf{L}(\theta_i \mid D)\mathbf{P}(\theta_i) 
\end{equation}

We have represented with $W_i(D)$ the unnormalized \textit{probability} of $\theta_i$, in anticipation of the notation to be used in the following chapter.
\begin{equation}
    \label{eq:unnormalized_p}
     W_i(D) \propto \mathbf{P}(\theta_i \mid D)
\end{equation}

Equations \ref{eq:bayes_prop_discrete} and \ref{eq:unnormalized_p} essentially make up proportionality \ref{eq:bayes_prop}, but with added emphasis on the discretization.

As we do this for all the $M$ grid points, we successively sum the results that we get for each. This normalization constant is to be used for all the points, so one need only calculate it once. Finally, we go back to get the normalized probabilities for all points, by dividing their respective values of equation \ref{eq:bayes_prop_discrete} by this common factor:
\begin{equation}
    \label{eq:discrete_normalization}
    w_i(D) \equiv \frac{W_i(D)}{\sum_{j=1}^M W_j(D)} =
    \frac{\mathbf{L}(\theta_i \mid D)\mathbf{P}(\theta_i) }
    {\sum_{j=1}^M \mathbf{L}(\theta_j \mid D)\mathbf{P}(\theta_j) }
\end{equation}

At this point, it is clear why normalization substitutes the calculation of the marginal probability: \textit{because it precisely amounts to calculating it}, apart from a constant factor $V/M$. This can readily be seen by comparing equation \ref{eq:numerical_marginal} with \ref{eq:discrete_normalization}'s denominator.
\begin{equation}
    \label{eq:marginal_approx}
    \mathbf{P}(D) \approx \frac{V}{M}\sum_{i=1}^M W_i(D)
\end{equation}

A single issue remains: the omission of this constant factor. But if one might be tempted to ask why we would leave it out, a more appropriate question would be, \textit{why would we not?}

As it turns out, this constant also appears in expression \ref{eq:numerical_expectation}, which we have deemed \textit{our ultimate goal}. So if we were to include it, we would constantly be \textit{dividing} equation \ref{eq:discrete_normalization} by it (multiplying its quotient), only to then \textit{multiply} it by the same factor in equation \ref{eq:numerical_expectation}. 

In the end, this \textit{oversight} is rather opportune, in the sense that assimilating the factor into the formalism simplifies the expectations we draw from it. Coincidentally (or not), it does so all the while imposing the familiar condition of normalization.

With this, when working with a a probability mass function, expectations take the form:
\begin{equation}
    \label{eq:discrete_expectation}
    \mathbb{E}_{\mathbf{P}( \theta \mid D)}\left[f(\theta)\right]
    = \int f(\theta) \mathbf{P}(\theta \mid D)\mathrm{d}\theta
    \approx \sum_{i=1}^{M} f(\theta_i) \cdot w_i(D)
\end{equation}

This is in essence a discretized version of equation \ref{eq:expectation}, which formalizes the approximation we considered in \ref{eq:numerical_expectation} by transitioning into an intrinsically discrete setting. Note that this is correct provided that the $w_i(D)$ are calculated according to \ref{eq:discrete_normalization}, whereas in the continuous domain $\mathbf{P}( \theta_i \mid D)$ represented single point evaluations of equation \ref{eq:bayesIII}. The difference is that now our normalization constant is an estimate of
\begin{equation}
    \frac{1}{P(D)\cdot M/V} = \frac{V}{M} \cdot \frac{1}{P(D)} 
\end{equation}

\noindent, which as we've seen takes care of the approximation in full, leaving only the target function evaluations.

Equivalently, we can describe the distribution using Dirac deltas to place \textit{impulses} at the grid points. These will map the expectation integrand $f(\theta)\mathbf{P}( \theta \mid D)$ at each chosen location $\theta_i$ to the corresponding evaluation at that point  $f(\theta_i)w_i(D)$, while disregarding all others - those (infinite) $\theta_X$s for which $\forall i\in \{1..M\} : \theta_X \neq \theta_i $. This converts \ref{eq:expectation} to \ref{eq:discrete_expectation} directly. 
\begin{equation}
    \label{eq:distribution_pulses}
    \mathbf{P}(\theta \mid D) \approx \sum_{i=1}^{M} w_i(D) \cdot \boldsymbol{\delta}(\theta - \theta_i)
\end{equation}

\begin{figure}[!ht]
    \centering
    \includegraphics[width=0.7\textwidth]{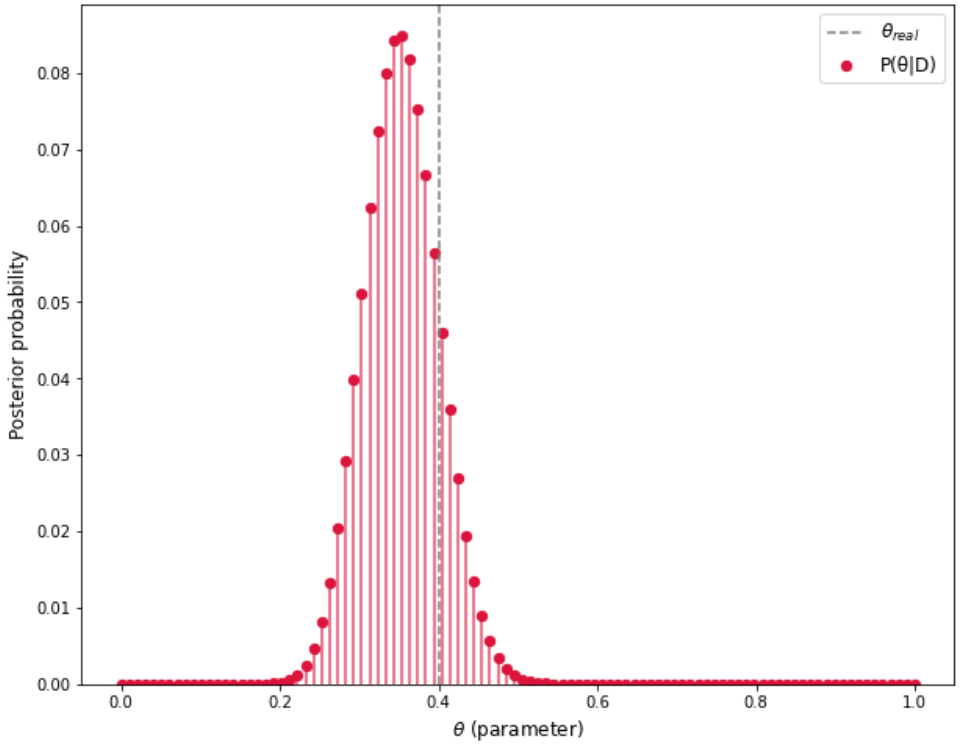}
     \caption{Bayesian posterior probability mass function for a binomial likelihood. The dashed line marks the real parameter used to generate the data. 100 trials were included in the data vector. }
     \label{fig:coin_pmf}
\end{figure}

When we are working in the discrete domain, we typically represent probabilities not with scattered points as in figure \ref{fig:coin_discrete}, but rather as the visually clearer alternative of a stem graph, as in figure \ref{fig:coin_pmf}. 

In contrast with figure \ref{fig:coin_normalized}, the probability mass function in \ref{fig:coin_pmf} has been normalized in accordance with \ref{eq:discrete_normalization}, which prompts a re-scaling of the y-axis by $V/M$. As said before, in our case $V=1$, and $M=100$ points were used; these features materialize as a compression by a factor of $100$.

With this we conclude our probe into approximate inference for now. In chapter \ref{cha:monte_carlo_posterior_sampling} we will take a more in-depth look at these methods, and elaborate on how their foundations can be built upon so as to outperform the basic grid arrangements we considered.

We will shortly proceed to discuss more realistic applications for the matters of this exposition, but before that one last topic merits a mention.

We assumed from the outset that we \textit{needed} to approximate the distributions, due to the obvious memory and power constraints. This was a sensible line of reasoning, and it \textit{is} the most general one. Nonetheless, it isn't necessarily the \textit{only} one: in some particular cases, the posterior is available analytically, dispensing with the need for numerical calculations entirely.

This happens for some classes of likelihood models, when paired with specific priors. For a few combinations of these two distributions, the posterior distribution is known to belong to the same family as the prior. In these circumstances, we call the latter a \textit{conjugate prior} for the likelihood function under consideration.

Such requirements may seem restrictive, because they are. Even so, a considerable amount of commonly occurring combinations of functions enable this approach, and its perks are enticing: we may abandon the resource-intensive techniques outlined above altogether, and simply take a known closed-form solution for a fraction of the effort.

This is made possible by the fact that we know in advance we can describe the posterior using the same parametric model as the prior. As such, we can generally compute the parameters that fully determine it efficiently. What is more, \textit{these parameters alone} contain all the information we could possibly want.

Clearly, these new parameters are distinct from the one(s) we are targeting with the learning process. They're just auxiliary constructs, used to characterize the consecutive distributions that depict our knowledge of the parameter that governs the probabilistic phenomenon under investigation (the $\theta$ from before). For this reason, and to avoid confusion, these somewhat tangential parameters are commonly termed \textit{hyper}parameters. For example, if we had a gaussian prior distribution, the hyperparameters would be the mean and variance; the parameter would simply be the independent variable.

We can think of the prior distribution of being molded by experimental results, yielding the posterior, whose shape has been altered to accommodate the data. This is done by tweaking the existing degrees of freedom; the extra structure that conjugate distributions provide reduces these degrees of freedom into a few variables. This enormously facilitates information storage and processing, because it is much easier to characterize a finite set of variables than it is a continuous distribution.

A well-known example is the normal distribution, which is self-conjugate. So if we start with a Gaussian prior, and the likelihood is Gaussian itself, then so too is the posterior.

Another case is when we have binomial likelihood as in our coin example, and the prior is a beta distribution. A beta distribution is described in the interval $\theta \in [0,1]$ by a probability density function which depends on two (hyper)parameters, $\alpha$ and $\beta$:
\begin{equation}
    \label{eq:beta_pdf}
    \boldsymbol{Beta}(\alpha,\beta) = \frac{\theta^{\alpha-1}(1-\theta)^{\beta-1}}{\mathrm{B}(\alpha,\beta)}
\end{equation}

\noindent, where $\mathrm{B}(\alpha,\beta)$ is the beta \textit{function}, which consists of the integral of the denominator and acts as a normalizing constant.

Let's say we start with some hyperparameters $\alpha_0$, $\beta_0$. It can be checked by direct calculation (by plugging equations \ref{eq:binomial} and \ref{eq:beta_pdf} into the Bayes' rule \ref{eq:bayesIII} and rearranging) that this yields a posterior given succinctly by:
\begin{equation}
    \label{eq:beta_posterior}
    \mathbf{P}(\theta \mid D) =
    \boldsymbol{Beta}\left(\alpha_0+s,\beta_0+(N-s)\right)
\end{equation}

It is convenient to remember from the probability density function in \ref{eq:binomial} that $s$ is the number of successes ($1$ outcomes), and $N$ is the total number of trials (so $(N-s)$ corresponds to the number of failures, or $0$ outcomes).

That means that all we have to do is count the number of $1$ (or tails) outcomes and keep track of the total number of coin tosses. Here our data can be summed up by two constants: $D=\{N,s\}$. In the end, we will have perfect understanding of our posterior, just by updating the hyperparameters according to these figures.

Of course, not all is perfect, and we find ourselves constricted in the choice of a prior, which must be a beta distribution itself (aside from the restriction we already placed on the likelihood). But often, and invoking if necessary the Bernstein–von Mises theorem, we can adjust the prior into taking the desired form; if so, this is a small price to pay considering the unparalleled convenience that it begets us.

In the case of the coin, we had picked a uniform prior, given by equation \ref{eq:flat_prior}. As luck would have it, that happens to be precisely a beta distribution, with parameters $\alpha=1$, $\beta=1$. This cognizance completes the necessary conditions for us to be two integer sums and a subtraction away from full possession of the \textit{exact} posterior distribution.

\begin{figure}[!ht]
    \centering
    \includegraphics[width=0.7\textwidth]{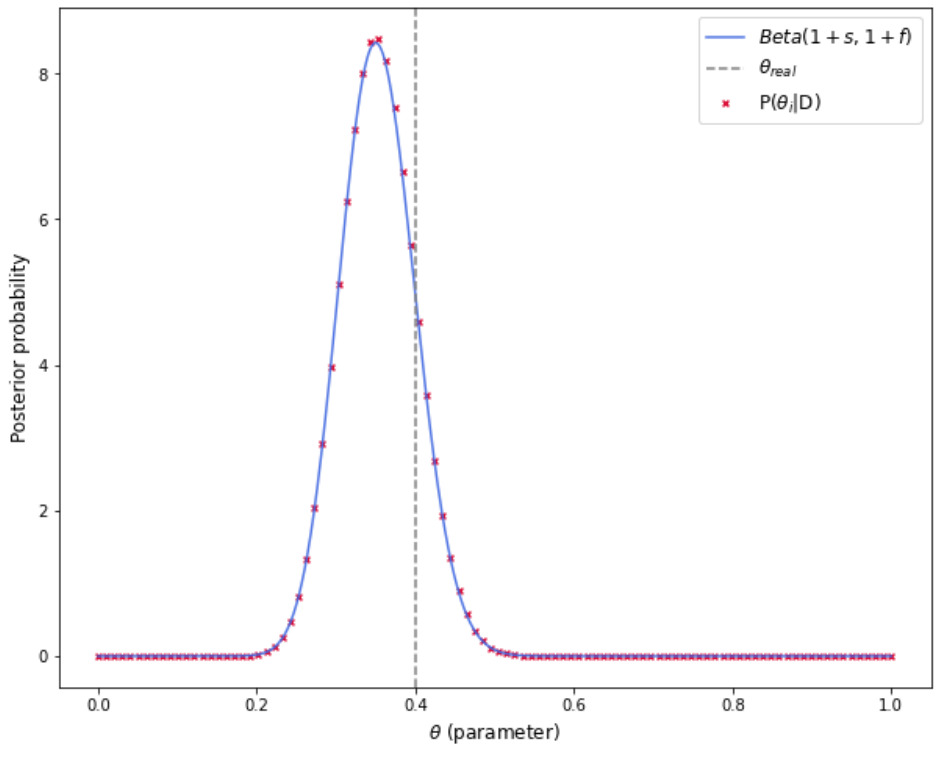}
  \caption{Bayesian conjugate posterior density for a binomial likelihood and a flat prior. The dashed line marks the real parameter used to generate the data, and the 'x's mark single point evaluations of the posterior. 100 trials were included in the data vector. }
  \label{fig:coin_conjugate}
\end{figure}

The posterior beta distribution is plotted in figure \ref{fig:coin_conjugate}, along with superposed discrete evaluations of the posterior using Bayes' equation directly (these are the same points as in figures \ref{fig:coin_discrete} and \ref{fig:coin_pmf}).

So not only does this allow us to forgo all the calculations laid out in section \ref{sub:estimation_in_practice}, but also yields an improved accuracy. Instead of performing $M$ likelihood evaluations and subsequent normalizations to get an approximate result, we perform a couple of elementary operations \textit{and} get a faultless one.

As a consequence, instead of the expectation values being approximated by equation \ref{eq:discrete_expectation}, they have closed forms. Most quantities of interest, namely all the estimators we mentioned before, can be looked up in a table and depend on the hyperparameters $\alpha$ and $\beta$ alone.

Naturally, more instances of conjugate distributions exist. Furthermore, strategies reminiscent of this one are still an option where conjugate distributions are not applicable: the posterior - and consequently expectations - can be approximated variationally. This is realized by assuming that it can be described by a parametrized distribution of choice, and optimizing its (hyper)parameters with respect to some measure of likeness to the actual posterior \cite{Blei_2017}. A similar selection was carried out implicitly in the case of conjugate priors, where the result was exact and determined by a quick data based calculation. 

In general such an assumption does not hold, wherefore the choice of a model introduces a bias. The problem can be alleviated by introducing enough variational freedom through additional hyperparameters, which yield a more customizable family of distributions and thus enhanced expressibility. This further demonstrates the ever present tradeoff between exactness and computational load (or tractability).

In the sample-based approaches we covered earlier, asymptotic exactness was generally assured, but the finite approximation error could be appreciable if not enough resources were employed. With variational inference, the error is mostly due to the limited descriptive power of the imposed model, whose complexity in turn determines the required resources.

Should the main goal be to lighten the computational cost, the latter option is a befitting choice. However, it can heavily weigh down the achievable accuracy. For this reason, we shall content ourselves with this brief overview of these alternative approaches, and instead focus on the \textit{general} approach. 

Working with the latter will bring us the benefits of a broad-spectrum solution, which preserves in full the flexibility that makes Bayesian inference attractive in higher complexity scenarios. Admittedly, it also presents more diverse intricacies. We will cover some of them in chapter \ref{cha:monte_carlo_posterior_sampling}; but still before that, we ought to examine one of the contexts that give rise to such scenarios.

\section{Application to quantum systems and general likelihood models}
\label{sec:application_to_quantum}
Finally, we would like to apply this inference technique to something a bit more useful than coin examination. By itself, determining the fairness of a coin is not the most rewarding endeavour. But this is where we benefit from the generality that we have claimed Bayesian inference enjoys: owing to it, measuring a qubit is essentially glorified coin tossing. On that account, we can characterize a quantum system much like we did a simpler one in section \ref{sec:bayesian_inference}.

This idea has been explored both theoretically and experimentally; here we will mostly follow \cite{Granade_2012,Ferrie_2011,Ferrie_2012}, but similar techniques have been approached in many other references \cite{Ferrie_2014,Wiebe_2014a,Wiebe_2014b,Sergeevich_2011,Wang_2017}.

The rules are now set by the Born rule, which yields the probability of finding a qubit in either basis state given its wavefunction. It states that the probabilities of these states are given by the squared modulus of their respective (complex) amplitudes. So if the wavefunction is given by 
\begin{equation}
    \label{eq:wavefunction}
    \ket{\Psi} = \alpha \ket{0} + \beta \ket{1}
\end{equation}

\noindent, then when measuring we will get each of the two outcomes with probabilities
\begin{align}
     \mathbf{P}(0)=|\alpha|^2\\
     \mathbf{P}(1)=|\beta|^2
\end{align}

As one might expect, imposing that $\Psi$ be normalized renders these two values redundant.
\begin{equation}
    \mathbf{P}(1)=1-|\alpha|^2
\end{equation}

This observation alone gives us a model quite alike the one of equation \ref{eq:binomial_likelihood}.
\begin{gather}
    \tag{\ref{eq:binomial_likelihood}}
    \mathbf{L}(\theta \mid D; M_{\text{coin}}) = \theta^{D}  (1-\theta)^{1-{D}}
    \\
    \label{eq:likelihood_qubit}
    \mathbf{L}(\alpha \mid D; M_{\text{qubit}})) = |\alpha|^{2D}  (1-|\alpha|^2)^{1-{D}}
\end{gather}

, where the annotations $M_{\text{coin}}$ and $M_{\text{qubit}}$ clarify which system each likelihood model pertains to.

We can point that one retrieves not the underlying parameter (amplitude) $\alpha$ from this information, but only its norm; this is of course a consequence of the projective measurement postulate of quantum mechanics. Aside from that, to shift from determining a coin's fairness to characterizing a pure quantum state, we need only identify the parameter $\theta$ with the squared norm of one of the wavefunction coefficients. This is because performing projective measurements on these systems is very much like sampling from a binomial distribution.

It is worth noting that quantum systems do \textit{not} in general behave like classical probabilistic phenomena. In fact, that they do not is ultimately attributable to the amplitude-versus-probability matter mentioned above. That the amplitude's phases at the time of measurement bear no impact the outcome does not mean that they aren't of consequence - which they are, due to the interference they give rise to as long as we're working in the quantum regime.

In the instant we measure in the computational basis, writing the wavefunction as in \ref{eq:wavefunction} tells the whole story as far as measurements are concerned. But leading up to that, the amplitudes can interfere with one another, be it constructively or destructively, to produce very different results than a weighted sum of the isolated basis states would have.

If our formulation seems unaffected by this, that's simply because we are implicitly describing the end result through the \textit{final} coefficients $\alpha$ and $\beta$, whose phases don't affect the outcome. Notwithstanding, all the ones coming before play a role in the algebra that produces it.

Moreover, this isn't the only subtlety that these coefficients hide. When speaking of coins we were targeting a constant variable, but that's no longer the case. The wavefunction could depend, in addition to the parameter(s) we're trying to find, on some external controls, such as the time of measurement; we could even change the measurement basis, which would condition the outcomes as well. In general, we could have some set of experiment controls $E$, such that:
\begin{equation}
    \label{eq:p0_controls}
     \mathbf{P}(0)=f(\theta ; E)\\
\end{equation}

We will from now separate random variables (here $\theta$, $D$) and non-random ones (here $E$, $M$) with a semicolon to avoid confusion; commas will be used to split arguments within the same \textit{class}. We prefer not to use the form $f(\theta \mid E, M)$, or in general $f(\theta \mid V)$ for $V$ a deterministic entity. While in English we might say that probabilities are \textit{conditioned} (dependent) on the experiment and model, we reserve the idea of conditioning to denote a dependence on \textit{random} variables, which are held fixed when taking conditional probabilities. Accordingly, we destine the vertical bar to standing between two random variables/events.

If the distribution of outcomes is given as in \ref{eq:p0_controls}, we must accommodate for it by adjusting the structure of the likelihood function we use to model the outcome distribution.
\begin{equation}
    \label{eq:likelihood_controls}
    \mathbf{L}(\theta \mid D; E, M) = f(\theta ; E)^{D} (1-f(\theta;E))^{1-{D}}
\end{equation}

The function $f$ is linked to the particular model $M$, mapping parameters to their associated conditional probabilities, given the experiment(s) and under $M$. We again separate random and non-random parameters using a semicolon, remembering that $\theta$ itself is treated as a random parameter.

Now, equation \ref{eq:bayesIII} too must be modified, to account for the likelihood's dependence on some experiment controls.
\begin{equation}
    \label{eq:bayes_controls}
    \boxed{\mathbf{P}(\theta \mid D; E) = \frac{\mathbf{L}(\theta \mid D; E)\mathbf{P}(\theta)}{\mathbf{P}(D ; E)}}
\end{equation}

We can also take the occasion to in addition make the model explicit, as we did to distinguish \ref{eq:likelihood_qubit} from \ref{eq:binomial_likelihood} and in equation \ref{eq:likelihood_controls} to mark the use of the function $f$ in the left-hand side.
\begin{equation}
    \label{eq:bayes_controls_model}
    \boxed{\mathbf{P}(\theta \mid D; E, M) = \frac{\mathbf{L}(\theta \mid D; E, M)\mathbf{P}(\theta ; M)}{\mathbf{P}(D ; E, M)}}
\end{equation}

The prior too signals dependence on the model, as it is itself an implementation choice; plus, the starting point for the parameters depends on the model's structure. Two different models may work on different scales, or even have different dimensions. The prior could be considered to depend on the controls too, but we assume that dependence to be absorbed into the model.

Finally, the marginal probability, being the integral of the nominator, also manifests a dependence on these new entrants. It is the prior probability of having obtained the datum $D$ given a choice of experiment $E$ and underlying model $M$.
\begin{equation}
    \label{eq:marginal_controls}
    \mathbf{P}(D ; E, M) = \int \mathbf{P}(D \mid \theta; E,M)\mathbf{P}(\theta ; M)\mathrm{d}\theta
\end{equation}

Of course, we don't necessarily have to consider data records of unit length, and we can condense longer data records into a single expression as we did in \ref{eq:bayes_datavector}.
\begin{equation}
    \label{eq:bayes_datavector_controls}
    \mathbf{P}(\theta \mid \vec{D};\vec{E})
    = \frac{\mathbf{L}(\theta \mid \vec{D}; \vec{E})\mathbf{P}(\theta)}{\mathbf{P}(\vec{D};\vec{E})}
\end{equation}

The data and experiment controls must have the same length and be written in the same order, that is, 
\begin{gather*}
    \vec{D}={D_1,D_2,\dots,D_N}
    \vec{E}={E_1,E_2,\dots,E_N}
\end{gather*}

\noindent means that experiment $E_1$ was performed and yielded datum $D_1$, etc. (We could of course implicitly incorporate everything into the data vector, making it a vector of tuples.)

When estimating quantum parameters, likelihood evaluations as per \ref{eq:likelihood_controls} amount to finding the probability distribution of outcomes for an experiment, given a parametrized quantum system. The experiment $E_j$ is a constant for each datum  $D_j$, and we bind the parameters to a target value $\theta_i$ at every evaluation to get $\mathbf{L}(\theta_i \mid D_j; E_j)$ - or, if we prefer, we put the data together and consider $\mathbf{L}(\theta_i \mid \vec{D}; \vec{E})$. Either way, we do this for multiple $\theta_i$ to approximate expectations as in equation \ref{eq:discrete_expectation}.

Such a parametrized description of the target system's behaviour is indispensable for the inference process, which cannot take place without a likelihood model; the only degrees of freedom bestowed on the solution are the parameters. As such, the existence of a fixed model has been assumed all throughout this discussion, to the point of being implicit in our early version of Bayes' rule (e.g. equation \ref{eq:bayesIII}).

However, such an assumption places severe restrictions on the scope of our analysis. In general we won't have one single, perfect model. What - if anything - can be done if we don't know the exact shape of the likelihood? Some alternative function $g(\theta ; E)$ could just as well replace the $f$ in equation \ref{eq:likelihood_controls}. Ideally, we would be able to tell which does best. Better yet, we would want to \textit{refine} the best one by increasing its complexity and evaluating whether its aptness to fit to the data is enhanced. All of these possibilities rely on the ability to compare models. 

So the question is, can this framework be extended to provide guidelines for model comparison?

The answer is, it doesn't need to be.

\subsection{Brief detour into model selection}
\label{sub:model_selection}

Not only is it possible to derive further generalization still within the Bayesian framework, but all the tools are already in our hands; we need only recognize and systematically apply them. More specifically, broader algorithms can achieved by branching the analysis so as to encompass several models. We can then rely on model selection rules based on constructions we've already discussed to choose between them.

The usual approach is to develop evolutionary strategies that construct and compare increasingly complex models, and it has recently been applied to quantum systems \cite{Gentile_2021}. An earlier such application, albeit somewhat \textit{less quantum} in nature (in the sense that model selection takes place within a classical clustering problem) is presented in \cite{Granade_2017}

How can we perform systematic model comparison? Bayes' theorem again provides an answer. This selection criterion is used and briefly discussed in e.g. \cite{Gentile_2021,Granade_2017}; a short but broader overview can be found in e.g. \cite{Ferrie_2014b}.

It should comes as no surprise that Bayes' rule can be used to reason about models also, bringing them from the sidelines to occupy the central role that in equation \ref{eq:bayesII} belonged to the parameters. 
\begin{equation}
    \label{eq:bayes_model}
    \mathbf{P}(M \mid D) = \frac{\mathbf{P}(D \mid M)\mathbf{P}(M)}{\mathbf{P}(D)}
\end{equation}

We can then quantify the relative merits of any two models $M_A$ and $M_B$ by means of the ratio of their posterior probabilities - in the words of \cite[chapter~5]{Jeffreys_61}, the proportionality of their chances:
\begin{equation}
    \label{eq:bayes_ratio}
    \frac{\mathbf{P}(M_B \mid D)}{\mathbf{P}(M_A \mid D)}= \frac{\mathbf{P}(D \mid M_B)\mathbf{P}(M_B)}
    {\mathbf{P}(D \mid M_A)\mathbf{P}(M_A)}
\end{equation}

For an objective assessment of their comparative predictive power, we choose $\mathbf{P}(M_A)=\mathbf{P}(M_B)$, getting:
\begin{equation}
    \label{eq:bayes_factor}
    \mathcal{B}_{B>A} \equiv \frac{\mathbf{P}(D \mid M_B)}{\mathbf{P}(D \mid M_A)}
\end{equation}

\noindent, where "$>$" denotes superiority.

The ratio in \ref{eq:bayes_factor} is termed a Bayes factor. It is called a factor because it transforms the prior odds into posterior odds. Notice that this quantity is entirely determined by the data: we make no \textit{a priori} distinction between models.

The magnitude of the ratio \ref{eq:bayes_factor} can be interpreted as how strongly the evidence supports one model (here $M_B$) over another (here $M_A$). Its significance can be gauged in a similar way to statistical hypothesis testing; scales have been proposed in \cite[appendix~B]{Jeffreys_61} and later \cite{Kass_1995}. In \cite{Granade_2017} this is applied to a quantum characterization example.

As to how we can compute it in practice, that's a discussion we've already had back in subsection \ref{sub:estimation_in_practice}. The elements of \ref{eq:bayes_factor} are simply the marginal probabilities of the data over each models' parameter space, as in equations \ref{eq:marginal} (where the model was implicit) and \ref{eq:marginal_controls} (where a single model was considered, and we also explicitly marked the experiment $E$):
\begin{equation}
    \label{eq:marginal_model}
    \mathbf{P}(D ; M_X) = \int \mathbf{P}(D \mid \theta_X; M_X)\mathbf{P}(\theta_X ; M_X)\mathrm{d}\theta_X
\end{equation}

This is where Bayes factors differ from traditional likelihood ratio tests, often used in hypothesis testing: maximization is replaced by marginalization over model parameters. This is convenient for two reasons. First, the marginal is at hand by virtue of the inference process (this has been illustrated in \ref{sub:estimation_in_practice} and will be discussed further in chapter \ref{cha:monte_carlo_posterior_sampling}). Second, we are exempted from optimization, which in addition to the cost is more susceptible to problems affecting generalization; integration tends to produce a more robust metric, in particular if noise is an issue.

Lastly, it naturally penalizes superfluous model complexity, privileging models with less structure if all else is equal \cite{Kass_1995,Vandekerckhove_2015}. This is opportune, given that model selection shouldn't unreservedly aim for the best goodness-of-fit, but rather factor in parsimony as well, seeking a happy medium. If we don't enforce the principle of parsimony (also known as Occam's razor), we risk overfitting owing to undue flexibility. This is related to the bias variance tradeoff and the idea of \textit{capacity} that is prevalent in machine learning.

Other approaches, such as the Akaike or Bayesian information criteria, artificially work in a penalty to disfavor an increase in parameters \cite{Ferrie_2014b}. By-default retention of the null hypothesis in frequentist hypothesis testing is another example of imposed bias against excessive structure \cite{Vandekerckhove_2015}.

In light of this, Bayes factors' inherent propensity towards simpler models has earned them the epithet of \textit{automatic Occam's razor} \cite{Kass_1995}. But where exactly does it come from?

It stems, of course, from the nature of the integral in \ref{eq:marginal_model}. We can shed some light on the reason why by taking a step back and  looking at what it stands for: instead of taking isolated likelihoods as a metric, we take their expectation value over the full prior.

This doesn't favor high likelihood densities per se, instead weighing them against space occupation. Should the parameter space be unnecessarily voluminous, a large part of it will have very low likelihoods associated. If so, the corresponding model will be penalized when taking the expectation in \ref{eq:marginal_model} as compared with a more frugal one which uses volume more sparingly. The gratuitous expansion of model space size for little or no benefit in terms of explanatory power provokes a penalty.

Put differently, the data tend to become more sparse relative to the working region as the latter expands. So if we add one parameter to our model, thereby increasing its dimension, this extra degree of freedom must actively contribute towards the interpretation of the data. If it does not, and the data fit is no better than it was before, its addition is unwarranted. This fact will be penalized in the Bayes factor analysis, where we take the average over a more \textit{diluted} prior. As a consequence, we will be led to keeping the simpler one, which does as good a job .

We can illustrate this with a simple example. Let's say we have a black-box that discards any inputs (i.e. the controls are irrelevant) and produces 0 or 1 outcomes. We start by proposing a model $M_1$, with a likelihood of the form
\begin{equation}
    \label{eq:likelihood_sectionI}
    \mathbf{L}(\theta \mid D; M_1) =
    \begin{cases}
    D, & \text{if } a \leq \theta \leq b\\
    1-D, & \text{otherwise}
    \end{cases}
\end{equation}

\noindent , with $a,b$ fixed and again $D \in \{0,1\}$.

We can interpret this as the description of a black-box controlled by some secret parameter. If the parameter lies between $a$ and $b$, it always outputs 1; if it doesn't, it always outputs 0.

Still, this doesn't fully determine our model; we need to define a domain for the parameter. Instead, we write: 
\begin{equation}
    \label{eq:likelihood_sectionII}
    \mathbf{L}(\theta \mid D; M_1) =
    \begin{cases}
    D, & \text{if } \theta \in [a,b]\\
    1-D, & \text{if } \theta \in [0,a[ \cup ]b,W[
    \end{cases}
\end{equation}

We furthermore assume a flat prior, as usual.
\begin{equation}
    \label{eq:m1_prior}
    \mathbf{P}(\theta ; M_1) = \frac{1}{W}
\end{equation}

 Here $a,b,W$ are constants with $0 \leq a \leq b \leq W$. The total prior width $W$ can be regarded as a hyperparameter that concretizes $M_1$.

Let's suppose our first call to the black-box produced a 1, and we want to tune our model so that it reproduces this behaviour. In this case, the support of the posterior is confined to the interval $[a,b]$, as this is the only region with non-null likelihood for $D=1$. This arrangement imitates what usually happens when performing inference: most of the posterior probability mass is contained within a region (a credible region, as discussed in subsection \ref{sub:estimators_and_expressions_of_uncertainty}), whose volume ideally decreases with the data collection if we fix the probability it should amass.

Recalling equation \ref{eq:marginal_model}, we can easily compute the marginal, or the \textit{evidence}, for this model:
\begin{equation}
    \label{eq:m1_evidence}
    \mathbf{P}(D=1 ; M_1) = \frac{b-a}{W}
\end{equation}

\begin{figure}[!ht]
    \centering
    \includegraphics[width=.8\textwidth]{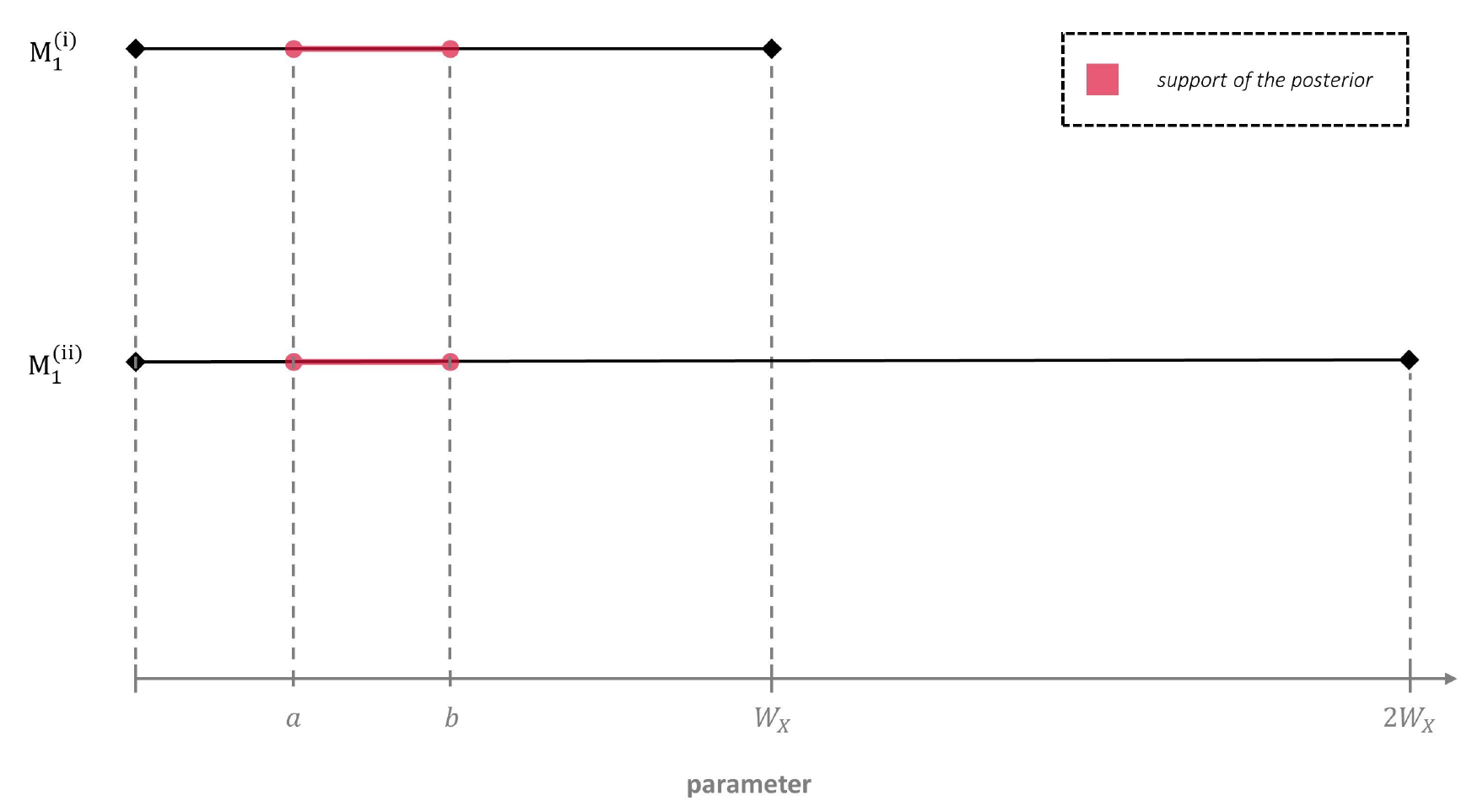}
    \caption{Illustration of the model $M_1$ from equation \ref{eq:likelihood_sectionII} for different values of $W$. The Bayes factor of $M_1^{(ii)}$ relative to $M_1^{(i)}$ is $\mathcal{B}=1/2$ for $D=1$.}
    \label{fig:model_1d}
\end{figure}

We can see that if we increase $W$, unnecessarily expanding the parameter space for no net increase in descriptive power for the data (in this case, datum), the marginal will decrease, effectively marking the model as \textit{less worthy} (figure \ref{fig:model_1d}). 

\begin{figure}[!ht]
    \centering
    \includegraphics[width=8cm]{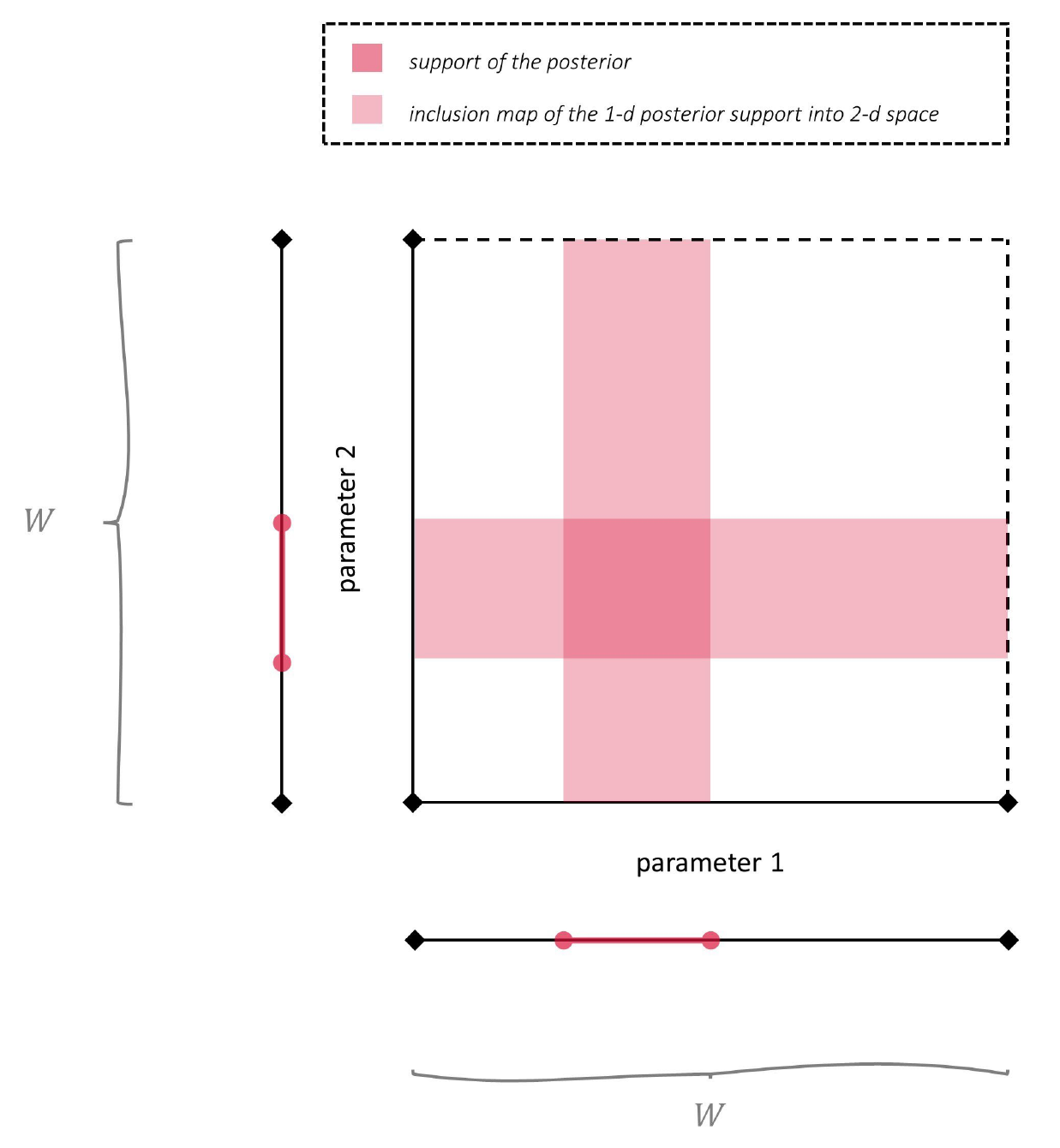}
    \caption{Illustration of the model $M_2$ from equation \ref{eq:likelihood_section_2d} as compared to its one-dimensional projections. Two such projections can be viewed on a scale in figure \ref{fig:model_1d} for different values of $W$. The space forms a square with side $W$.}
    \label{fig:model_2d}
\end{figure}

What if go further, and needlessly add \textit{an entire dimension}? We can lift the one-dimensional model up to two dimensions by taking the natural inclusion map of the likelihood \ref{eq:likelihood_sectionII} along each dimension and intersecting both (figure \ref{fig:model_2d}).

More formally:
\begin{equation}
    \label{eq:likelihood_section_2d}
    \mathbf{L}(\vec{\theta} \mid D; M_2) =
    \begin{cases}
    D, & \text{if } \{\theta_1,\theta_2\} \in [a,b]^2\\
    1-D, & \text{if } \{\theta_1,\theta_2\} \in [0,W[^2 \setminus [a,b]^2
    \end{cases}
\end{equation}

This is at least as expressive as the previous model. The posterior is symmetric relative to a 45° inclination line, and we could recover the previous posterior by projecting onto either of the axis the directional maxima orthogonal to it.

However, despite our having doubled the number of parameters, this doesn't fit the datum any better. As a matter of fact, the one-dimensional model already rendered a perfect description of what transpired, because any point with non-zero posterior probability would have reproduced it with a 100\% chance. Thus, increased dimensionality could never be of benefit to us.

We can see how this impacts the model's worth as measured by Bayes' factors. The same choice of a uniform prior now gives us
\begin{equation}
    \label{eq:m2_prior}
    \mathbf{P}(\vec{\theta} ; M_2) = \frac{1}{W^2}
\end{equation}

\noindent, and the evidence becomes:
\begin{equation}
    \label{eq:m2_evidence}
    \mathbf{P}(D=1 ; M_2) = \left( \frac{b-a}{W} \right)^2
\end{equation}

Other than in the scenario where all possible parameter specifications behave in the exact same manner (in which case all models are equally bad, because inference is useless), the two-fold increase in parameters brings a quadratic decrease in the evidence in favor of the model. A 2-dimensional model with a projection matching \ref{eq:likelihood_sectionII} could only have as large a marginal probability as the latter if the likelihood was independent of the additional parameter (with support on a vertical or horizontal strip, as the lighter colored sections of figure \ref{fig:model_2d}).

In the case of \ref{eq:likelihood_section_2d}, the ratio is:

\begin{equation}
    \label{eq:m2m1_bayes_factor}
    \mathcal{B}_{M_2>M_1} = \frac{b-a}{W} \leq 1
\end{equation}

Under these circumstances, we would reject the redundant parameter. This is due to the fact that the prior probability density is stretched thinner relative to the data-appointed plausible region, as a result of the tension between probability mass and parameter space volume. If we further generalize the likelihood to a cube and then hypercubes, the fraction of volume contained within the region will shrink exponentially with the dimension. 

Expectably, the decisions are more nuanced in realistic scenarios; this is an extreme example. Notwithstanding, it mimics the inner workings of Bayesian model comparison, and shows how it attributes higher scores to models which make better use of the space they inhabit. Encouraging simplicity in this manner is likely to improve generalization, making the winning model more apt to predict unseen data.

\subsection{Likelihood evaluations in the quantum case}
\label{sub:likelihood_evaluations_quantum}

Once models are out of the way, we are left with equation \ref{eq:bayes_controls}. The prior is predetermined, the marginal probability can be regarded as a normalization factor - and lastly, we have the likelihood function.

Likelihood evaluations are the key element when performing Bayes' updates, and their most challenging requirement. It is the experimentally determined likelihoods that will shape our final beliefs, by reworking the prior ones in consonance with our measurement results.

Conceptually, they consist of the probabilities that some hypothetical parameters $\theta_i$ would have responded like our target system did to our observations. That is, they correspond to the probabilities of sampling the conditional outcomes $D_j \mid E_j$ that we did (or the joint probability of having obtained the full data record $\vec{D} \mid \vec{E}$)  had the (unknown) real parameters been $\theta_i$. This describes the task of quantum simulation.

\begin{figure}[!ht]
    \centering
    \includegraphics[width=0.8\textwidth]{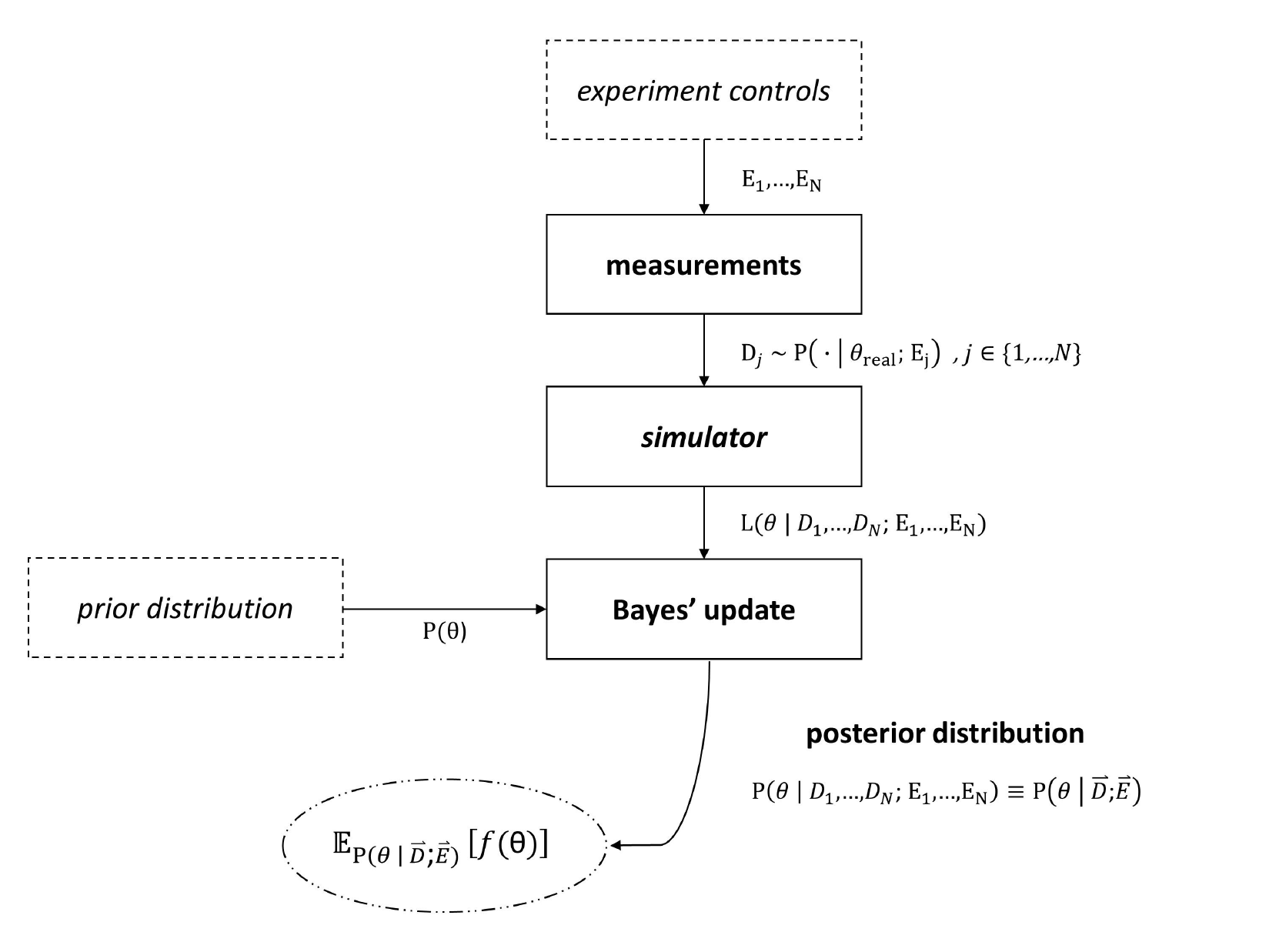}
    \caption{Diagrammatic representation of a quantum parameter estimation algorithm using Bayesian inference. Italics signal dependence on implementation choices, and dashed boxes denote one time only inputs. The tilde $\sim$ means "is distributed as" and the dot $\cdot$ is a stand-in for the object of the sampling.}
    \label{fig:offline_inference_diagram}
\end{figure}

Before proceeding to discuss this point, we observe that it settles the key differences between characterizing quantum system and characterizing a coin. A schematic overview of the inference process applied to the quantum case is presented in figure \ref{fig:offline_inference_diagram}. It illustrates the context that the application of equation \ref{eq:bayes_controls} in a quantum parameter learning problem demands. Similar diagrams (on which this one was based) can be found in \cite{Granade_2012} and \cite{Ferrie_2011}.

It is sometimes possible to estimate the probabilities by means of analytic expressions, derived from quantum mechanics. If so, apart from the measurements on the system under study, the inference protocol is as straightforward to implement as it was in the case of our coin. This is feasible for small quantum systems which are simple to analyse; such is the case of the examples we will approach here. But retrieving measurement statistics from a generic (parametrized) Hamiltonian is more often than not a rather challenging task.

Still more generally, if what we mean to characterize is an open quantum system - a realistic assumption -, the Schrödinger equation is substituted by a master equation, whose operators and constants we must discover to characterize the system. Evaluating likelihoods at some instant (this instant being an experimental control) requires simulating the time evolution produced by our formalism until then.

One possibility is to simulate this evolution classically. However, classical resources are known not to excel at simulating systems which manifest their non-classical nature. For this reason, the calculations soon become intractable, as the size of the quantum system of interest increases and with it the size of the classical simulator required to replicate it - it being that the latter scales exponentially with the former. 

Such circumstances prompt the use of quantum devices, which deliver our best known description of reality. This solution has been increasingly entertained in literature, ever since it was first advocated and discussed \cite{Feynman,Lloyd_1996}. For our particular purposes, they are interesting in that they supply efficient estimates of the likelihoods.

Let us briefly recap where we stand. We have only now stated that we would like to reconstruct the likelihood function for some parameter specification and experiment. It should be remembered that this is preceded by the measurement of a \textit{target} quantum system, which we have no control over. We mean to characterize that system; more specifically, we have a parametrized model describing it, and want to estimate the actual system parameters.

This can mean for instance that we are set on a fixed form Hamiltonian $\op{H}(\vec{\theta})$, which we believe describes the system's dynamics and whose real parameters $\vec{\theta}$ we would like to find. The description is to be concretized by establishing the numerical parameters $\vec{\theta}_{\text{real}}$ up to some tolerance. This settles our problem statement. In this context, quantum parameter estimation is often called Hamiltonian learning. 

All we need from the system is the ability to measure it (we would of course also appreciate it if it conformed to our model, but that responsibility falls on us). These measurements are described by experimental controls, such as the time, which we know give rise to a distribution of outcomes. We would like to know what distribution, because it is determined by the underlying parameters which we are attempting to estimate. 

Equivalently, knowing the exact distribution would inform us about the relevant parameters, as illustrated by equation \ref{eq:p0_controls} or \ref{eq:likelihood_controls}. Yet, all we get are individual samples - our data, which are comprised of measurement outcomes (paired with their respective controls). So we pick those samples, and we examine many possible configurations of our quantum model, expecting that the ones most likely to have behaved similarly under the same conditions correspond to the most accurate parametrizations. Conformity to the data \textit{is} accuracy for all we care, because the ultimate goal is to replicate the system's behaviour. Such an achievement enables the estimation of its properties by means of a well-tuned simulator, including the anticipation of future observations based on past ones.

Of course, if we only perform \textit{one} measurement, this doesn't convey very useful information. But as we gather more and more experimental data, we anticipate that it will. It also tends to perform better on limited data than just inverting expression \ref{eq:p0_controls} for a rough estimate of $\mathbf{P}(0)$ would, and is a more flexible approach (further exemplification will follow later in the document). 

It is when determining the merits of parametrization hypotheses that we need access to a quantum simulator. We want to mimic the real system; for that, we fill in the unknown (the parameters $\vec{\theta}$), and instruct our simulator to evolve accordingly. If dealing with a closed system, the evolution is dictated by the Schrödinger equation: we must approximate the dynamics induced by $\op{H}(\vec{\theta})$, e.g. by trotterizing it. The evolution time for the Hamiltonian dynamics must match how long the target system was allowed to evolve after some initialization (generally a projection).

Lastly, we replicate the exact same measurements we performed on our system, and assess how likely the outcomes would have been to match the registered data by repeating each and taking the relative frequencies. These constitute our likelihoods.

Note that, in spite of their likeness, there's an asymmetry between what is required of the target system as compared to the simulator; from the former we get binary samples, and from the latter we are assuming we get exact (or close enough, but certainly not binary) probabilities.

These ideas match in the limit where we take many samples (measurements), because we may then reconstruct the full probabilities of an outcome as the ratio of its occurrence to the total measurements. Or, meeting at the other end, if the simulator produces a single outcome. For this reason, the terms \textit{sampling} and \textit{simulation} are often used interchangeably in the domain of statistics. If they are not, a too strict view of simulation is to blame, for their difference is but a matter of degree.

Likewise, when speaking of quantum simulation one generally distinguishes between \textit{strong} and \textit{weak} simulation; the latter describes precisely the drawing of samples from the probability distribution, as opposed to its full specification. Their difference is quantitative and not qualitative, in spite of this quantitativeness' being rather formidable still (and quite possibly the difference between the simulation being computable and not). This emboldens one to ask, could weak simulation be used in the context of Bayesian inference? It can, and it has been \cite{Ferrie_2014}.

Still within quantum device based likelihood computations, more sophisticated strategies have been proposed, exploiting the structure of Bayesian learning to tune the protocol's degrees of freedom \cite{Wiebe_2014a,Wiebe_2014b}. Their underlying idea of adaptivity is related to the strategies to be covered in section \ref{sec:bayesian_experimental_design}.

This concludes our fundamental considerations on this topic. With them in hand, we can finally consider some real examples, corresponding to specific instances of the diagram in figure \ref{fig:offline_inference_diagram}.

\subsection{Quantum characterization examples: spin precession and phase estimation}
\label{sub:quantum_characterization_examples}

The main references for this section are \cite{Granade_2012,Wiebe_2016}, for the first and second illustrative cases respectively.

One simple yet elucidatory example is that of a single qubit oscillatory wavefunction whose frequency we mean to discover. This application has been widely studied, namely in \cite{Sergeevich_2011,Ferrie_2011,Granade_2012,Ferrie_2012,Wang_2017}.

We consider a qubit under the action of the following Hamiltonian:
\begin{equation}
    \label{eq:precession_hamiltonian}
    \op{H}(\omega) = 
    \frac{\omega}{2}\opsub{\sigma}{z}
\end{equation}

This class of Hamiltonians in particular is well-suited for the description of multiple physical phenomena of relevance, such as Larmor precession, Rabi flopping and Ramsey oscillations. We will elaborate further on chapter \ref{cha:experiments_quantum_hardware}, section \ref{sec:models_open_quantum_systems}; for now, we will be content with understanding the behaviour it induces.

That can be obtained by expanding the time evolution operator corresponding to \ref{eq:precession_hamiltonian} in its Taylor series, grouping the terms into even and odd functions, recognizing each as the cosine and sine series and using the identity $\opsub{\sigma}{i}^2 = \op{I}$ for $\opsub{\sigma}{i}$ in the Pauli group to get:
\begin{equation}
    \label{eq:precession_operator}
    \opsub{U}{\omega}(t) = e^{-i\op{H}t} = 
    \cos \left(\frac{\omega}{2}t \right) \op{I}
    + i\sin \left(\frac{\omega}{2}t \right) \opsub{\sigma}{z}
\end{equation}

Clearly, such an operator has no effect on a z-eigenstate, leaving basis states $\ket{0}$ and $\ket{1}$ unchanged apart from a global phase. For other initial states however, it will impact the  $\opsub{\sigma}{x}$ and  $\opsub{\sigma}{y}$ expectations.

In general, we can apply this to some initial state $\ket{\psi(0)}$,
\begin{equation}
    \ket{\psi(t)} = \opsub{U}{\omega}(t) \ 
    \ket{\psi(t=0)}
\end{equation}

\noindent, and use the Born postulate to extract outcome probabilities:
\begin{equation}
    \mathbf{P}(D \mid t) = \norm{\braket{D}{\psi(t)}} ^ 2
\end{equation}

\noindent for $\ket{D}$ some basis state corresponding to an observable's eigenstate. 

A straightforward choice of a starting state is the x-eigenstate. 
\begin{equation}
    \ket{+}=\frac{\ket{0}+\ket{1}}{\sqrt{2}}
\end{equation}

Since the identity maps it to itself and  $\opsub{\sigma}{z}$ is an inverter in the x-basis (mapping $\ket{+}$ to $\ket{-}$ and vice-versa), it is easy to see that:
\begin{equation}
    \label{eq:precession_+_wavefn}
    \ket{\psi(t)} = \opsub{U}{\omega}(t) \ 
    \ket{+} = \cos \left(\frac{\omega}{2}t \right) \ket{+}
    + i\sin \left(\frac{\omega}{2}t \right) \ket{-}
\end{equation}

\begin{figure}
\captionsetup[subfigure]{width=.9\linewidth}%
\begin{subfigure}[t]{.45\linewidth}
  \centering
  \includegraphics[width=\linewidth]{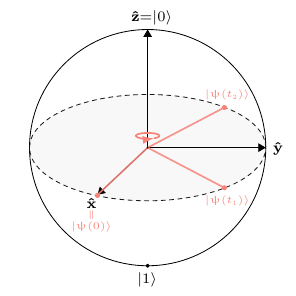}
    \caption{Spin precession starting at the $\ket{+}$ state. This illustrates the motion described by equation \ref{eq:precession_+_wavefn}. Here $0<t_1<t_2$.}
  \label{fig:bloch_precession_+}
\end{subfigure}
\begin{subfigure}[t]{.45\linewidth}
  \centering
  \includegraphics[width=\linewidth]{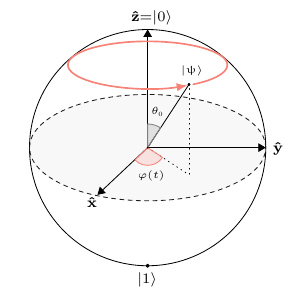}
  \caption{Spin precession in the general case. The angle $\theta_0$ is a constant, whereas the phase $\varphi(t)$ is time dependent.}
  \label{fig:bloch_precession_general}
\end{subfigure}
\caption{Bloch sphere representations of spin precession for a specific initial state and for a general one.}
\label{fig:bloch_precession}
\end{figure}

This describes a spin rotating about the z-axis. It can be visualized using the Bloch sphere (figure \ref{fig:bloch_precession_+}): the Bloch vector lies on the equatorial plane, and precesses about the vertical axis. Naturally, there is nothing unique about the z direction; we could modify \ref{eq:precession_hamiltonian} to make the Bloch vector rotate around other axes.

The result is that the $\opsub{\sigma}{z}$ expectation is constant (0), whereas those of $\opsub{\sigma}{x}$ and $\opsub{\sigma}{y}$  oscillate.
\begin{gather}
    \label{eq:precession_+}
    \mathbf{P}(+ \mid t) = \norm{\braket{+}{\psi(t)}} ^ 2
    = \cos^2 \left(\frac{\omega}{2}t \right)\\
    \label{eq:precession_-}
    \mathbf{P}(- \mid t) = \norm{\braket{-}{\psi(t)}} ^ 2
    = \sin^2 \left(\frac{\omega}{2}t \right)
\end{gather}

Thus by measuring on the x-basis, we can infer $\omega$. (Equivalently, we could measure in the y-basis or others, preferably orthogonal to the z-axis.)

For more general states, the operator preserves the $\opsub{\sigma}{z}$ expectation while inducing oscillations in the $\opsub{\sigma}{x}$ and $\opsub{\sigma}{y}$ components. This can most easily be seen by applying it to a general pure state in the Bloch form:
\begin{gather}
 \ket{\psi (0)} = 
 \cos(\theta_0 /2)\ket{0} + e^{i\varphi_0} \sin(\theta_0 /2)\ket{1} \\
 \ket{\psi (t)} = \opsub{U}{\omega}(t) \ket{\psi (0)} = 
 \cos(\theta_0 /2)\ket{0} + e^{i\varphi_0 + \omega t} \sin(\theta_0 /2)\ket{1}
\end{gather}

The initial angle with the z axis is fixed at $\theta_0$, whereas $\varphi (t) = \varphi_0 + \omega t$ (figure \ref{fig:bloch_precession_general}). For simplicity, we will consider the simpler case of $\ket{+}$ initialization.
 
All ingredients for inference are now set. We can identify the notation with the one used throughout the chapter:
\begin{equation}
    \label{eq:precession_summary}
    \dboxed{ \vphantom{gh} 
    \mathbf{P}(\theta \mid D;E) \leftrightarrow 
    \mathbf{P}(\omega \mid +;t) = \cos^2 \left(\omega t/2 \right)
    }
\end{equation}

Here the parameter is the precession frequency, the data are whether the measurement found the qubit at state $\ket{+}$ or $\ket{-}$, and the only experimental control is the evolution time that elapsed since the wave-function's collapse into the initial state. The likelihood in \ref{eq:precession_summary} contains all information, because the probability of $\ket{-}$ is complementary to that of $\ket{+}$.

\begin{figure}[!ht]
    \centering
    \includegraphics[width=0.7\textwidth]{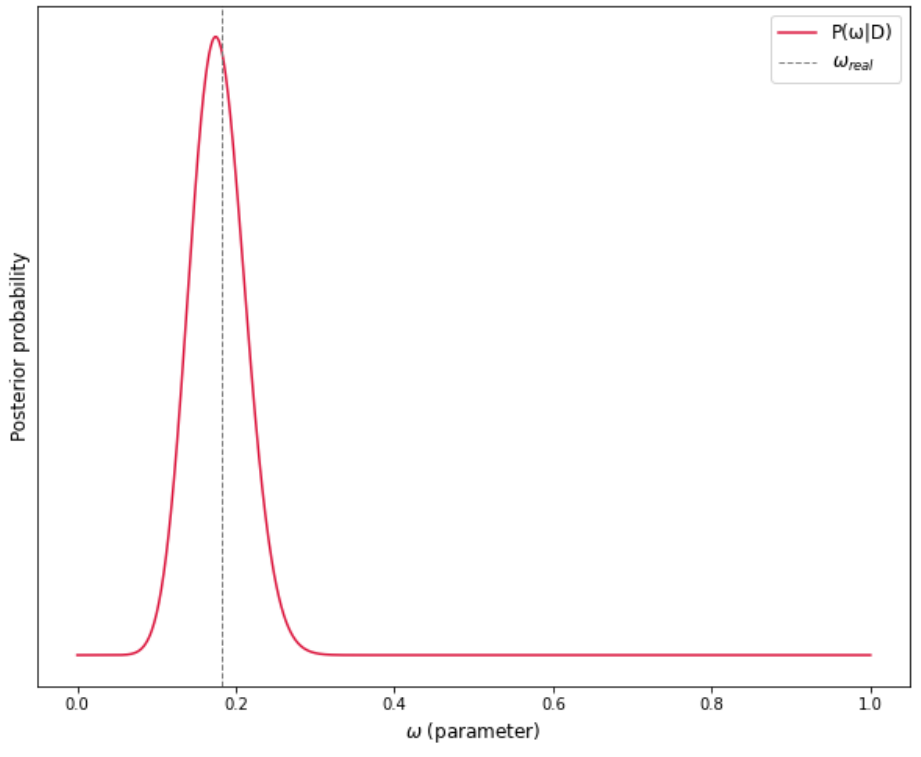}
    \caption{Bayesian posterior probability distribution for an oscillatory likelihood. The dashed vertical line mark the real parameter used to generate the data. 100 trials were contemplated in the data vector, using evenly spaced times on $[0,5]$.}
    \label{fig:posterior_precession}
\end{figure}

With this, everything is in place to learn $\omega$. An illustrative posterior for the likelihood in \ref{eq:precession_summary} is presented in figure \ref{fig:posterior_precession}. The times were chosen arbitrarily; we will get back to this choice later. For now, we'll be content with noting that the posterior looks similar to what it did in section \ref{sec:bayesian_inference} (e.g. figure \ref{fig:coin_bayes_vs_freq})

Another interesting example is phase estimation. The goal of a phase estimation is to estimate the phase $\phi$ of the eigenvalue $\lambda$ of a unitary $\op{U}$'s eigenstate $\ket{\psi}$. The two last elements are given. 
\begin{gather}
    \label{eq:eigen_equation}
    \op{U}\ket{\psi} = \lambda\ket{\psi}\\
    \label{eq:eigenvalue_phase}
    \lambda = e^{i\phi}
\end{gather}

The most well-known approach is to construct a circuit where we create superposition in an auxiliary register, apply to a second register initialized to $\ket{\psi}$ a sequence of multiples of $\op{U}$ controlled on the said register ($C \hyph \op{U}^\mathbf{m}$), and process the resulting state with an inverse Fourier transform (\gls{IQFT}). Then we measure to retrieve the phase information.

However, such a circuit is of prohibitive depth, considering the decoherence problems that plague today's (and tomorrow's) quantum devices. The main culprit is the Fourier transform, which requires a sequence of controlled rotations. This is doubly inconvenient, being \textit{a sequence} of \textit{controlled rotations}. Each of these rotations must be decomposed into a sequence of elementary gates, whose length scales oppositely to the required precision \cite{Kim_2018}. These groups of decompositions are to be executed mostly in a series, for the duration of which the system must not lose coherence.

Moreover, the precision is determined by the number of available qubits - a scant resource in quantum devices, and one which brings additional troubles (yet one more source of error). The result is hardly a reasonable thing to ask of near-term quantum devices, whose very limited coherence times will join a plethora of other sources of experimental error to produce essentially noise at the output. 

For this reason, alternative approaches to phase estimation have emerged, namely \gls{IPE} algorithms. They are typically based on a shorter quantum sub-routine, trusting on some level of classical processing to replace the troublesome quantum processing (the \gls{IQFT}).

\begin{figure}[!ht]
    \centering
    \includegraphics[width=10cm]{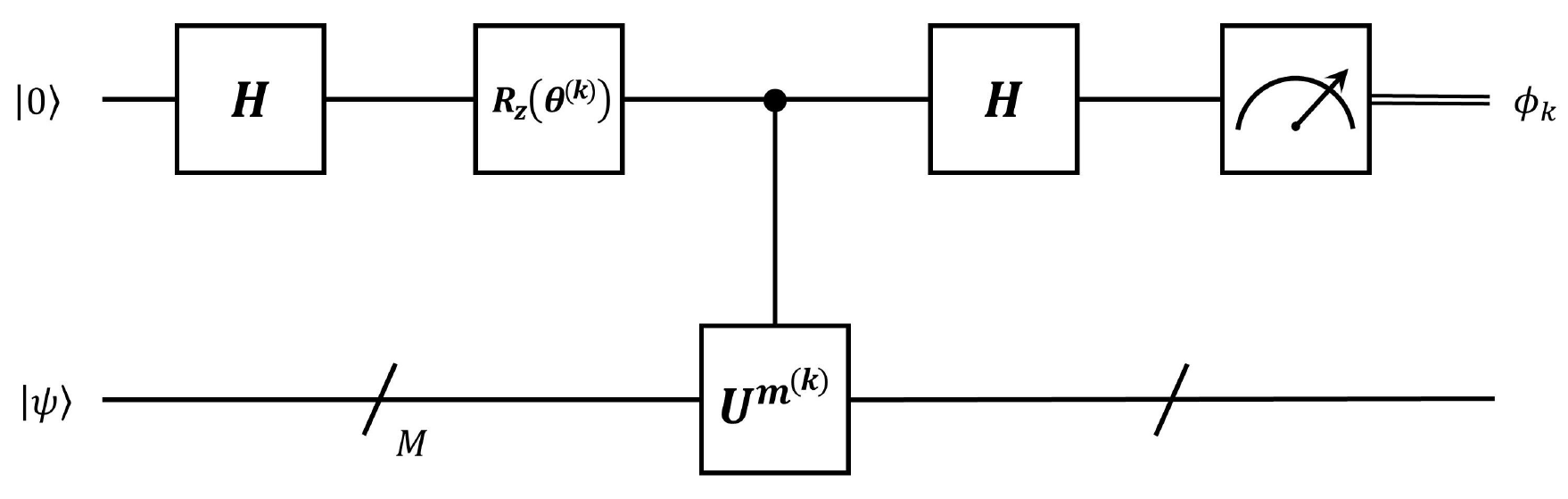}
    \caption{Iterative phase estimation circuit diagram.}
    \label{fig:phase_est_circ}
\end{figure}

This transfer of responsibility is achieved by shrinking the quantum circuit from a full algorithm to a fundamental measurement operation, which we plug into a classically controlled feedback loop. The typical quantum circuit used for this purpose is depicted in figure \ref{fig:phase_est_circ}. For a single qubit, it matches the traditional phase estimation circuit. 

Unsurprisingly, the circuit for iterative phase estimation resembles the key motif of its textbook version. We detach it from the circuit, alter it to our convenience - namely adding some degrees of freedom -, and feed it to a classical processor. This has a similar flavor to other \glsxtrshort{NISQ} era algorithms (such as the \glsxtrshort{QAOA}), where we take a shallow parametrized circuit that loosely reproduces the pattern of an exact quantum algorithm and throw in classical resources to (hopefully) compensate. The goal is to somehow extract the quantum advantage of the fundamental building blocks while placing most of the computational load elsewhere.

More specifically, the attractiveness of these iterative phase estimation protocols stems from three points. First, they are based on several more shallow circuits rather than one long one, which is beneficial when decoherence is a problem. Second, their precision is generally a function of the number of iterations rather than that of ancilla qubits. And lastly, they allow for the reduction of the ancillary register at the expense of extra rounds. This helps remedy the issues due to large circuit dimensions (both vertical and horizontal).

Several routes are possible within \gls{IPE}. A common approach is to obtain bits of the phase from last to first, using successive calls to the circuit in figure \ref{fig:phase_est_circ} for a pre-determined sequence of integers $m^{(k)}$ and adaptive angles $\theta^{(k)}$ \cite{Kitaev_1996,qiskit_textbook,Kitaev_2002,Svore_2014,Dob_ek_2007}. In this subsection we will prefer to use a superscript $^{(k)}$ to mark the iteration number, instead of a subscript which may cause confusion with binary notation and other subscripts (namely position indices). 

It is straightforward to see that the wavefunction of the upper qubit just before the Hadamard gate is:
\begin{equation}
    \frac{\ket{0} + e^{i(m^{(k)}\phi^{(k)}+\theta)}\ket{1}}{\sqrt{2}}
\end{equation}

\noindent, i.e. the eigenphase has been kicked back into the upper qubit $M$ times, as has the offset $\theta$, producing a relative phase. Clearly, the protocol exploits equations \ref{eq:eigen_equation} and \ref{eq:eigenvalue_phase}'s implication that $\ket{\psi}$ is also an eigenstate of $\op{U}^m$.
\begin{gather}
    \label{eq:eigen_equation_m}
    \op{U}^\mathbf{m}\ket{\psi} = \lambda^m\ket{\psi}\\
    \label{eq:eigenvalue_phase_m}
    \lambda^m = e^{im\phi}
\end{gather}

The final state will be, in the computational basis:
\begin{equation}
    \label{eq:phase_est_final_state}
    \frac{1}{2}
    \begin{bmatrix}
    1+e^{i(m^{(k)}\phi+\theta^{(k)})} \\
    1-e^{i(m^{(k)}\phi+\theta^{(k)})}
    \end{bmatrix}
\end{equation}

If the exponent's norm, 
\begin{equation}
    \label{eq:phase_est_xi}
    \xi \equiv m^{(k)}\phi+\theta^{(k)}
\end{equation}

\noindent, can be written in terms of a single digit binary fraction
\begin{equation}
    \label{eq:single_digit_frac}
    \xi = -2\pi(0.\xi_1)_2 = 2\pi\frac{\xi_1}{2^1} 
\end{equation}

\noindent, the probability of measuring 0 is
\begin{equation}
    \mathbf{P}(0 \mid \xi) = \begin{cases}
    1, \text{ if } \xi_1=0\\
    0, \text{ if } \xi_1=1\\
    \end{cases}
\end{equation}

In this case, the bit obtained by measuring precisely matches $\xi_1$, giving us $\phi$. These results are exact, apart from experimental error.

For longer $\phi$, still written as a binary fraction,
\begin{equation}
    \phi = -2\pi(0.\phi_1\dots\phi_M)_2 = 2\pi\left(\sum_{k=1}^\mathbf{m} \frac{\phi_k}{2^k} \right), \phi_k \in \{0,1\}
\end{equation}

\noindent, we can get any $\phi_k$ as the only digit in a fraction of the form \ref{eq:single_digit_frac}, and thus determine them all. For that, we leverage the degrees of freedom in equation \ref{eq:phase_est_xi}. We start by using $m^{(k)}$ to push the digits that are more significant than itself to the left of the radix point, rendering them an inconsequential integer factor for $2\pi$. After that, we erase the ones that are \textit{less} significant by performing a corrective rotation on the phase using $\theta^{(k)}$.

We don't know in advance what these corrections are; however, we do know that, for any digit, they are determined by the digits to the right of it. So we start with the rightmost digit $\phi_{M}$, which has none of those. In this case no correction is required. Then we move on to $\phi_{M-1}$, leveraging the freshly learned $\phi_{M-1}$ to tune our circuit, and we proceed similarly for all others as we accumulate more and more digits.

In short, we run the circuit from figure \ref{fig:phase_est_circ} $M$ times, getting all the $\phi_k$ in reverse order (from $\phi_M$ to $\phi_1$). We do so by iterating $k$ backwards from $M$ to $1$ and choosing at each iteration the controls that will carry out the corrections described above.
\begin{gather}
    m^{(k)} = 2^{k-1}\\
    \theta^{(k)} =  -2\pi (0.0\phi_{k+1}\phi_{k+2}\dots\phi_M)_2
    \label{eq:kitaev_theta_k}
\end{gather}

The iterator $k$, in addition to determining these controls, tags the digit being computed in the current iteration, $\phi_k$. Note that the phase correction $\theta^{(k)}$ depends on the digits inferred in all previous iterations $k+1,\dots,M$. That is, it considers as many digits as have been inferred so far (so if there are none, $\theta^{(k)}=0$; we've seen this happens for the edge case $k=M$).

It is this feedback mechanism what enables the direct extraction of the binary $\phi_k$ from the respective circuit's measurement outcome. At $k=M$ we infer $\phi_M$ with no prior knowledge; at $k=M-1$ we infer $\phi_{M-1}$ after using $\phi_M$ to rotate the phase; and so on. We can see that
\begin{equation}
    \label{eq:feedback_angle_m}
    \theta^{(k)} =  -m^{(k)} \phi_\text{curr}
\end{equation}

\noindent, where $\phi_\text{curr}$ contains the binary digits determined up to the previous iteration $k+1$, which are the $M-k$ least significant ones, in their proper positions.
\begin{equation}
    \phi_\text{curr} = (0.\underbrace{\vphantom{\phi_{M}} 0\dots0}_\text{k zeros}\phi_{k+1}\dots\phi_{M})_2
\end{equation}

That leaves $k$ placeholder zeroes corresponding to the digits to be filled in, all but one of which we eliminate when multiplying by $m^{(k)}=2^{k-1}$, retrieving \ref{eq:kitaev_theta_k}. This intuition will be alluded to later on.

Clearly, this strategy relies on accurate measurements of deterministic outcomes, as did the original protocol. Additionally, the maximum number of sequential applications of $\op{U}$ - and with it the necessary coherence time - scales rigidly and exponentially\footnote{Except if some additional structure can be exploited to get more efficient implementations of the required powers of $\op{U}$.} with the total number of iterations (as the original one did with the number of ancilla qubits). Thus, it is quite susceptible to experimental errors, especially when error corrected devices are not a reality (which they aren't). 
 
A more flexible approach is to regard circuit \ref{fig:phase_est_circ} not so much as a building block for a lineup that will output a solution, but rather as an entity on its own. It consists on no more than a system - one whose outcome distribution has a known dependence on the eigenphase $\phi$, in addition to some customizable parameters. This is the description of an inference problem.

Not unexpectedly, this can mean Bayesian inference. The resolution is up to the classical processing; we can apply whatever techniques we wish, namely machine learning ones. The Bayesian approach in particular has been proposed in \cite{Wiebe_2016} and subsequently implemented experimentally in \cite{Paesani_2017}. Finally, \cite{Lumino_2018} compares a few strategies, Bayesian and not, and in particular their error resilience.

This case differs from the precession example in that the target is a small circuit rather than a single qubit; but as before, its role is simply to relay the outcomes to a classical processing unit. Because the application of $\op{U}$ on $\ket{\psi}$ is of no material consequence, we must necessarily fall back on some auxiliary structure. We thus assemble a circuit that will impart some relevant information; this can be the again that of figure \ref{fig:phase_est_circ}.

We know that the structure of that circuit is sufficient, because it \textit{was} sufficient when neatly arranged into a a specific layout. We will simply control and process it a less orderly manner. This is even more reminiscent of other \glsxtrshort{NISQ} algorithms where optimization, feedback and repetition are joined together in an attempt to counter experimental limitations. The pairing of shallow circuits with more mature (classical) machine learning techniques is by nature quite robust. In particular, it is much more error resilient than routines ran entirely on non fault-tolerant quantum devices.

With this choice, we need only apply the Born rule to the final state \ref{eq:phase_est_final_state} to get our likelihood function corresponding to either of the basis states:
\begin{gather}
    \mathbf{P}(\phi \mid 0; m^{(k)},\theta^{(k)}) = 
    \frac{1+\cos \left(m^{(k)}\phi + \theta^{(k)} \right)}{2}\\
    \mathbf{P}(\phi \mid 1; m^{(k)},\theta^{(k)}) = 
    \frac{1-\cos \left(m^{(k)}\phi + \theta^{(k)} \right)}{2}
\end{gather}

\textit{Actually},
\begin{gather}
    \label{eq:phase_est_0}
    \mathbf{P}(\phi \mid 0; m^{(k)},\theta^{(k)}) = 
    \cos^2(\frac{m^{(k)}\phi + \theta^{(k)}}{2})\\
    \label{eq:phase_est_1}
    \mathbf{P}(\phi \mid 1; m^{(k)},\theta^{(k)}) = 
    \sin^2(\frac{m^{(k)}\phi + \theta^{(k)}}{2})
\end{gather}

 \noindent (using half angle formulas).
 
The likelihoods \ref{eq:phase_est_0} and \ref{eq:phase_est_0} and those from the precessing qubit example (\ref{eq:precession_+} and \ref{eq:precession_-}) look very quite alike. As a matter of fact, \textit{they are identical} if we identify the spanned angle $\omega t$ with $m^{(k)}\phi + \theta^{(k)}$ and convert between x and z basis.

This is, of course, because the two processes are equivalent. The effect of circuit \ref{fig:phase_est_circ} on the upper qubit is to pulse it into the x-basis, rotate it at the equatorial plane around the z axis, and then do an x-basis measurement (Hadamard transform + computational basis projection). The phase gate and the controlled $\op{U}$ both change the relative phase between basis states, affecting the Bloch coordinate $\varphi$ additively such that the accumulated phase is $m\phi + \theta$.

Predictably, the problem summary too nearly matches \ref{eq:precession_summary}.
\begin{equation}
    \label{eq:phase_est_summary}
    \dboxed{ \vphantom{gh} 
    \mathbf{P}(\theta \mid D;E) \leftrightarrow 
    \mathbf{P}(\phi \mid 0; m,\theta) = 
    \cos^2(\frac{m\phi + \theta}{2})
    }
\end{equation}

Here the parameter is the eigenphase, the data are the classical bits obtained upon measuring (which tell whether the qubit was found at state $\ket{0}$ or $\ket{1}$), and the experimental controls are two: the number of times $\op{U}$ is applied, plus the offset angle. The likelihood in \ref{eq:phase_est_summary} contains all information, because the probability of $\ket{0}$ is complementary to that of $D=\ket{1}$.

The nuance is that here our \textit{time} has discrete units $m$ (we would recover continuity if we took $\op{U}^\mathbf{m}=e^{-i\op{H}m}$)\footnote{A similar example is characterization of coherent gate error. For instance, applying an imperfect $\opsub{\sigma}{z}$ gate twice will rotate the Bloch vector by $(2\pi+2\epsilon)$ radians around the $z$ axis. This error $\epsilon$ will accumulate to cause an angle $2m\epsilon$ if the gate is applied an even number of times $2m$, which produces a similar oscillation of the probability of $\ket{+}$ as a function of $m$.}; also, there is now a controllable offset $\theta$, a freedom we could have added to the previous example by choosing a different initial state still within the x-y plane. 

Whatever they be, the existence of these variable controls is opportune: they allow us to tweak the outcome distribution. Generally, degrees of freedom in the likelihood function are welcome, albeit not strictly necessary. With them, we can make more out of each measurement by choosing appropriate controls. We are not simply passive agents who register and process data: we have the power to design the experiments that produce those data. But how do we use such a power?

\section{Bayesian experimental design}
\label{sec:bayesian_experimental_design}

When performing inference, not all measurements are made equal. The structure of the likelihood function itself impacts the achievable utility of each experiment: if it varies more slowly with the parameter(s), it will produce less precise parameter estimates. As an example, it is not hard to imagine that an exponential model exhibits reduced sensitivity as compared to a linear one.

Having chosen a model, the shape of the likelihood still changes depending on the controls. As a simple example, let's go back to our coin tosses. We had a coin, say coin A, with some probability of tails $p=\theta$, where $\theta$ was the target parameter. We can now imagine that we had access to a \textit{second} coin, coin B, in addition to the original one. We are told coin B's probability of tails $p=0.1\theta$ is a tenth of coin A's. That is,
\begin{gather}
    p_A = \mathbf{L}(\theta \mid 1; A) = \theta\\
    p_B = \mathbf{L}(\theta \mid 1; B) = 0.1\theta
\end{gather}

\begin{figure}
\centering
\captionsetup[subfigure]{width=.9\linewidth}%
\begin{subfigure}[t]{.45\linewidth}
  \centering
  \includegraphics[width=.95\linewidth]{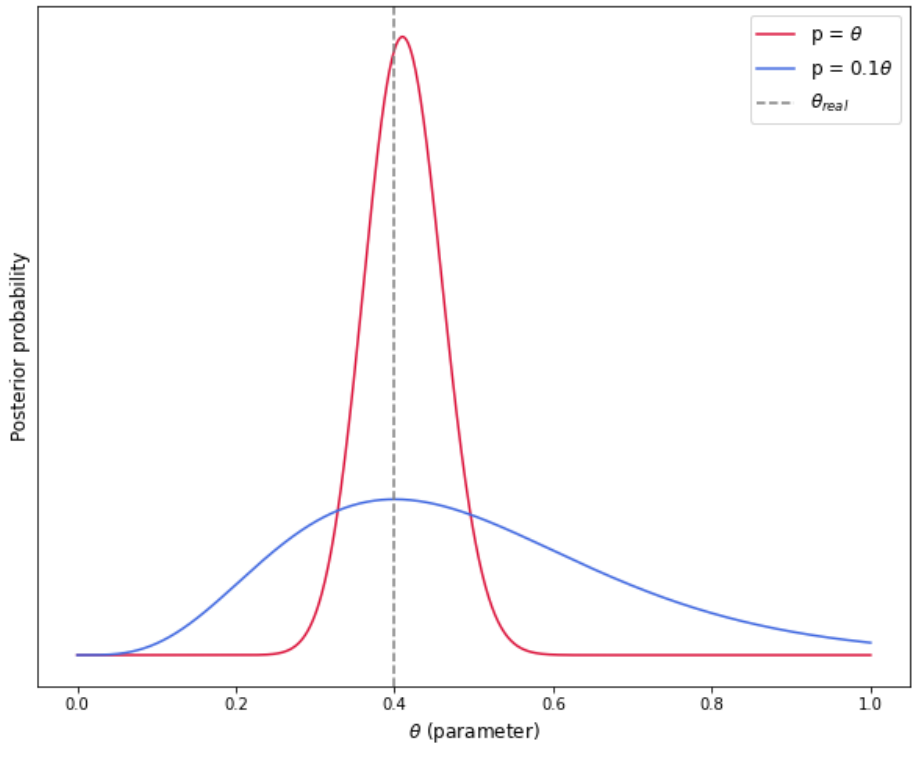}
  \caption{Posterior distributions for a simple binomial model, where 2 different multipliers of the parameter yield the samplable probability of success. The dashed line marks the real parameter used to generate the data. 100 trials were included in each data vector.}
  \label{fig:coin_prop}
\end{subfigure}
\begin{subfigure}[t]{.45\linewidth}
  \centering
  \includegraphics[width=.95\linewidth]{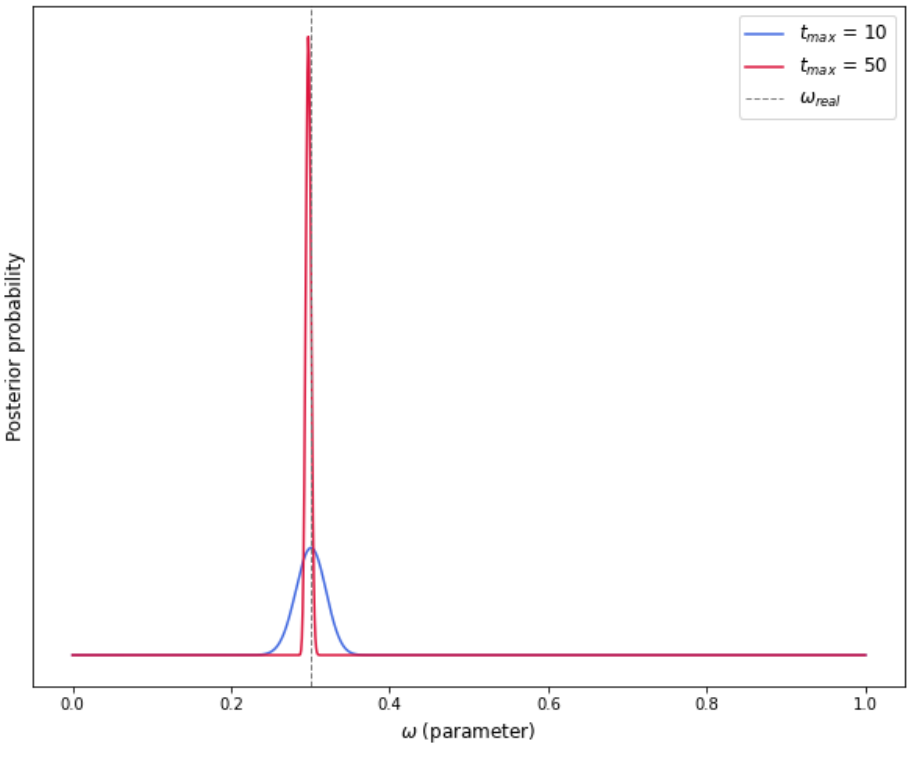}
  \caption{Posterior distributions for a time-dependent binomial model (equation \ref{eq:precession_summary}, for 2 sets of measurement times. The dashed line marks the real parameter. Each dataset included 100 measurements, with times evenly spaced up to $t_{max}$.}
  \label{fig:precession_tmax}
\end{subfigure}
\caption{Impact of the magnitude of the parameter's coefficient(s) in the shape of the likelihood for a simple binomial model (\ref{fig:coin_prop}) and a damped oscillator model (\ref{fig:precession_tmax}).}
\label{fig:experiment_coef}
\end{figure}

This introduces a new control: the coin pick. We can choose which coin to toss a few times before taking a guess at $\theta_\text{real}$. Which \textit{do we} choose? We may expect that coin B will do worse: it is harder to resolve $0.1\theta$ than it is $\theta$, as the latter manifests a stronger dependence on $\theta$. That is indeed the case. Figure \ref{fig:coin_prop} shows the posterior distributions resulting from 100 coin flips for each coin; the one with $p=0.1\theta$ has a flatter curve, with larger variance (signifying more uncertainty).

How about the oscillatory likelihoods of \ref{eq:precession_summary} and \ref{eq:phase_est_summary} - how should we pick the times? They show inside the sinusoidal function as a multiplying factor for the target parameter. For a fixed parameter value, longer times mean faster oscillations. We can anticipate that a Bayes' update will be more informative, and the posterior sharper, if the likelihood varies rapidly with the parameter.

Yet again, we have a point. The posterior distributions for two different sets of measurement times (spaced evenly on $[0,t_\text{max}]$ for two values of $t_\text{max}$) are depicted in figure \ref{fig:precession_tmax}. The longer times yield a visibly sharper curve\footnote{Although equation \ref{eq:precession_summary} is the one directly considered here, these results can be straightforwardly generalized to the case of \ref{eq:phase_est_summary}, where the \textit{times} are integers due to experimental constraints.}.

One more thing to notice is that both curves of \ref{fig:coin_prop} are narrower as compared to those of \ref{fig:precession_tmax}, testifying to the inherent differences between models mentioned earlier: the sinusoidal wave varies more rapidly than the proportionality (though naturally this depends on their frequency and slope). But once again, we will be holding the model fixed for each test case; it is ultimately determined by the target system, and we have already discussed how to select it. The experiments, on the contrary, can be chosen completely at our whim.

This is very promising: what we are observing is similar to what happened back in section \ref{sec:bayesian_inference} when we added more data (from figure \ref{fig:coin_bayes_vs_freq} to  \ref{fig:coin_bayes_vs_freq_sharper}). But here we are \textit{not} adding data, nor are we making any more Bayes' updates. We are simply changing the experiments we perform, and that alone has the power to change our results for the better - despite there being no extra processing whatsoever.

To make use of this fact, we need a way to identify the best experiments \textit{before} we perform them. How can we do that, if we don't know the posterior they would give rise to?

While it is true that we don't know the posterior in advance, we are from the outset in possession of the next best thing: our prediction of where it will stand, based on our current knowledge (the prior). We also don't know or how sharp a set of experiments will actually make it; but by the same token, we can compute how sharp a set of experiments makes it \textit{on average}. This is sufficient to make an informed choice, and stack the odds in our favour.

Interestingly, this discussion seems somewhat familiar. Back in subsection \ref{sub:estimators_and_expressions_of_uncertainty}, when talking about estimators, we spoke of their associated Bayes risk, which was the expectation over the data of the posterior averaged loss. We suggested we'd pick the estimator that we expected to minimize this risk.
\begin{equation}
    \tag{\ref{eq:bayes_riskIII}} 
    r(\hat{\theta}) = 
    \mathbb{E}_{\mathbf{P}(\vec{D}),\mathbf{P}(\theta \mid \vec{D})}
    \left[ 
    \mathcal{L}(\theta,\hat{\theta}(\vec{D})) 
    \right]
\end{equation}

With the choice of an estimator, the Bayes risk becomes a fixed quantity.
\begin{equation}
    \label{eq:bayes_risk_indep}
    r = 
    \mathbb{E}_{\mathbf{P}(\vec{D}),\mathbf{P}(\theta \mid \vec{D})}
    \left[ \mathcal{L}(\theta,\hat{\theta}_X(\vec{D}))  
    \right]
\end{equation}

Or does it? In the meantime, we have decorated Bayes' equation with some extra dependencies. As we are considering a fixed model, we'll refer back to equation \ref{eq:bayes_controls}. Adjusting \ref{eq:bayes_risk_indep} accordingly, we get:

\begin{equation}
    \label{eq:bayes_risk_exp}
    r(\vec{E}) = 
    \mathbb{E}_{\mathbf{P}(\vec{D};\vec{E}),\mathbf{P}(\theta \mid \vec{D}; \vec{E})}
    \left[ \mathcal{L}(\theta,\hat{\theta}_X(\vec{D};\vec{E}))  
    \right]
\end{equation}

The risk now depends \textit{on the experiments}. This is a formalization of what we saw in figure \ref{fig:experiment_coef}. In subsection \ref{sub:estimators_and_expressions_of_uncertainty}, we talked about several loss functions, and how they were typically a function of the deviation from the real value. When taking expectations over the posterior, they become measures of the spread of the distribution. The loss function we picked, the variance, clearly demonstrates this. If a set of experiment controls has a smaller risk relative to some reference, it is likely to make the posterior sharper.

When talking about experiments, we don't generally talk about \textit{risk}, but rather about \textit{utility}. Unlike an estimator, an experiment doesn't incur error; it is simply more or less informative. We can choose our utility function to be symmetric to the risk:
\begin{equation}
    U(\vec{E}) = -r(\vec{E})
\end{equation}

We will do just that and take the negative variance as our utility\footnote{The variance is a property of a specific posterior, so to get the final utility we must still average the variance over all possible ones (i.e. take the expectation over the data). The term utility is often used to refer both to the data-conditional utility and the expected utility. The subsequent double occurrence of the function $U$ in \ref{eq:expected_utility} with different arguments further attests to this fact.}, which was the result of our choice of a loss function in subsection \ref{sub:estimators_and_expressions_of_uncertainty}. Of course, alternative functions can be considered; another popular choice for estimation problems in physics is the information, measured by the entropy \cite{Ferrie_2011,Granade_2012}. Penalties can also be factored in, using e.g. the total evolution time if the cost of simulation is the main concern.

All tools are now at our disposal. For any group of experiments $\vec{E}$, we can obtain the posterior expected utility of a list outcomes by mock-updating the distribution, and then marginalizing the loss over the resulting posterior distribution $P(\theta \mid \vec{D})$. At the same time, we can obtain the current expected probabilities of the outcome list by marginalizing the likelihood over the prior. 

This is done for all datasets, after which we multiply each dataset's probability by the corresponding utility (both calculations depend on the experiments $\vec{E}$). Finally, we sum over all possible data vectors, which are $2^N$ for $N$ experiments with binary outcomes. With this, we get the expected utility of the vector of experiments:
\begin{equation}
    \label{eq:expected_utility}
    U(\vec{E}) = \sum_{\vec{D}}  \mathbf{P}(\vec{D};\vec{E}) 
    \cdot
    \mathbb{E}_{\mathbf{P}(\theta \mid \vec{D};\vec{E})}
    \left[ 
    U(\theta,\vec{D};\vec{E})  
    \right] 
    = \sum_{\vec{D}}  \mathbf{P}(\vec{D};\vec{E})  \cdot
    U(\vec{D};\vec{E})  
\end{equation}

\begin{figure}[!ht]
    \centering
    \includegraphics[width=\textwidth]{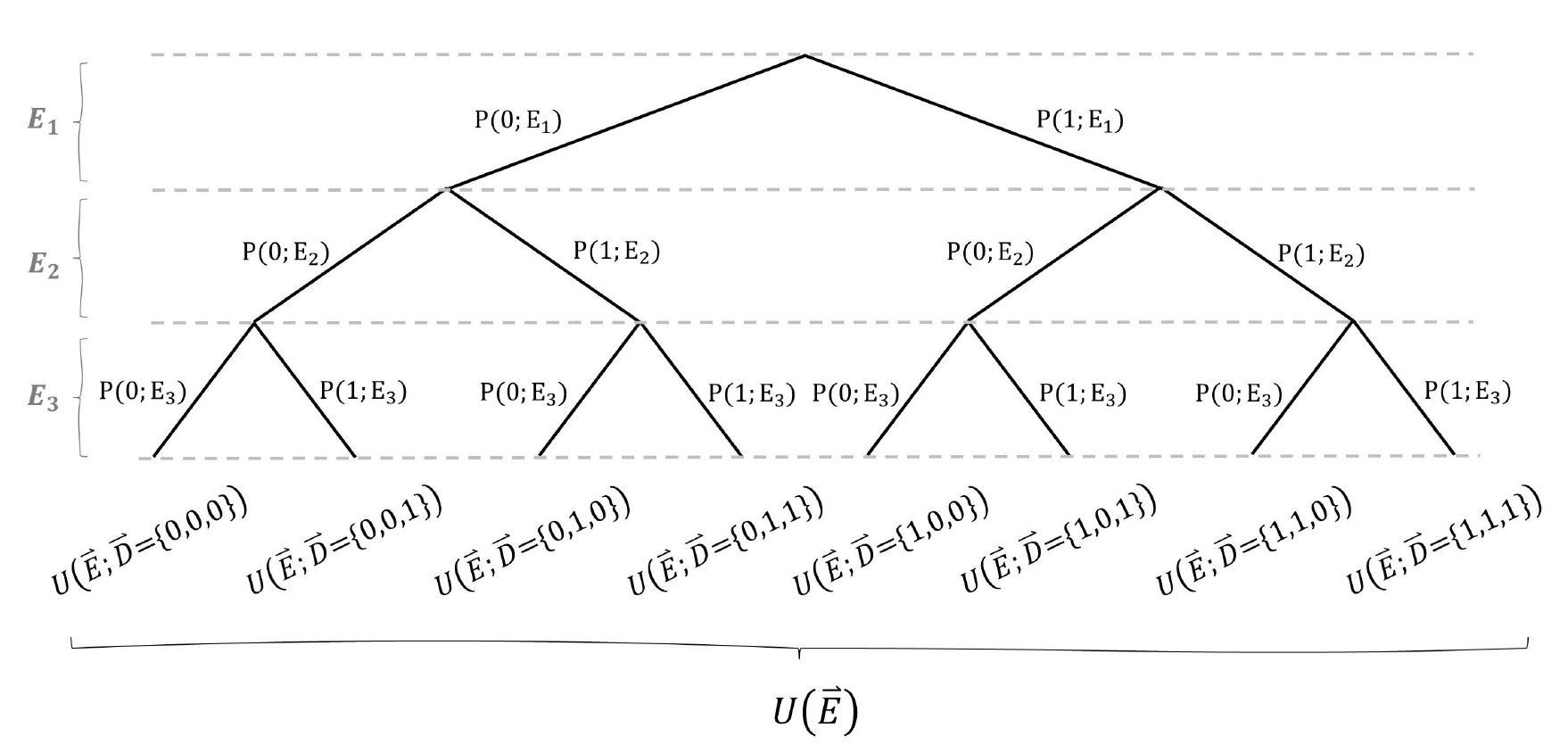}
    \caption{Binary tree illustrating the computation of the expected utility for a set of experiments $\vec{E}=\{E_1,E_2,E_3\}$. To each layer of branches corresponds one of the individual experiments, in chronological order from top to bottom. Each branch represents the occurrence of a specific datum, and is tagged with its likelihood. The leaves are marked with the utility of the data sequence traced out by the path of branches from the root to themselves. Using these calculations, the expected utility of $\vec{E}$ can be derived from \ref{eq:expected_utility}.}
    \label{fig:utility_tree}
\end{figure}

Figure \ref{fig:utility_tree} shows the operations required to compute the utility of a set of experiment controls. To make an optimal choice, this must of course be evaluated for multiple possibilities of $\vec{E}$.

Clearly, utility maximization is not all roses. In addition to computing the marginal probabilities of all $2^N$ data-vectors, we must compute their average posterior loss (e.g. the final variance). Each of these computations requires as much processing as the entire inference process did, because they are precisely a data-conditional \textit{look-ahead} into the posterior. And this is just for a single evaluation of the utility at a specified $\vec{E}$ - if we are to take on optimization, these the calculations need to be done all over again for many other possibilities, as they're evidently not re-usable for different choices of $\vec{E}$.

This exponential scaling with the total number of experiments is prohibitive, and soon becomes unfeasible. We're in the presence of a trade-off between the number of experiments and the processing cost. If measurements are an expensive resource, it may be interesting to take full advantage of each by being selective in the choice of controls. If not, however, we may be better off rejecting the processing overhead and running a larger number of arbitrary experiments. Between these two options, could there be hope for a more balanced compromise?

The issue is brought on by the construction of the annotated binary tree in figure \ref{fig:utility_tree}, which has height $N$. To a degree, it unavoidable that we undertake it; but a more lightweight strategy \textit{is} possible. We can exploit the parallel between the central Bayes' updates and the accessory ones by building the tree as we go, effectively intertwining its construction with the mandatory processing.

Such a coupling can be achieved by processing the data real-time, and forgoing the quest for a global maximum in favour of local optimality in what is called a \textit{greedy} algorithm. Equipped with adaptivity, we can move along the linear path that the data conduct us through. 

\subsection{Adaptive measurements for a greedy strategy}
\label{sub:adaptivity}

\begin{figure}[!ht]
    \centering
    \includegraphics[width=10cm]{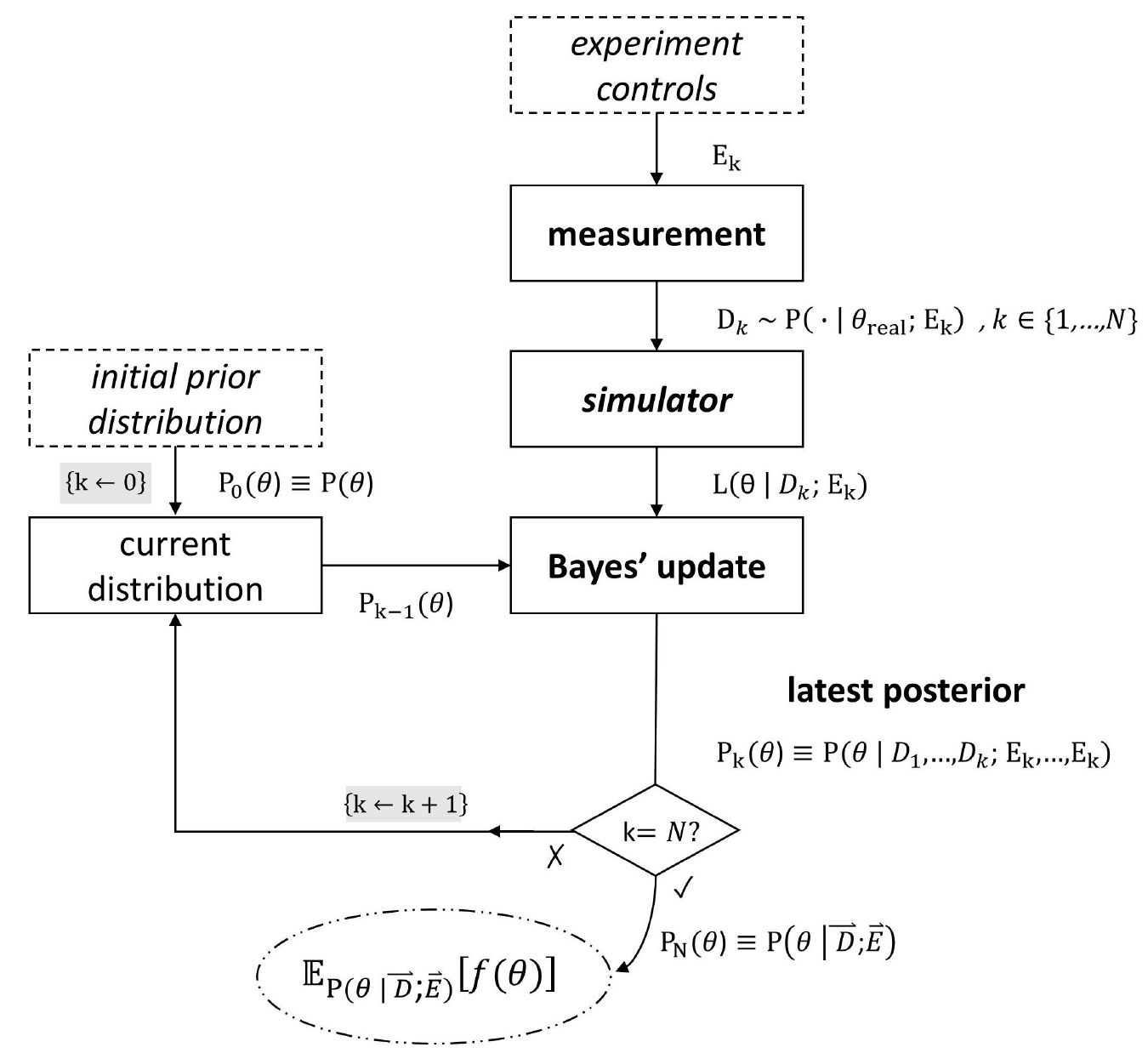}
    \caption{Diagrammatic representation of a quantum parameter estimation algorithm using Bayesian inference, based on iterative updates. Italics signal dependence on implementation choices, and dashed boxes denote one time only inputs.}
    \label{fig:inference_iterative}
\end{figure}

In sections \ref{sub:estimation_in_practice} and \ref{sec:application_to_quantum}, we claimed that whether we processed the data one by one or all at once was indifferent. And it is - as long as everything is equal. This \textit{everything} includes, of course, the experiments. Figure \ref{fig:inference_iterative} unfolds the process of \ref{fig:offline_inference_diagram}, representing the iterative version of the inference algorithm. These diagrams are again based on \cite{Granade_2012,Ferrie_2011}, as is the analysis itself. Under these conditions, the two diagrams are equivalent.

In spite of this equivalence, when going from the single datum updates of \ref{eq:bayes_controls} to the batch updating of \ref{eq:bayes_datavector_controls}, we actually did lose something: the passage through intermediate stages of knowledge. While it's fine to skip over them to our final conclusion if that's all we care about, we are now in a different scenario. After the collection of the first datum $D_1$, we already know a little bit more than we did to begin with; we have a first-iteration posterior, $\mathbf{P}(\theta \mid D_1;E_1)$. Then why not use this knowledge, instead of following a pre-determined route?

\begin{figure}[!ht]
    \centering
    \includegraphics[width=10cm]{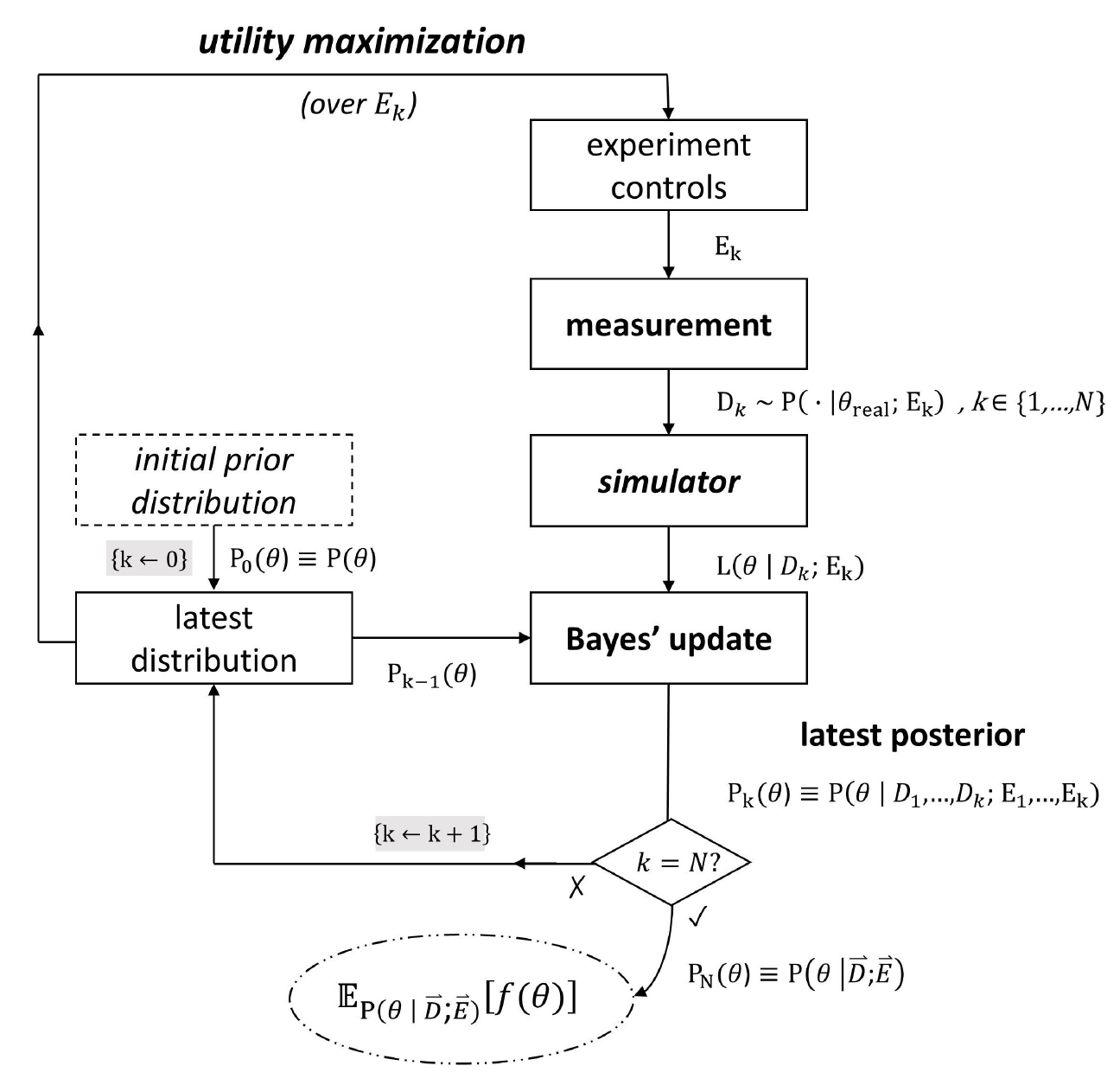}
    \caption{Diagrammatic representation of a quantum parameter estimation algorithm using Bayesian inference with local utility optimization. Italics signal dependence on implementation choices, and dashed boxes denote one time only inputs.}
    \label{fig:inference_local_opt}
\end{figure}

This can be done by introducing a feedback mechanism, as shown in figure \ref{fig:inference_local_opt}. We can delay deciding which second experiment $E_2$ to perform until the moment we collect $D_1$ and update the distribution according to it. \textit{That} is the full extent of our \textit{a priori} knowledge at the time we must perform the experiment $E_2$ - and we can put ourselves in a position to use it.

This is entirely applicable to all subsequent experiments too; for each, we use all the data gathered before it. The datum obtained with each experiment precipitates an update from the latest \textit{prior} to a transitory \textit{posterior}, which becomes the prior for the ensuing iteration. We had already mentioned that the posterior can be seen as an evolved version of the prior in equation \ref{eq:prior_to_posterior}.

The terms prior and posterior denote precisely the preexisting knowledge and its after-the-fact state, where the order is relative to the data-led updates. We are free to decide how phased the assimilation of the data should be; it can be broken up into chunks or regarded as a whole. Should we want to use it before finalizing the process, we can \textit{compile} it more often.

The only new imposition is that we switch to an online setting, and process the data as we go: in between measurements, we must update the distribution using the previous datum $D_{k-1}$ and employ local optimization to choose the next control $E_k$. The update is something we'd have to do eventually anyway, and the cumulative cost of the optimization will be smaller than it would if we were to seek a global optimum. 

\begin{figure}[!ht]
    \centering
    \includegraphics[width=12cm]{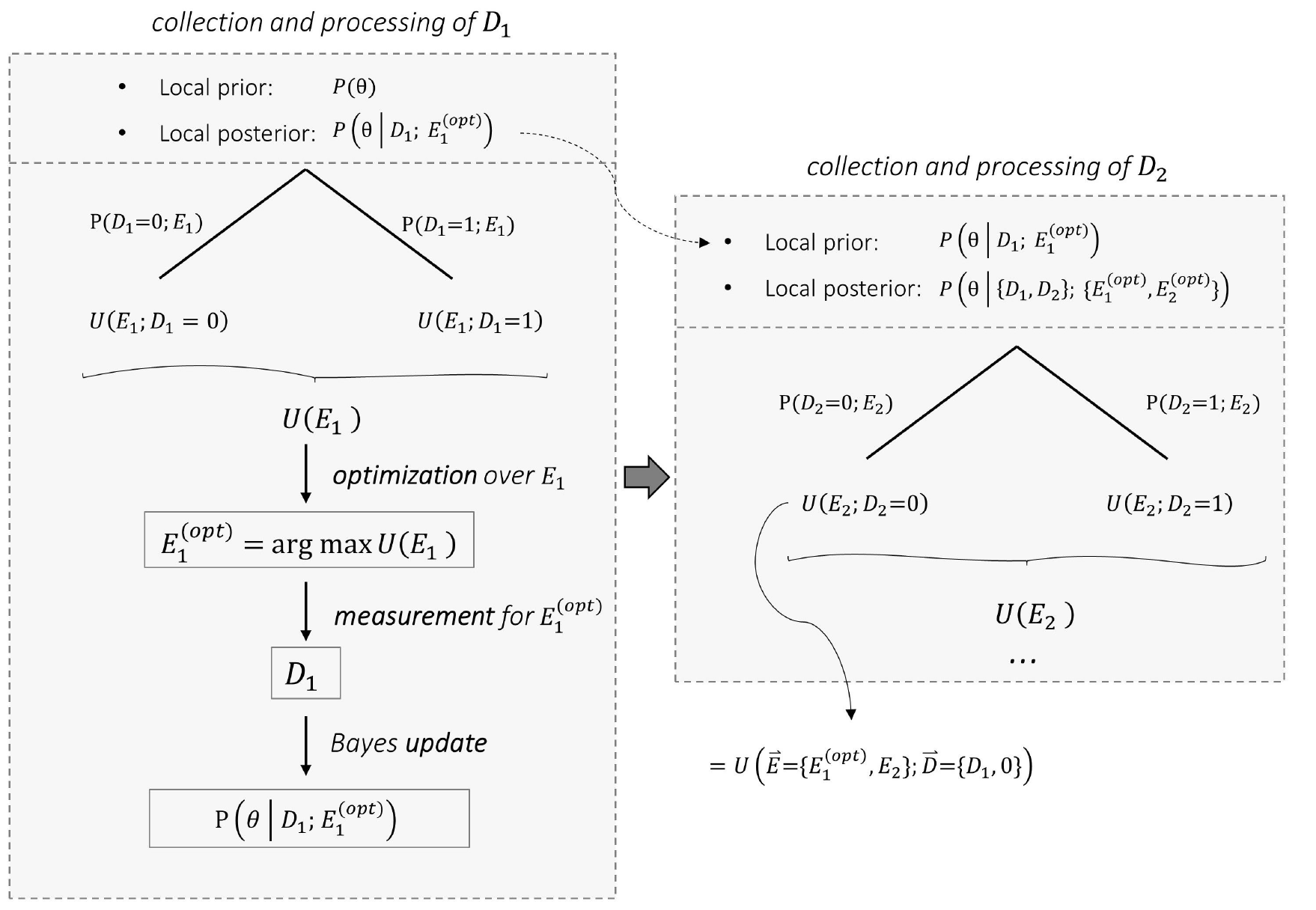}
    \caption{Binary trees illustrating the computation of the locally expected utility for two followed experiments, $E_1$ and  $E_2$. To each of the height-1 trees corresponds one experiment, in chronological order from left to right. Each branch represents the occurrence of a specific datum, and is tagged with its likelihood. The leaves are marked with the utility of the respective branch's datum. Using these calculations, the expected utilities of $E_1$ and $E_2$ can be derived from \ref{eq:expected_utility}. The equality annotated bottom right evidences that the latter calculation is dependent on $E_1$ and the effective $D_1$.}
    \label{fig:mini_trees}
\end{figure}

Looking at the binary tree of \ref{fig:utility_tree}, it is as if we travel down the branches as the inference advances, only ever using a single level look-ahead. Instead of concerning ourselves with the likes of the formidable tree in \ref{fig:utility_tree}, we consider a sequence of mini-trees of height 1, only ever thinking in terms of the experiment coming next and its two possible outcomes. Figure \ref{fig:mini_trees} shows two of these smaller trees. Their construction is necessarily sequential.

\begin{figure}[!ht]
    \centering
    \includegraphics[width=12cm]{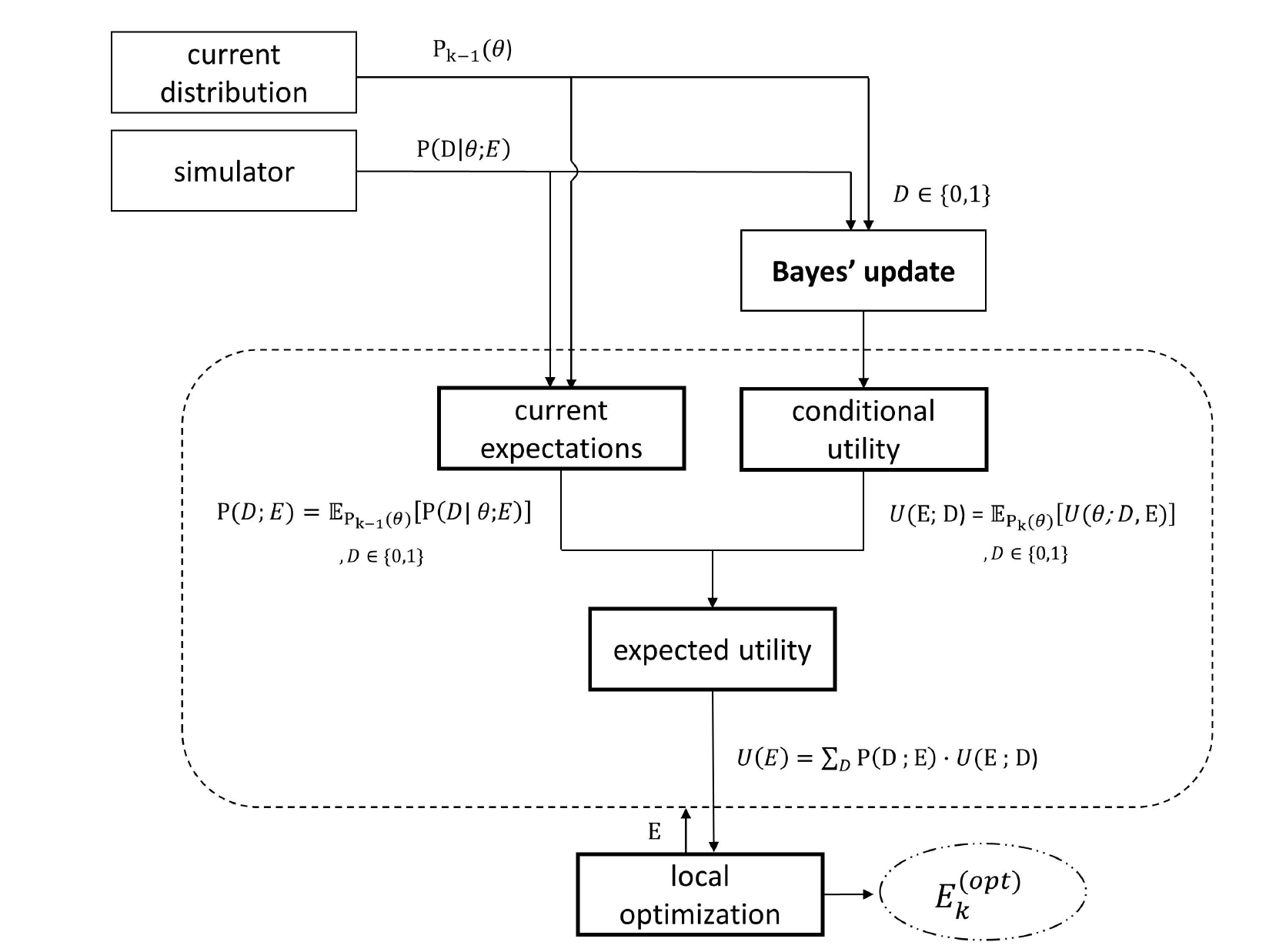}
    \caption{Diagrammatic representation of a local optimization routine in the context of Bayesian experimental design.}
    \label{fig:local_optimization}
\end{figure}

The feedback loop in \ref{fig:inference_local_opt} is oversimplified, and presupposes further processing, and access to a simulator (just as the main algorithm does). Figure \ref{fig:local_optimization} shows the required set up explicitly. The dashed section encloses the computation of the utility as per equation \ref{eq:expected_utility}.

In general, this greedy strategy will not prompt the same choices that the globally optimal one would have. However, it is a quite less demanding alternative, and has been shown (numerically) to produce good results in the case of a flat prior \cite{Ferrie_2011}. It may also be particularly interesting in the case of such an uninformative prior, since it particularly benefits from the inference process as it moves forward.

This type of measurement interdependence has been used to perform estimation with a precision scaling at the best limit allowed by quantum mechanics. As a matter of fact, adaptivity underpinned the first experimental realization of Heisenberg-limited phase estimation - filling in for the laborious state preparation at times thought to be irreplaceable \cite{Higgins_2007}. It may then be interesting to look at \textit{why} it allows reaching such a high precision regime. 

\subsection{Beating shot noise: quantum-enhanced, Heisenberg-limited estimation}
\label{sub:beating_shot_noise}

In metrology, one aims to achieve the best possible measurement tactics; in the case of a quantum system, it is to expect that they're be inescapably tied to a quantum description. For any estimation strategy, we don't arrive at the limits of describability until we overcome the shortcomings of our description. For this reason, a quantum problem begs a quantum solution, which promises the best known achievable accuracy. 

Here we will present a quick overview\footnote{The discussion is quite simplified; rigorous analysis requires more careful definitions, in particular of the resource count $N$ \cite{Zwierz_2010,Higgins_2007}.} of the limits quantifying the attainable bounds of this advantage, as they are quite linked to adaptive inference strategies.

For that end, we first consider an ensemble of $N$ independent probabilistic measurements on a qubit; as before considering experimental design, we would perform them at once, without much concern for their controls. They could be separated in space (different qubits) or in time (different passes through the same circuit). Based on these experiments, we get an estimate for some target quantity; for instance, the probability of finding the qubit in state $\ket{1}$. Writing each measurement outcome as $D_j$, that estimate is:
\begin{equation}
    \label{eq:sample_mean}
    \bar{D} \equiv \mathbb{E}\left[ D_j \right] = 
    \frac{\sum_{j=1}^{N} D_j}{N}
\end{equation}

This could be applied for instance to the frequency/phase estimation problems summarized in \ref{eq:precession_summary} and \ref{eq:precession_summary}; for some fixed experiment, we could determine the likelihood and then solve with respect to the parameter to get a ratio estimate.

Taking a finite number of random samples imposes a limit on the precision of this estimate; limited resolution is tantamount to observing the target quantity through noise. For our independent stochastic variables in particular, we consider gaussian noise: we can invoke the \gls{CLT} to say that the distribution of outcomes $D_j$ is approximately normal with some mean $\mu$ and variance $\sigma^2$.
\begin{equation}
    \label{eq:gaussian_noise}
    D_j \sim \mathcal{N}(\mu,\sigma^2)
\end{equation}

Equation \ref{eq:sample_mean} is the specific mean we get from a sample of $N$ results, and we are equating its obtention with sampling from a normal distribution \ref{eq:gaussian_noise}. This is despite the fact that the original variables are not normally distributed (this is justified by the \gls{CLT} under our suppositions).

In some cases, this sample mean might align perfectly with the exact value; but more often than not, it won't. To quantify the expected magnitude of this variability, we consider meta-statistics regarding this distribution of sample means.
\begin{equation}
    \label{eq:sample_mean_normal}
    \bar{D} \sim \mathcal{N}(\mathrm{M},\Sigma^2)
\end{equation}

This distribution has itself a mean, which by linearity of expectation and \ref{eq:gaussian_noise} is simply $\mu$.
\begin{equation}
    \mathrm{M} = \mathbb{E}\left[ \bar{D} \right] =
    \mathbb{E}\left[ \frac{\sum_{j=1}^{N} D_j}{N} \right] = 
    \frac{1}{N} \sum_{j=1}^{N} \mathbb{E}\left[  D_j \right] = 
    \frac{1}{N} \sum_{j=1}^{N} \mu = 
    \mu
\end{equation}

Centrally, we quantify its spread around the mean M by means of its variance $\Delta^2$:
\begin{equation}
    \label{eq:sample_mean_varianceI}
    \Delta^2 = \mathbb{V} \left[ \bar{D} \right] = \mathbb{V} \left[ \mathbb{E}[D_j] \right]
\end{equation}

The reason why we concern ourselves with the first two moments of the distribution only is that the mean is trivial and the variance suffices for our purposes; hence the representation as a Gaussian in \ref{eq:sample_mean_normal}.

By the properties of the variance, \ref{eq:sample_mean_varianceI} becomes:
\begin{equation}
    \mathbb{V} \left[ \bar{D} \right] =
    \mathbb{V} \left[ \frac{\sum_{j=1}^{N} D_j}{N} \right] =
    \frac{1}{N^2}\mathbb{V} \left[ \sum_{j=1}^{N} D_j \right]
\end{equation}

We can appeal to the fact that the measurements are independent, and so the variance of their sum is the sum of their variances.
\begin{equation}
    \frac{1}{N^2}\mathbb{V} \left[ \sum_{j=1}^{N} D_j \right] =
    \frac{1}{N^2} \sum_{j=1}^{N} \mathbb{V}  \left[ D_j \right]
\end{equation}

At which point we invoke \ref{eq:gaussian_noise} to get:
\begin{equation}
    \Delta^2 = \frac{1}{N^2} \sum_{j=1}^{N} \sigma^2 = \frac{\sigma^2}{N}
\end{equation}

This error arising from statistical fluctuations represents the so-called \gls{SQL}, or shot noise limit (a term borrowed from electronics, where it arises from discrete charges rather than discrete samples). The standard deviation of the sample means $\Delta$ quantifies our uncertainty, and the \gls{SQL} describes how it scales with the relevant resource (the number of shots, $N$).
\begin{equation}
    \label{eq:sql}
    \Delta^{(SQL)} \propto \frac{1}{\sqrt{N}}
\end{equation}

 The outcome variance $\sigma^2$ is of course independent of $N$: the total number of experiments could never influence their individual results.

Finally, we can dissect our assumptions. This is valid for \textit{independent} measurements; the calculations do not hold in other cases. In particular, better scaling can be achieved by exploiting state preparation and measurements in more thought-out schemes. If arbitrary quantum effects are used, the fundamental limit is Heisenberg's (\glsxtrshort{HL}):
\begin{equation}
    \label{eq:hl}
    \Delta^{(HL)} \propto \frac{1}{N}
\end{equation}

The scaling of \ref{eq:hl} represents the regime of \textit{quantum-enhanced} estimation, which offers the fundamentally optimal resolution \cite{Zwierz_2010}. Many of the proposals to achieve it consider introducing correlation by means of entanglement. The quantum Fisher information quantifies the extractable information for a state-observable pair, and can be used to directly determine the maximum attainable measurement accuracy (through the quantum Cramér-Rao bound). For separable states it is linearly bounded by the number of particles, whereas for general states this bound is quadratic \cite{T_th_2014}.

This has motivated the use of entangled states for parameter estimation \cite{Zwierz_2010}. However, the complexity required in the preparation of such states has often shown to be prohibitively challenging for the available technology \cite{Giovannetti_2004}, and the presence of noise sorely undermines their longevity \cite{T_th_2014}; such problems are further exacerbated by high qubit counts. These points have painted sub-shot noise estimation as something to be left for the future, especially in the case of large numbers of qubits.

A significantly more manageable alternative was proposed in \cite{Higgins_2007}, and relied precisely on a locally optimal Bayesian algorithm. The procedure sought to minimize the expected posterior variance at each step - effectively replacing intricate input states by sequential optimized reruns of a parametrized circuit. This greatly relieved the requirements, enabling the experimental demonstration of Heisenberg scaling.

\subsection{Empirical strategies for a precessing qubit}
\label{sub:precession_heuristics}

Even if so far we have spoken only of general estimation strategies, looking into the concrete structure of the likelihoods at hand may often prove advantageous. The adaptive measurements we surveyed produce good results, but are still quite demanding in view of the optimization to be carried out between measurements, especially when considering simulation based likelihoods. The processing overhead can be alleviated by problem-specific heuristics, which may afford favorable solutions at dramatically lower cost.

In this subsection, we will take a closer look at the particular case of our examples of section \ref{sub:quantum_characterization_examples}. Based on \cite{Ferrie_2012,Wiebe_2016,Granade_2012,Wiebe_2014a}, we will go over some adaptive heuristics that can be a proxy for full-on optimization or more invested approaches, posing a less demanding alternative. While at it, we will take the opportunity to get a more intuitive understanding of the mechanisms driving measurement optimality, and bring up some issues that become relevant if considering approximate inference. 

The strategies to be considered here are \textit{handcrafted}, in the sense that they are grounded in observed behaviour, be it theoretical or practical. More general (and powerful) approaches can be achieved by automating the reasoning; in some cases, pre-existing heuristics can even be used as a starting point, should any be available. As an example, \cite{Lumino_2018} relies on particle swarm optimization to create an \textit{offline adaptive}\footnote{This is in the sense that each measurement relies on the previous result, but virtually no online processing is required. The optimization is performed beforehand, after which the feedback is determined by a quick calculation.} phase estimation protocol. More recently, reinforcement learning has been used to create neural-network based heuristics; details can be found in \cite{Fiderer_2021}, which also presents an interesting performance comparison encompassing the approaches to be addressed here. However, these machine learning based methods are left out of this overview.

Having noted this, we will now introduce a few relevant schemes, starting with analytical considerations. For the squared-sinusoidal likelihoods in \ref{eq:precession_summary}, \cite{Ferrie_2012} analyses the behaviour of the risk under a normality assumption for the prior. Such an assumption enables the analytical treatment of the problem, which may assist in designing practical strategies for improving inference outcomes. A flat prior can be made approximately Gaussian by an update based on a small initial set of sensible predetermined measurements. 

Assuming the warm up phase has been realized, we seek optimality through adaptivity; the analytically tractable prior can be used to arrive at an asymptotic expression for the Bayes risk (again for quadratic loss) as a function of the experiment control. This control is the measurement time, a continuous variable. 

The authors find that this quantity oscillates between a constant maximum and a lower envelope, with a frequency that increases as the inference advances (i.e. with more measurements, as the spread of the distribution is reduced).

We consider a counter $k \in \{1..N\}$ starting out at the first adaptive measurement, after the prior has been warmed up to
\begin{equation}
    \mathbf{P}(\omega) = \mathcal{N}(\mu_0,\sigma_0)
\end{equation}

At any stage, the minimum of the envelope is given in terms of the latest posterior's variance  as
\begin{equation}
    r_k^{(opt)} \equiv r(t_k^{(opt)}) = (1-e^{-1})\sigma_{k-1}^2
\end{equation}

\noindent, and occurs for
\begin{equation}
    t_k = 1/\sigma_{k-1}
\end{equation}

\noindent, $\sigma_k$ being the standard deviation of the $k$th iteration posterior,$(k+1)$th iteration prior.

As the distribution becomes sharper, the minimum of the risk approaches this lower bound location, due to the oscillations' rapid variation with respect to the bottom of the envelope \cite[Appendix~B]{Ferrie_2012}. Thus, the optimal scaling of the uncertainty with the iteration number is asymptotically described by the successive envelope minima: 
\begin{gather}
    \sigma_k = \left( \sqrt{1-e^{-1}} \right)^{k}  \sigma_{0}
\end{gather}

This gives the rate at which we expect the uncertainty to shrink. These minima are associated with the times
\begin{equation}
    \label{eq:envelope_tmin}
    t_k = \frac{1}{\left( \sqrt{1-e^{-1}} \right)^{k-1}  
    \sigma_{0}}
    \approx \frac{1.26^{k-1}}{\sigma_0}
    , k \in \{1,\dots,N\}
\end{equation}

\noindent, i.e. $t_1=1/\sigma_0, t_2 = 1.26/\sigma_0,\dots$

This result is however only valid asymptotically, and does not hold for a finite number of experiments. Nonetheless, it may still be a useful insight; for instance, we could concentrate the search for an optimal solution around this lower bound.

We can even drop full-blown optimization altogether, and instead just browse around a vicinity of the envelope's minimum by comparing the utility of a few possible $t$s. We expect that the risk's minimum will be nearby, an expectation that only becomes stronger as we move forward.

In light of this, a possible solution is to compute the current standard deviation to obtain the time given by \ref{eq:envelope_tmin}, and then perturb its location for a number of attempts (e.g. by sampling from a Gaussian centered at it) to get a few tentative $t_j$. Then we could determine the expected utility of these selected $t_j$, following equation \ref{eq:expected_utility}, and choose the one among them which yielded the maximal value.
\begin{equation}
    t_k^{\text{(chosen)}} = \text{arg min}\big[U(t_j) \big]
\end{equation}

But we don't necessarily need to calculate the current standard deviation. We could pick pairs of values $(\omega_j,\omega_j')$ at random by sampling from the current distribution:
\begin{equation}
    \omega_j,\omega_j' \sim \mathbf{P}(\cdot \mid \vec{D}_{curr})
\end{equation}

\noindent, after which we take the reciprocal of their distance $t_j=1/|\omega_j-\omega_j'|$. This replaces $1/\sigma_\text{curr}$ while organically introducing variability. Also, the intuition of \ref{eq:envelope_tmin} can be generalized to higher dimensions \cite{Wiebe_2014a}, where this surrogate heuristic conveniently eschews the inversion of the covariance matrix.

This type of heuristic based guesses have been used in e.g. \cite{Granade_2012,qinfer}. In subsection \ref{sub:exp_ramsey_1d} they will be applied to a quantum computer, and \ref{sub:multi_cos} will attempt multi-modal generalizations. 

With this, we may significantly reduce the complexity of choosing good experimental controls, opting for an in-between solution where we neither choose them arbitrarily nor extensively optimize them. Computing the utilities still demands a number of likelihood evaluations, but we are in a position to decide how many (by adjusting the total number of attempts). Further, the utility can be made to account for experimental issues, namely by being partial to evolution times well within the coherence time of the system. This is pertinent when dealing with real devices, possibly more than idealized utility maximization; decoherence may significantly weigh down the accuracy of otherwise promising guesses.

What is more, concessions can easily be made, as experimental design can tolerate rougher calculations than the nucleus of the inference process. It is not of the utmost importance that we choose the absolute optimal experiments, so precision may be better left for where it matters most. The positioning of approximations in the inference process is discussed in \cite{Granade_2012}.

As a matter of fact, even settling on some apt guesses without ever considering the utility explicitly may bring better results than random measurements, for little to no added processing cost. By picking the asymptotic optimum \ref{eq:envelope_tmin}, we can hope to land in proximity to the actual optimum more than if we used indiscriminate measurements. Since as mentioned before the upper bound for the risk is flat, we never increase the maximum risk.

For these reasons, times inversely proportional to the standard deviation have been experimentally applied to the precession example in e.g. \cite{Wang_2017}, and also to phase estimation in \cite{Paesani_2017} after being proposed in \cite{Wiebe_2016}. In the latter case the times are rounded to an integer $m$, and there's one more parameter to choose: the feedback angle $\theta$, which is picked directly from the prior/latest posterior (apart from a multiplying constant dependent on the chosen $m$)\footnote{This seems inspired in the exact \gls{IPE} case of \ref{eq:feedback_angle_m} (note that there the iterations are in superscript to avoid ambiguous notation, whereas in other sections, namely this one, we opt for subscripts), of which \ref{eq:feedback_angle_bayes} is reminiscent. However, \cite{Lumino_2018} shows it is not optimal for inference, at least not under a (fairly applicable normality assumption).}. Summarizing, the phase estimation controls are obtained as (by this order):
\begin{gather}
    m_k \propto \left\lceil \frac{1}{\sigma_{k-1}} \right\rceil
    \label{eq:bayes_m}
    \\
    -\frac{\theta_k}{m_k} \sim \mathbf{P}_\text{curr}(\cdot) \label{eq:feedback_angle_bayes}
\end{gather}  

Since these results are not guaranteed to be optimal, the constant of proportionality of \ref{eq:bayes_m} can be optimized (e.g. by picking a set of hypothetical real values at random from the prior, performing a hyperparameter sweep, and choosing the hyperparameter instance that yielded the lowest loss or expected loss on average). The second equation \ref{eq:feedback_angle_bayes} just means that we sample a value from the current distribution (of the parameter, the phase $\phi$), which considers the data record up to the present iteration, $\mathbf{P}_\text{curr}(\phi) \equiv \mathbf{P}_\text{curr}(\phi \mid \vec{D}_\text{curr})$, and then multiply by the symmetric of the previously obtained constant $m_k$. Refer to figure \ref{fig:phase_est_circ} and/or box \ref{eq:phase_est_summary} for the meaning of these controls.

As an alternative, two different strategies are presented in \cite{Ferrie_2012}. One is based on locating the risk minimum with a correction to the asymptotic solution, while at the same time extending the Gaussian assumption to the successive posteriors (rather than restricting it to utility related considerations). This reduces the Bayes' update to a pair of rules that produce the mean and variance directly, thereby lessening the online processing load. This type of strategy is sometimes called \textit{offline adaptive} \cite{Lumino_2018}.

The second one is to choose the times to be exponentially spaced out (loosely emulating those of \ref{eq:envelope_tmin}, aside from a multiplying constant), i.e.
\begin{equation}
    t_k = C^k
\end{equation}

\noindent for some $C$ chosen beforehand (\textit{offline}, without making use of the data). The article proposes $C=9/8$ (for a prior domain normalized to $\omega \in ]0,1[$).

We started out with rigorous considerations, and gradually relaxed them to the point they're just some reasonable bids. If one is to only barely ground them in reason, why didn't we start with guesswork from the outset? Let us backtrack to the beginning of this section to try to get some insight on what useful experiments look like in practice, and why we can't identify them so easily. Along the way, we will consider the ways applied practice means can interfere with this usefulness, and how approximate inference may result in a more delicate interplay.

Considering the dependence illustrated by figure \ref{fig:experiment_coef}, the choice may have seemed easy enough: we can simply choose the controls whose gradient with respect to the parameters is larger in magnitude. They should be particularly useful, because they'll be quicker to tell us where in parameter space we should focus our attention.

So in the oscillator case, we might be tempted to snub shorter evolution times entirely. Of course, there's a limit to this; after a certain point decoherence and noise will take over, and we don't want our data collection process to take forever. But ideally, weren't it for those trade-offs, we may be inclined to  make the intervals as long as possible. This idea seems deceptively simple. Should we?

That's not really the case. Spectacularly sharp likelihoods aren't ideal so long as we can't perform exact inference, which we saw in section \ref{sub:estimation_in_practice} happens often. In the limit, our likelihood function could take the form of an infinitely tall and infinitesimally wide pulse at an exact value. It would be completely unambiguous, but that wouldn't do much for us, since we'd have vanishing probability of pinpointing its location in parameter space. 

In less extreme cases, it is still a point to consider: we have to select a finite set of points at which to evaluate the posterior, and if the function is very sharp we need a very dense grid to ever find non-zero probabilities. If we simply want a rough estimate, we may be better off relying on a flatter curve.

\begin{figure}[!ht]
    \centering
    \includegraphics[width=10cm]{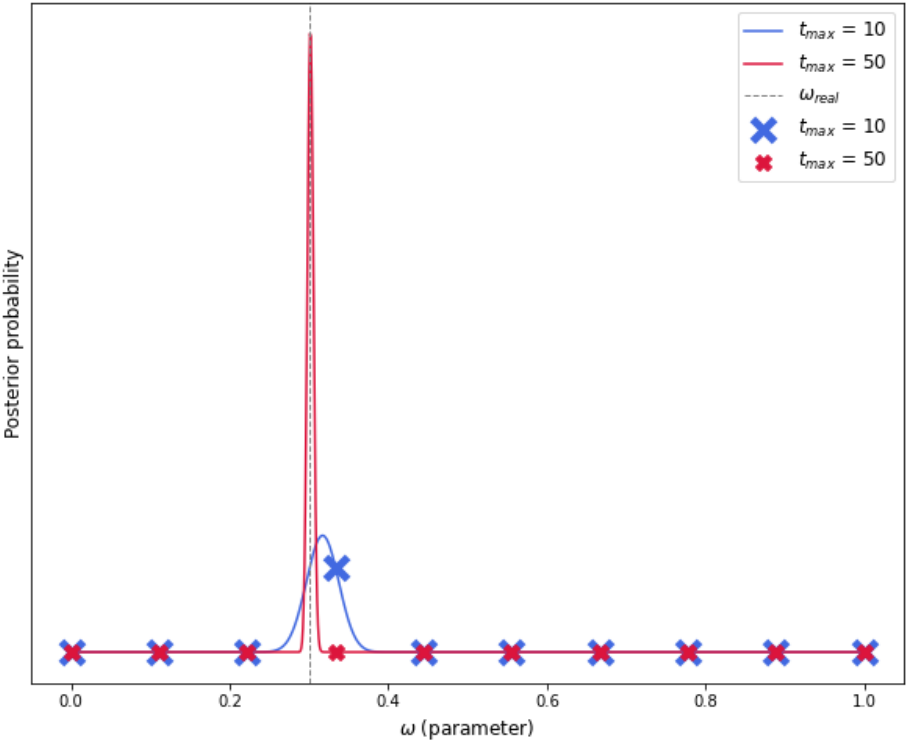}
    \caption{Posterior distributions for a time-dependent binomial model, for 2 sets of measurement times. The likelihoods take the undamped squared-sinusoidal form of \ref{eq:precession_summary}.  Both posteriors were evaluated on the same grid comprising 10 points (the 'x' markers), and are juxtaposed with smooth curves of matching color representing their respective continuous posterior distributions. The latter were approximated using 1000 points. The dashed line marks the real parameter used to generate the data. 100 measurements were included in each data vector, and the times were picked in constant increments up to the maximum $t_{max}$.}
    \label{fig:precession_tmax_points}
\end{figure}

This is illustrated in figure \ref{fig:precession_tmax_points}, where posteriors analogous to those of figure \ref{fig:precession_tmax} were evaluated on the same grid. The single point evaluations fail to provide any insight for the sharper curve, whereas in the case of the flatter curve one of the points succeeds in signaling mode proximity. 

It is clear that sharp likelihoods aren't the be-all and end-all of \textit{approximate} inference, in particular if resource economy is a concern. But save for that, should our idea inspire no more misgivings?

\begin{figure}[!ht]
    \centering
    \includegraphics[width=10cm]{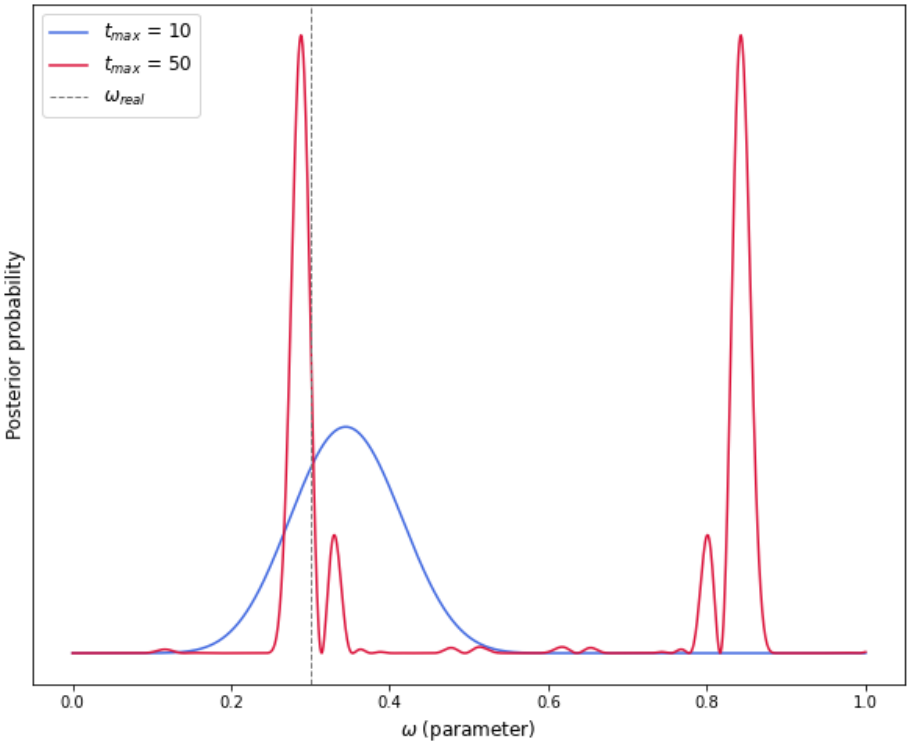}
      \caption{Posterior distributions for a time-dependent binomial model, for 2 sets of measurement times. The likelihoods take the undamped squared-sinusoidal form of \ref{eq:precession_summary}. The dashed line marks the real parameter used to generate the data. 10 measurements were included in each data vector, and the times were picked in constant increments up to the maximum $t_{max}$.}
    \label{fig:precession_tmax_ambiguity}
\end{figure}

That is still not the case. Elusive distribution peaks aren't the only caveat that it holds. As much as higher frequencies mean more information in a way, they also tend to bring ambiguity. This is demonstrated in \ref{fig:precession_tmax_ambiguity}, where as compared to \ref{fig:precession_tmax_points} we have reduced the number of measurements while preserving the intervals over which we space their corresponding times.

As the likelihoods associated to the data oscillate faster, their periodicity becomes more densely packed in parameter space. This gives rise to equivocal posteriors, where we are likely to observe multi-modality. This leads to irresolution: which mode do we pick if they have equal or similar predictive power for the results that we obtained? Our credible region is now much more disperse. The issue with \textit{high frequency information} is not unique to this problem or strategy; similar drawbacks appear when considering entangled states for phase estimation \cite{Berry_2009}.

In our case, incorporating lower times should help resolve this; clearly, there aren't as many modes as there could be in figure \ref{fig:precession_tmax_ambiguity}, and that's because we are considering a multitude of experiments. Each experimental time yields a datum, which produces an individual likelihood; only when we multiplying them all together do we get our final results.

In other words, our understanding doesn't spring from lone observations, but rather from their juxtaposition. If we add in a datum associated to a very long measurement time to the blue posterior, we don't expect that it will be immediately destabilized - most of the peaks will be cancelled out by lower probability events at the same parameter site.

This is evidenced by comparing the crimson curves of figures \ref{fig:precession_tmax_points} and \ref{fig:precession_tmax_ambiguity}; in the former we don't face the same problem as in the latter, despite using as long times. We strengthen the result through increasing the times more slowly; of course, this also means more data were necessary, and with them a higher processing cost. With this in mind, it is pivotal that we find a balance between reliable but shallow likelihoods, and sharp but undependable ones.

The problems represented in figures \ref{fig:precession_tmax_points} and \ref{fig:precession_tmax_ambiguity} are in a way related. They both suggest that we shouldn't unreservedly aim for long evolution times, even if dealing with a perfectly isolated system; and they both could be managed by progressive management of the search space.

In other words, we could profit from a fluid discretization of the posterior - one which deploys its grid\footnote{The term \textit{grid} is used loosely here.} points at strategic locations. This choice can draw on the previous one, using information that is already available by virtue of the inference process. In fact, each single-point evaluation of the Bayes rule quantifies the worth of the corresponding point (the coordinate(s) of a location in parameter space).

Shifting the disposition of evaluation sites via a serialization of the posterior sampling process would allow us to spare resources, by removing support from the unlikely regions. That way, we could concentrate them where they can actually be of benefit. As a bonus, we may to a degree evade ambiguity through preemptive action.

Ideally, we would start with a likelihood like the shallow one in figure \ref{fig:precession_tmax_points}, and then gradually transition (by dint of data collection) to the sharper one, while re-focusing our grid on the pertinent section of parameter space, where the posterior is non-null. Such a migration could populate this key section more densely, while abandoning its irrelevant surroundings. With that, a likelihood with ambiguity at the scale of \ref{fig:precession_tmax_ambiguity} should pose no issue: we’d have no coverage for the second peak, having eliminated it from the search space earlier on.

This reflection seems somewhat distant from the previous deliberation on utility. It stands closer to the details of how we represent distributions, in particular when dealing with analytically intractable inference and finite memory resources. While multi-modality or (faux) flatness \textit{would} be penalized if minimizing the expected posterior variance\footnote{Excessive sharpness in particular is technically associated with a low variance, but it is fair to assume optimization will operate our representation of the posterior. Since that's where the problem lies, this will reflect in the estimated variance.}, these problems may be more easily remediable via a tighter integration with the discretization methods. This is the subject of the next chapter.

\chapter{Monte Carlo methods for posterior sampling}
\label{cha:monte_carlo_posterior_sampling}

Up until now, we've been focusing on the core of the inference process. We were especially interested in how the allotted resources could be used to devise the best possible action plan. But all that is worth nothing, if we fail to seize the legacy of our perfect plan.

This was evident in figure \ref{fig:precession_tmax_points}, where a particularly good posterior produced particularly bad results. A chimerical dream of optimality is hardly any consolation when we've let it slip between our fingers. 

Granted, we could just reinforce the grid we started out with, increasing the number of evaluations from 5 to 10, or 100, or 1000. At some point, we would catch the mode; and we could pack the space as densely as our target precision demanded. However, this is somewhat precipitate: after all, we have four out of five points standing idly on the sidelines, relaying near zero probability. Why should we devote any more computational power to the problem, when we're squandering 80\% of it on undeserving calculations?

On this basis, this chapter concerns more resource-effective ways of sampling from the posterior. This is a determining factor for the success of Bayesian learning: it is these samples that provide a way of computing quantities of interest, as the sole carriers of the collected information.

An alternative to the grid based quadrature discussed in subsection \ref{sub:estimation_in_practice} is Monte Carlo integration, a simulation based approach which tends to scale better with dimension and complexity. This chapter overviews a selection Monte Carlo methods, starting with its simplest versions in section \ref{sec:monte_carlo}.

Because these algorithms are still unreliable for the task of characterizing a complex posterior, we will then move on to more advanced ones. These are grouped into two classes: \glsxtrfull{SMC} algorithms (section \ref{sec:sequential_monte_carlo}), which adaptively construct a dynamic grid, and \glsxtrfull{MCMC} algorithms (section \ref{sec:mcmc}), which sample from the posterior directly. Within each of these groups, a few structural variations are considered. 

For \gls{SMC}, the main degrees of freedom are the the source of dynamism (the sequence of distributions to be targeted) and the source of adaptivity (the mechanism for the strategic relocation of \textit{grid} points when transitioning from one target to the next). For the former we consider two possibilities: \glsxtrfull{SIR} with the prior as importance function, and \gls{SMC} with tempered likelihood estimation. For the latter we consider another two: Gaussian kernel smoothing and Markov kernels.

In the case of \gls{MCMC}, the variability lies with the construction of the transition kernels. Two variants of Metropolis-Hastings \gls{MCMC} are presented: random walk Metropolis and Hamiltonian Monte Carlo. Section \ref{sec:hmc_implementations} then goes into some implementation aspects concerning the latter: applicability, anti-correlation measures, and dynamic implementations (namely the no U-turn sampler extension). Finally, section \ref{sec:subsampling} discusses subsampling, overviewing variance reduction and pseudo-marginal methods with an emphasis on Hamiltonian Monte Carlo.

\section{Monte Carlo and importance sampling}
\label{sec:monte_carlo}

In section \ref{sub:estimation_in_practice}, we resorted to an elementary grid arrangement to perform numerical integration. For simple problems, a grid may actually be a viable choice - possibly with some improvements, such as the trapezoid rule and the like. But when working with large numbers of parameters and/or expensive likelihood evaluations, such a solution is a waste of resources in the best-case scenario, and a lost cause in the worst. In particular, creating a grid dense enough to capture features of high dimensional spaces is usually quite the thankless task; in these circumstances, a more efficient representation of the distribution is of the essence. 

This is due to that fact that as the parameter space's dimension grows, contributions to the expectation integral depend on a more complicated interplay between integrand (or probability) density and space volume. More specifically, the volume tends to accumulate away from the highest probability neighbourhood (or any neighbourhood for that matter).  As the parameters grow in number, the imbalance of volume grows exponentially with it. Indeed, this is the very reason why expectation integrals are far superior to optimization in high dimensional spaces: the latter blindly seeks probability density only, which means nothing by itself. These points are rooted in measure theory \cite{Betancourt_2014_talk}, which underpins statistics.

This discussion isn't unlike that of subsection \ref{sub:model_selection}, where we considered the effect of the dimension of an integral in its net value. In this context, this means that the region over which the integrand is to be accumulated must be selected carefully to achieve optimal balance between accuracy and evaluations. This is essential if one means to preserve the robustness of the expected value framework. 

To see how this fits the overarching narrative, we can refer back to our central equation. If we have collected a vector of data $\vec{D}$, we can compute the posterior expected value of a function as:
\begin{equation}
    \label{eq:expectation_batch}
    \mathbb{E}_{\mathbf{P}( \theta \mid \vec{D})}\left[f(\theta)\right]
    = \int f(\theta) \mathbf{P}(\theta \mid \vec{D})\mathrm{d}\theta
    \approx \sum_{i=1}^{M} f(\theta_i) \cdot w_i
\end{equation}

\noindent, where the $w_i = w_i(\vec{D})$ are our discrete-domain version of evaluations of the posterior $\mathbf{P}( \theta \mid \vec{D})$. We suppress the dependence on the data to lighten the notation.
\begin{equation}
    \tag{\ref{eq:discrete_normalization}}
    w_i \equiv \frac{W_i}{\sum_{j=1}^M W_j} =
    \frac{\mathbf{L}(\theta_i \mid \vec{D})\mathbf{P}(\theta_i) }
    {\sum_{j=1}^M \mathbf{L}(\theta_j \mid \vec{D})\mathbf{P}(\theta_j) }
\end{equation}

We have been trying to get a posterior with favorable properties to come by, so we could be confident in the expectations we compute (such as the mean, which we chose as our estimator). But our right-hand side calculation in \ref{eq:expectation_batch} reflects not this posterior, but rather its values at a discrete set of $M$ points.
\begin{equation}
    \{\theta_i\}_{i=1}^M
\end{equation}

Clearly, it is not sufficient that the posterior boasts good properties: it is also vital that these properties be relayed adequately by the $\theta_i$ of our choosing. Poorly chosen positions will produce negligible contributions, leading to inaccurate expectations.

Before proceeding to analyse and develop our algorithms,  we will dub two quantities to facilitate discussion. First, the parameter space points $\theta_i$ from $\{\theta_i\}_{i=1}^M$, which we use to discretize the distribution, are often called \textit{particles} (especially in the context of sequential Monte Carlo algorithms, which we will get to shortly). And second, their associated multipliers in \ref{eq:expectation_batch} are called their \textit{weights}, or importance weights.

Apart from this detail, we had already used these quantities in subsection \ref{sub:estimation_in_practice}, where we described a basic procedure to approximate expectations.  Algorithm \ref{alg:grid_expectation} outlines this first approach. It assumes single-parameter inference, though the generalization is straightforward.

\begin{algorithm}[ht!]
\caption{Grid-based algorithm for computing expected values.}

\textbf{Inputs}: prior distribution boundaries $\theta_L$ (left) and $\theta_R$ (right), data vector $\vec{D}$, number of particles $M$.

\textbf{Computes}: list of pairs $\{\theta_i,w_i\}_{i=1}^M$ to be used in \ref{eq:expectation_batch}.

\textbf{Assumes}: ability to evaluate the prior distribution and the likelihoods.

\begin{enumerate}
    \item Space out $M$ particles $\theta_i$ evenly in the interval $\theta \in [\theta_L, \theta_R]$ to get the fixed set $\{\theta_i\}_{i=1}^M$ (a grid). \label{stp:grid_space}
    \item For each particle $\theta_i \in \{\theta_i\}_{i=1}^M$, compute the unnormalized weight given the full dataset.
    \begin{equation}
        \tag{\ref{eq:bayes_prop_discrete}}
        W_i = \mathbf{L}(\theta_i \mid \vec{D})\mathbf{P}(\theta_i) 
    \end{equation}
    
    Additionally, accumulate these quantities to get the normalization constant $C$.
    \label{stp:grid_reweight}
    \begin{equation*}
        C = \sum_{i=1}^M W_i
    \end{equation*}
    
    \item For each $\theta_i \in \{\theta_i\}_{i=1}^M$, compute the normalized weights by dividing the former ones by the previously calculated normalization constant.
    \begin{equation}
    \tag{\ref{eq:discrete_normalization}}
    w_i = \frac{W_i}{C}
    \end{equation}

\end{enumerate}
\label{alg:grid_expectation}
\end{algorithm}

For one-dimensional integrals, improved efficiency can be achieved via interpolation, by using the well-known Newton-Cotes quadrature rules or other more sophisticated formulas. This can also be applied to higher dimensions (and so to multiple parameter estimation problems), through sequential application justified by Fubini's theorem.

However, this quickly becomes infeasible, due to exponential scaling on the number of dimensions: repeated application of \ref{eq:expectation_batch} along $d$ dimensions will require $M^d$ function evaluations. In that case, Monte Carlo (non-deterministic) methods tend to be preferable. They rely on random evaluations of the integrand, and their absolute error scales approximately with $M^{1/2}$, which is dimension-independent.

Their simplest version could be described by \ref{eq:expectation_batch} still, only the set $\{\theta_i\}_{i=1}^M$ would now be sampled from a uniform distribution. For the sake of clarity, it should be said that any integral (not just the ones corresponding to expected values) can be computed like so:
\begin{equation}
    \label{eq:monte_carlo}
    \int g(x)\mathrm{d}x \approx \frac{V}{M} \sum_{i=1}^{M} g(x_i)
\end{equation}

In our case, we have seen in section \ref{sub:estimation_in_practice} that normalizing the weights instead of dividing them by the marginal likelihood took care of the $\frac{V}{M}$ factor, because we were precisely dividing by $\frac{M}{V}\mathbf{P}(\vec{D})$.

Unfortunately, the idea of dimension-independence doesn't tell the whole story; these methods still suffer from increased dimensionality. In high-dimensional spaces, we expect that less of our evaluations will recount non-negligible probability, since the probability mass will tend to be more concentrated with respect to the total volume we must cover. This is likely to happen for our \textit{expectation} integrals, where we consider some parametrized distribution. Typically, vast sections of the parameter space will have near-zero probabilities; the more dimensions, the truer this is. The result is an inflation of the total space to explore; should particles fall in these regions, their weights will be insignificant, barely affecting the integral. Thus, the relative error will be amplified by the number of model variables, still frustrating our attempted estimates in complex cases.

The problem can be alleviated by using \textit{importance} sampling. Instead of sampling points from a uniform distribution, we sample them from one that is tailored to our specific integrand. Ideally, we would focus the $\theta_i$ in the regions that actually contribute to the integral; accumulating the integrand over null probability regions is, of course, futile. A sensible choice will reduce the variability in our estimate, making it more robust for finite $M$.

With that in mind, we would introduce a proposal distribution, also called an importance function, that we could sample from:
\begin{equation}
     \pi(\theta)
\end{equation}

\noindent, and use it to obtain our parameter locations:
\begin{equation}
    \theta_i \sim \pi(\cdot)
\end{equation}

In that case, equation \ref{eq:expectation_batch} must be altered to compensate for the points' relative importance, which is a trick to make the samples land around a critical zone and not literal \textit{importance}. That is, this importance is ascribed by our strategy, and not by the integrand. Consequently, if a point is very likely to occur under our choice, that likeliness should be discounted from its contribution; the intended value comes instead from the evaluation of the integrand, and is larger or smaller according to it alone.

For the general case of \ref{eq:monte_carlo}, this means:
\begin{equation}
    \label{eq:monte_carlo_importance}
    \int g(x)\mathrm{d}x \approx \frac{1}{M} \sum_{i=1}^{M} 
    \frac{g(x_i)}{\pi(x_i)}
    \quad , \ x_i \sim \pi (\cdot) 
\end{equation}

In hindsight, we can see that uniform sampling was just a particular case of this, with $\pi(x) = \frac{1}{V}$ (for $V$ the volume of the domain of integration, which should match that of the uniform distribution) yielding the fixed multiplicative constant $\frac{V}{M}$, which in that case was independent of the effective set of samples. 

As for our case, \ref{eq:expectation_batch} becomes:
\begin{equation}
    \label{eq:expectation_importance}
    W_i = \frac{\mathbf{P}( \theta_i \mid \vec{D})}{\pi(\theta_i)}
    \quad, \ \theta_i \sim \pi(\cdot)
\end{equation}

The normalization to get $w_i$ is as usual (i.e. divide each $W_i$ by the sum $\sum_{i=1}^{M} W_i$). Again, it is easy to see that for the particular case of a uniform distribution these contributions are as before, because the $\pi(\theta_i)=1/V$ factors cancel out when normalizing.

In algorithm \ref{alg:grid_expectation}, importance sampling would change two points:
\begin{itemize}
    \item In step \ref{stp:grid_space}, we would sample $\theta_i \sim \pi (\cdot)$ instead of arranging the particles on a grid.
    \item In step \ref{stp:grid_reweight}, we would compute the weights $W_i$ according to \ref{eq:expectation_importance} (for both occurrences).
\end{itemize}

As an example of importance sampling, we can think of estimating the fairness of a nearly balanced coin: we could sample the points from a Gaussian centered at $1/2$. Evaluating the posterior at those sites should be more fruitful than spreading them uniformly between $0$ and $1$, since we expect the probability to be low away from the center. Had we used uniform samples and small $M$, we would expect a significant discrepancy between any two different sets of samples, whereas a Gaussian should produce more consistent results; this testifies to the robustness of the estimation strategy.

How can we develop a \textit{general} approach to importance sampling, that will allow us to compute posterior expectations efficiently? This calls for some insight, but we have long decided to be neutral; our flat prior would suggest precisely uniform sampling. 

Overall, a nice importance function would resemble the integrand, $f(\theta) \mathbf{P}( \theta \mid \vec{D})$. We should remember that $f(\theta)$ can be any function of interest, such as the identity or the mean squared distance from the mean (whose expectations, as seen in subsection \ref{sub:estimators_and_expressions_of_uncertainty}, comprise our main targets, the mean and variance). What we want is to have a fairly generic group of $\theta_i$ - having registered these locations and their associated posterior weights $w_i$, we are in a position to retrospectively compute the expectation of whatever function of the parameters.

As such, the \textit{ideal} choice is actually objective: we would like to sample directly from $\mathbf{P}( \theta \mid \vec{D})$, in which case \ref{eq:expectation_importance} would reduce to perfectly distributed evaluations of $f$ with no multipliers at all. This manifests a dependence on the data-vector $\vec{D}$, and so assumes batch-processing. In such a context, we could use the results of our $N$ experiments to \textit{a posteriori} condition our proposal distribution:
\begin{equation}
    \label{eq:importance_sampling_data}
    \pi(\theta \mid \vec{D}) = \pi(\theta \mid \{D_1,\dots,D_N\})
\end{equation} 

This means that we can choose the posterior as the proposal distribution - though not quite. While we can resort to ready-made implementations to sample from some well-known families of functions, we don't know how to \textit{sample} from an arbitrary function just by evaluating it\footnote{\textit{Yet}; that's left for section \ref{sec:mcmc}.}.
 
What we \textit{can} do is to settle for a different strategy - one grounded on the data still, but aided by some extra contrivance. Expectably, this contrivance is again sequential processing.

\section{Sequential Monte Carlo}
\label{sec:sequential_monte_carlo}

\textit{Sequential Monte Carlo} (\glsxtrfull{SMC}) is an umbrella term for methods concerned with characterizing the succession of states in a stochastic process, based on consecutive partial observations. A review of such methods is presented in \cite{Doucet_2001}, which is the reference we'll follow for this section. They find use in a wide array of applications, notably Bayesian inference.

In that case, the target process easily finds a formalization in the theory of hidden Markov models. Such a model considers two concurrent processes: one is \textit{hidden}, and one is directly observable. The latter depending on the former, its direct observation amounts to making \textit{partial} observations regarding the hidden states\footnote{This \textit{partialness} could also be due to observing variables through some noise.}. These hidden states are of course the target parameter(s), while the \textit{visible} ones are the measurement outcomes; the \textit{emission probability} that functionally relates them is none other than the likelihood. The discrete time is set by the data collection, e.g. one datum corresponds to a time step.

Put simply, the purpose of \gls{SMC} is to sample from a sequence of distributions, in our case the successive posteriors. This is also called \textit{simulating} the successive posteriors. Just like quantum simulation provides the means for characterizing and predicting the behaviour of observable quantum systems, statistical simulation does so for observable statistical phenomena.

One possible application for sequential simulation is obvious: \textit{live} parameter learning, where we want either to capture the evolution of dynamic parameters or to track the estimates real-time. Clearly, this is our case, as we need to follow the data as they arrive in order to adapt the experiments accordingly.

However, these methods are as often used \textit{offline}, purely to aid at the post-processing stage. In that case, the goal is generally to travel across a sequence of slowly changing, increasingly complex distributions and arrive at the intended one, when it would be hard to do so directly. This is \textit{also} our case.

Our target sequence of distributions is comprised of the successive posterior distributions, with single-datum granularity. The $k$-th iteration posterior is:
\begin{equation}
    \label{eq:p_k}
    \mathbf{P}_k(\theta) \equiv \mathbf{P}(\theta \mid \{D_1,\dots,D_k\})
\end{equation}

\begin{figure}[!ht]
    \centering
    \includegraphics[width=0.7\textwidth]{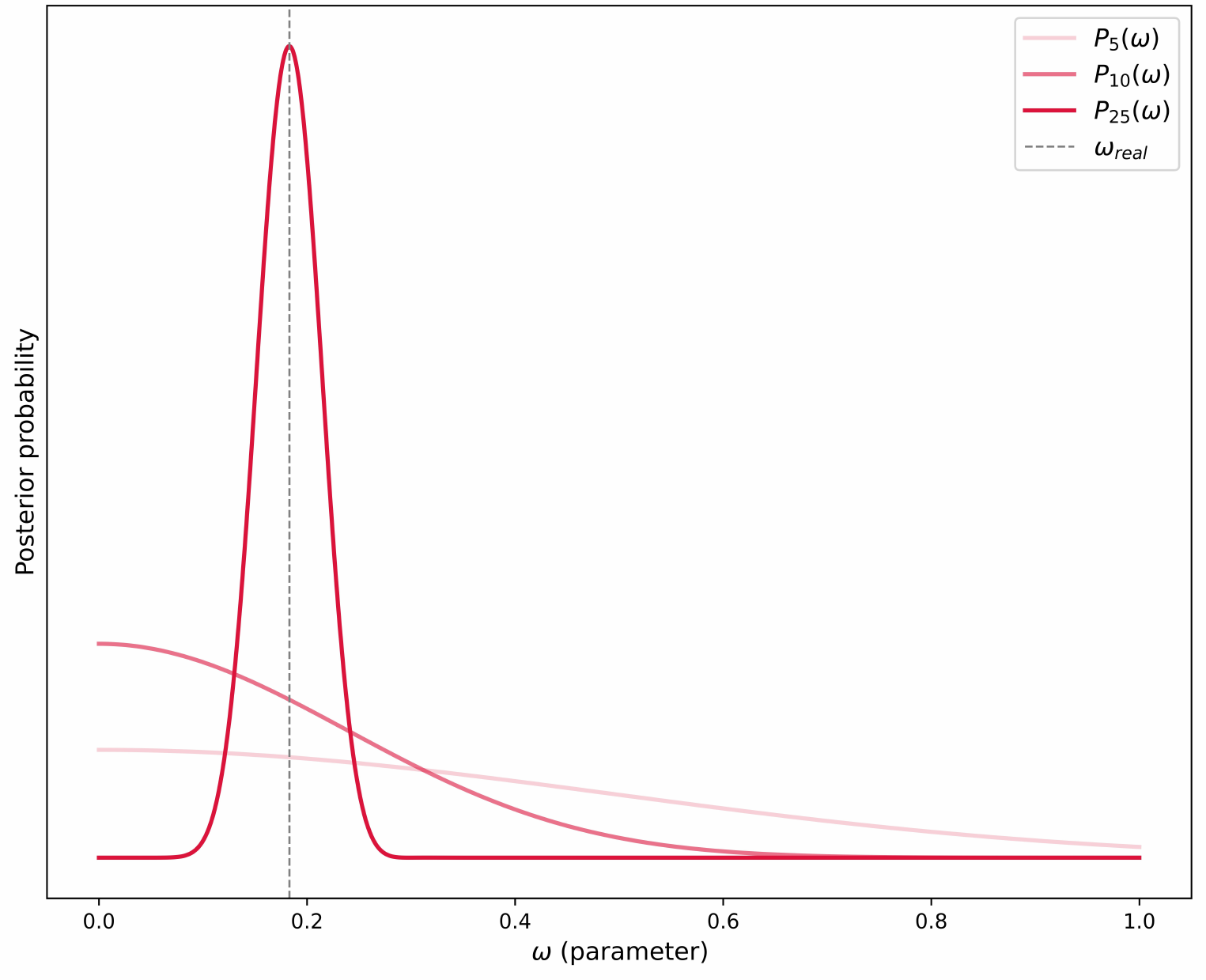}
    \caption{Evolution of the posterior probability distribution with the iteration number. The dashed vertical line marks the real parameter used to generate the data. The posterior is depicted for $k=5$, 10, and 25 accumulated data, according to equation \ref{eq:p_k}. The $D_k$ were obtained using exponentially increasing times $t_k=(9/8)^k$.}
    \label{fig:successive_likelihoods}
\end{figure}

A selection of such local posteriors is presented in figure \ref{fig:successive_likelihoods} for the same case as in the previous chapter (the precession example).

This local posterior should be available with no more information (i.e. not rely on $D_j$ for $j > k$). Again, one possible motivation for this is clear: we may want to choose the next experiment adaptively, or mean to get the up-to-date information as soon as possible. With it, we can get the current expectation for some function: 
\begin{equation}
    \label{eq:sequential_expectation}
    \mathbb{E}_{\mathbf{P}_k(\theta)} \left[f(\theta)\right]
    = \int f(\theta)\mathbf{P}_k(\theta)\mathrm{d}\theta
    \approx \sum_{i=1}^{M} f(\theta_i) \cdot w_i^{(k)}
\end{equation}

Another rationale in favor of adding data gradually is the fact that more attainable adjustments in the discretization should be due when transitioning between consecutive distributions, since a single datum isn't expected to bring about massive changes. This is made clear by figure \ref{fig:successive_likelihoods}, and will shortly be made still clearer.

To add to this fact, in the case of the previous chapter and figure \ref{fig:successive_likelihoods} extra flexibility is introduced by experimental freedom. That is, we can sort the data in increasing order of measurement times, to make the distribution evolve more smoothly; and we can increase them as fast or as slowly as we wish, in order to control the spread of the distribution through the iterations. If we maximize some utility function, we will be using our particle cloud to evaluate it, and so numerical precision will be minded by default.

Having outlined the appeal of a sequential sampling scheme, algorithm \ref{alg:sequential_expectation} lays out the associated procedure. It uses the same number of likelihood evaluations as \ref{alg:grid_expectation}, but it processes the data sequentially. 

\begin{algorithm}[!ht]
\caption{Sequential algorithm for computing expected values. A ${(k)}$ superscript signals dependence of the marked quantity on the first $k$ data and no more.}
\textbf{Inputs}: data vector $\vec{D}$ (to be used incrementally: $\{D_1\}$, $\{D_1,D_2\}$,...,$\{D_1,D_2,\cdots,D_N\}$), number of particles $M$.\\
\textbf{Computes}: successive lists of pairs $\{\theta_i,w_i^{(k)}\}_{i=1}^M$ for (by order) $k \in \{1,\dots,N\}$ to be used in \ref{eq:sequential_expectation}. \\
\textbf{Assumes}: ability to sample from the prior and to evaluate the likelihoods.\\
\textbf{Assures}: The $k$-th list is computed in the $k$-th iteration, using the datum $D_k$ and before processing $D_{k+1}$.
\begin{enumerate}
    \item Sample $\{\theta_i\}_{i=1}^M$ from the prior distribution and set $w_i^{(0)} = 1/M$ for all $i \in \{1,\dots,M\}$.
    
    \item For $k \in \{1,\cdots,N\}$:
    \begin{enumerate}
        \item For each particle $\theta_i \in \{\theta_i\}_{i=1}^M$, update the weight using the latest datum $D_k$.
        \begin{equation}
            \label{eq:update_weights}
           W_i^{(k)} = w_i^{(k-1)} \cdot \mathbf{L}(\theta_i \mid D_k)
        \end{equation}
        Additionally, accumulate these quantities to get the normalization constant $C$.
        \begin{equation*}
            C^{(k)} = \sum_{i=1}^M W_i^{(k)}
        \end{equation*}
        \label{stp:update_weights}
        \item For each $\theta_i \in \{\theta_i\}_{i=1}^M$, compute the normalized weights by dividing by the previously calculated normalization constant.
        \begin{equation}
            \tag{\ref{eq:discrete_normalization}}
            w_i^{(k)} = \frac{W_i^{(k)}}{C^{(k)}}
        \end{equation}
        \label{stp:normalize_weights}
        
    \end{enumerate}
\end{enumerate}
\label{alg:sequential_expectation}
\end{algorithm}

We now want to modify this to use sequential \textit{importance} sampling, in order to improve the positioning of the particles. Through such improvements, we hope to soothe our previous concerns relating the discretization of the distribution. 

In the context of \gls{SMC}, the problem we seek to treat is generally called the \textit{particle degeneracy} problem. As the inference process advances, the posterior probability becomes more and more concentrated with respect to the parameter space volume, leading to increasing disparity between the particle weights. As such, if we start out with $M$ uniform uncorrelated samples, our \textit{effective} number of samples will shrink down to a small few with time, due to particle collapse: most particles will cease to provide any useful information after a point.

This is due to the nature of inference: if all goes well, as the data pile up, the true posterior \textit{should} converge to an each time sharper peak around the real value. It is this fact what allows us to get results, and to trust them. But if we don't take heed when approximating the distributions, this same fact can be our undoing.

The achievable resolution will be determined by the disposition of the particles, as will how sharp a distribution we can capture. In general, it's safe to say that a coarse grid is a severe performance bottleneck, and undermines the credibility of the results - if they surface at all. What is more, this coarseness is relative, and bound to become a reality sooner or later.

In the case where we work with a fixed grid, after some precision threshold, the particles that fall closest to the peak are bound to amass all the weight, causing all others to yield negligible contributions (remembering that the weights are made to sum to one). Should we proceed further, we are likely to have a single particle be responsible for our estimates, as the others get sidetracked with (rather expensive) computations of zero.

Finally, if we add in still more data, even this resistant particle will find zero probability eventually. By then, we will encounter a zero-division indetermination when normalizing it, and be forced to stop. Even if one particle is lucky enough to fall very close to the asymptotic mode by chance - which we clearly should not rely on -, any estimates derived from it will be defective. In the end, if we take the inference process far enough, we're unlikely to be left with much more than the scant information that the particles we started out with are not plausible parameter specifications.

\begin{figure}[!ht]
    \centering
    \includegraphics[width=0.7\textwidth]{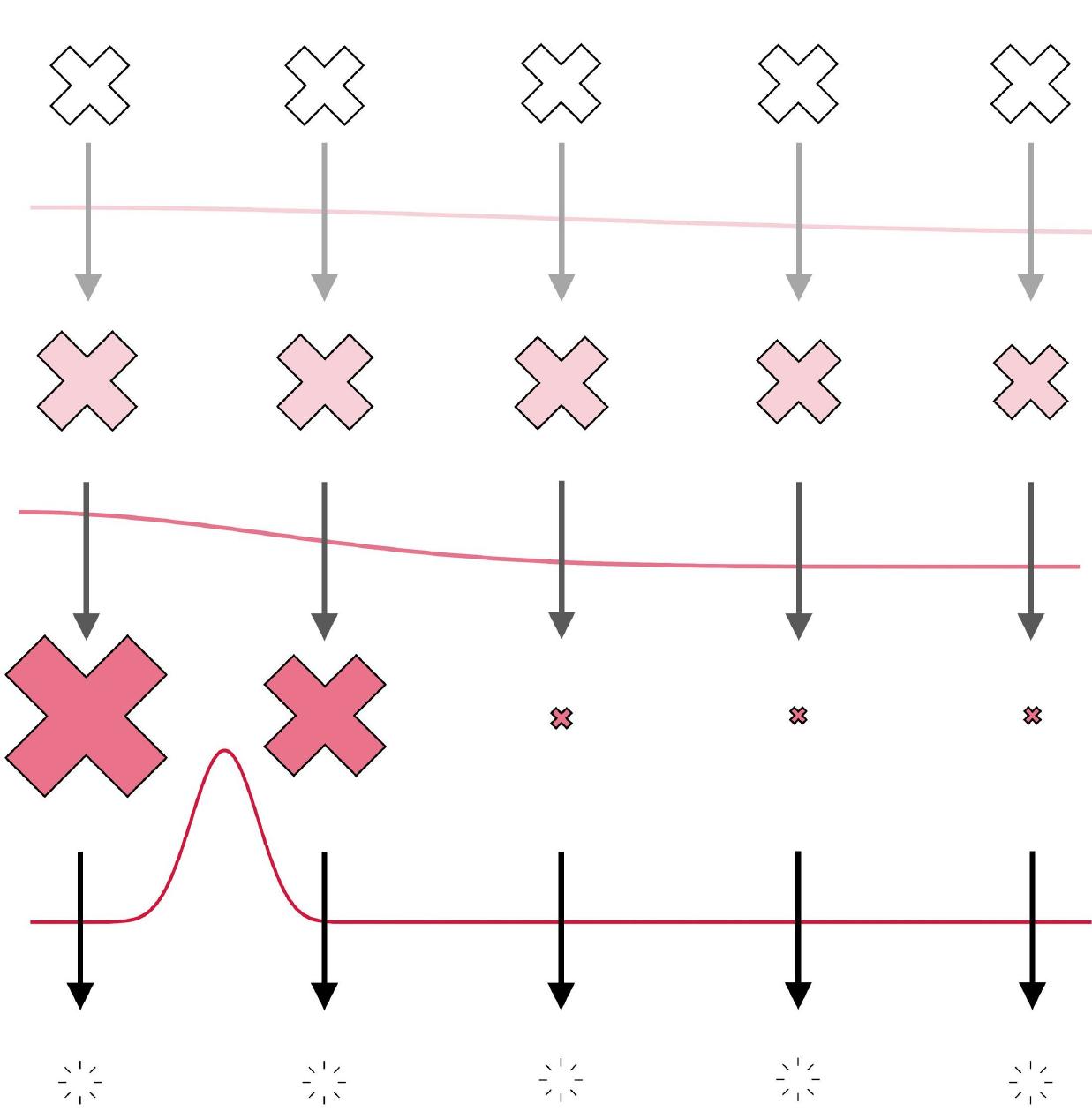}
    \caption{Illustration of the particle degeneracy/weight collapse problem, prompted by the particle cloud's evolution as the iterations (and with them the data) accumulate. The distributions are the same as in figure \ref{fig:successive_likelihoods}. The 'x' markers stand for particles, their sizes representing their weights.}
    \label{fig:particle_degeneracy}
\end{figure}

Figure \ref{fig:particle_degeneracy} illustrates this problem, using the posteriors displayed in figure \ref{fig:successive_likelihoods}. After the first curve, the weights are still fairly even; however, we can see that after the second the two leftmost particles have grown disproportionately, while the others wither away. This disparity foreshadows what will happen next: the grid is unfit to capture further information, because at all of its points the posterior probability is nearly zero (often it is for all purposes zero, due to arithmetic underflow).

It being that we proposed importance sampling as a possible remedy for this type of issue, we will try to incorporate it into algorithm \ref{alg:sequential_expectation}. For each iteration $k$, we can importance sample \textit{new} particles instead of recycling the previous ones. We can do so just before step \ref{stp:update_weights} of iteration $k$; and naturally, we can utilize the data supplied so far.
\begin{equation}
    \label{eq:importance_sampling_k}
    \theta_i^{(k)} \sim \pi (\cdot \mid \{D_1,\dots,D_k\})
\end{equation}

As a consequence, the weight calculations will be done from scratch at the new particle locations, and must be divided by the importance function instead of just being updated according to the latest datum (as before in step \ref{stp:update_weights}).
\begin{equation}
    \label{eq:importance_weights}
    w_i^{(k)} \propto \frac{\mathbf{P}_k(\theta_i^{(k)} \mid \{D_1, \dots, D_k\})}
    {\pi(\theta_i^{(k)} \mid \{D_1, \dots, D_k\})} 
\end{equation}

The values of the final weights are considered up to normalization for simplicity. Any scaling factor will be eliminated when normalizing, so determining the weights apart from a common normalizing constant is sufficient.

Even if we have some convenient importance function in mind, this is very demanding in terms of resources. We get the sequence of distributions we want and can use the past data to inform the importance distribution, but the price to pay is an immense cost as compared to our previous implementation. If before we took on the calculations for $N$ data, now we will do the same for 1 datum, then for 2, and so on until we \textit{finally} get to our last iteration - which alone is as costly as the full-length procedure previously was.

At this point, it is also not very clear why the situation that motivated our switch to a \textit{sequential} routine should be any better with it. Our problem was the importance function, or lack thereof. We started out with the purpose of ameliorating the case of \ref{eq:importance_sampling_data}, after observing that using the posterior as the importance distribution was infeasible and we had no other clues upon which to base it. What has changed as compared to then?

Two things have. At the time of sampling for the $k$-th iteration:
\begin{itemize}
    \item The available dataset at the time of importance sampling is not necessarily the full data-vector $\vec{D}$, but rather its $k$ first elements;
    \item Before computing the set $\{\theta_i^{(k)},w_i^{(k)}\}_{i=1}^M$, we have computed its precursor $\{\theta_i^{(k-1)},w_i^{(k-1)}\}_{i=1}^M$.
\end{itemize}

The first point means that the best importance function we could wish for at the $k$-th iteration is the posterior $\mathbf{P}_k(\theta)$ considering the first $k$ data, and not the final posterior. Here we encounter the same problem as before: we can't sample from the posterior.

However, thanks to the second point, we are \textit{much} closer. Equation \ref{eq:importance_sampling_k}, doesn't consider all the available resources at the time of its use. We can work in a dependence in all the preceding lists of particle-weight tuples, which will be all set by then. In particular, we can use the latest one, as we pointed out. Having just completed some iteration $k$, we have
\begin{equation}
    \label{eq:particle_cloud}
    \{\theta_i^{(k)},w_i^{(k)}\}_{i=1}^M
\end{equation}

In fact, this particle cloud is a weighted sample from $\mathbf{P}_{k}(\theta)$, as made clear by equation \ref{eq:sequential_expectation}. If we were in possession of uniform samples from $\mathbf{P}_{k}(\theta)$ to use as importance samples in the $(k+1)$th iteration, $\{\theta_i^{(k+1)}\}_{i=1}^M$, we could choose this $\mathbf{P}_{k}(\theta)$  as our importance distribution, and the weights to be calculated in \ref{eq:importance_weights} for the $(k+1)$th iteration would become simply:
\begin{equation}
    \label{eq:reweight_uniform}
    w_i^{(k+1)} \propto \frac{\mathbf{P}_{k+1}(\theta_i^{(k+1)})}
    {\mathbf{P}_{k}(\theta_i^{(k+1)})} \propto
    \mathbf{L}(\theta_i^{(k+1)} \mid  D_{k+1})
\end{equation}

That is, we would be a single datum away from the target, in which case this datum's likelihood is all that is needed for the iteration in question.

Were it not for the issues with reusing particle locations, we would just take the weighted sample and multiply the old weights by the likelihood, recovering algorithm \ref{alg:sequential_expectation}; this is enough to actualize the transition from $\mathbf{P}_{k}(\theta)$ to $\mathbf{P}_{k+1}(\theta)$. But we have seen the problem with that, so we want to proceed differently, somehow getting different - better - importance samples for the following operations.

Achieving this via the latest particle cloud would be doubly beneficial: on the one hand we would avoid repeating likelihood evaluations, and on the other the changes in the discretization would consist of smaller \textit{corrections} based on the information gathered since the previous alterations.

Essentially, we would want to base these smaller-scale corrections on \ref{eq:particle_cloud}, transferring the dependence of \ref{eq:importance_sampling_k} on the data to the particles in hand (except for the last datum). This leaves us with the last piece of the puzzle - we must figure out how to use these particles.

\subsection{Particle propagation and the Liu-West filter}
\label{sub:liu_west}

At this point, we are in possession of a particle cloud representing a weighted sample from the latest posterior, and from it mean to get a fresh sample for the following one. 
\begin{equation}
    \label{eq:importance_cloud}
    \theta_i^{(k+1)} \sim \pi (\cdot \mid 
    \{\theta_i^{(k)},w_i^{(k)}\}_{i=1}^M)
\end{equation}

More specifically, we want a new and improved set $\{\theta_i^{(k+1)}\}_{i=1}^M$, whose reweighting will be more fruitful than if we had reused the old locations. For that, we can \textit{recycle} poorly situated particles, summoning them to critical regions.

Equation \ref{eq:importance_cloud} raises two questions. First, the samples we have access to are weighted, whereas our importance samples up to this point were simple parameter vectors with no associated weights. We expect our importance density to place more particles where it is higher, but we don't want it to give us a high probability point and say it counts for 10.

Which takes us to our second point - we want \textit{variability}. We want to mobilize our particles away from the barren land that most witnessed, in order to sift through the interesting parts of parameter space. How can we make them decamp to those regions if all we have is a discrete set of old positions, some of which to top it off are given an inordinate amount of \textit{importance}?

As for the weight imbalance point, it can be solved easily. We just need to replace the weighted set of particles by an unweighted one (with uniform weights $w_i = 1/M$) according to the particle weights. For instance, we could divide the $M$ new particle locations by assigning $M \cdot w_i$ particles to each $\theta_i$. Since $M \cdot w_i$ will not in general be an integer value, the distribution is better preserved on average by drawing $M$ samples from a multinomial distribution with $w_i$ probability of getting $\theta_i$ ($w_i\leq 1$ given that $\sum_{i=1}^M w_i=1$). This is known as a bootstrap filter.

By doing this, we convert the weighted sample of particles to one where probability is conveyed by particle density. Naturally, even though this simplifies the problem, there is still an imbalance: the higher weight particle(s) will monopolize these new samples just as they did the weight. As a consequence, we expect to have multiple particles at some sites; what is more, these sites will be no more varied than they were before.

\begin{figure}[!ht]
    \centering
    \includegraphics[width=0.7\textwidth]{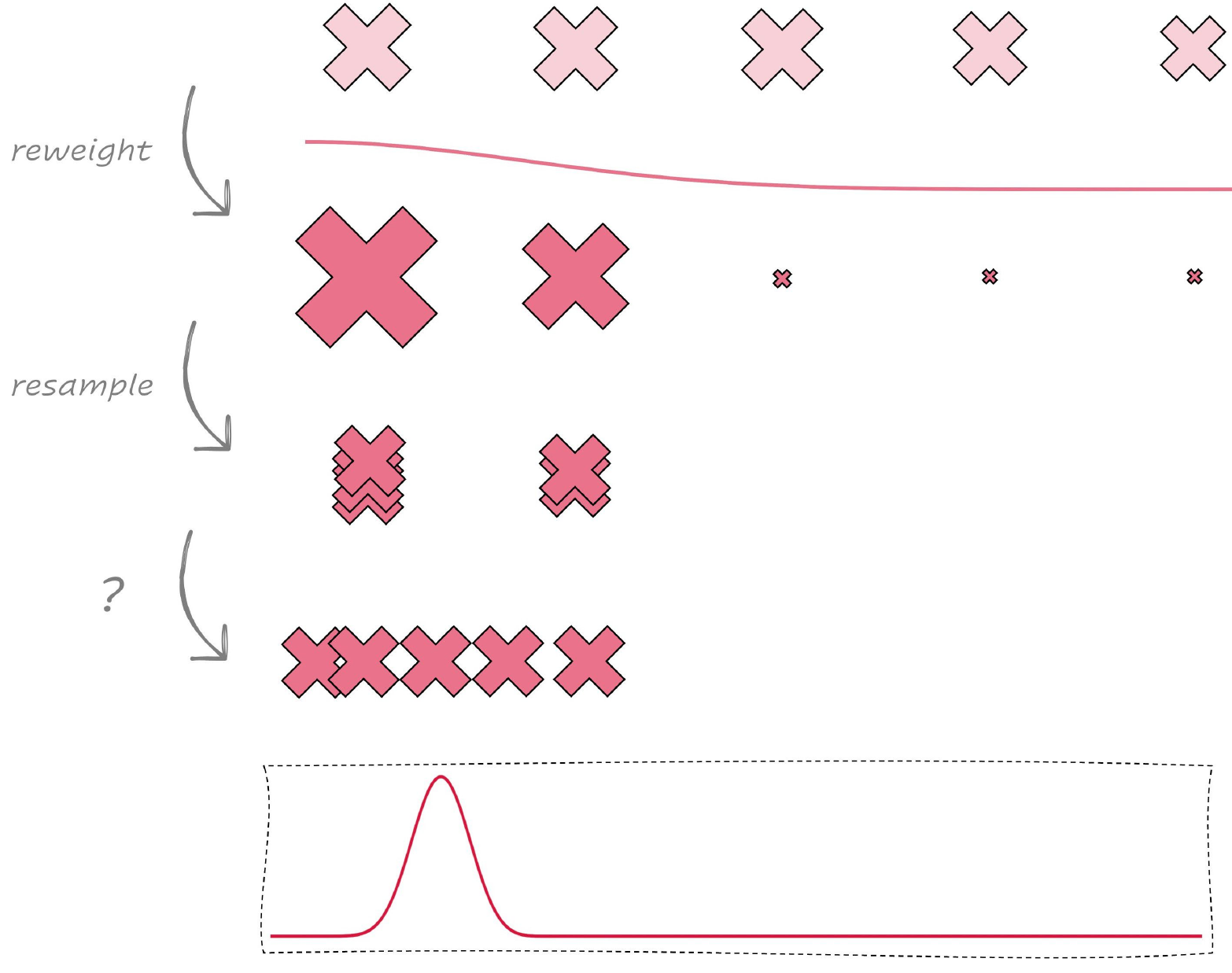}
    \caption{Illustration of the obtention of a weighted particle set and the necessary treatment to generate an efficient unweighted sample. This corresponds a segment of \ref{fig:particle_degeneracy}, with a change of the sampling techniques. The bottom clipping displays the following posterior, to be targeted by the darkest particle set. The 'x' markers stand for particles, their sizes representing their weights.}
    \label{fig:sir}
\end{figure}

Clearly, there's still something missing, as illustrated in figure \ref{fig:sir}. We have a weighted sample from the last iteration (e.g. the first, after we reweight the initial uniform samples according to their first datum likelihoods). Then we resample them with a bootstrap filter. Then what?

We must \textit{propagate} the particles somehow, but we can't do that if we adhere to our discrete set. No matter how well it represents the current distribution in terms of expectations, it will never be continuous like itself, and so it won't allow the particles to glide away from their original spots.

One possibility is to perturb the particles about their sites, using for example a Gaussian distribution. By allowing the particles to roam around, we are likely to place them at different points in parameter space, just as we wanted; and by doing so stochastically, we can expect that the overlaid ones will move into different positions.

This amounts to interpreting the weighted sample as a mixture of Gaussian distributions. Of course, we could just as well use other distributions to transform our particles into a smooth density. These distributions are commonly called kernels, and the method is called kernel smoothing. This is described in \cite{Liu_2001}, which is the main reference for the entirety of this subsection.

With Gaussian kernels, our smoothed distribution would be of the form:
\begin{equation}
    \label{eq:mixture_normals}
    \mathbf{P}_k(\theta) \approx \sum_{i=1}^M w_i^{(k)}
    \mathcal{N}\left(
    m_i^{(k)}, s_k^2 
    \right)
\end{equation}

We have already taken care of the weights by bootstrapping, so finalizing the kernel shape is all that's left. The reason why the Gaussian hyperparameters have been left vague is that they're tunable degrees of freedom. One obvious choice for the means is the previous particle location. After applying the bootstrap filter, we would take each sample's location as the mean for a Gaussian, sample from that Gaussian, and fix a particle at to this sample's coordinates for the $(k+1)$th iteration, instead of keeping it at its original site. We would proceed like such for all particles, after which we would have a more diverse sample.

Caution should be taken however in ensuring it doesn't get \textit{too} diverse, in the sense that we're injecting noise. The source is the Gaussians' spread, which we're yet to determine. Since we have no reason to distinguish between particles in this case, we choose a fixed standard deviation $s_k$ that depends only on the iteration.

Ideally, it would decrease as the iterations accumulate, as we don't want to toss the particles around too aggressively: we risk scattering them through the parameter space, thereby losing information (and possibly throwing them back into low probability regions). One natural way of informing this hyperparameter is to fall back on the latest particle cloud's standard deviation, namely using a proportion:
\begin{equation}
    s_k = h\sigma_k
\end{equation}

Here $h$ can itself be a function of the iteration number. This is also applicable in higher dimensions, where the covariance matrix would take the place of the variance. This only furthers the convenience of this idea, as it automatically considers the relative scale between parameters and their learning rates.

Even though $\sigma_k$ and $s_k$ are both standard deviations, they play very different roles. The former is associated with the \gls{SMC} estimate for the variance of $\theta$ given the distribution $\mathbf{P}_k$ after the reweighting,
\begin{equation}
    \sigma_k = \sqrt{\mathbb{V}_{\mathbf{P}_k}[\theta]}
\end{equation}

\noindent, which in turn estimates the expected mean squared error; the latter determines the distance between the new particles and the old ones (or its order of magnitude).

Together, they will determine the \textit{new} \gls{SMC} standard deviation, which comes to replace the former $\sigma_k$  and concerns the revitalized group of particles. Under our circumstances, it is given by:
\begin{equation}
    \sigma_k' = \sqrt{1+h^2} \cdot \sigma_k
\end{equation}

This affirms our concerns. Except for the case of $h=0$, where we stay by the bootstrap filter, we are increasing the uncertainty by dispersing the particles. While this may not seem very serious for small $h$, this over-dispersion will be amplified at each iteration we put this in practice, partially canceling out the efforts put into gathering and processing the data.

A technique called \textit{kernel shrinkage} has been proposed as a restorative. The idea is that to compensate for stirring the particles, we pull them slightly towards the \gls{SMC} mean. The more we agitate them, the more we impel them in the direction of the center, closer to the middle of the particle cloud. More specifically, to each particle we assign the kernel mean:
\begin{equation}
    \label{eq:lw_mean}
    m_i^{(k)} = a\theta_i^{(k)} + (1-a)\mu_k
\end{equation}

These means will preserve the variance for $a=\sqrt{1-h^2}$. Solving for $h$,
\begin{equation}
    \label{eq:lw_stdev}
    s_k = \sqrt{1-a^2} \cdot \sigma_k
\end{equation}

That is, in priming each particle location $\theta_i^{(k)}$ for the next iteration, we will be sampling from:
\begin{equation}
    \theta_i^{(k+1)} \sim \mathcal{N}(m_i^{(k)}, s_k^2)
\end{equation}

Putting everything together, we have a closed form for our smoothed density of \ref{eq:mixture_normals}, and we can sample fresh particles from it for the following iteration. We can do so before step \ref{stp:update_weights} of the $(k+1)$th iteration, as mentioned earlier, or after step \ref{stp:normalize_weights} of the $k$-th iteration.

The choice may vary depending on the source, seemingly with an inclination toward the latter option, which we will therefore opt for. However, this is a matter of taste, since either way the samples are based on the same information and available when needed. The difference is otherwise not very relevant; fresh samples are unlikely to be a necessity in the end of the last iteration as in the beginning of the first one.

With this option, at the end of each iteration we get new samples for the following one according to the aforementioned smoothed density:
\begin{equation}
    \theta^{(k+1)} \sim \sum_{i=1}^M w_i^{(k)}
    \mathcal{N}\left(
    m_i^{(k)}, s_k^2 
    \right)
\end{equation}

This describes the Liu-West \textit{resampler} or \textit{filter} (\glsxtrshort{LWF}). It is named after its two authors Liu and West, who both developed the underlying ideas and presented this particular algorithm \cite{Liu_2001}. The \textit{filter} designation comes from the fact that \gls{SMC} samplers are ordinarily called \textit{particle filters}, due to how they carry through particle selection. 

The method is summarized in algorithm \ref{alg:liu_west}. No bounds are enforced for the particles' motion, though one could easily exclude points without posterior support; the particles are not expected to stray much within this scheme, and are unlikely to move away from the originally covered region at all.

\begin{algorithm}[!ht]
\caption{Liu-West resampling algorithm. The calculations assume $\theta$ to be one-dimensional, but the generalization is straightforward.}
\textbf{Inputs}: weighted sample $\{\theta_i,w_i\}_{i=1}^M$, resampling parameter $a$.\\
\textbf{Computes}: new equally weighted positions.\\
\textbf{Assumes}: Ability to sample from multinomial and Gaussian distributions.\\
\textbf{Assures}: The new sample has the same mean and variance as the original one.

\begin{enumerate}
        \item Compute the current mean and variance of the particle cloud.
        \begin{multicols}{2}
        \noindent
        \begin{equation}
            \mu_k = \sum_{i=0}^{M} w_i \theta_i\\
        \end{equation}
        \begin{equation}
            \sigma^2 = \sum_{i=0}^{M} w_i
            \theta_i^2-\mu^2
        \end{equation}
        \end{multicols}
        
        Define
        \begin{equation}
            s = \sqrt{1-a^2} \cdot \sigma
        \end{equation}
    \item For $i \in \{1,\dots,M\}$, get a sample $\theta_i'$:
    
        \begin{enumerate}
            \item Sample a particle $\theta_s$ from the original set $\{\theta_i\}_{i=1}^M$ with probabilities given by their weights $\{w_i\}_{i=1}^M$ (i.e. a multinomial distribution), and calculate
            \begin{equation}
                \tag{\ref{eq:lw_mean}}
                m_i = a\theta_s + (1-a)\mu_k
            \end{equation}
            \item Sample $\theta_i'$ from a gaussian centered at $m_i$
            \begin{equation}
                \theta_i' \sim \mathcal{N}(m_i, s^2)
            \end{equation}
        \end{enumerate}     
    \end{enumerate}
\label{alg:liu_west}
\end{algorithm}

\gls{SMC} samplers making use of this resampling strategy have often been adopted for quantum characterization examples. The method is presented in detail in \cite{Granade_2012}, and applied and/or briefly discussed in many others (e.g. \cite{Wiebe_2014b,Granade_2017}). It has also been used in experimental implementations \cite{Wang_2017}. A strong point in favour of the Liu West resampler is that it doesn't require likelihood evaluations, which is rarely the case for other methods.

On the flip-side, it only preserves the first two moments of the distribution (mean and variance, in our case representing the estimate and uncertainty respectively), which is not always sufficient. In particular, it is utterly unfit for multimodal posteriors, except for the uninteresting case where $a=0$. This also means it may converge to wrong values in the case of redundancy, which it won't be able to recover from. Furthermore, it can't be expected to yield very accurate expectations. These problems may be relieved by using more particles and resampling less often, but this is a precarious solution; not only does it defeat the purpose of using such a lightweight method, it is also bound to fail in more complex scenarious. A more sophisticated resampling strategy will be mentioned in subsection \ref{sub:mcmc_smc}.

The behaviour with respect to other moments depends on the hyperparameter $a$, which should be adapted to the application, namely the characteristics of the likelihood function. This quantity negotiates a trade-off between variety and preservation. For $a=1$, we are back to the bootstrap filter, which draws random samples (with replacement) from the original particle locations. At the other extreme, choosing $a=0$ would amount to making a more drastic normality assumption: we would ignore the individual particle locations and sample from a Gaussian with the same mean and variance as the particle ensemble\footnote{It's worth noting that under such an assumption more lightweight strategies may be suitable, eliminating the need to store the entire particle cloud. For instance, \cite{Wiebe_2016} proposes a \gls{GRF} approach to phase estimation, where samples are produced as for $a=0$ and the reweightings (Bayes' updates) are replaced by a acceptance-rejection step. In that case, the particles can be treated sequentially and discarded after their contributions to quantities of interest have been accumulated, greatly easing the memory requirements. The mean and variance can be calculated from 4 common counters instead of $M$ particles. Results of the application of this strategy can be found in subsection \ref{sub:phase_estimation}.}.

High values of $a$ are customarily picked so as to mostly preserve the structure of the particle cloud. Most references above choose $a=0.98$, with the original authors applying an even higher value ($0.995$), which is fitting for models with pronounced non-Gaussianity. For others, a lower value may improve convergence; $0.9$ is suggested in \cite{Wiebe_2014b} for that reason. 

One more thing to observe is that the filtering interrupts the parallelism of the \gls{SMC} algorithm. While the particles can be separately reweighted (up to a constant factor) and propagated, the probabilities for the multinomial sampling step, which must be determined between these two tasks, depend on the whole particle group. In contrast, algorithm \ref{alg:grid_expectation} could be executed in parallel up to the normalization step, and algorithm \ref{alg:sequential_expectation} would be serialized only by the demands of the application; the normalizing constant's use can be deferred until an expectation needs to be computed.
 
Finally, it should be noted that not all sources define the scope of the term \textit{resampling} equally. It aims to emphasize the fact that we are reworking the sample we had to begin with, but what it entails may vary. In our case, equation \ref{eq:mixture_normals} suggests that we join the multinomial sampling together with the normal perturbations into a unified resampling step of which the bootstrap filter is a particular case.

Nevertheless, just as often the \textit{resampling} and \textit{particle propagation} or \textit{move} steps are split up in \gls{SMC} algorithms, the former referring exclusively to bootstrapping. In other words, \textit{resampling} can refer to the multinomial sampling alone \cite{South_2019,Gunawan_2020}, or it can integrate the variety-introducing element as well \cite{Granade_2012}. The first case commonly occurs when the fusion isn't as natural as it is in the case of our mixture density assumption (subsection \ref{sub:mcmc_smc}), which for convenience we sample from in two steps.

\subsection{Sequential importance resampling}
\label{sub:sir}

We are now nearly all set for our sequential importance (re)sampling algorithm, having elected a resampling strategy. But before adopting this new-found strategy, it is convenient to ask whether we actually want to use it.

In truth, it depends. Even leaving aside the chosen resampler's characteristics, we may not want to put it into action at every step of the algorithm. Instead of processing one datum at a time to get a weighted sample and immediately resampling, we may want to ponder whether to do so. Depending on the situation, and in particular on the effect of the fresh data, we could instead add a sequence of a few data before altering the particle locations; for that sequence, the weights would be updated directly as in algorithm \ref{alg:sequential_expectation}.

Such a wait is justified when these data have caused slow variations in the distribution, making the particles as-is fit to convey its picture. In that case, spending resources to shuffle the particles around is unnecessary - and possibly even harmful, since the overreliance on circumstantial evidence may result in their unsubstantiated relocation.

This is made evident by looking at figure \ref{fig:particle_degeneracy} (along with \ref{fig:successive_likelihoods}). After the first curve, which includes not one but \textit{five} data, resampling is uncalled for. The weights are still fairly even, which indicates a still representative sample. Only after the second curve - where 5 more data have been added, making up a total of 10 - have the weights become unbalanced.

Somewhere between the $5^{th}$ and the $10^{th}$ distributions, we may have benefited from a resampling step, which would have helped prevent the sample's being virtually reduced to a meager two particles; notwithstanding, to include one such step at every iteration may be excessive.

Motivated by this, we may want to opt for a \textit{dynamic} approach, where we decide whether to resample depending on the current particle cloud's state. A commonly used diagnostic criterion for weight disparity is the \gls{ESS}, or an approximation thereof. This is a widely used criterion, and is presented in e.g. \cite{Martino_2017}.

As the name suggests, the \gls{ESS} tells us how many exact samples our $M$ imperfect samples would amount to, thus assessing the statistical efficiency of the weighted sample. In general, it penalizes correlation; in our case, this is naturally expressed by the weights, as the dominance of some samples means they \textit{count for several} while being only one. 
\begin{equation}
    \widehat{\text{ESS}} = \frac{ \left( \sum_{i=1}^M W_i \right)^2}
    {\sum_{i=1}^M W_i^2} = \frac{1}{\sum_{i=1}^M w_i^2}
\end{equation}

With this definition, 
\begin{equation}
    1 \leq \widehat{\text{ESS}} \leq M
\end{equation}

After the initialization, we have $M$ particles with uniform weights, so $\widehat{\text{ESS}}=M$. The same holds immediately after a resampling step. For all other possible ways of assigning the weights, the \gls{ESS} is smaller: non-uniformity introduces a penalty. 

It is easy to see that if $M_X$ of the $M$ particles have among them uniform weights $1/M_X$, and the other $(1-M_X)$ have zero weights, the \gls{ESS} is $M_X$, which makes sense since only the non-zero weight particles contribute to the estimates. For instance, if all but two particles have collapsed, and those two particles have each a weight of $1/2$, the sample size is 2, as should be. In general, the more drastic the weight inequality, the lower the \gls{ESS}. This measures the progressive deterioration of our initially even sample: a low sample size is symptomatic of particle degeneracy.

Generally, some threshold is chosen for the \gls{ESS}, below which a resampling step is triggered. This threshold is often given in terms of the original particle number, $M_\text{thr}=\epsilon M$. In \cite{Doucet_2009,Granade_2012} and many others a value of $\epsilon=0.5$ is suggested, though sometimes more conservative values are imposed.

It should be noted that the \gls{ESS} is not always controlled through adaptive resampling. There are alternative ways of keeping it around or above a target value, such as choosing the sequential distributions adaptively based on a \textit{look-ahead} into prospective posteriors \cite{Gunawan_2020}. Such alternatives do however tend to be more resource-intensive, whereas this one barely adds to the overall cost of the algorithm.

Finally, we point out that alternative resampling criteria may be used, e.g. the entropy of the weights \cite{Doucet_2009}.

Algorithm \ref{alg:sir} lays out the steps of the \gls{SIR} algorithm. It also goes by the name of sequential importance sampling (SIS) or a more wordy sequential importance sampling and resampling (SISR). 

\begin{algorithm}[!ht]
\caption{Sequential importance resampling (\gls{SIR}) algorithm (with the prior distribution as importance function). A ${(k)}$ superscript signals dependence of the marked quantity on the first $k$ data and no more.}
\textbf{Inputs}: data vector $\vec{D}$ (to be used incrementally: $\{D_1\}$, $\{D_1,D_2\}$,...,$\{D_1,D_2,\cdots,D_N\}$), number of particles $M$, resampling parameter $a$, threshold effective sample size $M_\text{thr}$.\\
\textbf{Computes}: successive lists of pairs $\{\theta_i^{(k)},w_i^{(k)}\}_{i=1}^M$ for (by order) $k \in \{1,\dots,N\}$ to be used in \ref{eq:sequential_expectation}. \\
\textbf{Assumes}:  ability to sample from the prior and to evaluate the likelihoods, access to a resampling routine.\\
\textbf{Assures}: The $k$-th list is computed in the $k$-th iteration, using only the datum $D_k$ and the particle history (namely the last generation of particles).
\begin{enumerate}
    \item Sample $\{\theta_i^{(0)}\}_{i=1}^M$ from the prior distribution and set $w_i^{(0)} = 1/M$ for all $i \in \{1,\dots,M\}$. Set $\{\theta_i^{(1)}\}_{i=1}^M=\{\theta_i^{(0)}\}_{i=1}^M$.
    
    \item For $k \in \{1,\cdots,N\}$:
    \begin{enumerate}
        \item For each particle $\theta_i^{(k)} \in \{\theta_i^{(k)}\}_{i=1}^M$, update the weight using the latest datum $D_k$.
        \begin{equation}
           W_i^{(k)} = w_i^{(k-1)} \cdot \mathbf{L}(\theta_i^{(k)} \mid D_k)
        \end{equation}
        Accumulate these quantities to get the normalization constant $C^{(k)}= \sum_{i=1}^M W_i^{(k)}$.

        \item For each $\theta_i^{(k)} \in \{\theta_i^{(k)}\}_{i=1}^M$, compute the normalized weights $w_i^{(k)}=W_i^{(k)}/C^{(k)}$.
        \item Calculate the approximated effective sample size
        \begin{equation}
            \widehat{\text{ESS}}_k = \left( \sum_{i=1}^M \left[ w_i^{(k)}
            \right]^2\right)^{-1}
        \end{equation}
        
        \item If $\widehat{\text{ESS}}_k < M_\text{thr}$, resample the particles $\{\theta_i^{(k)},w_i^{(k)}\}_{i=1}^M$ to get a new set $\{\theta_i^{(k)},1/M\}_{i=1}^M$.
        
        \item Set $\{\theta_i^{(k+1)}\}_{i=1}^M=\{\theta_i^{(k)}\}_{i=1}^M$. 

    \end{enumerate}

\end{enumerate}
\label{alg:sir}
\end{algorithm}

In spite of the fact that the exposition has been focused on a particular instance of the \gls{SIR} algorithm, the method allows far more generality. There are multiple things that could have been chosen differently, such as the sequence of distributions, the means of monitoring the sample's representativeness, the resampler itself, and others.

In particular, we took the importance distribution to be the prior (a tentative latest posterior), and highlighted the convenience of this choice. Even though this is the most widely used strategy, it should be mentioned that others exist - and more importantly, they do while still enabling recursion in the weight computations. That is, a different choice won't necessarily involve reanalysing the accumulated data at each iteration.

The mechanism behind this is fairly unsurprising: it suffices to take a slightly broader view of what we just did. Just like we used the particle cloud to leave only the last datum's likelihood in equation \ref{eq:reweight_uniform}, we can use it to \textit{deduct} the prior probability from the calculations, even if we don't want to appoint it directly as the importance distribution.  We can formalize our particle representation of the prior by means of a transition equation:
\begin{equation}
    \label{eq:transition_eq}
    \mathbf{P}(\theta^{(k+1)} \mid \theta^{(k)})
\end{equation}

This integrates our Markovian model, dictating the evolution of the states. In our case, it encodes our particle-based representation of the current distribution: the prior probability at the $(k+1)$th iteration. The point is that this estimate allows us to cancel out the contributions of previous data (within our model; outside of it, this annulment is approximate).

With this, we can jump from one iteration to the next and maintain some continuity, whatever be our importance function. This is done (as before) by attributing an individualized importance distribution to each particle - for instance, by placing a kernel at its position. 
\begin{equation}
    \theta_i^{(k+1)} \sim \mathbf{P}(\cdot
    \mid \theta_i^{(k)})
\end{equation}

In the following iteration, since the new particles are generated from the old ones and exist in one-to-one correspondence, the weights can be reused. That of each particle is inherited by its offspring, and the new weights can be calculated - \textit{updated} - as:
\begin{equation}
    \label{eq:transition_weights}
    w_i^{(k+1)} \propto w_i^{(k)} 
    \frac{\mathbf{L}(\theta_i^{(k+1)} \mid D_{k+1}) \cdot \mathbf{P}(\theta_i^{(k+1)} \mid \theta_i^{(k)})}
    {\pi(\theta_i^{(k+1)} \mid \cdots)}
\end{equation}

\noindent, where the importance function can be conditioned both on the data and on the particles.

In our case, it precisely corresponded to \ref{eq:transition_eq}, yielding:
\begin{equation}
    w_i^{(k+1)} \propto w_i^{(k)}\cdot 
    \mathbf{L}(\theta_i^{(k+1)} \mid D_{k+1})
\end{equation}

For uniform weights (e.g. after applying a bootstrap filter), this reduces back to equation \ref{eq:reweight_uniform}. If furthermore we use gaussian kernels in the transition density, we can recover our previous resampling strategy. If on the other hand we interpret the discrete sample of the prior more literally and use Dirac-delta kernels (as per \ref{eq:distribution_pulses}), this takes us back to the basic reweighting we started out with in algorithm \ref{alg:sequential_expectation}, with no repositioning. Other approaches can be found in the literature; for instance, \cite{Daviet_2016} suggests using a (leave-one-out) \glsxtrfull{KDE} as the transition density for the weight updates, despite not using kernel smoothing when refreshing the particle locations. If the particle cloud is a poor representation of the previous distribution (due to failure to approach equilibrium), this may allow it to recover for the subsequent iterations.

In conclusion, the \gls{SIR} algorithm systematizes approximate inference in a flexible manner, without overt application-dependent restrictions (namely, without assumptions about the distributions, excluding those imposed by the employed resampler). By bringing the state of knowledge up to speed as soon as possible, it at once enables adaptive experimental design - or other \textit{online} assessments - and conducts the discretization so as to preserve its aptness, effectively revising it according to the turns of events. In doing so, it tackles the issues brought about by finite numerical precision, averting the unavoidable fate of any \textit{a priori} discretization: powerlessness in the face of ever-growing information.

 \subsection{Tempered likelihood estimation}
\label{sub:tle}

In the previous section, we considered targeting a sequence of increasingly narrow distributions. The benefits were evident: adaptivity played a pivotal role in helping the particle cloud keep up with this narrowness, acting as a stand-in for a much larger amount of resources. 

It was this breaking up of the inference process what allowed us to bring a resampler into action when needed. As we saw, this isn't necessarily always. Sometimes, progress happens too slowly to warrant it, so we may want to perform several updates before acting on it. That was considered when conceiving the algorithm; yet, we failed to consider the opposite extreme. What if progress happens too \textit{fast} to act on it?

This is even more dangerous, because taking a too big leap may lead to particle degeneracy and taint the results. Unfortunately, it is also harder to solve, because we don't have as much freedom at the other end of the spectrum. We considered adding data one by one, which is the slowest achievable pace in this setting. However, this definition of the minimum time-step is rather arbitrary, in the sense that we don't necessarily exert control over the effect each datum induces.

Generally, the data don't make guarantees about the mildness of their alterations to the parameter space landscape. If in two different problems there is a ten-fold difference in an individual datum's impact, in one of them the resampling schedule is ten times more coarse-grained. This may well mean that resampling in between data isn't enough.

In the precession example, the effect of each datum could be manipulated with relative ease, since we could choose the evolution time in a continuum and it affected the posterior in a predictable manner. Sadly, this is not always the case. Often the likelihood function is much too complex to anticipate how it will behave, and the freedom in choosing experimental controls isn't enough to moderate its influence on the distribution. 

Moreover, other hard to solve intricacies can arise. Under the presence of multimodality, be it intrinsic or due to redundancy, there is a danger of localized particle depletion. We can think of a process that randomly switches between two unknown regimes $A$ and $B$, corresponding to two different parameter specifications and thus two different asymptotic modes. Further, by hypothesis, the processes induce close to mutually exclusive observations, conceivably because the modes are far apart.

Let's say that we collect a first datum while regime $A$ was operative, and that it presents evidence in favor of a region containing the parameter vector associated with $A$. The particles stationed close to that of $B$ will get zero or near zero weights, and are likely to be driven away from it. As a result, even if we next get a datum that would uphold $B$, there won't be support in the distribution to register that. Clearly, our approach is ill-suited to this type of scenario, and will do very poorly. 

Again, this didn't present much of an issue in the precession frequency case: not only was multi-modality periodic and unlikely to ever exclude the true parameter, it also tended to arise for longer times, which we could handle with care. But naturally, we can't expect this to come to pass in other applications; and even in this one, such a heedless choice may trap the sampler in local minima.

Such problems can be solved by choosing the sequence of target distributions more carefully. A popular solution is known as \textit{tempered} or \textit{annealed} likelihood estimation (\gls{TLE}), presented in detail in \cite{Neal2001} and briefly described in e.g. \cite{Del_Moral_2006,South_2019}. In this approach, we consider a sequence of $S$ target distributions (not counting the prior) that are nearly proportional to powers of the posterior:
\begin{equation}
    \mathbf{P}_s(\theta) \propto \mathbf{L}(\theta \mid \vec{D})^{\gamma_s}\mathbf{P}(\theta)
    \quad , \ 0 \leq \gamma_s \leq 1
\end{equation}

The goal being to smooth the function and the prior being smooth already, making it more level through an exponent on $\mathbf{P}(\theta)$ would serve no purpose; as such, this treatment is reserved for the likelihood, which is the challenging item. 

The exact distributions are naturally given by:
\begin{equation}
    \mathbf{P}_s(\theta) =
    \frac{\mathbf{L}(\theta \mid \vec{D})^{\gamma_s} \mathbf{P}(\theta)}{\int \mathbf{L}(\theta \mid \vec{D})^{\gamma_s} \mathbf{P}(\theta) \mathrm{d}\theta} 
    \quad, \ s \in [0..S]
\end{equation}

The $\gamma_s$ are the tempering or annealing coefficients, also called the (inverse) \textit{temperatures}. They must satisfy:\begin{equation}
    \gamma_0 = 0 < \gamma_1 < \dots < \gamma_S = 1
\end{equation}

For $s=0$ we have our prior distribution as before, $\mathbf{P}(\theta)$. For $s=1$ we have our posterior distribution (also as before), $\mathbf{P}(\theta \mid \vec{D}) \propto \mathbf{L}(\theta \mid \vec{D})^{\gamma_s} \mathbf{P}(\theta)$.

The $s$ indices are chosen to avoid confusion with the $k$s from before. In that case, a datum $D_k$ was assigned to  each iteration $k$. Now, \textit{all} $D_k$ for $k \in \{1,\dots,N\}$ are used at \textit{each} iteration, and the sequential distributions differ solely in the exponentiation of their common likelihood function. 

Using algorithm \ref{alg:sir} as a reference, the only thing that changes is the weight updating step. As before, to jump from the $s$-th iteration to the $(s+1)$th one, we must multiply by $\mathbf{P}_{s+1}(\theta)$ and divide by $\mathbf{P}_s(\theta)$ (to compensate for the samples' being distributed according to the latter, as in equation \ref{eq:monte_carlo_importance}). 
\begin{equation}
    w_i^{(s+1)} \propto w_i^{(s)}\cdot  
    \frac{\mathbf{P}_{s+1}(\theta_i^{(s+1)})}
    {\mathbf{P}_{s}(\theta_i^{(s+1)})} \propto
    w_i^{(k)}\cdot \mathbf{L}(\theta_i^{(s+1)} \mid \vec{D})^
    {\gamma_{s+1}-\gamma_s}
\end{equation}

Again, we don't need to worry about scale factors, which cancel out when normalizing. We only ever need to evaluate each of the distributions up to a constant.

With this, the distribution starts out flattened, and is gradually made less so; by slowly increasing the exponent to $1$, we lift the posterior up to its final form. The tempering coefficients being real numbers, they can be chosen continuously unlike before, and they have a direct impact on the distribution's shape.

The term \textit{annealing} takes a similar sense as in \textit{simulated annealing}, an approximate optimization strategy which in turn borrows its name from metallurgy. In common they have the fact that they concern some \textit{particles}, physical or otherwise, that undergo random motion. In the beginning this randomness is more yielding, and they are allowed to move around quite freely; but as time passes, they are more and more bound to the desirable sites they stumble upon. In the case of the algorithms, the point is that exploration should be comprehensive at first, then progressively scale down.

In the other two cases, this is done by decreasing a \textit{temperature}, the reduction in kinetic energy resulting in decreased mobility. In annealed likelihood estimation, it's almost as if we have an each time stronger potential energy (associated with the likelihood), which pulls the particles towards higher probability locations - more the more advanced the iteration is.

As for the idea of \textit{tempering}, it is perhaps even more suitable. We can think of tempering a sauce, where we want to add in an ingredient that doesn't combine easily. In our case this ingredient is the likelihood, whereas the prior is the premade sauce. If we add in the likelihood all at once (to get the posterior), we will likely observe poor mixing. Instead, we add just a little bit of the likelihood, folding it in as we go; that is, consolidating the distribution through the iterations by refining how we discretize it. Although these distributions will at first still resemble the prior more than the posterior, each tempering step makes the two parts slightly more compatible, until we can finally mix them together in full.

In short, this new scheme ensures that the distribution evolves consistently and controllably at all points, whereas the old one was merely based on an empirical idea. That being said, it was a \textit{very} convenient idea: it fixed the number of likelihood evaluations for the reweightings at $N \cdot M$ ($N$ is the number of data, and $M$ that of particles), which for \gls{TLE} becomes $N \cdot M \cdot S$ for some number of iterations $S$. Also, the latter algorithm is clearly unsuited for online estimation.

Regardless, it is quite more robust and flexible than sequential updating; for that reason, it is a fairly popular strategy for dealing with target posteriors that are difficult to simulate, namely in high-dimensional spaces and in the presence of multimodality. In addition, a more meticulous strategy may increase the accuracy of the estimates produced by the samples - namely those regarding the sought-after marginal likelihood, or equivalently the model evidence (subsection \ref{sub:model_selection}).

Since it tends to be applied in more demanding contexts, the \gls{TLE} approach is usually paired with more sophisticated move steps than the ones mentioned so far. In particular, \gls{LWF} is unfit for multimodal distributions, one of the main motivating scenarios. Though work-arounds can be used to solve this, such as an intermediate clustering step \cite{Granade_2017}, they can be computationally demanding while not addressing the underlying problem. In fact, failure in the face of multimodality is purely a symptom betraying an overall context of insufficient property preservation. This same context is likely to bias estimates, including those of the model evidence - hence, a high accuracy variety introducing mechanism is decisive for the reliability of expectations. The most widely used strategy will be presented later in subsection \ref{sub:mcmc_smc}.

These alternatives do however tend to be quite more computationally demanding, which is amplified by the complexity of the distributions they target in \gls{TLE}. More often than not, they require likelihood evaluations of their own, to be performed independently for each particle. Expectably, this aggravates the exigency of the protocol, further widening the gulf between the two algorithms.

All of the aforementioned requirements may be to a degree relaxed by making use of subsampling strategies, which can take advantage of the increased regularity of the sequential distributions; this will be considered in subsection \ref{sec:subsampling}. Unbiased estimators coupled with variance reduction techniques may reduce the number of necessary likelihood evaluations while largely preserving accuracy, making the method at once more efficient and more parallelizable \cite{Gunawan_2020}.

\newpage
\subsection{General SMC schemes and estimation of the model evidence}

Estimators for the model evidence, or the marginal probability, have been a recurring theme. The reason for this is clear. They aren't simply a normalizing constant, but rather one that assesses a model's worth - including \textit{comparative} worth, with all that entails. This brings up an important matter: at some point along the chapter, a means of calculating it was lost.

In subsection \ref{sub:estimation_in_practice}, we found that the marginal, $\mathbf{P}(\vec{D})$, could be approximated using the sum of the unnormalized weights:
\begin{equation}
    \sum_{i=1}^M W_i(D) \equalhat \frac{M}{V} \mathbf{P}(\vec{D}) 
\end{equation}

This provides instant access to one of the most important estimates in Bayesian inference, dispensing with the extra calculations - including likelihood evaluations - that appealing to \ref{eq:expectation_batch} would. Algorithm \ref{alg:grid_expectation} is compatible with such a calculation, and so too is algorithm \ref{alg:sequential_expectation}. But for algorithm \ref{alg:sir} and the likes of it, the same no longer holds. 

In the first one, the normalizing constant was directly available, because it was calculated at once. In the second, we normalized at each step, so the cumulative product of multiplying factors would do. In the third, things are no longer so clear.

Algorithm \ref{alg:sequential_expectation} already presents a minor complication: at each iteration, we must sum the likelihoods used to update the weights in equation \ref{eq:update_weights}, and not the resulting weights themselves. Each datum's probability - apart from a constant factor - is calculated at the corresponding step by removing the influence of the previous iterations. This is of course unless we can delay normalizing until the end, in which case we'd be back to algorithm \ref{alg:grid_expectation}. Normalization is only necessary when evaluating expectations.

When we move on to \gls{SIR}, even this is out of reach. The problem lies with the resampling stage. By transforming weight into particle density, it disrupts the scheme. While this \textit{disruption} is intentional, in this context it means that the weights no longer stand on their own as uniform samples. After the first resampling step, we can't \textit{remove the influence} of the previous iteration anymore, because it is ingrained on the particle cloud. The available information is fused with the particles' attributes (locations and weights); if the original idea was to use nothing but their weights, their locations now contain irreplaceable information.

Simply put, we would want to calculate the expectations over the prior, but can only average over the current distribution $\mathbf{P}_k(\theta)$. And this is in the case of algorithm \ref{alg:sir}, where the sequence of target distributions is the same as in \ref{alg:sequential_expectation}; for other sequences of distributions, the weights don't even begin to resemble the data's probabilities.

To see if this can be salvaged, we can start with a generic view of \gls{SMC} where we consider the normalizing constant explicitly. The target is always a sequence of probability distributions; let's say, of length $T+1$. They can be written as:
\begin{equation}
    \mathbf{P}_t(\theta) = 
    \frac{\eta_t(\theta)}{Z_t} \quad, \ t \in \{0,\dots,T\}
\end{equation}

\noindent, having defined the normalizing constant:
\begin{equation}
    Z_t \equiv \int \eta_t(\theta) \mathrm{d}\theta
\end{equation}

This covers all possibilities, including of course the previously examined cases. In algorithm \ref{alg:sir} the $\eta_t(\theta)$ were given by:
\begin{equation}
    \eta_t(\theta) \equiv \mathbf{L}(\theta \mid \{D_1,\dots,D_t\})
    \mathbf{P}(\theta)
\end{equation}

While for \gls{TLE}:
\begin{equation}
    \eta_t(\theta) \equiv \mathbf{L}(\theta \mid \vec{D})^{\gamma_t}
    \mathbf{P}(\theta)
\end{equation}

In general, the reweightings abide by:
\begin{equation}
    w_i^{(t)} \propto w_i^{(t-1)}\cdot 
    \frac{\eta_{t}(\theta_i^{(t)})}
    {\eta_{t-1}(\theta_i^{(t)})}
\end{equation}

\noindent, since the $Z_t$ constants (this constancy being of course relative to $\theta$) are rendered irrelevant by normalization. More explicitly,
\begin{gather}
    \label{eq:reweighting_eta}
    W_i^{(t)} = w_i^{(t-1)}\cdot 
    \frac{\eta_{t}(\theta_i^{(t)})}
    {\eta_{t-1}(\theta_i^{(t)})}\\
    w_i^{(t)} = \frac{W_i^{(t)}}{\sum_{j=1}^M W_j^{(t)}}
\end{gather}

With this, expectations can be computed at each step as:
\begin{equation}
    \label{eq:expectation_eta}
    \mathbb{E}_{\mathbf{P}_t(\theta)}\left[f(\theta)\right]
    = \int f(\theta) \mathbf{P}_t(\theta)\mathrm{d}\theta
    \approx \sum_{i=1}^{M} f(\theta_i^{(t)}) \cdot w_i^{(t)}
\end{equation}

If a resampling stage does not take place at the end of the $t$-th iteration,the $(t+1)$th particle set is unchanged: $\{\theta_i^{(t+1)}\}_{i=1}^M=\{\theta_i^{(t)}\}_{i=1}^M$. If on the contrary a resampling mechanism is activated, $\{\theta_i^{(t)}\}_{i=1}^M$ is recast to get $\{\theta_i^{(t+1)}\}_{i=1}^M$. This preparation aims to introduce variety for the following iteration, and preserves $\mathbf{P}_t(\theta)$ (or a selection of relevant attributes). 

Thus, taking into account the fact that resamplers typically have nice convergence properties, we may as well write expectation \ref{eq:expectation_eta} as:
\begin{equation}
    \label{eq:expectation_eta_res}
    \mathbb{E}_{\mathbf{P}_t(\theta)}\left[f(\theta)\right]
    \approx \sum_{i=1}^{M} f(\theta_i^{(t+1)}) \cdot w_i^{(t)}
\end{equation}

With this, the main ideas for a generic \gls{SMC} inference algorithm are established.

Our interest lies in the normalizing constants $Z_t$. As long as the final distribution is the posterior $\mathbf{P}(\theta \mid \vec{D})$, then $Z_T$ is the marginal.
\begin{equation}
    Z_T = \mathbf{P}(\vec{D}) = \int 
    \mathbf{L}(\theta \mid \vec{D})
    \mathbf{P}(\theta) \mathrm{d}\theta
\end{equation}

At the other end, $Z_0$ is just the integral of the prior, which is 1 for any valid distribution.
\begin{equation}
    Z_0 = \int \mathbf{P}(\theta) \mathrm{d}\theta = 1
\end{equation}

Thus we can write:
\begin{equation}
    \mathbf{P}(\vec{D}) = \frac{Z_T}{1} = \frac{Z_T}{Z_0}
\end{equation}

Which in turn can be written as:
\begin{equation}
    \label{eq:ratio_evidence}
    \mathbf{P}(\vec{D}) = \frac{Z_T}{Z_0} = \prod_{t=1}^{T} \frac{Z_{t}}{Z_{t-1}}
\end{equation}

\noindent, since all the intermediate terms cancel out when expanding.

Now we need to get the individual terms, or some estimate thereof:
\begin{equation}
    \widehat{\left( 
    \frac{Z_{t}}{Z_{t-1}}
    \right)}
\end{equation}

This is written in terms of two consecutive distributions, which is promising since we were initially displeased with their closeness. Developing the expression further, we can write the term corresponding to the $t$-th iteration as an average over the $(t-1)$th distribution, which is closer to what we wanted:
\begin{gather}
    \left( 
    \frac{Z_{t}}{Z_{t-1}}
    \right)
    = \frac{\int \eta_t(\theta) \mathrm{d}\theta}{Z_{t-1}}
    = \int \frac{\eta_t(\theta)}{Z_{t-1}} \mathrm{d}\theta
    \nonumber \\
    = \int \frac{\eta_t(\theta) }{\eta_{t-1}(\theta)}
    \cdot \frac{\eta_{t-1}(\theta)}{Z_{t-1}} \mathrm{d}\theta
    = \int \frac{\eta_t(\theta)}{\eta_{t-1}(\theta)}
    \mathbf{P}_{t-1}(\theta)\mathrm{d}\theta
\end{gather}

At this point we fall back on \ref{eq:expectation_eta_res}.
\begin{equation}
    \left( 
    \frac{Z_{t}}{Z_{t-1}}
    \right)
    = \int \frac{\eta_t(\theta)}{\eta_{t-1}(\theta)}
    \mathbf{P}_{t-1}(\theta)\mathrm{d}\theta
    \approx \sum_{i=1}^{M} \frac{\eta_t(\theta_i^{(t)})}
    {\eta_{t-1}(\theta_i^{(t)})} 
    \cdot w_i^{(t-1)}
\end{equation}

Recognizing \ref{eq:reweighting_eta}, this becomes:
\begin{equation}
    \label{eq:ratio_estimator}
    \left( 
    \frac{Z_{t}}{Z_{t-1}}
    \right)
    \approx \sum_{i=1}^{M}  W_i^{(t)}
\end{equation}

The finalized estimate is obtained by plugging \ref{eq:ratio_estimator} into \ref{eq:ratio_evidence}.
\begin{equation}
    \label{eq:marginal_smc}
    \mathbf{P}(\vec{D}) 
    \approx \prod_{t=1}^{T} \left( \sum_{i=1}^{M}  W_i^{(t)} \right)
\end{equation}

As it turns out, the marginal probability can still be estimated using nothing but sums of unnormalized weights - only now some extra work is required to see why and how.

Also, in this case there isn't a missing $V/M$ factor, which is made clear by the fact that our derivation is correct. It was based on approximate expectations, which correspond to perfectly proper estimates, though \ref{eq:marginal_smc} may make it seem like they are equivalent for the single step, single datum case.

Looking at the final formula only, the fundamental difference in this regard is brought about by the fact that at the first step we normalized after sampling from the prior, which exerts an influence in the product of \ref{eq:marginal_smc}. If we had just sampled from the (flat) prior, the starting weights would be $1/V$ and sum to $M/V$. Since we normalized too, by attributing uniform weights, they are $1/M$ and sum to $1$. Of course, the correctness holds for any choice of a prior distribution, as the proof is clearly inclusive.

This restores the ability to obtain the model evidence basically \textit{for free} along with the adoption of \gls{SMC}. Because the act of determining the sums in \ref{eq:marginal_smc} is intrinsic to the algorithm (e.g. the $C^{(k)}$ in algorithm \ref{alg:sir}), one need only retrofit it with a calculation of their cumulative product.

\section{Markov Chain Monte Carlo and the  Metropolis-Hastings algorithm}
\label{sec:mcmc}

\glsxtrfull{MCMC} comprises a family of methods that rely on Markov chains to produce samples from a probability distribution. In its simplest form, the construction of such a chain rests only upon the ability to evaluate the probability at specific points; more advanced methods may require gradient information as well. For this section, the main references are \cite{Andrieu2003,Speagle_2020,Betancourt_2018}.

This ties into what has been discussed so far. It had been noted that, if we could sample from the posterior directly, then to compute its associated expectations those samples would be all we needed from it. It would suffice to evaluate the function whose expected value is to be calculated at those samples. This can be seen as an alternative to \ref{eq:expectation_batch}, and a special case of \ref{eq:monte_carlo_importance} - which for the case of posterior expectations becomes:
\begin{equation}
    \label{eq:importance_expectation_posterior}
    \mathbb{E}_{\mathbf{P}( \theta \mid \vec{D})}\left[f(\theta)\right]
    = \int f(\theta) \mathbf{P}(\theta \mid \vec{D})\mathrm{d}\theta
    \approx \frac{1}{M} \sum_{i=1}^{M} \frac{f(\theta_i) \cdot 
    \mathbf{P}(\theta_i \mid \vec{D})}{\pi (\theta_i)}
    \quad , \ \theta_i \sim \pi (\cdot) 
\end{equation}

The choice of the posterior as importance distribution materializes our idea.
\begin{equation}
    \label{eq:posterior_mcmc}
    \mathbb{E}_{\mathbf{P}( \theta \mid \vec{D})}\left[f(\theta)\right]
    \approx \frac{1}{M} \sum_{i=1}^{M} f(\theta_i)
    \quad , \ \theta_i \sim \mathbf{P}(\theta \mid \vec{D})
\end{equation}

This is a somewhat familiar idea, since strategies reminiscent of this one were used all throughout section \ref{sec:sequential_monte_carlo}. First, we used a similar approach to cancel out the most part of the influence each distribution had in its corresponding weight updates, effectively removing the overlap of information with their predecessors. And second, we ultimately used this very approach, it being precisely the end goal of \gls{SMC}; the last iteration produces samples from the posterior. Accordingly, \ref{eq:sequential_expectation} is just an execution of \ref{eq:posterior_mcmc} (for $k=N$ and uniform weights).

There, we adopted a group of particles that we collectively propagated, allowing them to \textit{communicate} at several points while traversing through a sequence of distributions that little by little morphed into the posterior. The difference is that now we will pick a single \textit{particle} (which isn't usually called a particle, but rather a state) and take it on a lone trip, with the purpose of of producing samples (consecutive states) from a fixed distribution. Naturally, this distribution is the posterior.

Both methods are sequential in a sense, but the sense is not the same. In \gls{SMC}, the fundamental sequentiality comes from targeting a sequence of progressive distributions; the samples from the posterior are produced all at once in the last step, as are those from all intermediate distributions in their respective steps. In \gls{MCMC}, a Markov chain is sequential in the sense that it produces samples one after the other, but they all respect the same distribution. There isn't a series of distributions, only one of samples.

To sample from the posterior without slowly warming up to it as we did before, we must build a personalized Markov chain. This describes a stochastic process that travels through a sequence of states, in our case points in parameter space. The transition from one state to another is probabilistic; importantly, it depends only on the original state. In other words, the chain has no memory. Its current state fully determines the probability distribution of states it may transition to next. We write the probability of moving from a current state $\theta^{(i)}$ to some prospective $\theta^{(i+1)}$ as:
\begin{equation}
    T(\theta^{(i+1)} \mid \theta^{(i)})
\end{equation}

The shift of the $i$ index to a superscript illustrates the contrast between \gls{SMC} and \gls{MCMC}. 

Thus, if the chain is at state $\theta^{(i)}$, the following state can be obtained by sampling:
\begin{equation}
    \theta^{(i+1)} \sim T(\cdot \mid \theta^{(i)})
\end{equation}

For our purposes, we need a chain whose equilibrium distribution is the posterior distribution. For that, it must of course be invariant under the posterior. That is, if we have a group of samples from the posterior and perform a Markov transition on each (using parallel Markov chains), we should get another such group. If we instead perform multiple transitions on the same sample, the history of states will also constitute a sample from the posterior, and we can in theory obtain expected values by averaging over the registered trajectory of the Markov chain.
\begin{equation}
    \label{eq:expectation_mcmc}
    \mathbb{E}_{\mathbf{P}( \theta \mid \vec{D})}\left[f(\theta)\right]
    \approx \frac{1}{M} \sum_{i=1}^{M} f(\theta^{(i)})
    \quad , \ \theta^{(i)} \in \mathbf{t}_\text{\tiny MC}
\end{equation}

A sufficient condition for having some distribution $\mathbf{P}_X(\theta)$ as stationary distribution of the Markov chain is that of \textit{detailed balance}, or reversibility:
\begin{equation}
    \label{eq:detailed_balance}
    T(\theta' \mid \theta) \mathbf{P}_X(\theta)
    = T(\theta \mid \theta') \mathbf{P}_X(\theta')
\end{equation}

Essentially, this says that the probability of observing a transition from state $\theta$ to state $\theta'$ or vice-versa should be the same. In our case $\mathbf{P}_X(\theta) = \mathbf{P}(\theta \mid \vec{D})$. It's interesting to note the similarity between this equation and that which gave rise to Bayes' rule back in chapter \ref{cha:quantum_parameter_estimation}. 

The condition \ref{eq:detailed_balance} can easily be achieved using rejection sampling. We divide our transition distribution into a proposal $q$, and an acceptance/rejection step $a$:
\begin{equation}
    T(\theta' \mid \theta)  
    = q(\theta' \mid \theta) \cdot a(\theta',\theta)
\end{equation}

We can preserve the freedom of choosing the proposal distribution by tailoring the acceptance probability to it, effectively performing a \textit{ correction}. 
\begin{equation}
    \label{eq:acceptance_probI}
    a(\theta',\theta) = 
    \frac{q(\theta \mid \theta') \mathbf{P}_X(\theta')}
    {q(\theta' \mid \theta) \mathbf{P}_X(\theta)}
\end{equation}

With this, it is easy to see that equation \ref{eq:detailed_balance} holds.
\begin{gather}
    T(\theta' \mid \theta) \mathbf{P}_X(\theta)
    = q(\theta' \mid \theta) a(\theta',\theta) \mathbf{P}_X(\theta)\\
    = q(\theta' \mid \theta) 
    \frac{q(\theta \mid \theta') \mathbf{P}_X(\theta')}
    {q(\theta' \mid \theta) \mathbf{P}_X(\theta)}
    \mathbf{P}_X(\theta)
    = T(\theta \mid \theta') \mathbf{P}_X(\theta')
\end{gather}

A caveat is that $a$, being a probability, mustn't be higher than one, and the ratio in \ref{eq:acceptance_probI} guarantees no such thing. We can patch this by having an exception for the case where the ratio exceeds unity, bringing it down to $1$. 
\begin{equation}
    \label{eq:acceptance_probII}
    a(\theta',\theta) = \min \left(1,
    \frac{q(\theta \mid \theta') \mathbf{P}_X(\theta')}
    {q(\theta' \mid \theta) \mathbf{P}_X(\theta)}
    \right)
\end{equation}

This solution is called the \glsxtrfull{MH} algorithm \cite{Metropolis_1953,Hastings_1970}; there are other methods still within \gls{MCMC}.

An interesting thing to note is that the theoretical framework is unaffected by constant factors, which cancel out in the ratio as well as in all the equations. In particular, the probability distribution doesn't need to be normalized. This is a crucial point, because when we let go of the multi-particle arrangement of \gls{SMC} we lost our means of calculating the normalizing constant for the posterior. This way, we can still formulate the transitions by evaluating the likelihood and the prior only. 

Similarly, we lost some ease in assessing convergence and robustness when changing methods. In \gls{SMC}, the congruity of the particle cloud conveniently briefed us on where it stood, whereas an individual chain offers no such measure. This too can be fixed, though less directly, by starting multiple chains at different locations and running them in parallel. The ensemble can then used to extract statistics\footnote{Similarly, if the intel provided by the particle cloud in \gls{SMC} is insufficient, the particles too can be split into groups that evolve independently of each other, an their outcomes compared \cite{Daviet_2016}.}. In particular, their proximity is used to evaluate consistency: large discrepancies should raise suspicion. Such a scheme resembles \gls{SMC}, but it is completely parallelizable because the particles don't interact . 

Having laid out these ideas, we are ready to construct a theoretically sound Markov chain. The procedure is described in algorithm \ref{alg:mcmc}. 

\begin{algorithm}[!ht]
\caption{Metropolis-Hastings algorithm for posterior sampling in Bayesian inference.}
\textbf{Inputs}: starting state $\theta^{(0)}$.\\
\textbf{Computes}: sequence of states
$\mathbf{t}_\text{MC} = \{\theta^{(i)}\}_{i=1}^M$ to be used in \ref{eq:expectation_mcmc}.\\
\textbf{Assumes}: Ability to evaluate the prior, the likelihoods and a predefined proposal density $q(\theta' \mid \theta)$, and additionally to sample from the latter and from a binomial distribution.\\
\textbf{Assures}: The samples are distributed according to the posterior distribution as  $M \rightarrow \infty$.
\begin{enumerate}
    \item For $i \in \{1, \cdots, M\}$:
    
    \begin{enumerate}
        \item Sample a proposal
        \begin{equation}
            \theta^* \sim q(\cdot \mid \theta^{(i-1)})
        \end{equation}
        \item Compute the acceptance probability
        \begin{equation}
            a(\theta^*,\theta^{(i-1)}) = \min \left(1,
            \frac{q(\theta^{(i-1)} \mid \theta^*) 
            \mathbf{L}(\theta^* \mid \vec{D}) \mathbf{P}(\theta^*)}
            {q(\theta^* \mid \theta^{(i-1)}) 
            \mathbf{L}(\theta^{(i-1)} \mid \vec{D}) \mathbf{P}(\theta^{(i-1)})}
            \right)
        \end{equation}
        \item With probability $a(\theta^*,\theta^{(i-1)})$, set
        \begin{equation}
            \theta^{(i)} = \theta^*
        \end{equation}
        Otherwise, keep $\theta^{(i)} = \theta^{(i-1)}$.
    \end{enumerate}
\end{enumerate}

\label{alg:mcmc}
\end{algorithm}

As long as the stationary distribution is the correct one, the  Markov chain will converge as desired\footnote{This is an oversimplified idea; some transitions may cause the chain to become faulty due to complex geometrical attributes of the target distribution \cite{Betancourt_2018}. This may frustrate ergodicity, a characteristic which underpins \gls{MCMC} - thus invalidating estimates.}. As such, the starting point $\theta^{(0)}$ can be chosen at random, though it will have an effect on convergence. Likewise, we are free to choose a proposal distribution $q(\theta' \mid \theta)$, but it strongly affects the performance of the algorithm.

The chain \textit{will} eventually reach its steady state, or be close enough - at which point we get exact samples from the posterior, or accurate enough ones. Nonetheless, what this \textit{eventually} offers is scant assurance. If care is not taken, runtime until convergence may be too long for its fruits to ever be reaped. Even if this is not true, in practice two provisions may be considered for accuracy's sake:
\begin{itemize}
    \item A number of initial samples is to be discarded, because the chain won't have reached equilibrium yet. This is called a \textbf{burn-in} period.
    \item We may want to keep only every $L$th sample, discarding the ones in between. This is called \textbf{thinning}. The point is that consecutive samples will be correlated due to the localized proposals; this correlation can be made negligible at or after lag $L$ (for large enough $L$). Correlated samples may bias calculations. This is particularly relevant if a \textit{compact} set of samples is desired, due to e.g. memory or post-processing constraints \cite{Link_2011}.
\end{itemize}

Both of these points are greatly impacted by the proposal distribution. The time required to reach the stationary regime (or get close enough), called the mixing time, greatly depends on it, as it determines how fast the chain explores the space. For the same reason, it influences the serial correlation between samples. If it causes minuscule perturbations, the chain will mix slowly; however, countering this isn't as simple as opting for bold proposals.

Ideally, the transitions \textit{should} produce distant moves; but since the ratio $\mathbf{P}(\theta^*)/\mathbf{P}(\theta^{(i-1)}$) is a factor in the acceptance probability, unchecked distant proposals may suffer from high rejection rates, which is self-defeating since the chain will stall. This is of course not applicable if the proposals are well founded and can systematically find regions of high probability, but that how to generate such proposals is an open problem.

By now, it is clear that the single most important implementation detail is the proposal distribution, which takes a similar role to our resampler from before. It can affect convergence speed, computational efficiency, and even correctness. The following two sections we will go over two well-known approaches: random walk Metropolis, and Hamiltonian Monte Carlo. Both are mostly based on \cite{Betancourt_2018}, one of the main references listed for this section.

\subsection{Random walk Metropolis}
\label{sub:rwm}

A common simplification of the Metropolis-Hastings algorithm is called simply the \textit{Metropolis} algorithm. It describes the original method, which Wilfred Hastings later generalized it to the general case. The only difference is that the proposal distribution is chosen to be symmetric, i.e. 
\begin{equation}
    q(\theta' \mid \theta)  = q(\theta \mid \theta')
\end{equation}

With this, the acceptance ratio from \ref{eq:acceptance_probII} becomes:
\begin{equation}
    \label{eq:acceptance_prob_metropolis}
    a(\theta',\theta) = \min \left(1,
    \frac{\mathbf{P}_X(\theta')}
    {\mathbf{P}_X(\theta)}
    \right)
\end{equation}

This has a nice interpretation: each point's chances of being picked grow in proportion with the probability the target distribution assigns to it. Naturally, if we want to sample from a distribution, we mean to favor precisely the points that it favors, and that too with the insistence that it favors them.

In its earliest version, the algorithm utilized gaussian proposals centered at the starting state, which are obviously symmetric. 
\begin{gather}
    \label{eq:metropolis_proposal}
    \theta' \sim \mathcal{N}(\theta, \sigma) \\
    q(\theta' \mid \theta)  = \mathcal{N}(\theta' \mid \theta, \sigma)
\end{gather}

The standard deviation is often tuned to maintain the acceptance ratio around a pre-defined value, though more sophisticated methods can be employed \cite{Gelman_2007}.

Along with \ref{eq:acceptance_prob_metropolis}, \ref{eq:metropolis_proposal} defines a protocol which we will call \glsxtrfull{RWM}. It is based on stochastic exploration corrected for probability density. Thinking of the behaviour of such a strategy in response to an increase in dimensions, we can see how it balances the tension between probability density and parameter space volume (mentioned in section \ref{sec:monte_carlo}). On the one hand, the Gaussian generalizes to a multivariate normal distribution, which relays the behaviour of high dimensional spaces through the way the tails take up increasing volume. On the other, the correction of \ref{eq:acceptance_prob_metropolis} always privileges higher probability regions. Together, they will negotiate the intended compromise, focusing on the region that produces significant contributions to the integral.

On the downside, equation \ref{eq:metropolis_proposal} describes a random walk: the direction and speed of exploration are not customized to the target function. Due to this fact, the chain is likely to move in small steps, and further to often retrace them since there is no element of continuity between transitions (which there could still be in a Markovian process: it can by achieved via the target function's geometry rather than past states). 

This tends to worsen with increased dimensions. As the region that yields consequential contributions to the expectations becomes more and more concentrated and the volume outside of it grows, uninformed proposals become each time more inadequate. Either they are bold and likely to fall outside the narrow non-negligible probability region (wherefore they are rejected), or they are conservative and slow down the chain.

For simple problems, \gls{RWM} may still perform well within a reasonable span of time. Unfortunately, its issues can quickly get out of hand, bringing unacceptable inefficiency. In that case, the differential structure of the posterior distribution may aid the production of more educated guesses. This is a commonly exploited resource in more advanced \gls{MCMC} methods, namely Hamiltonian Monte Carlo\footnote{Another common approach is Langevin Monte Carlo, also called the Metropolis-adjusted Langevin algorithm, which is a particular case of Hamiltonian Monte Carlo \cite{Barbu_2020}.}.

\subsection{Hamiltonian Monte Carlo}
\label{sub:hmc}

Hamiltonian Monte Carlo (\gls{HMC}), originally called Hybrid Monte Carlo (also HMC) \cite{Duane_1987}, comes to solve the problems that a naive random walk brings, and reinstate the efficacy of \gls{MCMC} for high-dimensional spaces. This is approached at length in references \cite{Betancourt_2018}, which contains an intuitive overview on \gls{MCMC} methods with a focus on Hamiltonian Monte Carlo, and \cite{Neal_2011}, which is a more detailed review of \gls{HMC}. Finally, the reference manual and source code of the statistical inference platform \textit{Stan} \cite{Gelman_2015,Stan} presents most if not all of the state-of-the-art \gls{HMC} techniques, along with select bibliography in the first case. 

We have suggested that gradient information may come in handy, but its application is not so direct: we don't simply want to steer towards high probability regions, and so it necessitates some correction. This correction should be based on differential geometry, but a physical analogy provides a more accessible explanation. 

It is inspired in the Boltzmann distribution, which assigns occupation probabilities to states based on their energy. For some state $i$ with energy $E_i$ and at temperature $T$, the occupation $p_i$ is:
\begin{equation}
    p_i \propto e^{\frac{-E_i}{kT}}
\end{equation}

\noindent, $k$ being the Boltzmann constant.

We regard our possible $\theta$s as states, and our goal is that the chain finds them with probability $\mathbf{P}_X(\theta)$. Thus the energy our \textit{physical system} has in association with the target distribution is given by:
\begin{equation}
    \mathbf{P}_X(\theta) \propto e^{\frac{-U(\theta)}{kT}}
\end{equation}

We take this as a potential energy, regarding $\theta$ as a position - hence the use of the letter $U$. We then choose:
\begin{equation}
    \label{eq:hmc_potential}
    U(\theta) = -\ln \Big( \mathbf{P}_X(\theta) \Big)
    = -\ell(\theta)
\end{equation}

This energy is the negative of the log-likelihood $\ell(\theta)$, a commonly used construction by virtue of its computational convenience - it turns products of small (positive) numbers into sums of large (negative) numbers. The energy offset is irrelevant, as is a multiplying constant in the probability; the implementation is insensitive to them, as it should. This choice of reference produces an always negative energy, $U(\theta) \in ]-\infty,0]$ for $\mathbf{P}_X(\theta) \in [0,1]$. 

Additionally, we introduce an auxiliary \textit{momentum} variable, or vector - it should have as many dimensions as there are parameters. It supplies a \textit{kinetic energy}, which will help us explore the parameter space in a more orderly fashion. The effect is that now we work on a phase space with double the dimensions as compared to the parameter space.
\begin{equation}
    \theta \rightarrow (\theta, p)
\end{equation}

We also introduce a distribution over the momentum, which may be conditioned on the position: $\mathbf{P}(p \mid \theta)$. The joint position/momentum distribution must admit the original distribution as marginal distribution. That way, we can transform samples from this augmented distribution into samples from the target distribution simply by discarding the momentum component.
\begin{equation}
    \mathbf{P}(\theta,p) = \mathbf{P}(p \mid \theta) \mathbf{P}_X(\theta)
\end{equation}

Reformulating the probability of occupation, we can define a (classical) Hamiltonian that gives the total energy of the system.
\begin{gather}
    \mathbf{P}(\theta,p) = e^{-H(\theta,p)}\\
    \label{eq:hmc_hamiltonian}
    H(\theta,p) = U(\theta) + K(p,q) 
    = -\ln \Big( \mathbf{P}_X(\theta) \Big)
    - \ln \Big( \mathbf{P}(p \mid \theta) \Big)
\end{gather}

While the position distribution is given by the problem, the momentum distribution can be picked at will for the implementation. If we forgo position dependence and borrow the kinetic energy from classical mechanics, we get\footnote{The one-dimensional case is considered for simplicity, but the generalization is straightforward: $K = p^TM^{-1}p/2$. The mass $M$ must be a symmetric positive-definite matrix. Other elements can be similarly generalized.}:
\begin{equation}
    \label{eq:kinetic_hmc}
    K = \frac{p^2}{2\mathcal{M}} 
    \rightarrow \mathbf{P}(p \mid \theta) 
    = e^{-\frac{p^2}{2\mathcal{M}} } 
\end{equation}

With a corrective offset in the kinetic energy\footnote{This offset is the logarithm of the Gaussian's normalizing constant, which depends on the mass, but it doesn't affect the dynamics as long as $\mathcal{M}$ is independent of $\theta$.}, the momentum distribution associated with \ref{eq:kinetic_hmc} becomes an easy-to-sample-from Gaussian, with mean $0$ and (co)variance given by the mass. This is called an Euclidean-Gaussian kinetic energy \cite{Betancourt_2018}. The mass must be tuned according to the problem, but is often chosen to be diagonal (commonly a multiple of the identity matrix). This will be discussed later in this subsection. For now, we have:
\begin{equation}
    p \sim \mathcal{N} (\cdot \mid 0, M)
\end{equation}

It is important that we can sample from the momentum distribution, because we don't want it to be a hindrance to the Markov chain's journey. Here it is trivial to sample a momentum vector from the correct distribution before even beginning to construct the chain. This way, we can start the chain with a perfect sample on the momentum side, and no burn-in period is required as far as that side is concerned. This is in contrast with the position distribution, where we can't sample a starting $\theta$ directly from the target distribution; the inability to do so is the very reason why we use \gls{MCMC}. In that case, the starting state generally falls far from the mode(s), and it is the chain's task to conduct it to higher probability regions.

We then start the process at some starting position $\theta(t=0)$, which is chosen as for any Markov chain, and with some momentum $p(t=0)$, which is sampled directly from the selected momentum distribution. The objective is to advance the chain while preserving the augmented distribution; or equivalently, while preserving the energy given by the Hamiltonian of equation \ref{eq:hmc_hamiltonian}. We can do so by having the \textit{state}, regarded as a mechanical system, undergo Hamiltonian dynamics.

\begin{figure}[!ht]
    \centering
    \includegraphics[width=15cm]{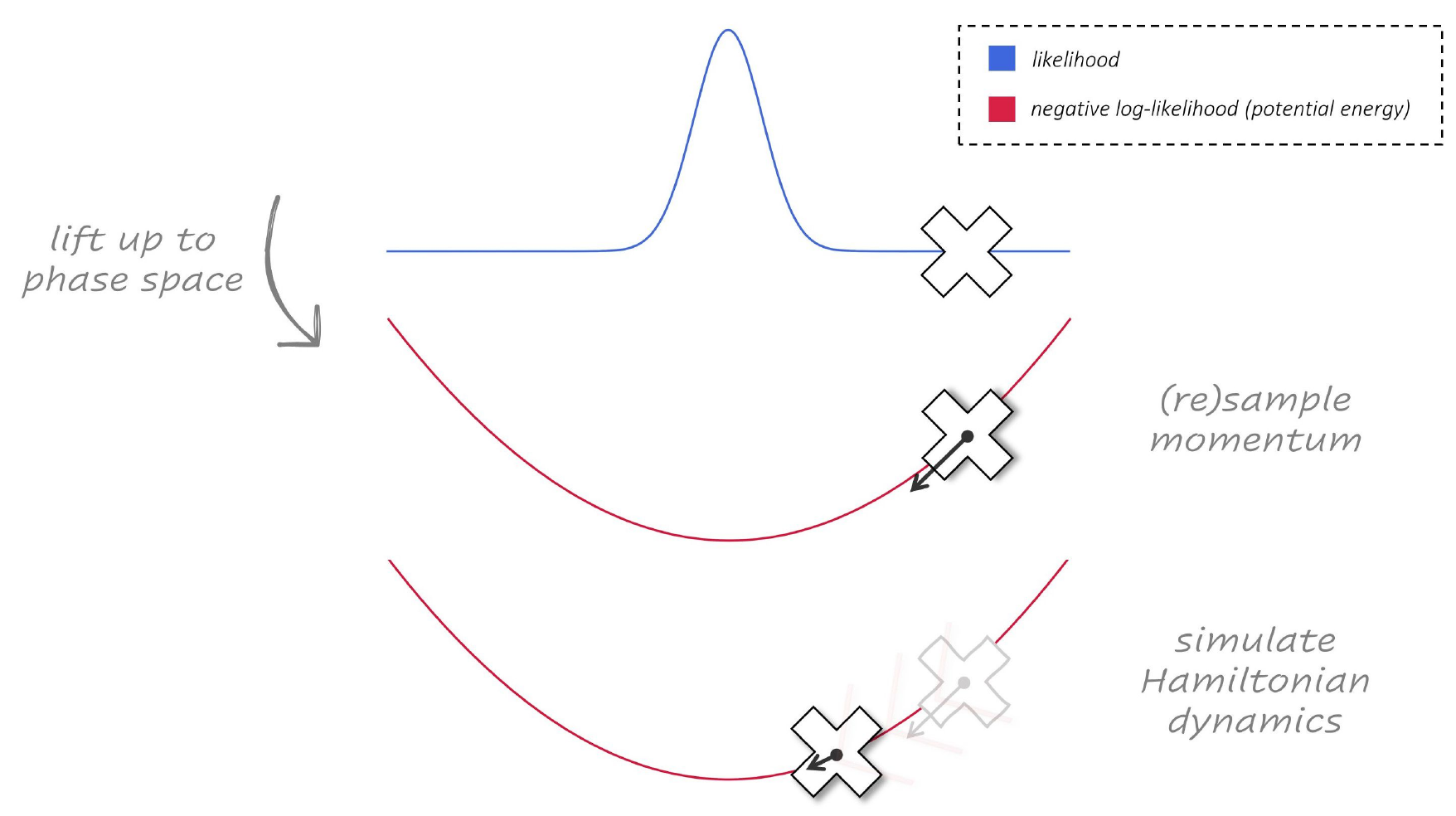}
    \caption{Illustration of the process behind a \glsfmtshort{HMC} proposal (for a gaussian distribution).}
    \label{fig:hmc_proposal}
\end{figure}

This completes the necessary structure for the construction of a \gls{HMC} proposal, which comes to replace the Gaussian sampling of random walk Metropolis. It makes use of the improvised momentum to slide along the potential determined by equation \ref{eq:hmc_potential}, as the augmented system explores the extended space (a \textit{mechanical} phase space) that is built from the original one (the parameter space). This is depicted in figure \ref{fig:hmc_proposal}.

Simulating the evolution prompted by a Hamiltonian is conceptually simple: it is described by Hamilton's equations, and we are in possession of all of their elements. Assuming the same position-independent kinetic energy as before, they take the form:
\begin{equation}
\label{eq:hamiltons_eq}
\begin{cases}
\begin{aligned}
    \dfrac{\dd \theta}{\dd t} &= \dfrac {\partial H}{\partial p}\\
    \dfrac{\dd p}{\dd t} &= - \dfrac{\partial H}{\partial \theta}
\end{aligned}
\end{cases}
\leftrightarrow
\begin{cases}
\begin{aligned}
    \dfrac{\dd \theta}{\dd t} &= \dfrac{\partial K}{\partial p}\\
    \dfrac{\dd p}{\dd t} &= - \dfrac{\partial U}{\partial \theta}
\end{aligned}
\end{cases}
\leftrightarrow
\begin{cases}
\begin{aligned}
    \dfrac{\dd \theta}{\dd t} &= \dfrac{p}{\mathcal{M}}\\
    \dfrac{\dd p}{\dd t} &= - \dfrac{\partial U}{\partial \theta}
\end{aligned}
\end{cases}
\end{equation}

To evolve the Markov chain, we must simulate the dynamics this system induces for some evolution time $\Delta t$.
\begin{equation}
    \Big(\theta(0),p(0) \Big) \rightarrow 
    \Big(\theta(\Delta t),p(\Delta t) \Big)
\end{equation}

For simplicity we write as before
\begin{equation}
    (\theta,p) \rightarrow 
    (\theta',p')
\end{equation}

If we can do execute this perfectly, the Hamiltonian will be preserved, as will the position and momentum distributions, producing perfect samples. But as per usual, we won't, and must resort to approximate numerical methods. In that case we can regard the last state of the trajectory as an elaborate proposal $\theta' \sim q(\cdot \mid \theta)$, and perform a Metropolis-Hastings step to account for discretization errors. The acceptance rate is as before (defining $v \equiv (\theta,p)$):
\begin{equation}
    \label{eq:acceptance_prob_hmcI}
    a(v',v) = \min \left(1,
    \frac{q(v \mid v') \mathbf{P}(v')}
    {q(v' \mid v) \mathbf{P}(v)}
    \right)
\end{equation}

This is a problem. Unless there is no movement, the probability of recovering the initial state of the trajectory $(\theta,p)$ when starting at its final state $(\theta',p')$ is null, just like a sliding object will never revert its motion to go back uphill. Hamiltonian dynamics are deterministic, fixing the ratio of proposal probabilities at $0/1$ - which sets the  acceptance probability to zero. For any $(\theta_0,p_0)$, there is only one possible final state $(\theta_F,p_F)$, so any other point has null probability of being proposed.

Fortunately, the correction is straightforward: it suffices to invert the momentum before making the proposal $(p' = -p(\Delta t))$. Should the trajectory have started at $(\theta',-p)$, it would have been the exact inverse of the one starting at $(\theta',-p)$. This renders the proposal symmetric ($1/1$), in which case we're back to the Metropolis case:
\begin{equation}
    \label{eq:acceptance_prob_hmcII}
    a(\theta',\theta) = \min \left(1,
    \frac{\mathbf{P}(v')}
    {\mathbf{P}(v)}
    \right)
    = \min \left(1,
    \frac{e^{-H(v')}}
    {e^{-H(v)}}
    \right)
    = \min \Big(1, e^{H(v)-H(v')} \Big)
\end{equation}

This acceptance probability resembles the Boltzmann factor, depending on the difference between energies. The more the new state lowers the energy as compared to the old one, the more likely it is to be accepted. Also, we can see that the sign of the momentum doesn't matter as long as the kinetic energy's dependence on it is quadratic, so although the sign flip is required for the sake of theoretical soundness no practical adjustments are necessary other than using \ref{eq:acceptance_prob_hmcII} and not \ref{eq:acceptance_prob_hmcI} to compute the acceptance probability (it being that the latter would mean setting it to $0$).

With this, we get exact samples from the augmented density, whose projection into the original space obeys our target distribution. What is more, if the energy is approximately preserved, which it should be for reasonable integration methods, we will have that $a(\theta', \theta)\approx 1$: an acceptance rate nearing 100\%. This is how the methodology affords efficient exploration of the space, producing the ideal combination of distant proposals and high acceptance probabilities - of course, at the cost of the transition kernel's higher complexity.

Here the \textit{reasonability} of a given integrator may take an unusual form as compared to other applications. In particular, it is important that the integrator be a symplectic map, or equivalently a canonical transformation. This means that it should preserve the form of Hamilton's equations, though not necessarily the energy itself. This means that even if it doesn't exactly obey our ideal Hamiltonian, it acts under the action of a \textit{perturbed} one that approximates it. This prevents the trajectory from drifting - it instead oscillates around the correct one, and its fluctuations average out unbiasedly instead of building up for long integration times. 

An example of a symplectic integrator is the elementary leapfrog integrator, which interleaves momentum and position updates to advance the system while evaluating the gradients at midpoints. The algorithm is presented in \ref{alg:leapfrog}.  Note the time reversibility: if we start at the final point and integrate backwards in time, we will retrace our steps all the way back to the initial point.

\begin{algorithm}[!ht]
\caption{Leapfrog integration for solving Hamilton's equations in the context of Bayesian posterior sampling.}
\textbf{Inputs}: initial position $\theta(0)$ and momentum $p(0)$, integration stepsize $\epsilon$, mass $\mathcal{M}$, integration time $\Delta t$.\\
\textbf{Computes}: sequence of position $\theta(t)$ and momentum
$p(t)$ states for $0<t \lessapprox \Delta t$, where the two instances of $t$ are matched for the last state.\\
\textbf{Assumes}: Ability to evaluate the prior and the likelihoods, and additionally their gradients.\\
\textbf{Assures}: Preserves the form of Hamilton's equations, though not necessarily the Hamiltonian itself.
\begin{enumerate}
    \item Define
    \begin{gather}
        U(\theta) = \mathbf{L}(\theta \mid \vec{D}) \mathbf{P}(\theta)\\
        \theta_{x}, p_{x} = 
        \theta(x \cdot \epsilon), p(x \cdot \epsilon) 
    \end{gather}
    \item Advance the momentum by a half timestep. 
    \begin{equation}
        p_{1/2} = p_0 - \frac{\epsilon}{2} \cdot \nabla U(\theta_0)
    \end{equation}
    \item For $k \in \{1, \dots, L\}$, with $L \equiv \text{round}(\Delta t / \epsilon)$:
    \begin{enumerate}
            \item Advance the position by a timestep.
            \begin{equation}
                \theta_{k} = \theta_{k-1} + \epsilon \cdot \frac{p_{k-1/2}}{\mathcal{M}}
            \end{equation}
            \item If $k \neq L$, advance the momentum by a timestep.
            \begin{equation}
                p_{k+1/2} = p_{k-1/2} - \epsilon \cdot \nabla U(\theta_k)
            \end{equation}
    \end{enumerate}
    \item Advance the momentum by a half timestep.
    \begin{equation}
        p_{L} = p_{L-1/2} - \frac{\epsilon}{2} \cdot \nabla U(\theta_L)
    \end{equation}
\end{enumerate}
\label{alg:leapfrog}
\end{algorithm}

The leapfrog algorithm can be generalized to higher order, though the increase in accuracy may not justify the cost. Even though this method is very simple, it beats others whose smaller but coherent errors may prove disastrous for the Hamiltonian trajectory.

Another benefit is that even though they are not immune to problems that frustrate exploration (and thus accuracy), these problems are easy to recognize. This is because the same geometrical features - namely high curvature - that bias estimators will affect numerical stability, causing the trajectory to diverge. Such problems can be easily identified so compensatory strategies can be employed. Overall, their simplicity and theoretical grounds facilitate the task of assessing their effectiveness. 

The integrator's performance depends on some hyperparameters, namely the stepsize $\epsilon$ and the path length $L$. Together, they determine the integration time $\epsilon \cdot L$. For a fixed integration time, the stepsize determines how many iterations $L$ must be performed, which defines the cost of the algorithm; in particular, the number of evaluations of the likelihood and its gradient scale inversely with $\epsilon$. Because the accuracy and stability of the integration depends strongly on this same $\epsilon$, which determines the roughness of time discretization, a good balance should be sought when choosing it.

Having discussed the integrator, we can finalize the main details of \gls{HMC}. The procedure is laid out in algorithm \ref{alg:hmc}. One may note that an important detail is missing in its construction: it doesn't explicitly cater to constraints (parameter space boundaries). They're left to be controlled by the fact that their prior probability will be zero, which will cause a minimum overflow. A way to treat restraints is to consider the effect of attributing an infinite potential energy to out-of-bounds regions, and handle it within the integration \ref{alg:leapfrog}. This leads to the position ($\theta$) advancement step being followed by a \textit{rebounding} motion (the position is mirrored relative to the boundary and the momentum is negated) if any positional boundary has been crossed \cite{Neal_2011,Daviet_2016}.

\begin{algorithm}[!ht]
\caption{Hamiltonian Monte Carlo algorithm for posterior sampling in Bayesian inference.}
\textbf{Inputs}: starting state $\theta^{(0)}$, mass $\mathcal{M}$, integration time $\Delta t$.\\
\textbf{Computes}: sequence of states
$\mathbf{t}_\text{MC} = \{\theta^{(i)}\}_{i=1}^M$ to be used in \ref{eq:expectation_mcmc}.\\
\textbf{Assumes}: Ability to evaluate the prior and the likelihoods, to sample from gaussian and binomial distributions, and to integrate Hamilton's equations.\\
\textbf{Assures}: The samples are distributed according to the posterior distribution as $M \rightarrow \infty$,.
\begin{enumerate}
    \item Define
        \begin{gather}
            H(\theta,p)
            = -\ln \Big( \mathbf{L}(\theta \mid \vec{D}) 
            \mathbf{P}(\theta) \Big)
            - \frac{p^2}{2\mathcal{M}}\\
            v^{(i)} \equiv (\theta^{(i)},p^{(i)})
        \end{gather}
    
    \item For $i \in \{1, \dots, M\}$:
    
    \begin{enumerate}
        \item Sample the initial momentum to define the initial state.
        \begin{equation}
            p^{(i-1)} \sim \mathcal{N}(\cdot \mid \mu=0, \sigma^2=M)\\
        \end{equation}
        
        \item Simulate Hamilton's equations for $H(\theta^{(i-1)},p^{(i-1)})$ for a time $\Delta t$ and define the proposal according to the final state.
        \begin{equation}
            v^* \equiv (\theta^*,p^*) \equiv \Big(\theta(\Delta t),-p(\Delta t) \Big)
        \end{equation}
        \item Compute the acceptance probability.
        \begin{equation}
            a(v^*,v) = \min 
            \Big(1, \exp \big( H(v^{(i-1)})-H(v^*)\big) \Big)
        \end{equation}
        \item With probability $a(v^*,v^{(i-1)})$, set
        \begin{equation}
            \theta^{(i)} = \theta^*
        \end{equation}
        Otherwise, keep $\theta^{(i)} = \theta^{(i-1)}$. The momentum can be discarded.
    \end{enumerate}
\end{enumerate}
\label{alg:hmc}
\end{algorithm}

Note that all that is used from the generated phase space trajectory of \ref{alg:leapfrog} is the last state, which is proposed as a state for the actual trajectory of the Markov chain. There exist some alternative approaches that don't treat intermediary states simply as an auxiliary construct but rather as proposals by themselves. More advanced strategies may drop the Metropolis correction entirely and instate some other scheme that safeguards correctness. A short overview is left for subsection \ref{sub:trajectory_generation}. In order to get valid proposals for such a scheme, the leapfrog algorithm \ref{alg:leapfrog} must be adapted to produce valid points, by synchronizing the position and momentum coordinates. This is more commonly known as the velocity Verlet method. It could be done by evaluating the momentum at integer steps as well, e.g. moving the edging half-steps inside the loop and doing 3 updates per iteration instead of 2. Alternatively, one could drop midpoint evaluations, or at least change the role of the momentum and position since the latter has more expensive updates (log-gradient evaluations).

As compared to random walk Metropolis, well-tuned \gls{HMC} only presents random walk behaviour between energy levels. This is induced by the random choice of a momentum variable that lifts the position coordinates up to a double-dimensional (as compared to the original parameter space) phase space, joining the position coordinate to establish the energy\footnote{Even this random walk aspect can be lessened by only partially refreshing the momentum between transitions \cite{Neal_2011}.}. Other than that commencing step, the exploration of the energy level set is deterministic - fixed by the integrator - and informed by the geometry of the target distribution, which brings improved efficacy.

The energy transition distribution is induced by the sampling of the momentum, and should be adapted to the marginal energy distribution of the target distribution. This is similar to how the \gls{RWM} proposal should be tuned to the distribution itself. On the other hand, the efficacy of the exploration for a fixed energy depends on the integration method and time. As for the integration time, there is a trade-off between completeness of exploration and cost, so the goal would be to keep it \textit{just long enough} to characterize the energy level set before the returns diminish. On the opposite extreme, its being too small may not bring gains as compared to exploration via a random walk. This will be further discussed in subsection \ref{sub:trajectory_generation}.

As for the mass, its inverse should resemble the covariance of the target distribution. Intuitively, this makes sense: the sharper the distribution, the \textit{heavier} the system should be, so that the moves are less adventurous. In the general case, the mass is a matrix, and it rotates and rescales the augmented parameter space, reshaping the structures that must be traversed. Naturally, this means that it may help smooth the way for the integrator.

For a Gaussian distribution, optimal results can be achieved with a Gaussian kinetic energy, but in the general case that is no longer true. As such, a position dependence may be worked in to better treat local features of the distribution's landscape \cite{Girolami_2011}.

All of these considerations were based on \cite{Betancourt_2018}, which presents an accessible but thorough introduction to \gls{HMC} and the main aspects concerning its implementation. It additionally contains illustrations that aid the visualization of many of the points approached here (and more).

Despite its ability to propose distant moves with high acceptance probabilities, \gls{HMC} is not without its shortcomings. Namely, its performance is strongly affected by the tuning of the parameters $\epsilon$, $L$ and $M$ (and possibly others in more advanced implementations), which is a complex task. Furthermore, it is far harder to implement than \gls{RWM}, and rather computationally demanding - though this is in terms of cost per iteration, which per se doesn't matter much. Also, it is not well suited to multi-modal distributions, and can get trapped in isolated energy minima, especially if the modes are far apart and/or differ in height (since even if the chain gets there the acceptance probability will be low due to the difference in energy). However, not only is this weakness common to most other approaches to \gls{MCMC}, it also can be mitigated via improvements to the basic \gls{HMC} algorithm \cite{Betancourt_2015b,Neal_2011,Lan_2014,Gu_2019,Betancourt_2018_talk} or by embedding the chain in an interacting ensemble (as in \gls{SMC}). 

The circumstances under which the penny plain version  of \gls{HMC} shines the most are high dimensionality, hard-to-resolve (elongated and/or sharp) shapes, and near normality - all of which greatly benefit from the fictitious momentum variable. Due to this enhancement, \gls{HMC} chains are capable of accomplishing geometric ergodicity for a wider range of functions than \gls{RWM} or other simplistic approaches, in which case the expectation estimates abide by a \gls{CLT} instead of achieving correctness only in the asymptotic limit (as for a usual Markov chain) \cite{Betancourt_2018_talk}. Even if this does not hold, progress tends to occur on a much faster time scale for a geometric method as compared to random walk based exploration. Further, even in a worst case scenario \gls{HMC} offers an interesting perk: powerful diagnostic tools. This is owing to the fact than when it does fail, it tends to fail quite spectacularly, facilitating fault detection and correction \cite{Betancourt_2018_talk}, as mentioned regarding the construction of the numerical trajectory.

These advantages of \gls{HMC} will be demonstrated in chapter \ref{cha:applied_examples} by means of an illustrative set of examples.

\subsection{Markov kernels within SMC}
\label{sub:mcmc_smc}

Despite the alluring simplicity of their underlying principle, \gls{MCMC} methods have many drawbacks of their own. Coming near their promised correctness presents many hurdles, since theoretical ergodicity often doesn't transfer well to practice. Not only is the stationary regime hard to reach, but also hard to recognize. 

For these reasons and others, they do worse in many aspects as compared to \gls{SMC}. Namely, they:
\begin{itemize}
    \item Are not suitable for online estimation;
    \item Tend to require more likelihood evaluations (especially when considering \gls{SMC}-\gls{SIR} with the prior as importance function and likelihood free resamplers like Liu West's);
    \item Don't deal well with complex features of the likelihood landscape, such as multimodality - especially when the modes are sharp and distant from one another (Markov chains are likely to get stuck in the one closest to the starting point for the duration of any realistic runtime);
    \item Do not offer as effortless a way of computing of the marginal likelihood/model evidence;
    \item Can be hard to tune.
\end{itemize}

As such, they can often lag behind as a competitor to \gls{SMC}. However, they don't have to be a competitor - they can be an ally. Their invariance properties make them perfect to refresh the particle cloud while exactly preserving the distribution it represents. For that reason, \gls{MCMC} kernels are widely used within \gls{SMC} for particle propagation, \cite{Del_Moral_2006,South_2019,Gunawan_2020,Rios_2013}. After performing the reweighting and multinomial sampling steps as before, the uniformly weighted particles are moved in parallel using one or more \gls{MCMC} transitions. A \gls{RWM} kernel is perhaps the most popular approach, but \gls{HMC} may significantly enhance performance. When using \gls{HMC} moves, \gls{SMC} is often termed sequential Hamiltonian Monte Carlo (\gls{SHMC}).

As compared to \gls{LWF}, this scheme preserves the entire distribution and not just its first two moments. This automatically solves the issues in capturing multimodality, by getting directly to the root of the problem. 

The aid is reciprocal, and the \gls{SMC} framework also complements the Markov transitions well in several ways. When they're used as variety-introducing mutations, a burn-in period is not required, because the particle cloud is already a Monte Carlo approximation of the desired distribution before they are performed. If it does not approximate it exactly, these moves should aid convergence, since the Markov chain's stationary distribution is precisely the target.

Discarding lag samples is also unnecessary, given that the distribution is characterized by weighted particle density and not successive Markov chain states. However, correlation between samples is still relevant, despite not being strictly necessary for formal correctness in the asymptotic limit (infinite number of particles). It dictates how effectively variety is introduced in the particle positions, which strongly impacts how well covered the parameter space is. Correlation between samples can be used for deciding by how many Markov moves each particle should be displaced, in order to assure an effective exploration \cite{Gunawan_2020}.

The two simplifications mentioned above are direct benefits of shifting responsibility from sequential states to parallel particles, but there are others. For instance, the particle cloud structure can provide an insight into the tuning of the \gls{MCMC} kernels \cite{Buchholz_2021}, whose efficacy strongly depends on their parameters. This is particularly critical when using Markov transitions which are very sensitive to the parameters and where choosing them is a challenging problem, as with \gls{HMC}. For instance, the mass matrix should resemble the covariance of the target distribution, which can be approximated by that of the particle cloud.

Finally, the fact that \gls{SMC} is more supple with respect to the targeted densities brings added flexibility, relaxing the \gls{MCMC} requirement that the target density be a constant function of the parameters.

Thus constructed \gls{SMC} samplers enjoy, unlike before, absolute formal correction (asymptotically), in the sense that the resampling step doesn't introduce error. They can also be interpreted as \textit{interacting} Markov chains. This is perhaps the most common implementation of particle filters of this sort, and it is popular across various fields of knowledge. For that reason, it is known by many different names, both more specific and more general: interacting Metropolis algorithms, Feynman-Kac particle models (mostly in measure-theoretic literature), etc. \cite{Del_Moral_2004,Del_Moral_2014}.

It can also be seen as a genetic algorithm which alternates between mutation and selection phases. For the duration of the \textit{mutation} periods, the particles are allowed to evolve independently through Markov transitions. The \textit{selection} stage is performed based on a probabilistic survival of the fittest criterion: the most apt particles \textit{reproduce}, and the least apt ones \textit{die}.

All the advantages of exact \textit{mutations} come at the cost of extra likelihood evaluations. Given that a finite number of particles is unable to characterize in full a continuous probability density, it was expected that to thoroughly preserve it consulting the particle cloud would not be enough. The extra required resources are significant, making the propagation step the most costly part of the algorithm. For the scheme of algorithm \ref{alg:sir}, which is particularly frugal with respect to likelihood evaluations, moving a particle at some step $k$ will demand at least twice as many resources as have been used for all the reweightings of the particle up to that point ($2k$). And this is for \gls{RWM} mutations - \gls{HMC} requires on top of that $L\cdot k$ gradient evaluations, $L$ often being of the order of the hundreds or more and the gradients being rather expensive for large datasets and/or complex functions.

The use of Markov kernels also makes the algorithm more memory-intensive, because all the data must be preserved throughout the runtime (to be used at least in the move steps). While in the case of tempered likelihood estimation this was already the case, this still brings some added difficulties, namely if meaning to exploit parallelism \cite{Gunawan_2020}. While parallel Markov chains are in theory \textit{embarrassingly parallel}, large shared datasets make them less so due to the overwhelming memory requirements.

All of this may be dispensable if the mean and variance offer a sufficient description of the distribution, but not for more complex targets or if the accuracy of expectations is of concern. In particular, the model evidence estimator is sensitive to the propagation strategy. Further, not only do Markov steps in themselves improve the accuracy of this estimator, they can also can be structured in a way to enable recycling the proposals for the construction of a different and more robust one. This strategy is proposed in \cite{South_2019}.

Either way, the \gls{SMC} samplers described here have been employed to deal with complex statistical functions, including tempered inference with \gls{HMC} move kernels \cite{Daviet_2016,Gunawan_2020} (the most demanding combination of all alternatives reviewed here). Cost doesn't matter \textit{per se}, and for many applications the value of these costly but dependable strategies is completely unmatched by simpler but unavailing ones.

\section{Implementations of Hamiltonian Monte Carlo}
\label{sec:hmc_implementations}

The previous section contained superficial considerations on the tuning of \gls{HMC}'s parameters (namely the mass matrix and integration step size and path length), but mostly left out structural modifications to the algorithm or its target, instead focusing on the \textit{standard} version of \gls{HMC}. This section will present a short survey of a few such modifications, namely alternative takes on generating a trajectory, selecting a sample, and determining the integration path length. Before that, a short aside on features \gls{HMC} struggles with, which can themselves be resolved by more sophisticated implementations.

\subsection{Conditions for applicability}
\label{sub:applicability_hmc}

The most determining factor for the efficiency of \gls{HMC} is whether it is well-suited for the target function. A few cases where it is not have been briefly mentioned, such as multi-modality. But one needn't go so far to watch \gls{HMC} fail.

Let us take for instance the example of section \ref{sub:quantum_characterization_examples}. It is a rather simple example, where we can expect the posterior to be approximately normal after just a few data. The distribution was uni-modal, and univariate; \gls{HMC} is known to excel in much more complex scenarios. Yet, it doesn't in this one, and upon implementing it we would find it to be no more effective than \gls{RWM}.

\begin{figure}
\centering
\captionsetup[subfigure]{width=.9\textwidth}%
\begin{subfigure}[t]{.45\textwidth}
  \centering
  \includegraphics[width=\linewidth]{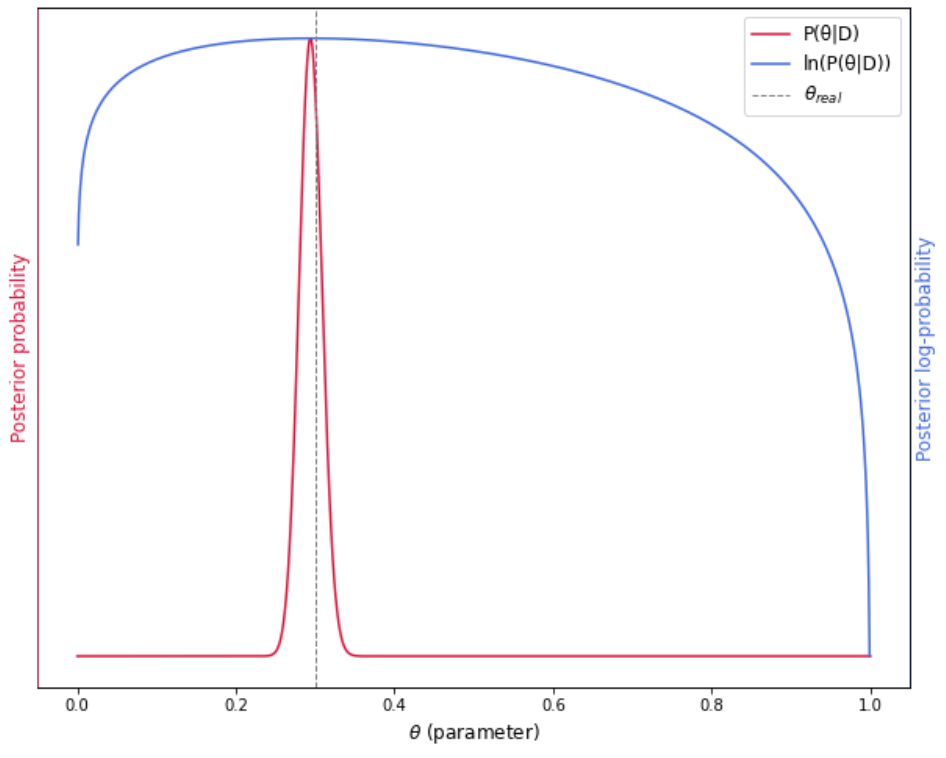}
  \caption{Binomial likelihood model (the example explored throughout sections \ref{sec:bayes_rule} and \ref{sec:bayesian_inference}). 100 trials were contemplated in the data vector.}
  \label{fig:loglikelihood_coin}
\end{subfigure}
\begin{subfigure}[t]{.45\textwidth}
  \centering
  \includegraphics[width=\linewidth]{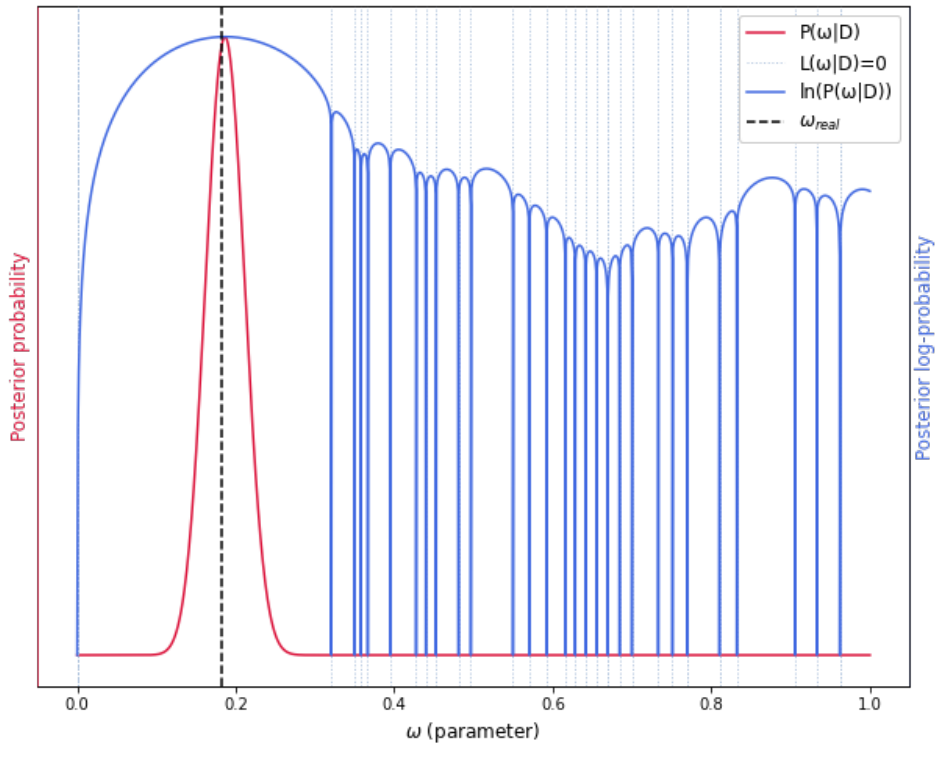}
  \caption{Squared-cosine likelihood model (the example from subsection \ref{sub:quantum_characterization_examples}). 50 measurements were contemplated in the data vector, using evenly spaced times on $[0,10[$. The thin dotted vertical lines (in blue) mark the log-likelihood asymptotes, where the likelihood is null.}
  \label{fig:loglikelihood_precession}
\end{subfigure}
\caption{Posterior probability (crimson) and log-probability (blue) for two different likelihood models and flat priors. The dashed vertical lines mark the real parameter used to generate the data $D$.}
\label{fig:loglikelihoods}
\end{figure}

The reason why can be seen in figure \ref{fig:loglikelihoods}, which shows the likelihood and log-likelihood for the \textit{coin} model (constant binomial function) and for the precession model (time-varying sinusoidal likelihood).  

In the first case (subfigure \ref{fig:loglikelihood_coin}), the log-likelihood is a smooth curve that \gls{HMC} can easily sail through - as much as it may not be the most resource-effective method for such a simple target. But in the second case (subfigure \ref{fig:loglikelihood_precession}), the log-likelihood looks grim for \gls{HMC}.

The problem is that the likelihood function (given in \ref{eq:precession_summary}) has roots. This should if anything be a \textit{nice} attribute: it may allow for ruling out some hypotheses based on a single datum. However, the resulting discontinuous log-likelihood is a problem for \gls{HMC}\footnote{Though some ways to deal with discrete or discontinuous likelihoods have recently been proposed \cite{Nishimura_2020}, they are certainly not tolerated by construction.}.

Our \gls{HMC} algorithm assumed the \textit{potential energy} to be continuously differentiable within some boundaries, which means that no points should have null likelihoods. That they do creates potential barriers that are overridden only if the discretized dynamics happens to leap over them by chance. As a consequence, the periodic roots of the precession example then create a minefield of potential wells that trap the particle in.

If these discontinuities affected the likelihood landscape at single points, we would with unbounded probability never encounter them; unfortunately, their neighbourhoods have a disruptive effect as well due to how the energy tends to infinity. This is aggravated by limited numerical precision; after a point the trajectory will become unstable, as the energy and its gradient overflow. When they do, the particle won't even be able to slide back into representable probability territory.

Even though the particles may be able to escape by resampling induced relocation or luck, if one is to rely on such methods much cheaper alternatives exist than \gls{HMC}, in particular confessedly luck-based methods (random walks). If there aren't many such asymptotes, a possible way of escaping these regions would be to screen for vanishing acceptance rates and perform a \gls{RWM} step after \gls{HMC} if they're found. Obviously, such a sequence is posterior-invariant if each step is, and by slightly moving the particle one may hope to introduce some chance that it crosses the potential wall and is positioned more favorably to Hamiltonian dynamics. This is attempted in section \ref{sub:phase_estimation}. However, it may lead to expensive proposals often being constructed just to be (with near unit probability) discarded and replaced by cheaper ones, which is rather unproductive.

\subsection{Sampling a trajectory and state}
\label{sub:trajectory_generation}

This section considers variations in how the \gls{HMC} proposals are picked. It is based on \cite{Betancourt_2018,Hoffman_2011}.

In the \gls{HMC} implementation of algorithm \ref{alg:hmc}, the last state obtained by integrating was proposed as sample, and all others ignored. A Metropolis-Hastings acceptance/rejection step was then performed for correction, which was necessary due to the simulation of the Hamiltonian dynamics' not being exact. In that case, \gls{HMC} could be seen as a particular (and remarkably creative) case of the Metropolis-Hastings algorithm.

The reason why the last state is generally adopted as proposal is that it tends to be more distant to the initial point, meaning we expect to get less correlated samples. However, this isn't a perfect rule, nor is it the only possible one. Clearly, the energy conserving properties of the symplectic integrator apply to the full trajectory. As such, any other state in the trajectory could be proposed, and accepted or rejected based on its own coordinates.

As a matter of fact, a sample from the trajectory will asymptotically correspond to a sample from the augmented target distribution (i.e. as the path length goes to infinity). Accordingly, we could pick a point from the trajectory uniformly at random, and perform the corresponding Metropolis correction. 

While this is a valid strategy, it is not the best, because each trajectory point is assigned equal probability. This isn't an issue in what respects correctness, as long as the correction is performed, but we may get unnecessarily low acceptance probabilities when there would have been better proposals in the generated trajectory. In this scenario, we would with considerable probability be forced to discard the point, and the construction of the entire trajectory would be in vain.

An alternative is to propose the points with probabilities proportional to the augmented probability density at their phase space locations. From here on, this is what is intended with the phrase "sampling a point from the trajectory" unless otherwise specified. This could be done using for example slice sampling, or simply multinomial sampling. It requires extra target evaluations, but also allows for devising more efficient sampling schemes.

The most basic approach would be to generate the whole trajectory while keeping track of all points, and then sample one of them using some sampling strategy of choice. We must in this case register not only the points' positions, but also their momentum coordinates, so as to be able to evaluate the augmented target density at each of them. As this scheme by itself safeguards the invariance of the target distribution under the Markov moves, the Metropolis correction is no longer necessary (see \cite{Betancourt_2018} for a formal analysis). As a matter of fact, this is no longer a Metropolis scheme.

The fact that all states must be kept in memory is highly discouraging, but better tactics can be used that improve memory management all the while performing equivalently. This can be done by choreographing the trajectory expansion in such a way that we can sample from it along its construction. The key thing to note is that if we have some initial point and want to integrate the Hamilton's equations for some number of discrete time-steps $L$, picking a trajectory uniformly at random from all ($L$) possible ones and \textit{then} sampling a point from that selected trajectory amounts to sampling from all reachable points under the same conditions.

Picking a trajectory at random from all that could contain the initial point given a fixed trajectory length is straightforward: it suffices to simulate the dynamics incrementally, successively choosing between forward or backwards time evolution (i.e. we would use positive or negative time-steps in the leapfrog integration). We would start at the current extremes of the trajectory (last/latest in time when integrating forward, first/earliest in time when integrating backwards) and integrate until the desired length was achieved.

Essentially, we would be progressively appending extra states to the previously obtained trajectory, producing a new and longer one.  These states can be added one or a few at a time in an additive scheme, but for faster progression a multiplicative scheme can be used. A common choice is to append a new trajectory as long as the last one, effectively doubling the trajectory length at each iteration.

In that case, we could generate a trajectory $A$ of length $W$ and keep a representative sample from it, along with the cumulative target density associated with the ensemble of points (which we will call its weight). Then we would pick at random a direction for the ensuing simulation, and take the corresponding extreme of trajectory A as the initial state for a second trajectory $B$. Finally, we would build trajectory $B$ until its length matched that of $A$. After this, we could concatenate those trajectories, and keep the sample from $A$ or $B$ with probabilities proportional to the weights of the entire trajectories. This provides a sample from the concatenated trajectory $A+B$.

In the first iteration, the $W$ above would be one. That means that each trajectory would contain a single point; the first one would be constituted by the initial state, and the second one by one of the two possible adjoining states in phase space (given the integrator parameters, namely the stepsize, the evolution is deterministic; the only random choice is the direction for the time evolution). In this case, the trajectories' weights would be simply the extended density at their only constituents, and the samples representing them would be the only possible one for each. As such, this base case corresponds exactly to sampling a point from the whole two-state trajectory. Joining the two parts together, we would get a length-two trajectory, and a sample representing it. Summing up the individual trajectories' weights, we'd also get a weight for the compound trajectory. 

Then we would integrate again; backwards in time with 50\% probability, and forward with another 50\%.  This integration would start one of the two points obtained so far, depending on their relative time positioning. If choosing a doubling scheme, we would advance by two integrator steps. After that, we would proceed similarly for the length four trajectory. With the additive one-by-one scheme, the new trajectory would always contain a single point, and its weight would always be the density at that point. In either case, this would be repeated until the intended path length was reached.   

So far we have assumed sampling proportional to the trajectory weights, but a higher efficacy can be achieved by introducing bias. This is another benefit of progressive sampling, that wouldn't be possible in the first two schemes outlined above (picking a state uniformly and performing a Metropolis correction, and sampling from the whole trajectory after building it). It is possible to attribute higher probability to the newest trajectory's sample, reducing correlation - and that too while preserving correctness, as exposed in \cite{Betancourt_2018}. 

In the scheme proposed therein, should the newest trajectory outweigh the former or their weights match, the newest trajectory's representative sample is kept, resulting in a proposal which is further away from the initial point (which obviously always falls within the oldest/previous trajectory segment). Even if it doesn't, the new trajectory will be favored over the old one. The Markov chain's relevant properties remain unharmed, because this bias is cancelled out if we consider all possible initial points: should the initial point have been in the newest trajectory instead, the success probability would be inverted with respect to the trajectories. 

\subsection{Dynamic implementations: the no U-turn sampler}
\label{sub:nuts}

We have seen that the opportunity to introduce anti-correlation measures makes it beneficial to build the trajectory as we go, until it achieves some \textit{desired length}. This length is the tunable $L$ of subsection \ref{sub:hmc}. It had been noted that it is a critical parameter for the performance of \gls{HMC}, it being that the trajectory should stop short of dwindling returns. However, a preliminary calibration is speculative at best, given that we don't know what the trajectory will look like. Why not then judge this in the light of its current state, if we are to be sitting at its frontier? Such is the idea presented in \cite{Betancourt_2018,Hoffman_2011}, which we will again follow.

When constructing the trajectory along the way, one can opt for a \textit{dynamic} scheme, where the number of integration time-steps is chosen adaptively based on some termination criterion. This can bring sizeable gains, since even perfect static tuning is in general sub-optimal: because the random momentum resampling at each Markov chain state lifts the sample up to a random energy, the ideal path length varies at each step, even for the same position coordinates.

In general, to different initial phase space states correspond different energies. Depending on the energy, a longer or shorter evolution may be required to properly explore the associated phase space region. If using a one-size-fits-all approach for the path length selection, we will be under-exploring some energy level sets (falling short of the returns they might otherwise have contributed with), and over-exploring others (wasting computational resources for little benefit). It is then clear it would be beneficial to introduce some adaptivity in the extent of this exploration.

This can be realized by defining a criterion with whose satisfaction the trajectory draws to a close, finalizing the state transition. A dynamic criterion can be contingent on the current standing of the trajectory, and is usually made to depend in its endpoints - earliest and latest in time, respectively $(\theta^{(-)},p^{(-)})$ and $(\theta^{(+)},p^{(+)}$). A well advised choice is the no-U-turn termination criterion (with which \gls{HMC} is called the No U-turn sampler, or \gls{NUTS}) \cite{Hoffman_2011,Stan,Betancourt_2018}:
\begin{gather}
    \text{Terminate if:} \nonumber \\
    p^{(+)} \cdot (\theta^{(-)} - \theta^{(+)}) > 0
    \text{ AND/OR }
    - p^{(-)} \cdot (\theta^{(+)} - \theta^{(-)}) > 0
\end{gather}

Each of these conditions signals that the velocity at an endpoint of the trajectory has the same direction\footnote{In the sense that the vectors make an acute angle, not that they're in perfect alignment.} as the distance vector from that endpoint's position to the other's. That is, integrating further from that extreme while keeping the other one fixed is expected to bring the endpoints closer together - or in other words, a U-turn is made. This is a rough indication that most of the exploration has taken place, and continuing it will bring reduced returns. Note that when integrating backwards the effective momentum is negated, since rewinding time reverts the direction of the velocity.

Some sources suggest taking the disjunction of the conditions \cite{Hoffman_2011}, while others propose the conjunction \cite{Betancourt_2018}. The former option may be too aggressive, since it stops as soon as any of the points moves towards the other (which may be moving away still faster); by contrast, the latter ensures that the endpoint are indeed coming towards each other, which may be too conservative. Which is best will depend on the application. 

An important point to note is that this criterion can't be applied naively, under penalty of forfeiting correctness. To ensure reversibility, any dynamic termination scheme must be implemented carefully.

In order for the Markov chain to admit the target distribution as stationary distribution, any two states in every trajectory must generate that trajectory with equal probability. Sampling from all the phase space points contained in it with probabilities proportional to the extended density then preserves the augmented target, and in particular the original target.

In static schemes, the trajectory generated by any initial state can easily be picked at random from all possibilities by randomly switching between time directions, and this principle will be respected. However, when considering dynamic schemes where adaptivity is employed in the choice of path length things get somewhat trickier. Instead of expanding our trajectory a fixed amount, we expand it until a requisite is satisfied. The problem is that we can't neglect all the ways this requisite \textit{could have been} satisfied, had the construction gone differently. In other words, the termination criterion might have cheated other trajectory elements out of their chances of reconstructing the trajectory. A point belonging to a trajectory it had no chances of originating is a clear violation of our condition.

If for instance we have some current trajectory, append a point to an extreme, and that point fulfills the termination criterion with any point other than its opposite extreme, that would mean that had the trajectory started at it, it would have stopped earlier. The aforementioned reversibility principle would then be breached, shattering the theoretical foundations of the Markov chain and rendering the construction invalid for our purposes. More specifically, this defies the validity of any estimator obtained by simulation.

This can be corrected by scanning through newcomer points that could have matched pairwise to provoke an early termination, had the trajectory started somewhere else; they are intruders, and should be excluded. That is, one must ensure that \textit{any} point that we welcome into our trajectory would have been able to, as a starting point, recreate it from end to end. Here an exponential expansion strategy performs much better than an additive one: it allows for saving on these correction checks, because it is more selective in the trajectories it allows \cite{Betancourt_2018}.

To any trajectory built through a doubling scheme corresponds a balanced binary tree, so for any path that we want to concatenate with our previous trajectory we need only check its balanced binary sub-trees (of which there are as many as nodes). By contrast, a one-by-one addition scheme would require any new arrival to be compared with every single one of the older states.

Independently of the strategy, if for any new state in the to-be-appended trajectory the termination criterion would have brought the process to a halt, the construction must be at once interrupted. The entirety of the new trajectory is to be discarded, and the old one to be taken as is. This is the downside of a bold expansion scheme: if considering a single addition at a time, we would never discard more than one point, whereas when doubling we may have to discard as many as we actually use (the latest correct trajectory). 

Although it may sound unreasonable that we'd be asked to dispose of a trajectory packed with potential samples, the method is still less wasteful than a one-size-fits-all approach, and is commonly used with a doubling progression scheme. 

An additional point to note is that when biasing sample selection (as discussed in the previous section) within such a scheme, one may choose to introduce bias in the construction of the entire trajectory (including all the sub-trees when constructing a new stretch) or only when appending a new trajectory to the old one - that is, when doubling. The original paper \cite{Hoffman_2011}, which coined the term \gls{NUTS}, suggests opting for the latter option. Details for the efficient implementation of all the ideas mentioned here (and others) can be found in that paper, as well as performance assessments. 

The algorithm was implemented along with other sample picking strategies, and the results are shown for a Rosenbrock function in subsection \ref{sub:rosenbrock}. The no U-turn sampler performs better even when matching the static and dynamic strategies for total states covered, i.e. including the discarded ones.

\section{Subsampling strategies}
\label{sec:subsampling}

Hamiltonian Monte Carlo is a very powerful method, and one which has enjoyed a lot of success - but it is also quite computationally costly. Monte Carlo methods are in general quite computationally demanding, but this is particularly true of \gls{HMC}. In inference, likelihood evaluations are the most expensive resource, and \gls{HMC} requires both itself and its gradient to be evaluated at each point of the phase space trajectory brought about by the (discretized) Hamiltonian dynamics. It is then clear why subsampling strategies may be of interest when scaling up complexity (namely dimensions and/or dataset size) - the data batch size may determine whether \gls{HMC} is an efficient or even viable solution in view of the available memory and processing resources. As an added benefit, it may increase efficacy if employing parallel processing, which may be too memory-intensive if the full dataset is to be available to each worker \cite{Gunawan_2020}.

Unfortunately, naive subsampling strategies can be catastrophic for the performance of \gls{HMC}, debasing the very features that make it so appealing \cite{Betancourt_2015}. As such, subsampling strategies must be developed either carefully, or not at all. This section will present a shallow view of a few proposed solutions in the literature. The main references are \cite{Betancourt_2015,Chen_2014,Quiroz_2018,Dang_2019,Gunawan_2020}. Subsection \ref{sub:control_variates} presents a simple and fairly general estimator, whereas \ref{sub:sg_hmc} and \ref{sub:shmc_ecs} outline the ideas of \gls{SG}-\gls{HMC} and \gls{ECS}-\gls{HMC} respectively. The latter is more prudent approach, relying on a pseudo-marginal method to refresh the subsample so a \gls{HMC} (within Gibbs) step can be performed in a seemly manner.

\subsection{Unbiased estimators and variance reduction with control variates}
\label{sub:control_variates}

We start by discussion likelihood estimators, based on \cite{Quiroz_2018,Dang_2019,Gunawan_2020}. Ideally we would use the exact likelihood and log-likelihood, which depend on all data:
\begin{gather}
    L(\theta) = \prod_{k=1}^{N} L_k(\theta \mid D_k) \equiv
    \prod_{k=1}^{N} L_k(\theta) \\
    \ell (\theta) = \sum_{k=1}^{N} \ell(\theta \mid D_k)
    \equiv \sum_{k=1}^{N} \ell_k(\theta)
\end{gather}

However, this may be infeasible or impractical for large datasets, which motivates the use of mini-batches sampled from the original data record $\vec{D}$. That is, we would like to construct an unbiased estimator based on a subsample of $m<N$ observations.

If $\widehat{\ell}_m(\theta)$ is an unbiased estimator of the log-likelihood based on subsampling $m$ data and $\widehat{\sigma}^2_m(\theta)$ estimates its standard deviation, we can estimate the likelihood by:
\begin{equation}
    \label{eq:likelihood_estimator}
    \widehat{\mathbf{L}}_m(\theta) = 
    \exp \Big( \widehat{\ell}_m(\theta) 
    - \frac{1}{2} \widehat{\sigma}^2_m(\theta)
    \Big)
\end{equation}

The motivation is that, by linearity of expectation, an unbiased estimator can easily be found for the log-likelihood, and $\widehat{\mathbf{L}}(\theta)$ will be unbiased too if $\widehat{\ell}_m(\theta)$ is normal (which is likely for large $N$ as per the \gls{CLT}) and $\widehat{\sigma}^2_m(\theta)$ is exact. The variance of the estimator corrects for bias; if it is approximate (a more realistic assumption), the estimator in \ref{eq:likelihood_estimator} is approximately unbiased.

As for the log-likelihood estimator, the starting point is:
\begin{equation}
    \label{eq:naive_subsampling_ll}
    \widehat{\ell}_m(\theta)  = 
    \frac{N}{m} \sum_{j=1}^{m} \ell_{u_j}(\theta) 
\end{equation}

\noindent, where the $u_j$ are subsampling indices sampled at random: $u_j \in \{1,\dots,N\}$. Once we fix the $m$ elements of $\vec{u}$, $\widehat{\ell}_m(\theta)$ is a value that estimates $\widehat{\ell}(\theta)$. Evaluating it has cost $\mathcal{O}(m)$.

Clearly, $\mathbb{E}\big[\widehat{\ell}_m(\theta)\big]=\ell_m(\theta)$ by linearity of expectation. However, subsampling tends to increase the variance in the estimate. A well-known variance reduction technique for Monte Carlo estimates is the use of control variates\footnote{Of course, other variance reduction techniques exist, some of which have been successfully employed in \gls{HMC} \cite{Li_2019}.}. They exploit the fact that the expected value is unaltered if we add a term and cancel out its expectation:
\begin{equation}
    \widehat{\ell}_m(\theta)  = 
    \frac{N}{m} \sum_{j=1}^{m} \ell_{u_j}(\theta) 
    + C \Big(
    \frac{N}{m}\sum_{j=1}^{m}q_{u_j}(\theta) - \sum_{k=1}^{N}q_k(\theta)
    \Big)
\end{equation}

The $q_k(\theta)$ are auxiliary random variables called \textit{control variates}. Even though the expectation is unchanged, the variance is not, and can be minimized through the choice of the $q_k$ and $C$. If we have some way of approximating $\ell_k(\theta)$, one way of achieving this is by choosing $q_k(\theta) \approx \ell_k(\theta)$ and setting $C=-1$ to get the so-called (for obvious reasons) \textit{difference estimator}:
\begin{equation}
    \label{eq:difference_estimator}
    \widehat{\ell}_m(\theta)  = 
    \sum_{k=1}^{N}q_k(\theta)
    + \frac{N}{m} \sum_{j=1}^{m} 
    \Big( \ell_{u_j}(\theta) 
    - q_{u_j}(\theta) \Big)
\end{equation}

The $q_k(\theta)$ can be Taylor series around some reference value $\theta^*$ for the distribution\footnote{This assumes unimodality (unless provisions are made, such as clustering the data to form  multiple centroids $\theta^*_k$ to serve as references).}. For \gls{MCMC} within \gls{SMC}, the reference $\theta^*$ can be the current \gls{SMC} mean $\bar{\theta}$, and for \gls{MCMC} it can be determined in a preliminary processing phase. In both cases, the coefficients for the Taylor expansion can be calculated once using the full dataset and are then used for all parallel particle updates/transitions (for \gls{SMC}/\gls{MCMC} respectively), signifying $\mathcal{O}(1)$ cost (i.e. no likelihood evaluations, only distances with respect to $\theta^*$). This implies that \ref{eq:difference_estimator} has cost $\mathcal{O}(m)$ just like \ref{eq:naive_subsampling_ll}.

Finally, $\widehat{\sigma}^2_m(\theta)$ too can be obtained from the subsample as the variance of the differences in \ref{eq:difference_estimator} (between parenthesis). This clears the requirements for the computation of the estimator in \ref{eq:likelihood_estimator}, which can be used in any sampling strategy, namely any \gls{MCMC} and \gls{SMC} method (for both the weight updates and the particle propagation steps in the latter case). In \gls{HMC}, the potential energy would become:
\begin{equation}
    \widehat{U} = -\ln \Big( \widehat{\mathbf{L}}_m(\theta) \Big)
\end{equation}

We don't use $\widehat{U}=-\widehat{\ell}_m(\theta)$ directly because we are interested in the unbiasedness of the target distribution\footnote{Though one \textit{could} ignore this and opt for a more biased but simpler estimator based on a sub-product, which may still do well depending on the implementation details and structure of the data.}, the potential energy being simply a derived construct. The \gls{HMC} updates require the log-gradient through $-\nabla \widehat{U}$, which can can be obtained from the sample as well by differentiating \ref{eq:difference_estimator} and its variance. Most of these calculations are laid out in \cite{Quiroz_2018}, which additionally shows that the error in expected values associated with the perturbed posterior decreases as $\mathcal{O}\big(1/(Nm^2)\big)$. In particular, this is the expected convergence rate of the calculated mean and variance for the posterior distribution.  

For instance, letting $m=\mathcal{O}(\sqrt{N})$ means that their error decreases quadratically with the total number of observations, which is very favorable. However, this is assuming that $\theta^*$ is the posterior mode based on the full dataset. For \gls{SMC}, there are convenient statistics are at hand via the particle cloud, whose mean we can use directly; for \gls{MCMC}, auxiliary methods such as gradient descent become necessary. In that case, \textit{stochastic} gradient descent based on a fixed subsample may be a less purpose-defeating assumption, but it slows down convergence. If the gradient is also based on $m=\mathcal{O}(\sqrt{N})$ observations, the rate becomes $\mathcal{O}(N^{-1/2})$. 

Being in possession of an estimator, we can simply use it instead of the complete likelihood (or its byproducts) when performing inference. However, this doesn't necessarily guarantee a good cost-to-performance ratio; improvements in the estimator \textit{do} bring improved results - but this is as compared to a crude approximation. When the reference is the fully proper algorithm, this is not necessarily so.

In the case of \gls{HMC}, it's particularly tempting to subsample the gradients, which are responsible for the lion's share of the cost. The subsampling indices could be sampled only at the beginning of each trajectory (picking a new $\vec{u}$ for each iteration of algorithm \ref{alg:hmc}), or within the trajectory itself (picking a new $\vec{u}$ for each iteration of algorithm \ref{alg:leapfrog}). 

If we do this but impose the usual Metropolis corrections based on the full dataset, which is in itself costly, we correct the bias that we are introducing. However, we are likely to dramatically lower the acceptance rate, since the energy is no longer stable. If we skip this correction to avoid processing all the data, the trajectory will be biased - just how much depends on the quality of the estimators, the features of the data, and the details of the implementation. Resampling the subsampling indices at each step of the trajectory performs better than not, but still not satisfactorily so \cite{Betancourt_2015}. As such, this approach severely compromises the scalability that \gls{HMC} prides itself on. In this context, some revisions of the algorithm have been suggested.

\subsection{Stochastic gradient Hamiltonian Monte Carlo}
\label{sub:sg_hmc}

Reference \cite{Chen_2014} adopts an intuitive view of the dynamics behind \gls{HMC} in an attempt to remedy the downfalls of subsampling. They point out that using a subsampled gradient amounts to using a noisy gradient, which by the \gls{CLT} they write as:
\begin{equation}
    \label{eq:stochastic_gradient}
    \Delta \widehat{U}(\theta) \approx 
    \Delta U(\theta) + \mathcal{N}(0,V)
\end{equation}

\noindent, where $V$ is the (co)variance of the stochastic gradient noise and may depend on $\theta$. This defines a diffusion matrix that shows an additional dependence on the stepsize, $B=\epsilon V/2$, which in turn determines the unwanted random force that affects the dynamics.

\begin{figure}[!ht]
    \centering
    \includegraphics[width=9cm]{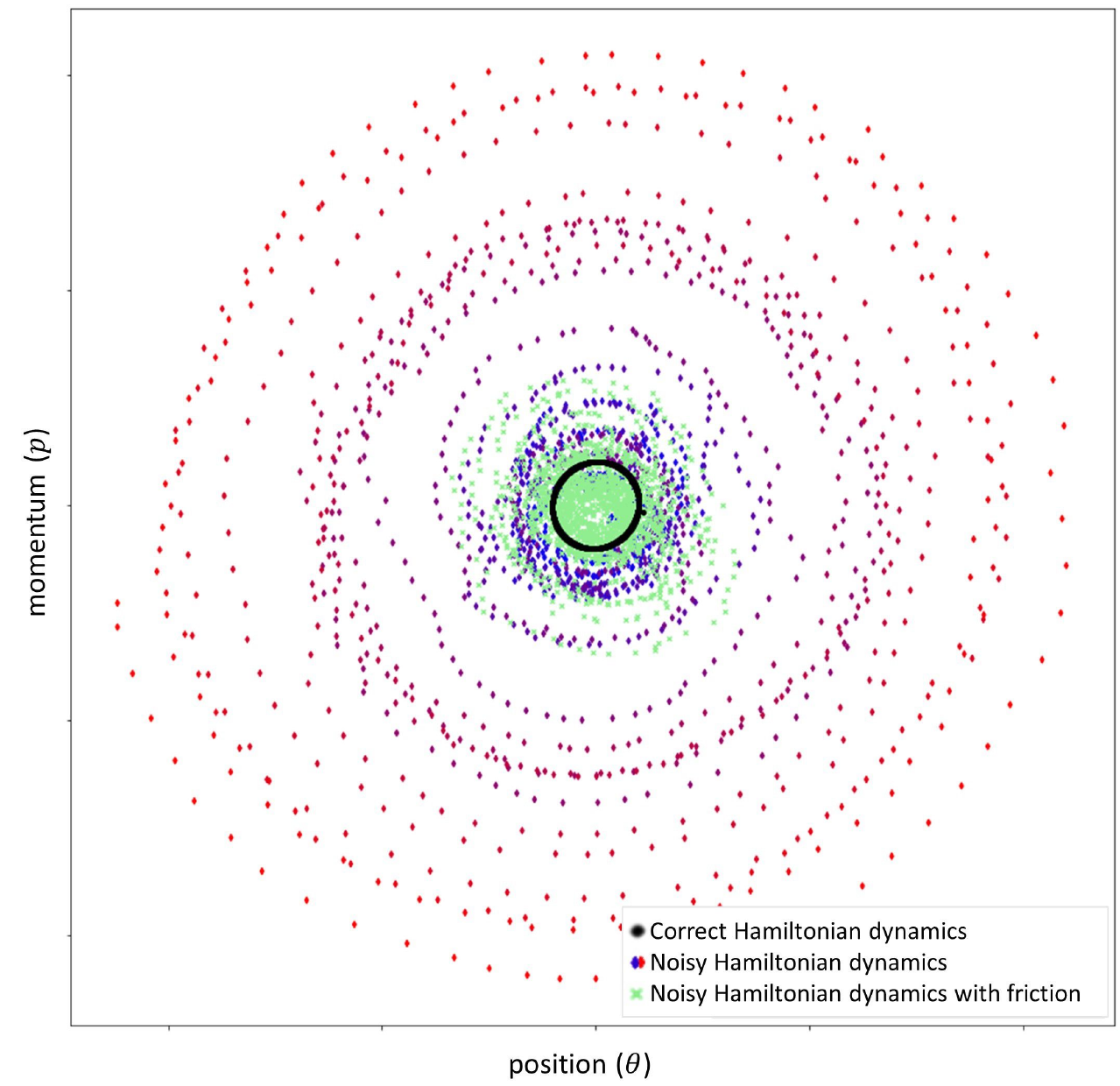}
    \caption{Effect of noise on the phase space trajectory for a harmonic oscillator model ($U=\theta^2/2$). The trajectory with correct dynamics is plotted in black, that with a noisy gradient in green, and that with a noisy gradient compensated by a friction term in a gradation from blue to red (by ascending order of elapsed time). This example is based on an identical one from \cite{Chen_2014}.}
    \label{fig:sghmc}
\end{figure}

It is shown that the dynamics associated with \ref{eq:stochastic_gradient} fail to preserve the target distribution, leading to divergent trajectories. The potentially catastrophic impact of noise in the dynamics is demonstrated in figure \ref{fig:sghmc} for a simple oscillator. 

To mend this problem, the authors suggest introducing a friction term inspired in the Langevin equation (a stochastic differential equation initially used to describe Brownian motion \cite{Tong_2012}): 
\begin{equation}
    \dfrac{\dd p}{\dd t} = - \dfrac{\partial \widehat{U}}{\partial \theta}
    -\gamma \frac{p}{\mathcal{M}} 
\end{equation}

This presents a modification of \ref{eq:hamiltons_eq}. For $\gamma=B$, equilibrium is restored. In general, the exact noise model won't be known exactly, but rather approximately; the algorithm can be adapted accordingly. The variance $V$ can be estimated using the inverse of the Fisher information matrix based on a data mini-batch. All details can be found in \cite{Chen_2014}. Results of implementing the algorithm are shown in subsection \ref{sub:multi_cos}.

The authors suggest omitting the Metropolis corrections entirely to spare costs, and instead controlling the quality of the sampler (namely the bias) through the discretization of the dynamics (by keeping the integration stepsize small). However, there is then a sensitive trade-off between bias and integration cost for a fixed evolution time. Overall, even though it helps control divergence, this scheme may undermine the efficiency of \gls{HMC}, especially in high-dimensional scenarios \cite{Betancourt_2015,Dang_2019}.

\subsection{Hamiltonian Monte Carlo with energy conserving subsampling}
\label{sub:shmc_ecs}

A possible pathway to a more correct subsampling \gls{HMC} strategy is an extended target density. We will consider an approach presented in \cite{Quiroz_2018} for \gls{MCMC} in general, in \cite{Dang_2019} for \gls{HMC} in particular, and still more particularly in \cite{Gunawan_2020} for \gls{SMC} with \gls{HMC} move steps. The idea is to have the target distribution encompass the subsampling indices, as the random variable that they are:
\begin{equation}
    \label{eq:HMC_ECS_target}
    \mathbf{P}_X(\theta,\vec{u}) 
    = \widehat{\mathbf{L}}_m(\theta,\vec{u})\mathbf{P}(\theta)\mathbf{P}(\vec{u})
\end{equation}

When performing \gls{HMC} steps, only the posterior is targeted; that is, the Hamiltonian dynamics operate for a constant $\vec{u}$ \footnote{Auxiliary variables have been attributed a momenta of their own in \gls{HMC} \cite{Alenlov_2019}, but typically they're continuous, as in latent variable models. \gls{HMC} isn't made for discrete domains like that of $\vec{u}$, though accommodations can be made \cite{Nishimura_2020}}. Accordingly, when we further augment the distribution with a momentum variable, it respects $\theta$ only, and matches it in dimension. Thus, within \gls{HMC}:
\begin{equation}
    \label{eq:HMC_ECS_target_p}
    \mathbf{P}_X(\theta,\vec{u}) =
    \exp \Big(-\widehat{H}_{\vec{u}}(\theta,\vec{p}) \Big)
    \mathbf{P}(\vec{u})
\end{equation}

This doesn't mean that $\widehat{H}_{\vec{u}}(\theta,\vec{p})$ doesn't depend on $\vec{u}$: it does, because now the potential energy is $\widehat{U}(\theta) = -\ln \widehat{\mathbf{L}}_m(\theta,\vec{u}) -\ln \mathbf{P}(\theta)$. However, this vector is fixed within the Hamiltonian, its variability being relegated to this vector's own distribution $\mathbf{P}(\vec{u})$.

Still, $\vec{u}$ must be updated, or else the extension serves no purpose. That can be done with a simple Metropolis step (using a uniform proposal distribution, which is clearly symmetric too), as long as $\theta$ is held fixed. Updating each parameter by conditioning it on all others is a strategy called \textit{Gibbs sampling}, a special case of the Metropolis-Hastings algorithm where we sample each variable individually while sweeping through the parameter vector \cite{Andrieu2003}.

Here the separation is between $\theta$ and $\vec{u}$. We want to sample $\vec{u}$ according to $\mathbf{P}_X(\theta,\vec{u})$ conditionally on $\theta$; we can do so with a \gls{MH} step. This strategy is called Metropolis-within-Gibbs, or more descriptively one-variable-at-a-time Metropolis. We fix $\theta$, sample new indices uniformly, and accept them with probability:
\begin{equation}
    \label{eq:pm_mh}
    a(\vec{u}\, ',\vec{u}) = \min \left(1,
    \frac{\widehat{\mathbf{L}}_m(\theta,\vec{u}\, ' )}
    {\widehat{\mathbf{L}}_m(\theta,\vec{u})}
    \right)
\end{equation}


Note that $\theta$ is fixed, corresponding in practice to the latest $\theta$ iterate. This is an instance of \textit{pseudo-marginal} Metropolis-Hastings (\gls{PM}-\gls{MH}), a version of \gls{MH} where the likelihood isn't available analytically and a non-negative unbiased estimator takes its place. This preserves correctness (the chain's invariant distribution is retained), earning marginal methods the title of \textit{exact approximations} \cite{Warne_2020}. 

A common example comes from latent variable models whose integrals can't be solved analytically: importance sampling can be used to get a Monte Carlo estimator of the likelihood (this requires only pointwise evaluations of the integrand, and is for that reason often called a \textit{likelihood-free} method). An alternative would be to target the joint distribution of parameters and latent variables. This is less efficient than the \textit{marginal} methods (such as the ones considered so far), which target the intended distribution directly; as such, pseudo-marginal methods attempt to bring the approach closer to them \cite{Andrieu_2009,Alenlov_2019}. 

In our subsampling case, we introduce not evaluation sites but subsampling indices, since we are truncating a sum and not an integral. Either way, this amounts to extending the target density to encompass some extra random variables, now the auxiliary samples used to estimate the likelihood \cite{Andrieu_2009,Tran_2016,Warne_2020,Quiroz_2018}. They can be refreshed at each iteration; this refreshment is factored in when computing the joint acceptance rate \cite{Andrieu_2009,Tran_2016} (as it is in \ref{eq:pm_mh}). That there is variability in the compound \gls{MH} step for even fixed $\theta$ - equivalently, that the distribution is extended - slows down convergence, how much depending on the variance of the estimator (hence the attempted reduction through e.g. control variates) \cite{Warne_2020}.

One may wonder whether the samples can't simply be picked anew at each iteration to instance the estimator and then discarded. This is sometimes done, and given the self-explanatory name of Monte Carlo within Metropolis (in the case of a Monte Carlo likelihood estimator). However, in that case, absolute validity is only achieved in the case of perfect \textit{estimation}; otherwise the scheme is rather an \textit{inexact approximation} whose accuracy grows with the number of samples \cite{Andrieu_2009}.

A possible complication of \ref{eq:pm_mh} is index stagnation. Should the estimator significantly overestimate the likelihood associated with some $\vec{u}$, others will suffer from low acceptance, \textit{jamming} the sampler and affecting validity. To solve this, correlation can be introduced between consecutive proposals, by updating only a segment of the previous index set in what is called a \gls{bPM} strategy \cite{Tran_2016}. This should make the acceptance rate more stable, preventing the monopoly that could happen otherwise.

After the index updating step, a \gls{HMC} step can be performed while holding $\vec{u}$ fixed (targeting \ref{eq:likelihood_estimator}). The \gls{HMC} moves preserve the energy because the \gls{MH} correction step targets precisely the energy according to which the Hamiltonian dynamics are simulated, guaranteeing high acceptance as well as correctness.

In short, \gls{ECS} is achieved by wrapping the \gls{HMC} transitions in a Gibbs update, in which a block Metropolis-Hastings step is used to pick the subsampling indices (conditioned on the latest particle location) before the usual \gls{HMC} step (conditioned on the indices). This amounts to a pseudo-marginal approach where the posterior is augmented to include some supplementary sample identifiers, the joint density being left invariant by the composite - Gibbs - Markov transition comprised of two \gls{MH} steps (one of which is \gls{HMC}).  Marginalizing over the indices $\vec{u}$ should yield the intended $\theta$ samples.

This approach was applied to \gls{TLE}-\gls{SMC} (with \gls{HMC}-\gls{ECS} for particle propagation); the results are shown in subsection \ref{sub:multi_cos}. All of the implementation details can be found in \cite{Gunawan_2020}.
\chapter{Applied sampling and inference examples}
\label{cha:applied_examples}

This chapter presents a few test cases for the methods presented in chapters \ref{cha:quantum_parameter_estimation} and \ref{cha:monte_carlo_posterior_sampling}. The tests on actual quantum devices are left for chapter \ref{cha:experiments_quantum_hardware}.

Section \ref{sec:sampling_problems} applies statistical simulation techniques to hard-to-sample-from density functions. These functions are non-negative, describing probability distributions up to a constant - but not Bayesian posterior distributions. They consist of a 6-dimensional Gaussian (subsection \ref{sub:6d_gaussian}), a Rosenbrock function (\ref{sub:rosenbrock}), and a smiley-like structure built from a pre-determined set of points (\ref{sub:smiley_kde}). Their complexity is due to high dimensionality, sharpness and multimodality respectively. All the subsections test \gls{MCMC} as described in section \ref{sec:mcmc}. Two instances of \glsxtrlong{MH} are considered: \glsxtrlong{RWM}, and \glsxtrlong{HMC}. 

The performances of static, progressive and dynamic variations of \gls{HMC} (as per subsection \ref{sub:trajectory_generation}) are compared in subsection \ref{sub:rosenbrock}. Everywhere else, a standard static last-state
implementation of \gls{HMC} (algorithm \ref{alg:hmc}) is implicit. Subsection \ref{sub:smiley_kde} concerns a smiley-shaped trimodal distribution, to which \gls{SMC} is applied in addition to the other methods. Two different Markov kernels are tested: \gls{RWM} and \gls{HMC}. 

In section \ref{sec:inference_problems}, sampling methods are instead used in inference problems, using data generated according to the intended likelihoods. Subsection \ref{sub:phase_estimation} applied Gaussian rejection filtering and \gls{MCMC} (\gls{RWM} and \gls{HMC}, section \ref{sec:mcmc}) to phase estimation (subsection \ref{sub:quantum_characterization_examples}). Subsection \ref{sub:sampling_precession} targets the precession example of subsection \ref{sub:quantum_characterization_examples} using the same \gls{MCMC} methods, and additionally \gls{SMC}-\gls{SIR} with \gls{LWF} (subsections \ref{sub:sir} and \ref{sub:liu_west}) as well as with \gls{HMC} move steps (this being typically abbreviated as \gls{SHMC}). The example is easily made to result in multi(bi)-modal posteriors, the nemesis of the Liu-West kernel shrinkage method - effectively orchestrating the ruin which befalls it in a wide array of realistic applications.

Finally, subsection \ref{sub:multi_cos} generalizes the precession likelihood to higher dimensions, to test the methods in more challenging settings. Results are shown for \gls{SG}-\gls{HMC} (subsection \ref{sub:sg_hmc}) within \gls{SMC}-\gls{SIR}, and for \gls{TLE}-\gls{SHMC}-\gls{ECS} (subsections \ref{sub:tle}, \ref{sub:hmc}, \ref{sub:shmc_ecs}) with control variates (section \ref{sub:control_variates}). Finally, some heuristics for choosing likelihood hyperparameters (\textit{experiment controls}) are tested.

\section{Sampling problems}
\label{sec:sampling_problems} 

These examples are based on \cite{Daviet_2016}. 

\subsection{6-dimensional normal distribution}
\label{sub:6d_gaussian}

\begin{figure}[!ht]
    \centering
    \includegraphics[width=11cm]{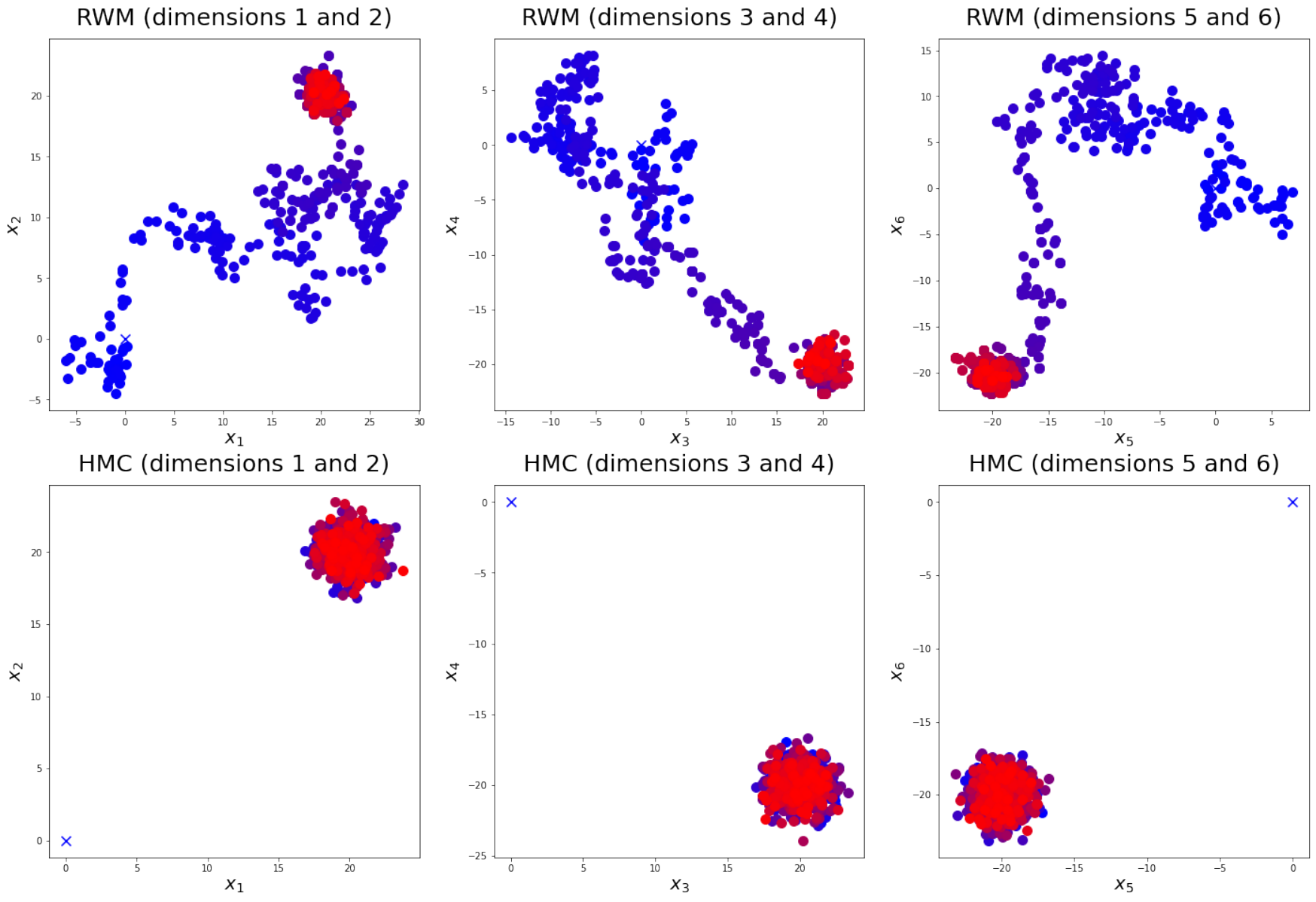}
    \caption{Trajectories generated using \glsfmtshort{RWM} (top) and \glsfmtshort{HMC} (bottom) to sample from a multivariate normal distribution with $6$ dimensions/parameters. Both chains started at the origin and were evolved for $10^3$ states. These states are plotted as points gradating from blue to red (from earlier to later iterations), whereas the starting point is marked with an 'x'. The dimensions are plotted pairwise for convenience.}
    \label{fig:hmc_gaussian}
\end{figure}

The target for this subsection is a multivariate normal distribution. This example showcases the benefits of \gls{HMC} as compared to \gls{RWM} or other basic \gls{MH} algorithms. The distribution was chosen to have mean $\Vec{\mu}=[20,20,20,-20,-20,-20]$ and identity covariance. The results are shown in figure \ref{fig:hmc_gaussian}.

The \gls{RWM} proposal variance was such that the mean acceptance rate was $42\%$, whereas for \gls{HMC} it was $100\%$ with $L=30,\epsilon=0.05,\mathcal{M}=\opsub{I}{6}$. \gls{HMC} manages a golden combination of distant proposals and very high acceptance probabilities, whereas \gls{RWM} struggles a fair bit more to reach the mode. It should be mentioned that they're not matched for resources; each \gls{HMC} step is by nature more expensive, due to how it is informed by the target's geometry. However, larger and larger distances will render \gls{RWM}'s wandering less and less productive, hence it is bound to become less cost-effective be it sooner or later. 

Furthermore, not only does the \gls{HMC} chain mix faster, there is also less correlation between consecutive samples, as can be seen from both the unimportance of the starting point and the uniform red color in figure \ref{fig:hmc_gaussian}: the latest iterations alone cover the mode thoroughly and evenly, which can't be said of \gls{RWM}. This reduces the  number of samples necessary to comply with accuracy requirements for the expectations.

The performance of \gls{HMC} is especially impressive in this example: its proposals are particularly good if the target itself is Gaussian, the usual Gaussian distribution over the momentum is chosen, and the covariance matrices of these distributions are alike \cite{Betancourt_2018}. While in practice sampling from a Gaussian isn't a great achievement, by the \gls{CLT} many posteriors will approach normality for large enough datasets. Thus, the virtues that shine through in this particular case can be expected to manifest in others.

\subsection{Rosenbrock function}
\label{sub:rosenbrock}

The target for this subsection is a Rosenbrock function as defined as in \cite{Daviet_2016} (where it seems to be exponentiated as compared to its usual form):
\begin{equation}
    \label{eq:rosenbrock}
    g(x,y) = \exp \Big[ \frac{1}{8} \big(-5(y-x^2)^2-x^2 \big) \Big]
\end{equation}

\begin{figure}[!ht]
    \centering
    \includegraphics[width=12cm]{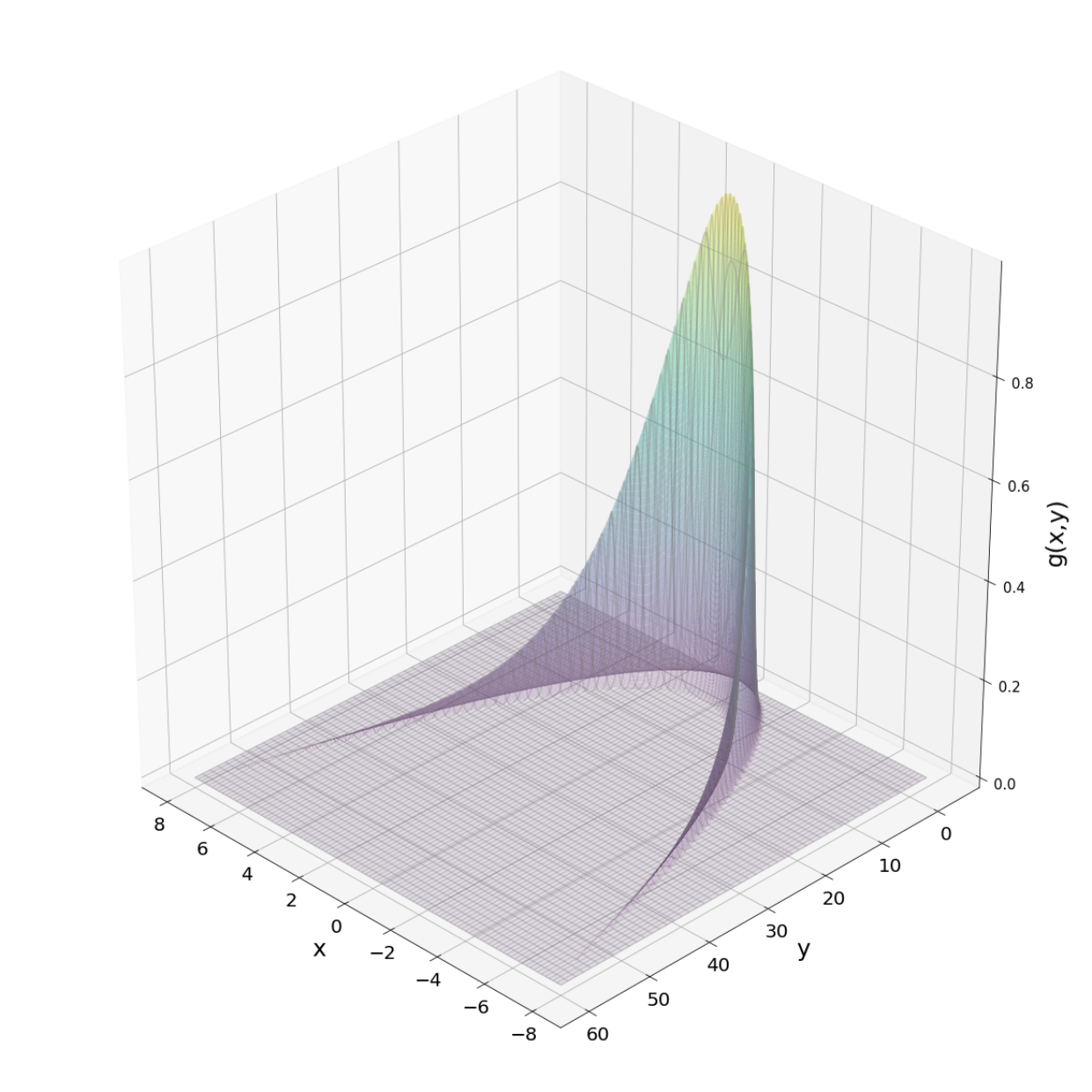}
    \caption{3-dimensional plot of the Rosenbrock function from equation \ref{eq:rosenbrock}.}
    \label{fig:rosenbrock_3d_plot}
\end{figure}

The resulting density resembles a narrow smile with sharp corners. Its shape is plotted in figure \ref{fig:rosenbrock_3d_plot}. The key thing to note other than its overall form is that the corners of the \textit{mouth} reach all the way up to $y \approx 55$, with a probability density that decreases rapidly with increasing $y$. This feature is especially hard to resolve, because Markov chains struggle to explore such narrow and low probability regions.

\begin{figure}[!ht]
\captionsetup[subfigure]{width=\textwidth}%
\begin{subfigure}[t]{.5\textwidth}
  \centering
  \includegraphics[width=.8\textwidth]{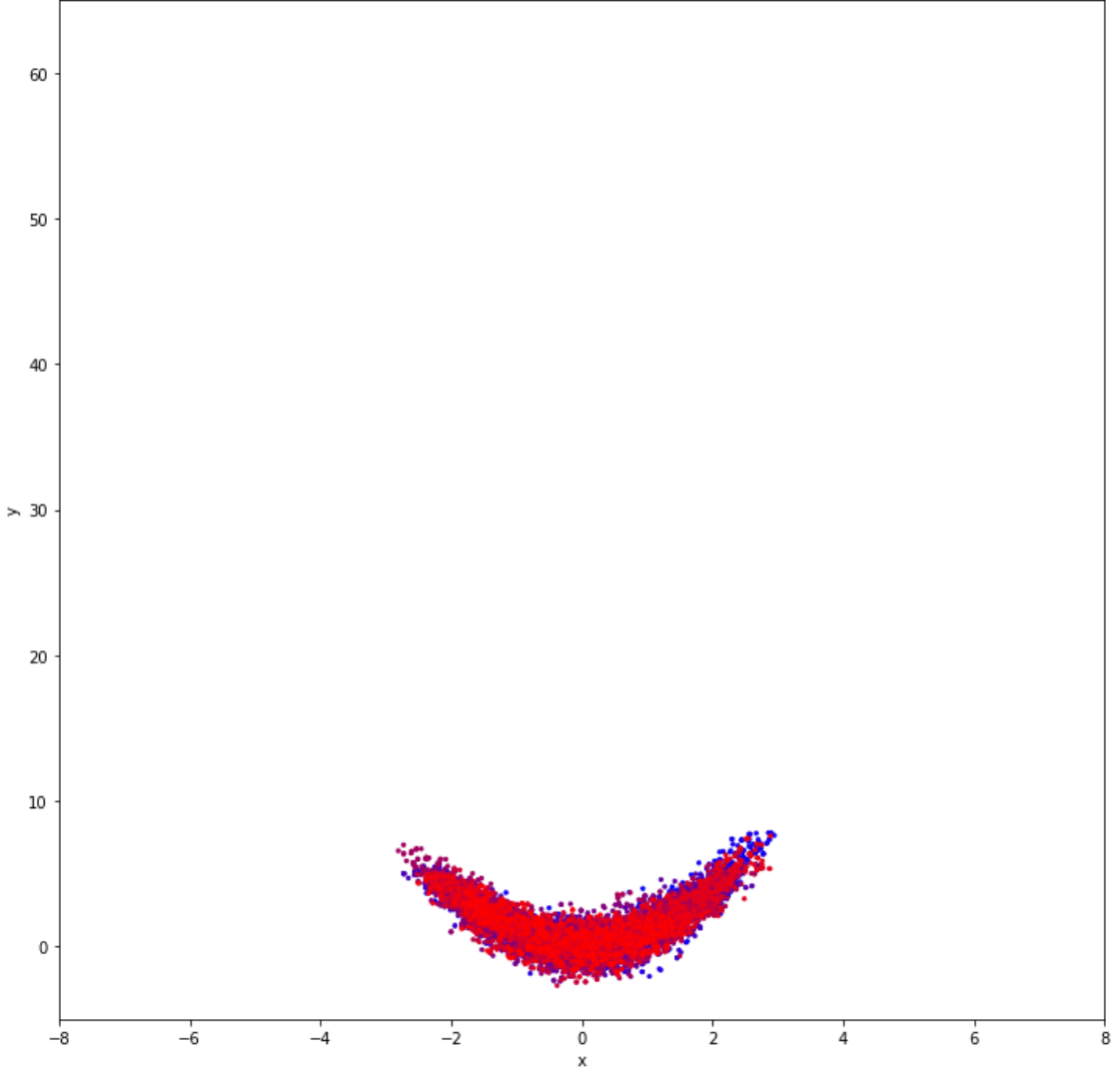}
  \caption{}
  \label{fig:rosenbrock_rwm}
\end{subfigure}%
\begin{subfigure}[t]{.5\textwidth}
  \centering
  \includegraphics[width=.8\textwidth]{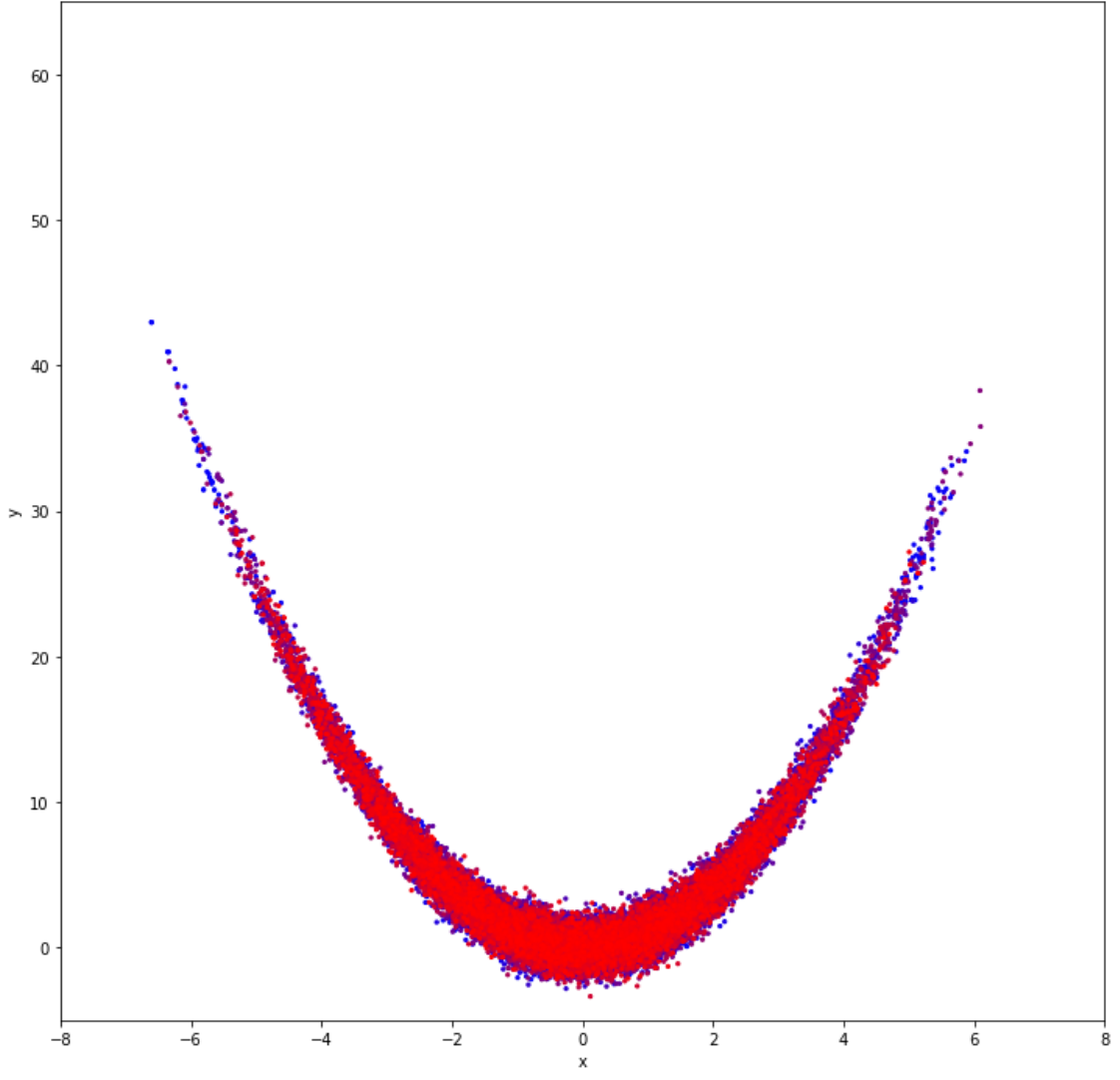}
  \caption{}
  \label{fig:rosenbrock_last}
\end{subfigure}%
\caption{Trajectories generated using \glsfmtshort{RWM} (\ref{fig:rosenbrock_rwm}) and \glsfmtshort{HMC} (\ref{fig:rosenbrock_last}) to sample from a (bivariate) Rosenbrock function. Both chains started at the origin and were evolved for $10^4$ states. These states are plotted as points gradating from blue to red (from earlier to later iterations).}
\label{fig:rosenbrock_rwm_hmc}
\end{figure}

The results of using \gls{RWM} and \gls{HMC} to sample from this function are shown in figure \ref{fig:rosenbrock_rwm_hmc}. The \gls{RWM} proposal standard deviation was fixed at $\text{Cov}_\text{prop}=\texttt{diagonal}(0.15)$, yielding a $68\%$ acceptance rate, whereas for \gls{HMC} the acceptance rate was $98\%$ with $L=2^6,\epsilon=0.1,\mathcal{M}=\opsub{I}{2}$. 

Clearly, \gls{HMC} achieves a resolution that \gls{RWM} does not: the momentum is vital for climbing up the elongated corners of the shape. Owing to it, \gls{HMC} comes much closer to reaching the regions with largest values of $y$ and non-null density (which occur at $y \approx 55$), reaching roughly $y \approx 40$ while \gls{RWM} fails to achieve even $y = 10$. This result could be made still better by increasing $L$ or the number of transitions.

\begin{figure}[!ht]
\captionsetup[subfigure]{width=\textwidth}%
\begin{subfigure}[t]{\textwidth/3}
  \centering
  \includegraphics[width=\textwidth]{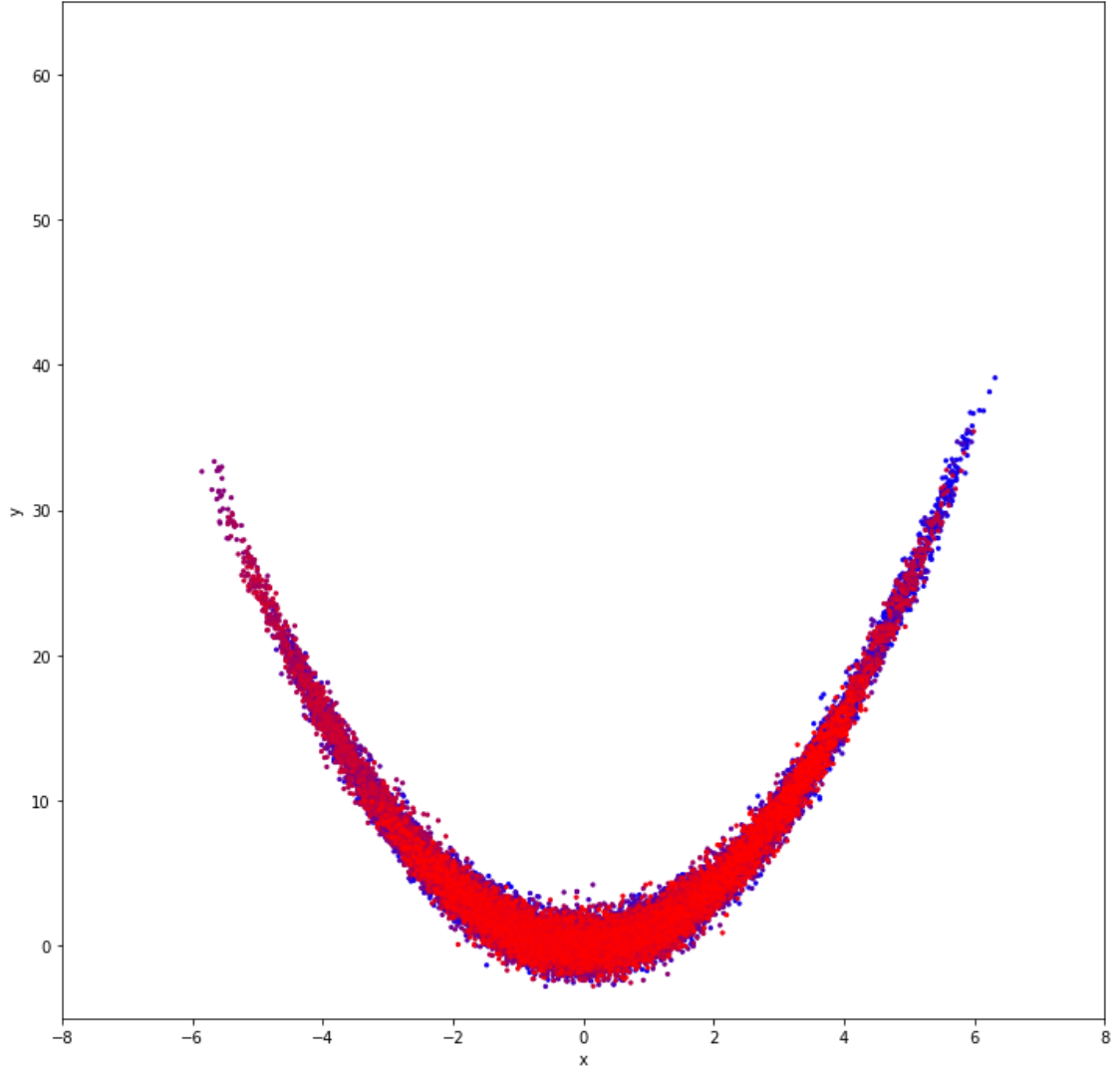}
  \caption{}
  \label{fig:rosenbrock_uniform}
\end{subfigure}%
\begin{subfigure}[t]{\textwidth/3}
  \centering
  \includegraphics[width=\textwidth]{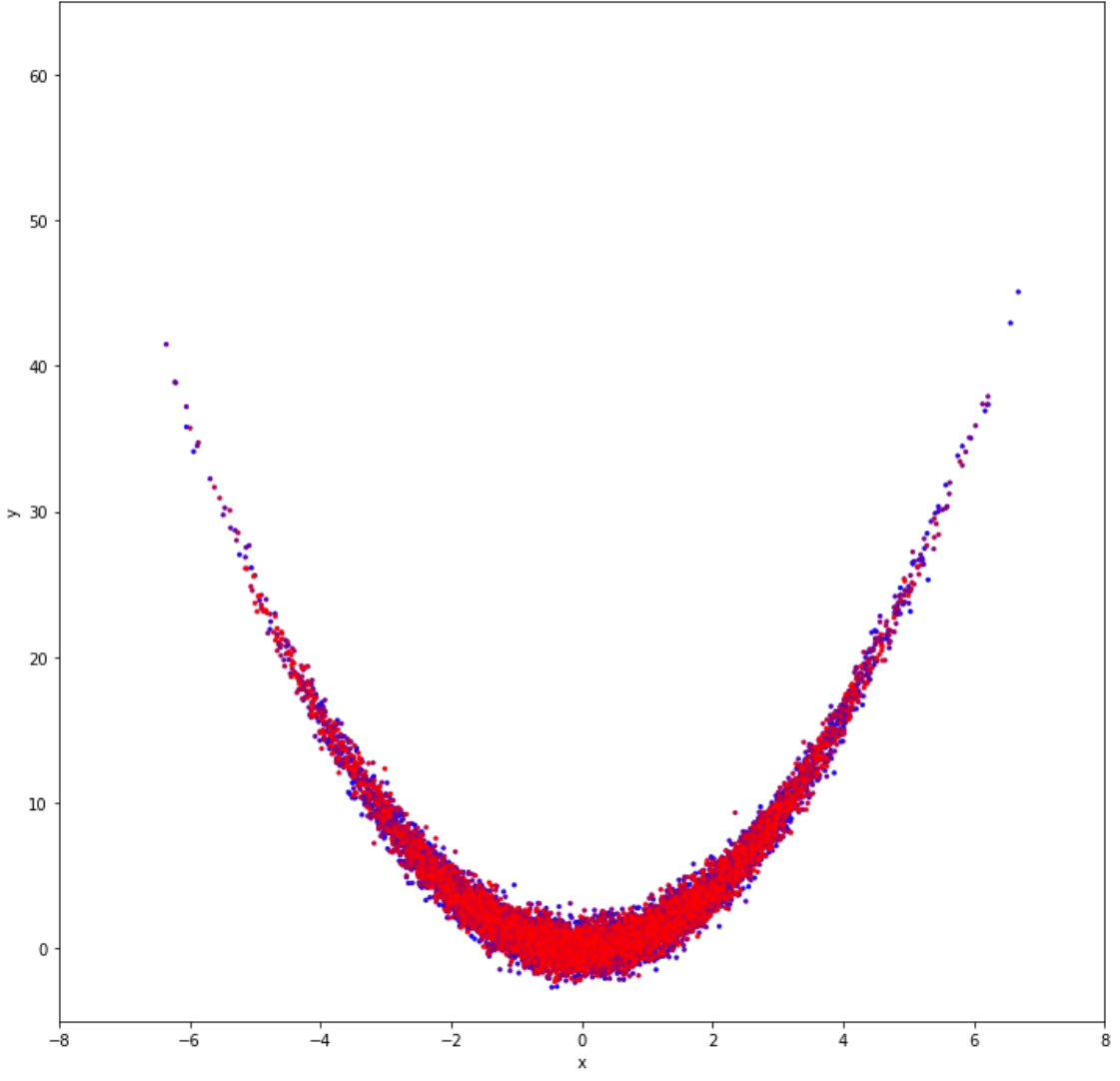}
  \caption{}
  \label{fig:rosenbrock_biased}
  \end{subfigure}%
\begin{subfigure}[t]{\textwidth/3}
  \centering
  \includegraphics[width=\textwidth]{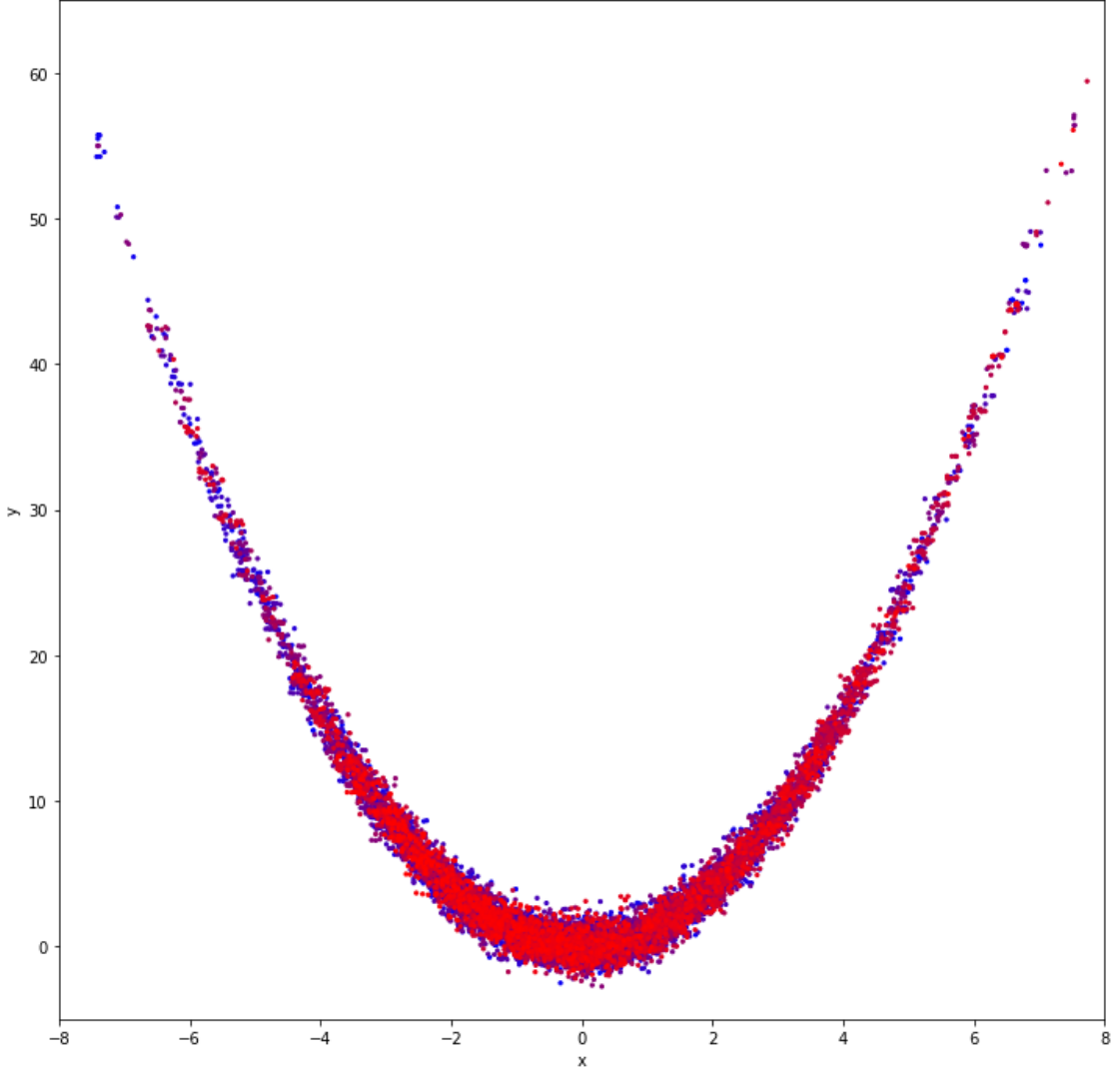}
  \caption{}
  \label{fig:rosenbrock_nuts}
  \end{subfigure}%
\caption{Trajectories generated using different variants of \glsfmtshort{HMC} to sample from a (bivariate) Rosenbrock function. They differ (both between themselves and from the standard case of \ref{fig:rosenbrock_last}) in how the trajectories are generated and their samples picked. The sample selection methods are: uniform sampling (\ref{fig:rosenbrock_uniform}), biased progressive sampling (\ref{fig:rosenbrock_biased}) and \glsfmtshort{NUTS} with partly biased progressive sampling (\ref{fig:rosenbrock_nuts}). Figures \ref{fig:rosenbrock_uniform} and \ref{fig:rosenbrock_biased} used static implementations with $L=2^6$, whereas \ref{fig:rosenbrock_nuts}'s dynamic termination produced on average $L=2^{5.4}$. All chains started at the origin and were evolved for $10^4$ states. These states are plotted as points gradating from blue to red (from earlier to later iterations).}
\label{fig:rosenbrock_hmc_traj}
\end{figure}

Another way of improving it without increasing any iteration number is to refine the algorithm. Figure \ref{fig:rosenbrock_last} used the best known \gls{HMC} variant, where the last state of the trajectory is deterministically appointed as the proposal, and a probabilistic \gls{MH} accept/reject step is performed (algorithm \ref{alg:hmc}). Figure \ref{fig:rosenbrock_hmc_traj} shows the results of using the variants discussed in subsections \ref{sub:trajectory_generation} and \ref{sub:nuts} - namely uniform sampling, biased progressive sampling and the \gls{NUTS}. The best performing one is the \gls{NUTS}, which exhibits a rather impressive ascent to the edges all the way up to the maximal values at $y \approx 55$ (though they're visibly less populated than the main body, in conformity with figure \ref{fig:rosenbrock_3d_plot}).

In these cases, a doubling progression scheme was used for generating the trajectories. In the static biased strategy, bias was introduced both when generating and concatenating trajectories, as this was the option that worked better. For the \gls{NUTS}, however, only when joining trajectories were the probabilities skewed, as in the original implementation of \cite{Hoffman_2011}. The difference relative to the strategy exposed therein is that multinomial sampling was used instead of slice sampling for picking the proposals.

The \gls{HMC} parameters $\epsilon,\mathcal{M}$ were the same as in the previous case (figure \ref{fig:rosenbrock_last}). Due to using a dynamic termination criterion, the \gls{NUTS} doesn't allow for pre-determining the path length $L$. For assessing the performance, the average generation trajectory length per transition $L_\text{avg}$ was computed, which determines the total cost. For a fair comparison, all other methods were chosen to have $L>L_\text{avg}$; the \gls{NUTS} still outperformed them. An important point is that this $L_\text{avg}$ includes \textit{all} the computed transitions, counting those that had to be discarded to guard reversibility. By contrast, the other methods always use all the generated states; again, the \gls{NUTS} outperforms them regardless. This proves what was claimed in subsection \ref{sub:nuts}.

Another subtlety mentioned in that section was the use of a conjunction or a disjunction of the 2 U-turn conditions. For figure \ref{fig:rosenbrock_nuts}, an \texttt{OR} was used, but the \texttt{AND} was also tested. It increased $L_\text{avg}$ to $2^{7.7}$ while barely enhancing the exploration, so in this case the more restrictive condition worked best.

\subsection{Smiley kernel density estimate}
\label{sub:smiley_kde}
\begin{figure}[!ht]
\captionsetup[subfigure]{width=.4\textwidth}%
\begin{subfigure}[t]{.5\textwidth}
  \centering
  \includegraphics[width=.85\textwidth]{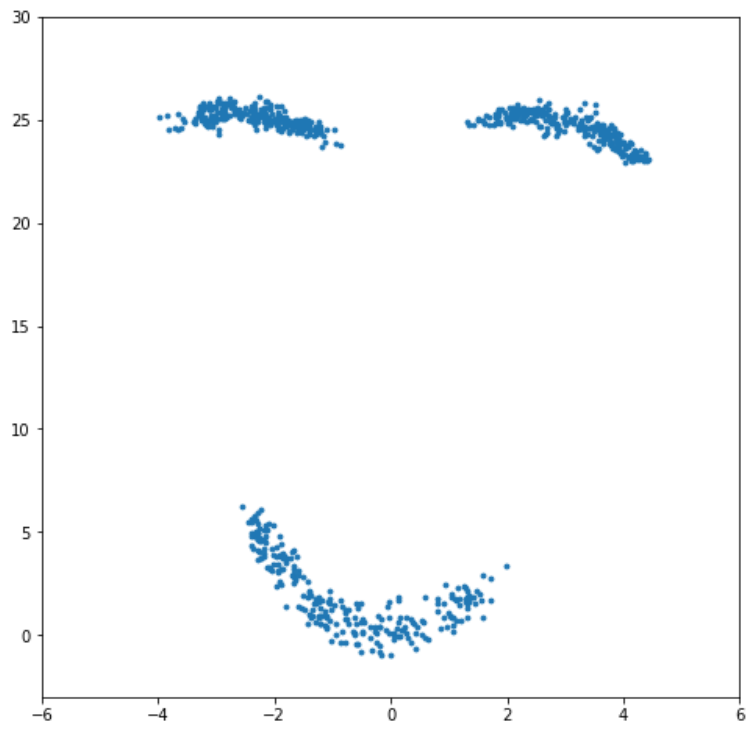}
  \caption{}
  \label{fig:smiley_og}
\end{subfigure}%
\begin{subfigure}[t]{.45\textwidth}
  \centering
  \includegraphics[width=.9\textwidth]{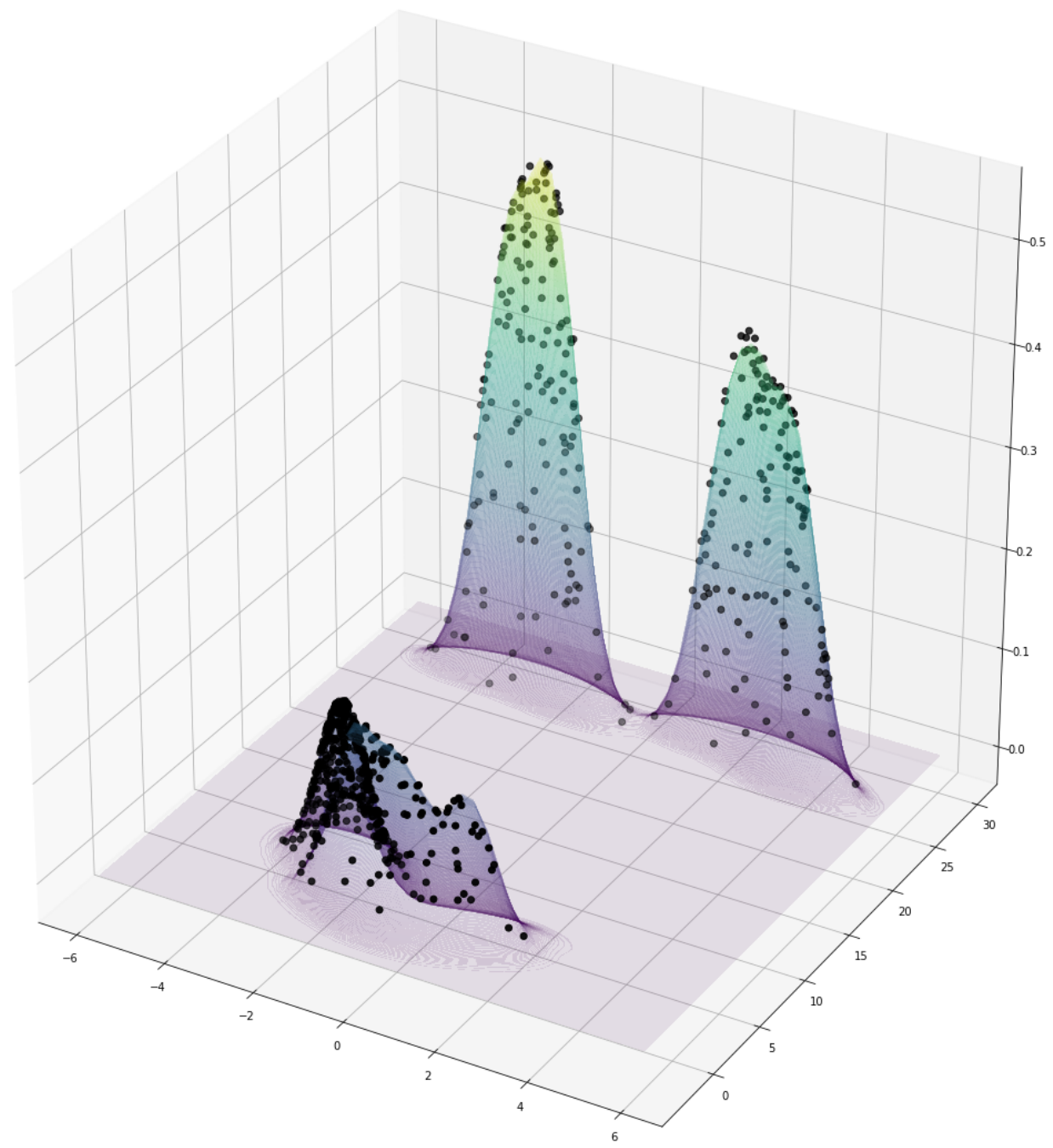}
  \caption{}
  \label{fig:smiley_3d}
\end{subfigure}%
\caption{Target bivariate smiley function (\ref{fig:smiley_og}), generated using $768$ gaussian kernels, and 3-d scatter of the particle cloud produced by \glsfmtshort{SHMC} (\ref{fig:smiley_3d}) plotted over a surface plot of the target.}
\label{fig:smiley_hmc} 
\end{figure}

The target for this subsection is a smiley face. A smiley-like density can be formed by joining together three functions similar to the one of the previous section, and rotating or flattening them to taste. Instead of using that density directly, we sample $768$ points from it and smooth them into a continuous density. This smoothing was achieved by a gaussian \gls{KDE} (figure \ref{fig:smiley_og}).

The goal is to make the data divisible: by only considering a fraction of the points, we can make the function shallower. This provides a way of creating a sequence of targets for \gls{SMC} without having to consider inference. The data can be divided into a list of chunks, and that list traversed at the rate of an item per iteration. The target density at iteration $i$ is the data contained by the first $i$ elements (counting from 1). The $768$ points were split into $8$ data-chunks, 7 containing $100$ points and $1$ containing the leftover $68$. The first iteration of \gls{SMC} then targets the first one, the second iteration targets the first one plus the second one, and so on until the eight iteration.

Note that in principle obtaining the problem-statement set of points is in itself a hard sampling problem. However, it can be made easier by splitting the probability density into 3 local densities and starting 3 separate Markov chains around the center of each eye and the mouth. That was the adopted approach, with $268$ samples attributed to each feature; \gls{HMC} chains were used, with some burn-in and lag samples discarded for better coverage. When testing, the information of the modes' number and locations were withheld from the samplers, whose particles were initially placed by a normal distribution centered at $[0,12.5]$ and with diagonal covariance \texttt{diag}(10,20).

Figure \ref{fig:smiley_3d} shows the result of applying \gls{SHMC} in these conditions, and with $768$ points. For the \gls{SMC} reweightings, a leave-one-out \gls{KDE} of the particle cloud was used in the denominator instead of the previous target. This idea from \cite{Daviet_2016} was mentioned in section \ref{sub:sir}, and aided coverage. The \gls{HMC} parameters were $L=20,\epsilon=0.05,\mathcal{M}=\opsub{I}{2}$, producing a $99\%$ acceptance rate, and resampling was performed at each step. The sampler succeeded in finding all modes and covering them satisfactorily. 

\begin{figure}[!ht]
\captionsetup[subfigure]{width=.9\textwidth}%
\begin{subfigure}[t]{\textwidth/4}
  \centering
  \includegraphics[width=.95\textwidth]{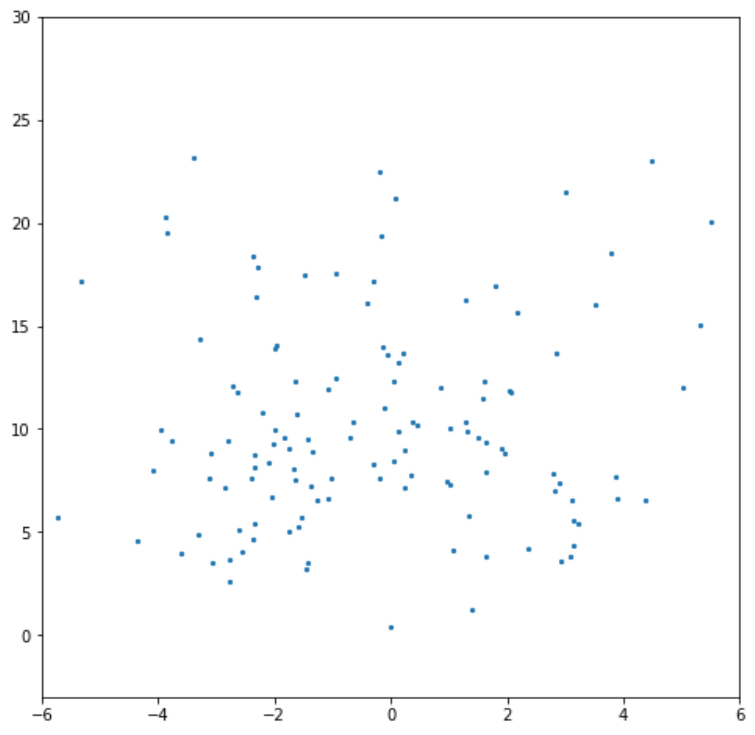}
  \caption{}
  \label{fig:smiley_RWM}
\end{subfigure}%
\begin{subfigure}[t]{\textwidth/4}
  \centering
  \includegraphics[width=.95\textwidth]{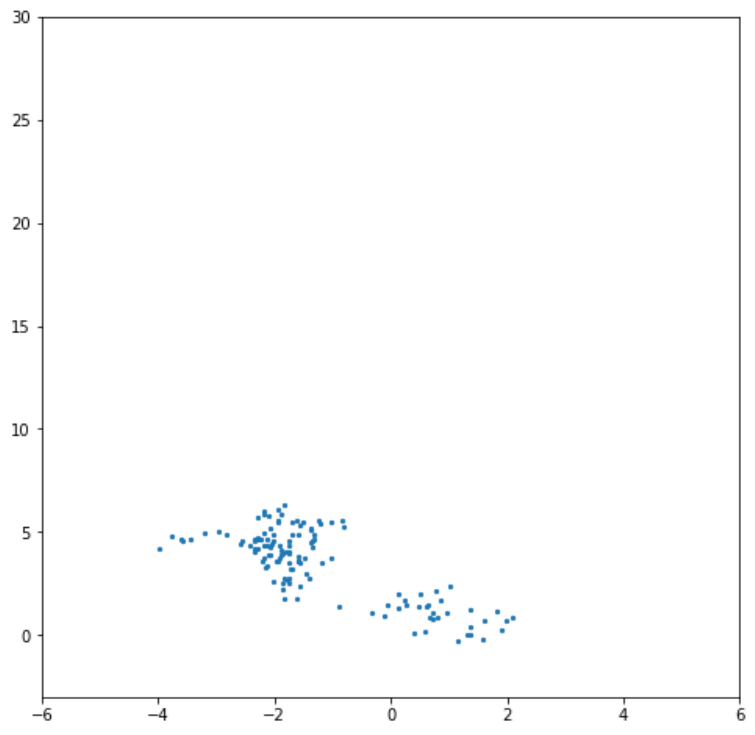}
  \caption{}
  \label{fig:smiley_SMC_RWM}
  \end{subfigure}%
\begin{subfigure}[t]{\textwidth/4}
  \centering
  \includegraphics[width=.95\textwidth]{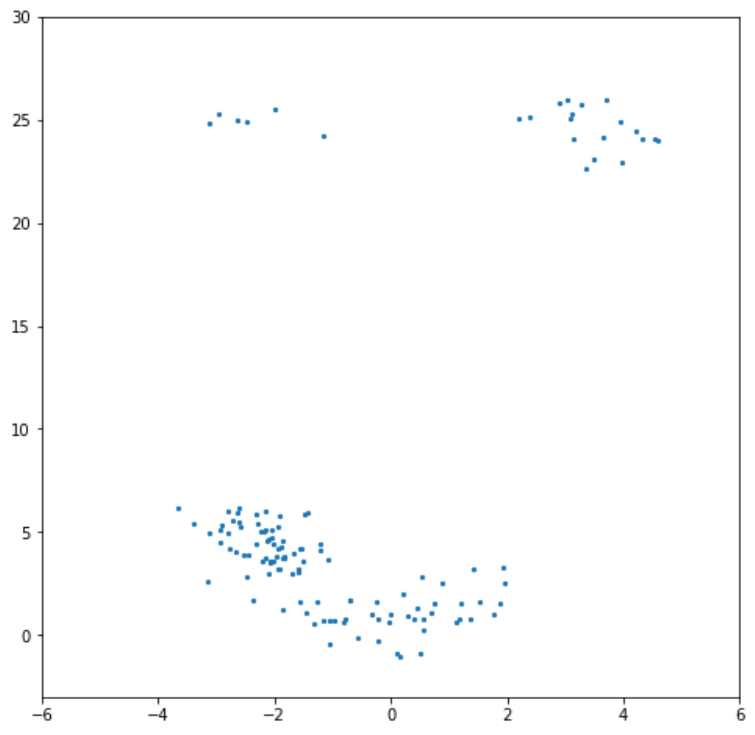}
  \caption{}
  \label{fig:smiley_HMC}
  \end{subfigure}%
  \begin{subfigure}[t]{\textwidth/4}
  \centering
  \includegraphics[width=.95\textwidth]{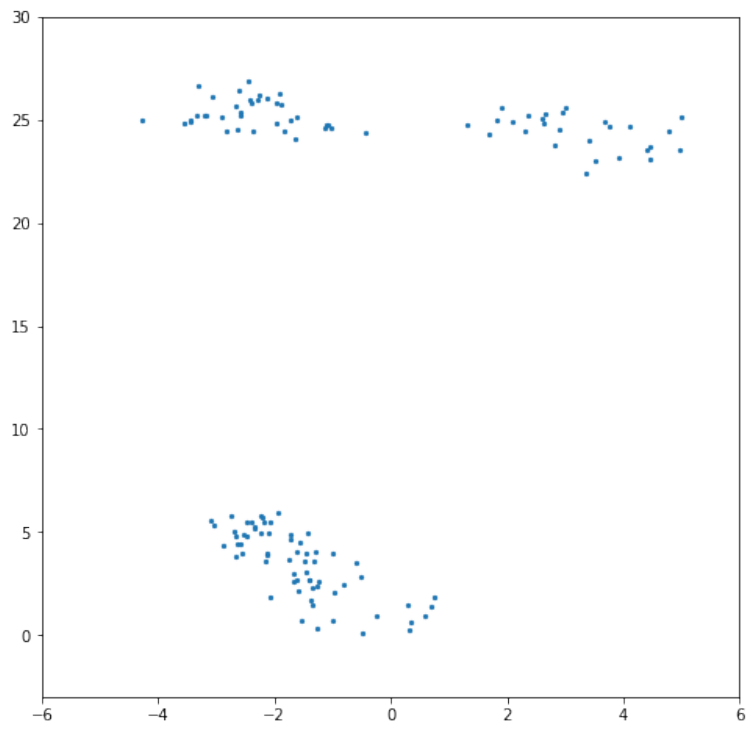}
  \caption{}
  \label{fig:smiley_SHMC}
  \end{subfigure}%
\caption{Trajectories generated using different \glsfmtshort{SMC}/\glsfmtshort{MCMC} sampling methods to sample from a (bivariate) smiley \glsfmtshort{KDE}. Results are shown for 4 methods: parallel \glsfmtshort{RWM} chains (\ref{fig:smiley_RWM}), \glsfmtshort{SMC} with \glsfmtshort{RWM} moves (\ref{fig:smiley_SMC_RWM}), parallel \glsfmtshort{HMC} chains (\ref{fig:smiley_HMC}), and \glsfmtshort{SHMC} (\ref{fig:smiley_SHMC}).}
\label{fig:smileys}
\end{figure}

Other methods were tested, namely \gls{SMC} and parallel Markov chains; figure \ref{fig:smileys} presents a selection of results. All methods used $128$ particles for $8$ steps, and the same prior (a multivariate gaussian). The particle reduction is significant, making the problem at once more challenging for the sampler and less challenging for the processing unit. Note that \gls{MCMC} typically relies on longer evolutions, but this wouldn't make for an equitable comparison: for the parallel methods, all steps are performed on the final density, so even when matching them for iterations they're already more costly than \gls{SMC}. 

The conditions for \gls{HMC} were as before, producing $96\%$ acceptance. \gls{RWM} used $\sigma_\text{prop}=0.05$, producing $78\%$  and $66\%$ acceptance within and outside \gls{SMC} respectively. The difference between methods is clear on two fronts: the coupling of \gls{SMC} greatly benefits the particle positioning and mode evenness as compared to non-interacting chains, and the insight of \gls{HMC} seems to be an as strong asset. 

\section{Inference problems}
\label{sec:inference_problems}

\subsection{Phase estimation}
\label{sub:phase_estimation}

This example is based on \cite{Wiebe_2016}, for the idea of Bayesian (iterative) phase estimation in general as for the implementation of \gls{GRF} and adaptive heuristics in particular. All of this was explored in subsection \ref{sub:quantum_characterization_examples}.

\begin{figure}[!ht]
\captionsetup[subfigure]{width=.9\textwidth}%
\begin{subfigure}[t]{.5\textwidth}
  \centering
  \includegraphics[width=\textwidth]{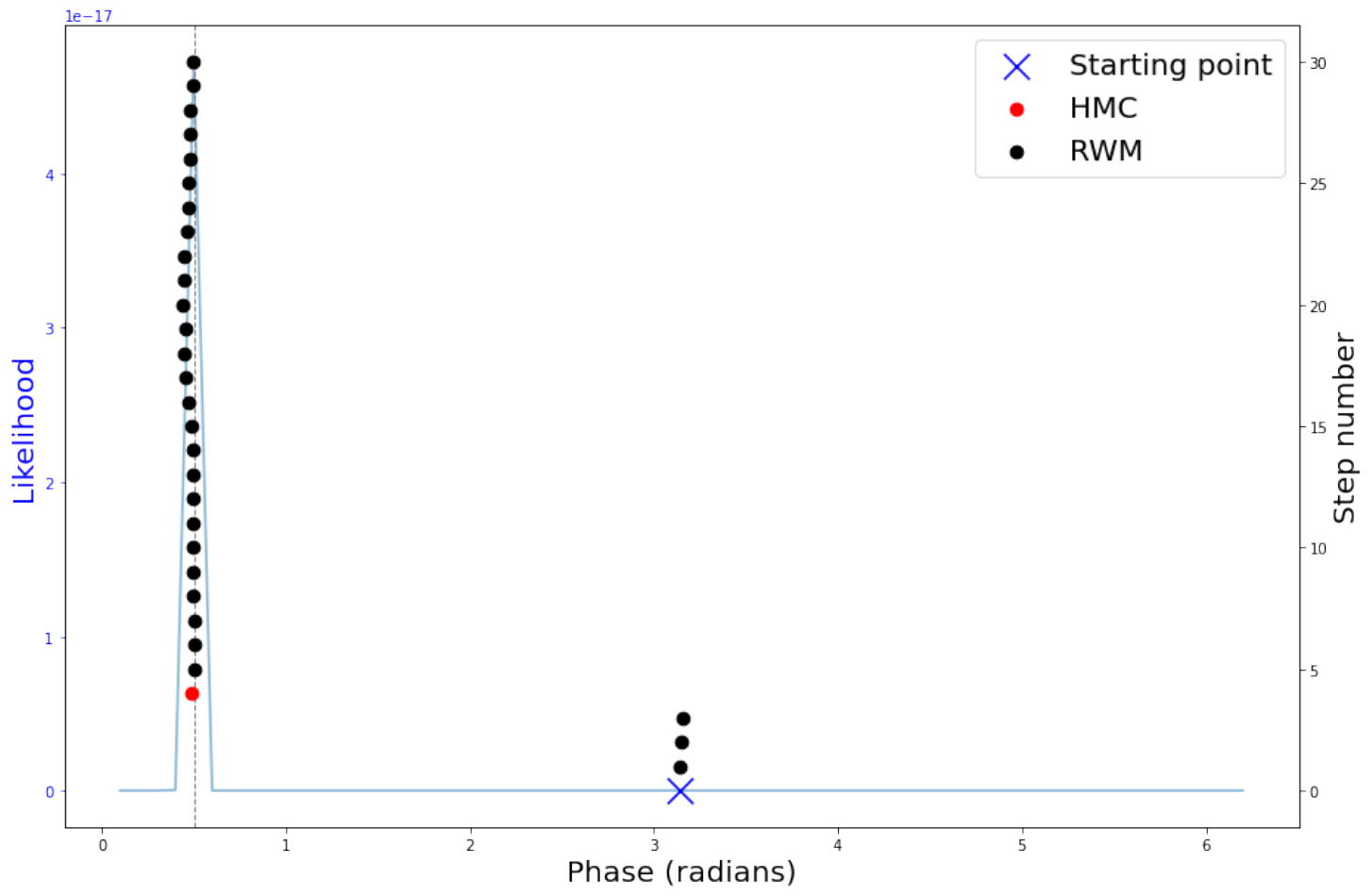}
  \caption{Trajectory generated by \glsfmtshort{MCMC} with \glsfmtshort{RWM} (black dots) and \glsfmtshort{HMC} (red dots) transitions. The \glsfmtshort{HMC} proposals prompted essentially all bigger leaps, but were most often unavailing. The experiment controls were chosen semi-arbitrarily in advance, and all the data are used at each step.}
  \label{fig:ipe_hmc}
\end{subfigure}%
\begin{subfigure}[t]{.5\textwidth}
  \centering
  \includegraphics[width=\textwidth]{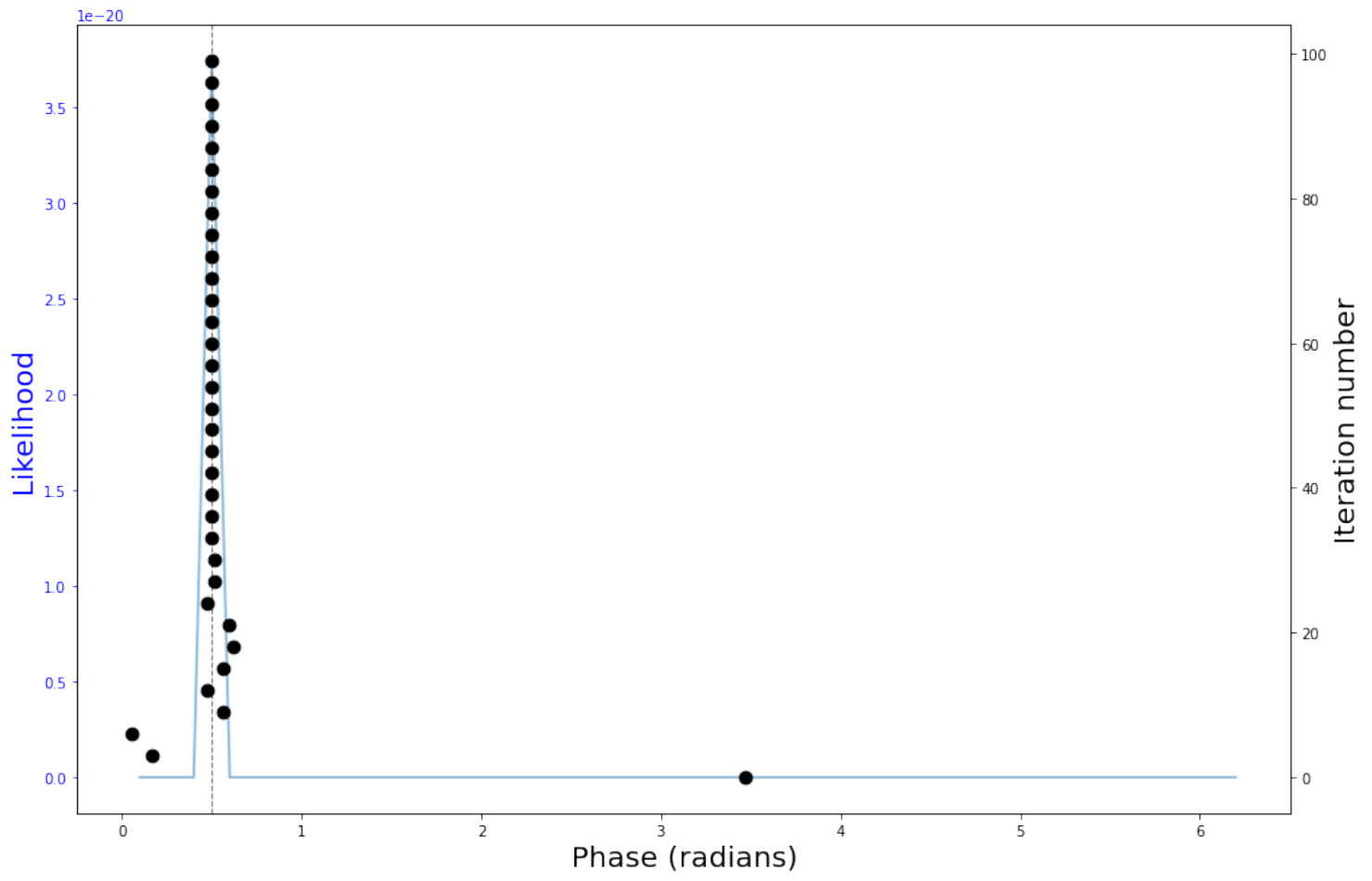}
  \caption{Evolution of the distribution means through the iterations of \glsfmtshort{GRF}. The experiment controls were chosen adaptively. Each iteration $i$ uses only the $i$th datum and the $(i+1)$th iteration's results.}
  \label{fig:ipe_grf}
\end{subfigure}%
\caption{Evolution of the sampling process through the iterations for \glsfmtshort{MCMC} with \glsfmtshort{RWM} and \glsfmtshort{HMC} move steps (\ref{fig:ipe_hmc}) and for \glsfmtshort{GRF} (\ref{fig:ipe_grf}). Both consider $100$ experiments total, but their design and use are different. The horizontal coordinate is the phase (angle) value. The points are positioned vertically and upward according to the iteration number, e.g. the topmost point corresponds to the final result. The likelihoods based on the full datasets are plotted in blue in a second axis for visualization.}
\label{fig:ipe_graphs}
\end{figure}

Figure \ref{fig:ipe_graphs} shows the results of applying \gls{MCMC} and \gls{GRF} to learning a phase set at $\phi_\text{real}=0.5$ using $100$ measurements and a flat prior on $[0,2\pi[$. The parameter space is periodic. \gls{GRF} works under a normality assumption, where the distributions (the successive posteriors) are represented in the period-long interval by two quantities: the arithmetic mean and variance of the particle locations. With this, some distributions over the phase are bound to be misrepresented. For instance, if the distribution is sharp and centered at $2\pi$, this description will relay a broad gaussian centered at $\pi$ due to a bisection at the borders. To account for this, a wrapped normal distribution (on the unit circle) can be considered. A clever alternative fix is suggested in \cite{Wiebe_2016}; the authors propose screening for this type of problem using the variance and treating it as a special case using $\pi$ shifts. That was the option opted for here.

For \gls{GRF}, the experimental controls were chosen adaptively according to section \ref{sub:precession_heuristics}; the data were considered one at a time (per iteration), meaning $100$ steps. For \gls{MCMC}, all combinations of $M \in [1..10]$ and $\theta \in [0..9] \cdot \pi/5$ were used, and $30$ transitions were realized. Note that even though this corresponds to fewer iterations, each of them considers all the data, in contrast with \gls{GRF} where a single datum is contemplated per step (along with the previous results). This makes \gls{GRF} more lightweight at the expense of correctness. It also enables adaptivity, bringing more informative data: after roughly $20$ measurements, the results are already quite decisive, whereas with so few data \gls{MCMC} typically fails to converge.

The problem of applying \gls{HMC} to this type of problem was discussed in \ref{sub:applicability_hmc}. To manage that, a \gls{RWM} step was performed after the \gls{HMC} one whenever the acceptance probability for the latter fell below $0.01$.  This resulted in only $3.3\%$ \gls{HMC} steps in spite of \gls{HMC} acceptance's nearing $100\%$. In other words, all points attempted \gls{HMC}, but for $96.7\%$ this resulted in vanishing probability of moving; a select few had near unit probability of relocation. It is clear that the \gls{HMC} transitions are the driving force behind the bold transitions; all big leaps that the chain carried off was produced by \gls{HMC}. Notwithstanding, this is a hard problem to manage, especially for longer evolutions, and requires lavishing resources on the proposals for rare benefit. Even though it can be controlled by fine-graining the integration further, the effect is limited and resource-intensive.

\subsection{Precession and bimodality}
\label{sub:sampling_precession}

This example is based on \cite{Granade_2017}. It has been mentioned in subsection \ref{sub:quantum_characterization_examples}.

Figure \ref{fig:ipe_graphs} shows the results of using \gls{SMC}-\gls{SIR} with \gls{LWF} and \gls{MCMC} to learn a frequency $\omega_\text{real}=0.5$ using $100$ measurements and a flat prior on $\omega \in ]0,10]$. The times were chosen in increments of $0.08$, decided with a parameter sweep. Note that what matters is not the units, but only the relative scale between times and frequencies. They're always found paired up in the form $\omega \cdot t$.

\begin{figure}[!ht]
\captionsetup[subfigure]{width=.9\textwidth}%
\begin{subfigure}[t]{.5\textwidth}
  \centering
  \includegraphics[width=\textwidth]{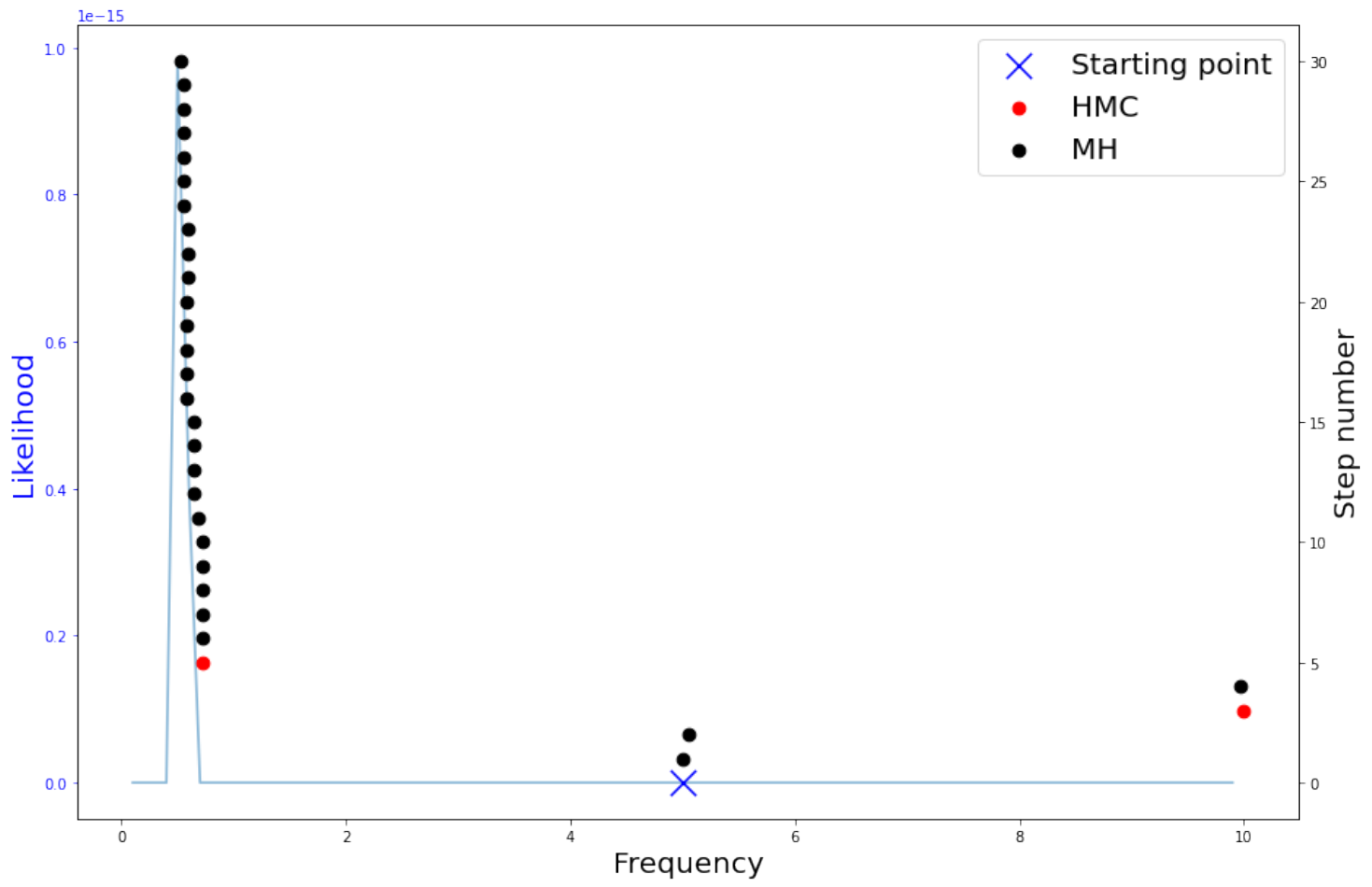}
  \caption{Trajectory generated by \gls{MCMC} with \gls{RWM} (black dots) and \gls{HMC} (red dots) transitions. All data are used at each step. \gls{HMC} proposals prompted essentially all bigger leaps, but were most often unavailing.}
  \label{fig:precession_hmc_rwm}
\end{subfigure}%
\begin{subfigure}[t]{.5\textwidth}
  \centering
  \includegraphics[width=\textwidth]{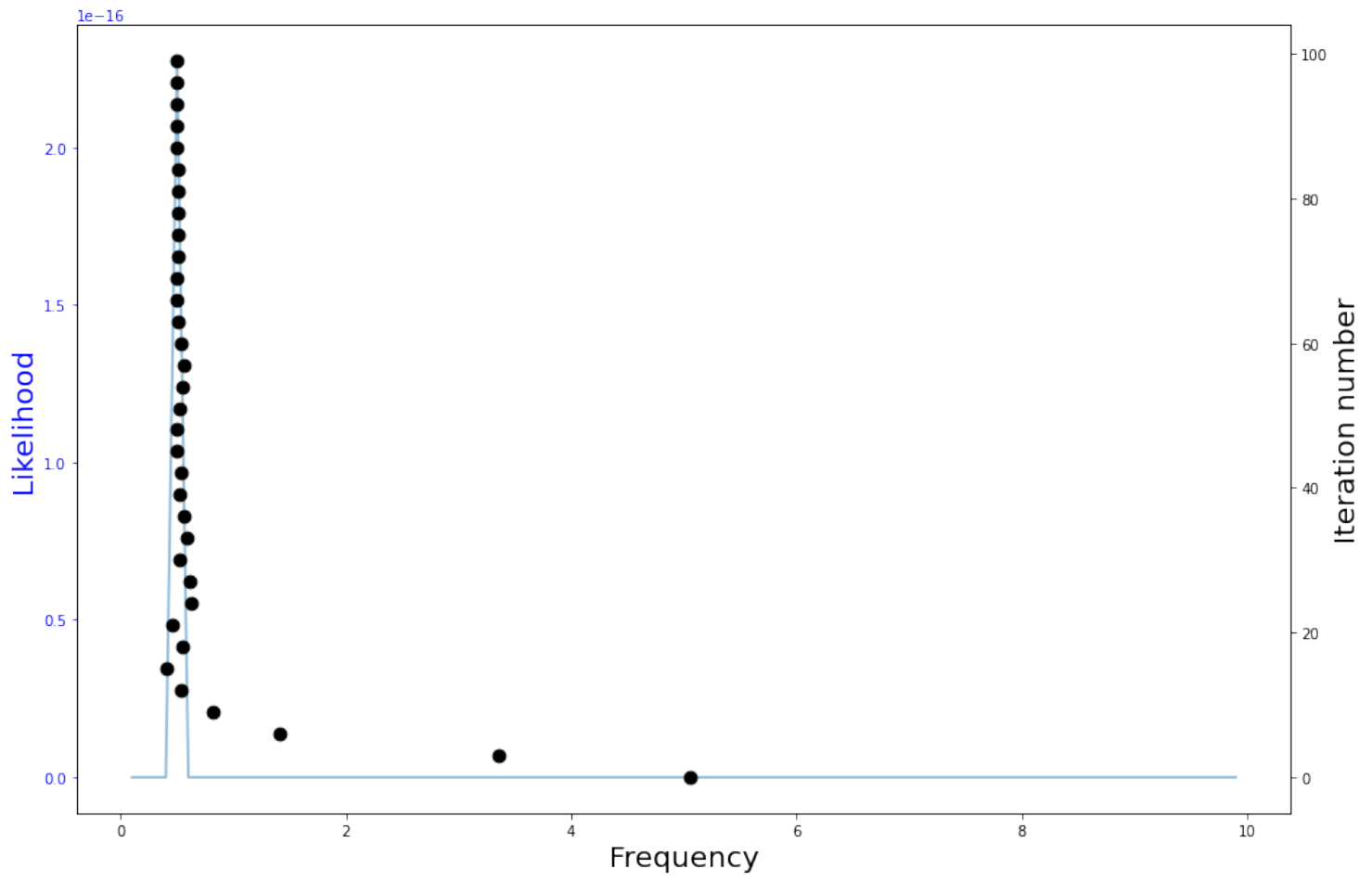}
  \caption{Evolution of the distribution means through the iterations of \gls{SIR} with \gls{LWF}. Each iteration $i$ uses only the $i$th datum and the $(i+1)$th iteration's results.}
  \label{fig:precession_smc_lw}
\end{subfigure}%
\caption{Evolution of the sampling process through the iterations for for \glsfmtshort{MCMC} with \glsfmtshort{RWM} and \glsfmtshort{HMC} move steps (\ref{fig:precession_hmc_rwm}) and \glsfmtshort{SMC} with \glsfmtshort{LWF} (\ref{fig:precession_smc_lw}). Both consider $100$ experiments total, and the same set of pre-determined controls. The points are positioned vertically and upward according to the iteration number, e.g. the topmost point is the final result; the horizontal coordinate is the phase (angle) value. The likelihood is plotted in blue on a second axis for visualization.}
\label{fig:precession_graphs_dots}
\end{figure}

\gls{MCMC} was implemented in the same conditions as section \ref{sub:phase_estimation}, resulting in roughly $100\%$ mean \gls{HMC} acceptance but only $6.7\%$ \gls{HMC} steps. Clearly, we face the same problem as before. Both used 100 data; \gls{MCMC} used 30 steps, and \gls{SMC} was implemented as in algorithm \ref{alg:sir}, with 100 particles and the \gls{LWF} parameter $a=0.98$. Despite using more steps and particles, it actually uses fewer likelihood evaluations than \gls{MCMC}: $T=100 \cdot 100$ for the $100$ reweightings of $100$ particles. Had \gls{MCMC} ran for 100 iterations, it would use as many with \gls{RWM} alone, whereas \gls{HMC} would add to that $LT$ gradient evaluations. Naturally, performing both at some iterations accumulates their costs, which end up higher even for $30$ steps (and would still even if \gls{RWM} were dispensable, especially if considering that the cost of evaluating the likelihood is usually much less significant than that of computing its gradient). Also, \gls{MCMC} by itself is less robust, especially for higher frequencies (or equivalently longer evolution times) and/or broader priors, as well as incompatible with online processing (unlike \gls{SIR}).

\begin{figure}[!ht]
\captionsetup[subfigure]{width=.9\textwidth}%
\begin{subfigure}[t]{.5\textwidth}
  \centering
  \includegraphics[width=0.8\textwidth]{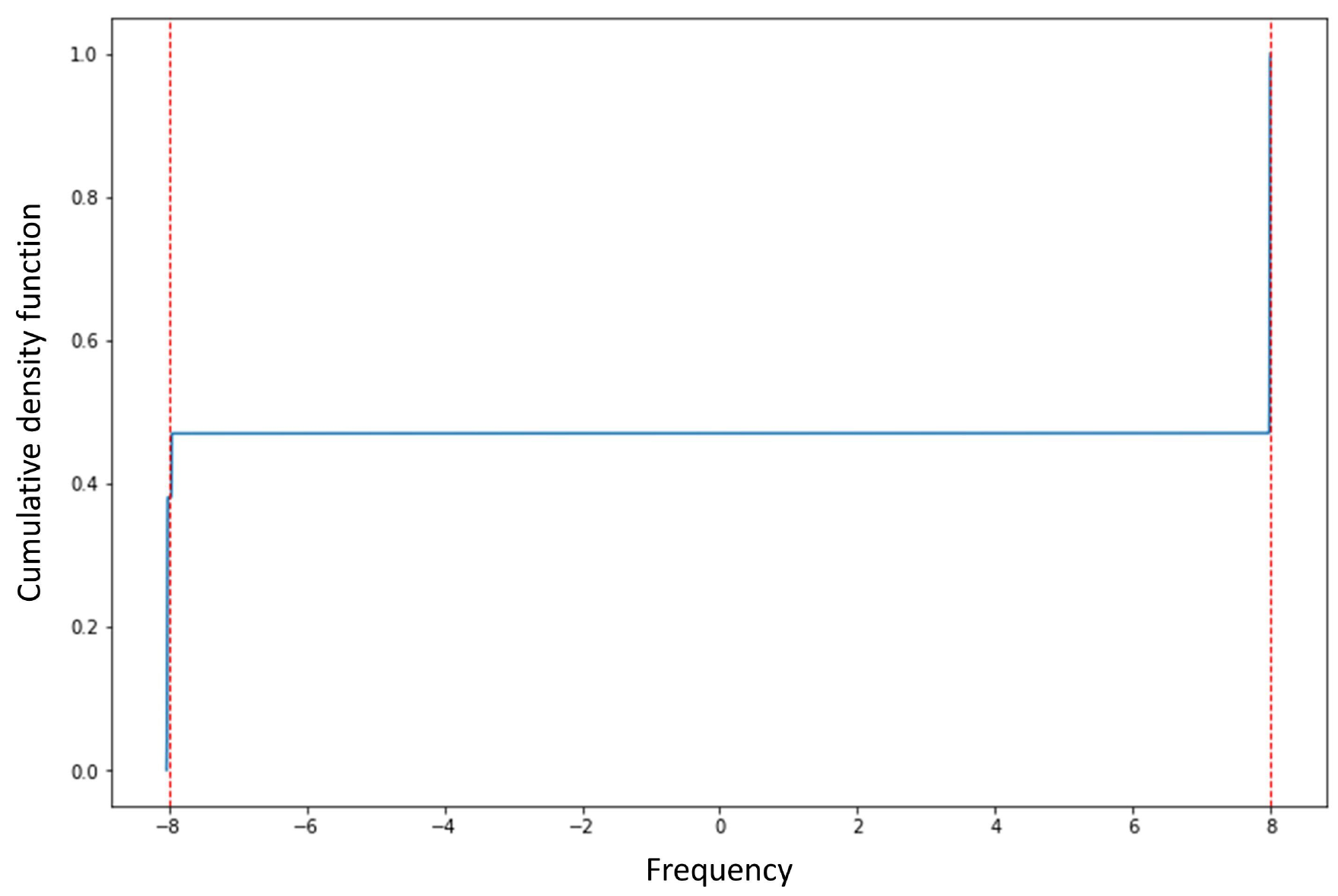}
  \caption{Distribution produced by \glsfmtshort{SHMC}.}
  \label{fig:shmc_cdf}
\end{subfigure}%
\begin{subfigure}[t]{.5\textwidth}
  \centering
  \includegraphics[width=0.8\textwidth]{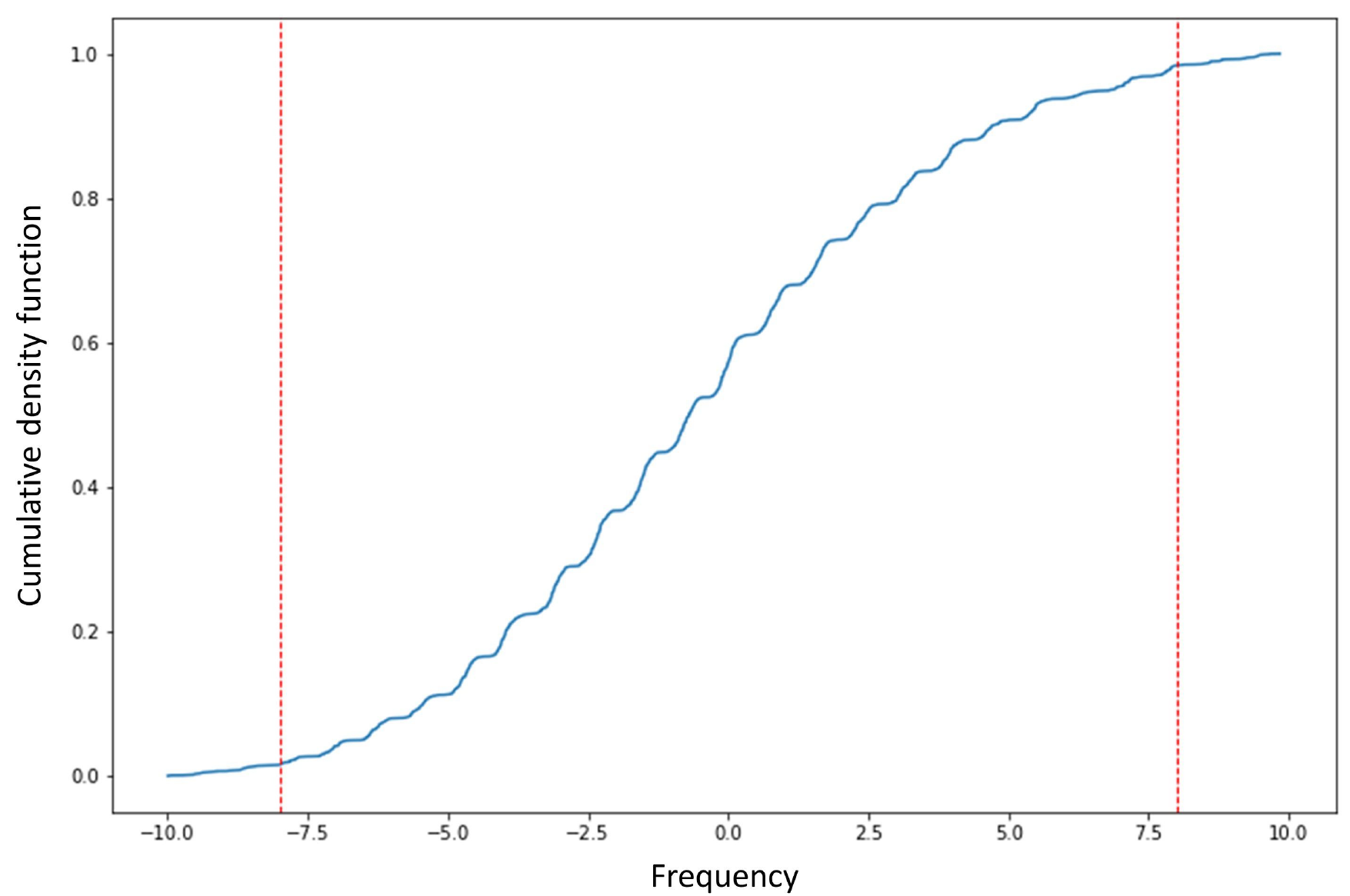}
  \caption{Distribution produced by \glsfmtshort{SMC} with \glsfmtshort{LWF}.}
  \label{fig:lw_cdf}
\end{subfigure}%
\caption{Cumulative density functions obtained via  \glsfmtshort{MCMC} with \glsfmtshort{RWM} and \glsfmtshort{HMC} move steps (\ref{fig:shmc_cdf}) and \glsfmtshort{SMC} with \glsfmtshort{LWF} (\ref{fig:lw_cdf}), for a prior with support over negative frequencies. The modes are marked with dashed vertical red lines.}
\label{fig:precession_cdf}
\end{figure}

As discussed in \ref{sub:mcmc_smc}, Markov chains can be made more robust by coupling. We now switch from \gls{MCMC} per se to Markov transitions within an \gls{SMC} scheme, and compare this with \gls{SMC}-\gls{LWF} (both with \gls{SIR}). The advantages can be made obvious by extending the domain to encompass negative frequencies. Due to the cosine function's (formula \ref{eq:precession_summary}) symmetry, this gives rise to muldimodality, in which case \gls{LWF}'s performance is disastrous (as was to be expected). This was given as a motivation for seeking alternative methods in the reference cited for this section, \cite{Granade_2017}. The issue can be seen clearly in figure \ref{fig:precession_cdf}.

The conditions are the same as before, except for the fact that $2000$ particles were used instead of $100$ to make for smoother curves and the frequency was fixed at $\omega=8$ (or equivalently $\omega=-8$).

Finally, we present results for \gls{TLE} (section \ref{sub:tle}). We test \gls{TLE}-\gls{SHMC}, both based on the full data and with energy conserving subsampling and control variates (\gls{TLE}-\gls{SHMC}-\gls{ECS}, section \ref{sec:subsampling}). In the latter case, the likelihood evaluations were based on a subset of the data for both the reweightings and \gls{HMC}. The subsampling indices for each iteration and particle were chosen in \gls{bPM}-\gls{MH} steps (for $3$ blocks, inducing roughly $67\%$ correlation). 

\begin{figure}[!ht]
    \centering
    \includegraphics[width=6.5cm]{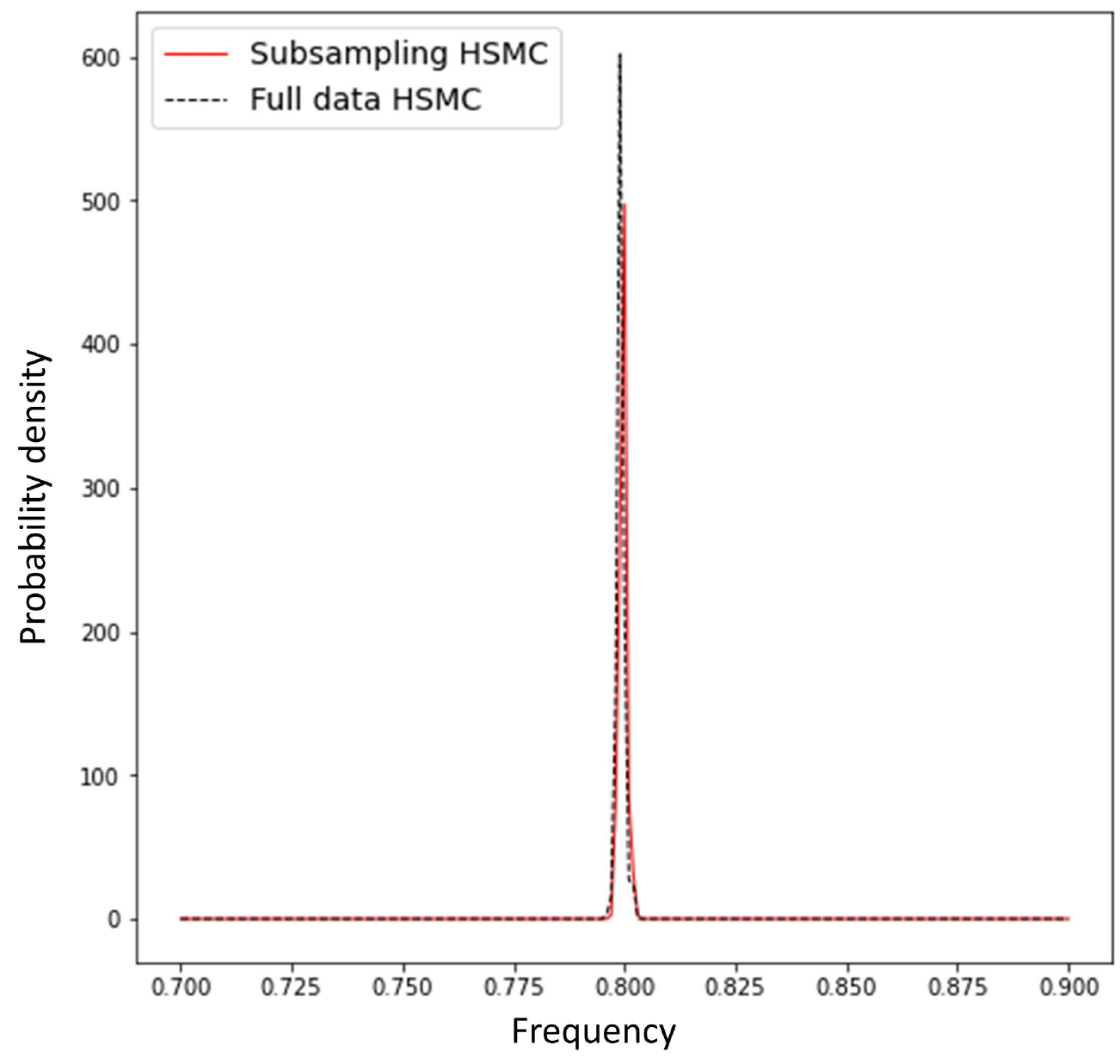}
    \caption{Kernel density estimates of the posterior resulting from full data \glsfmtshort{SHMC} and from subsampling 50/400 observations (\glsfmtshort{SHMC}-\glsfmtshort{ECS}).}
    \label{fig:ecs_kde}
\end{figure}

The applied control variates being gradient based (Taylor expansions), they didn't work well due to high target curvature. This can be controlled in several ways. One option would be to keep the evolution times short, but this is inefficient because they bring little information after a point. Another one would be to narrow down the prior. This can be done by a warm up phase on the first (lower times) observations if prior knowledge is
insufficient to justify it. Chapter \ref{cha:experiments_quantum_hardware} will show the warming up alternative; for these tests, a less general prior was chosen. The posterior distributions are presented in figure \ref{fig:ecs_kde} as \gls{KDE}s in the domain of the prior. The real frequency was set at $\omega = 0.8$.

 The $400$ measurement times were chosen randomly up to $t_\text{max}=100$, and when subsampling $50/400$ were evaluated. Apart from that, all conditions were matched for the two methods. The tempering coefficients were $10$ and evenly spaced. The resulting standard deviations were $\sigma_\text{FD} = 8.2 \times 10^{-4}$ and $\sigma_\text{subs} = 8.6 \times 10^{-4}$ for the full data and subsampling cases respectively.

\subsection{Multivariate function (2 and 4-dimensional estimation)}
\label{sub:multi_cos}

This section considers a multi-parameter generalization of the two previous examples, with the purpose of testing methods in more challenging settings. This can be achieved by choosing a vector $\Vec{\omega}$ containing $dim$ parameters $\omega_d$ and choosing a still binomial ($D \in \{0,1\}$) likelihood:
\begin{equation}
    L(\Vec{\omega} \mid D) = \frac{1}{dim} \sum_{d=1}^{dim} \cos^2 (\frac{\omega_d t}{2})
\end{equation}

Clearly, this introduces redundancy, resulting in multimodality. The number of modes reflects in how many ways the frequencies can be distributed by the dimensions, which are symmetric: the parameters all play identical roles. In other words, there are as many modes as there are possible orders for the parameters. This total is given by the number of permutations (without repetitions) of the dimensions, which is the factorial $dim!$.  

\begin{figure}[!ht]
\captionsetup[subfigure]{width=.9\textwidth}%
\begin{subfigure}[t]{\textwidth/3}
  \centering
  \includegraphics[width=.95\textwidth]{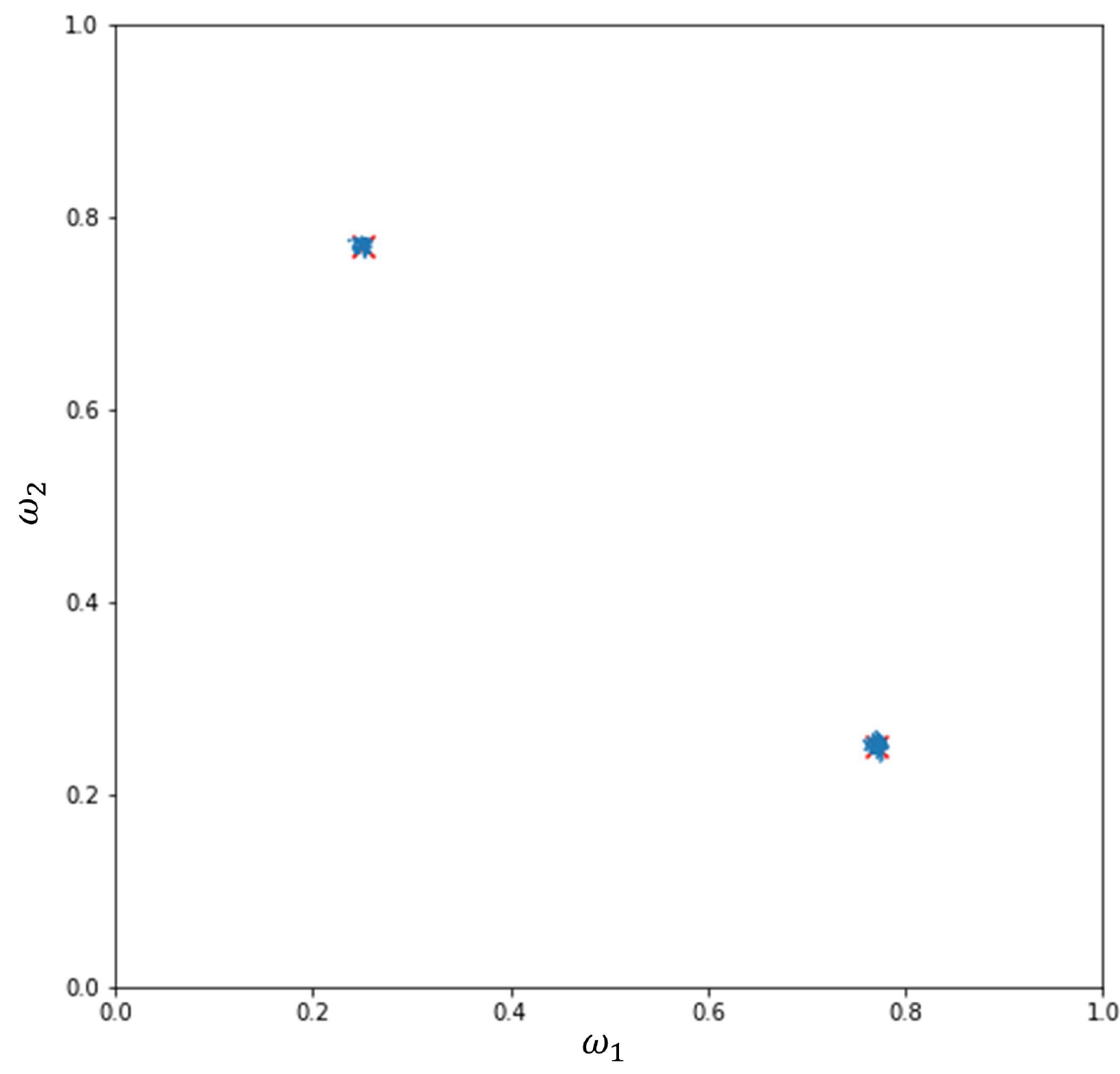}
  \caption{}
  \label{fig:2d_resampling}
\end{subfigure}%
\begin{subfigure}[t]{\textwidth/3}
  \centering
  \includegraphics[width=.95\textwidth]{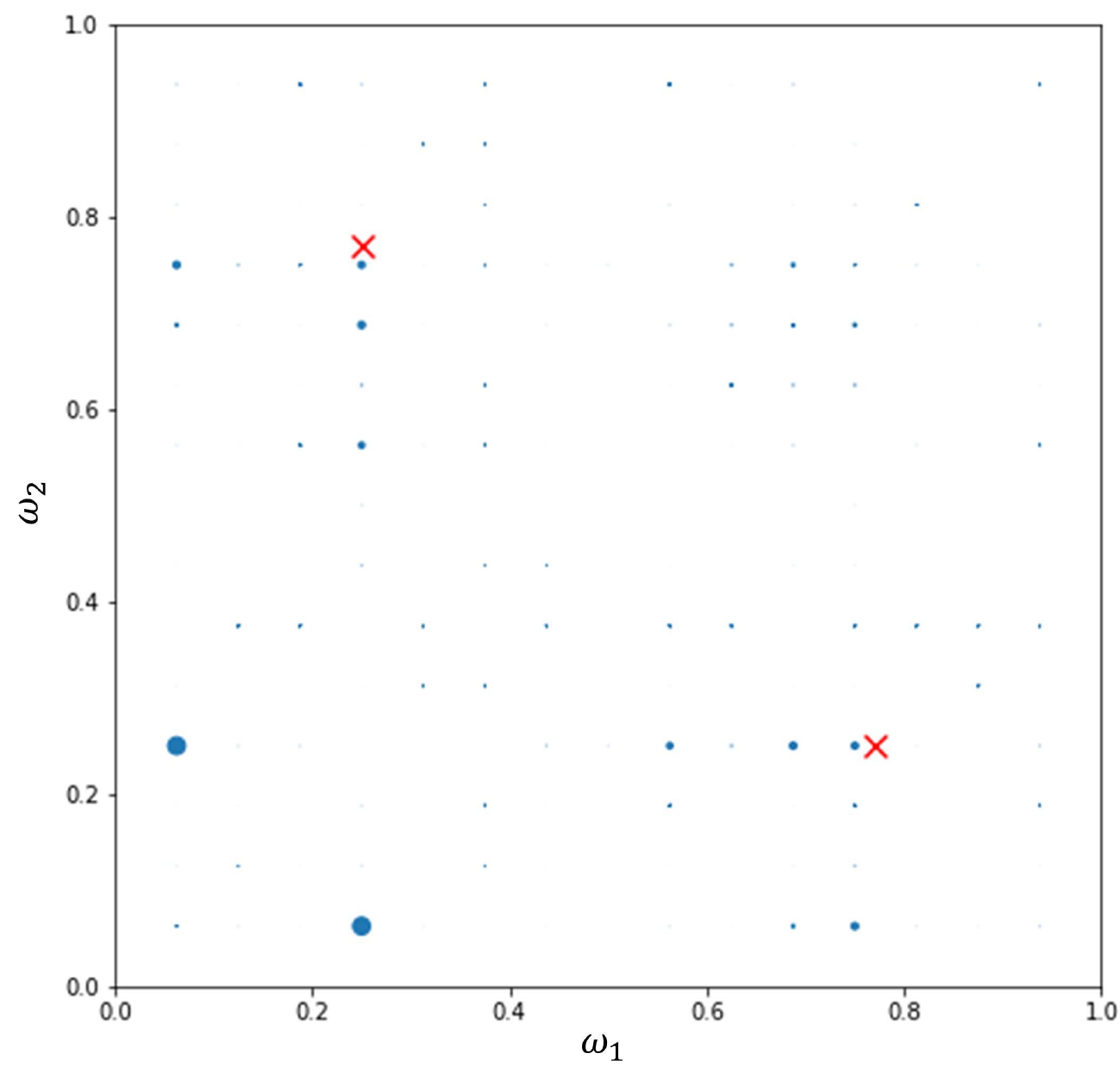}
  \caption{}
  \label{fig:2d_grid_matched}
  \end{subfigure}%
\begin{subfigure}[t]{\textwidth/3}
  \centering
  \includegraphics[width=.95\textwidth]{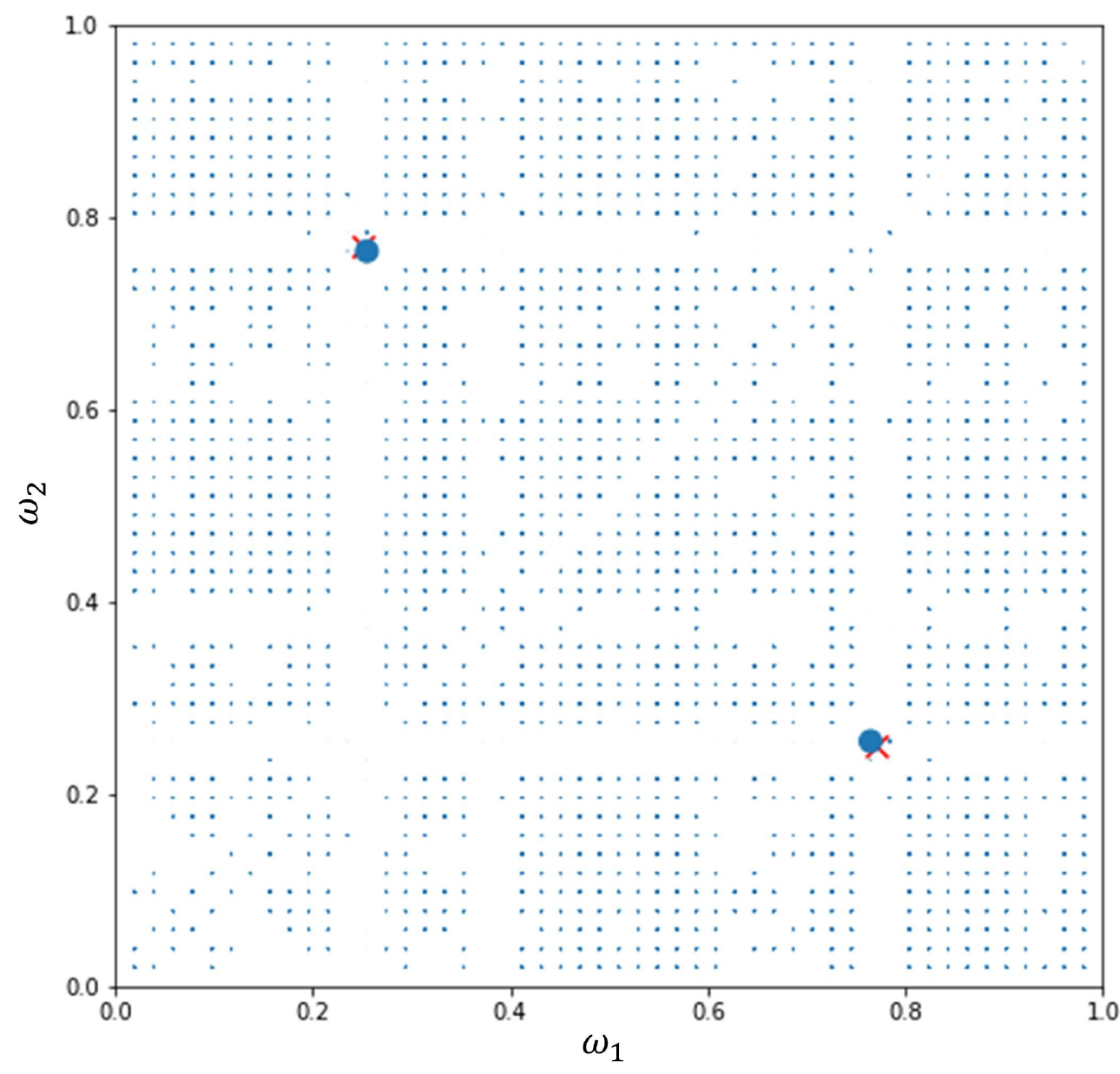}
  \caption{}
  \label{fig:2d_grid_denser}
  \end{subfigure}%
\caption{Results of 2-dimensional inference using \glsfmtshort{SIR} with \glsfmtshort{RWM} propagation for $15^2$ particles (\ref{fig:2d_resampling}), a grid for $15^2$ particles (\ref{fig:2d_grid_matched}), and a grid for $50^2$ particles (\ref{fig:2d_grid_denser}). The modes are marked with red 'x' markers.}
\label{fig:2d_cos}
\end{figure}

First, we can use this to illustrate the imperativeness of resampling in a more realistic setting than that of chapter \ref{cha:experiments_quantum_hardware}. Figure \ref{fig:2d_cos} shows how a particle number that  converges to the right modes with the help of resampling fails miserably without. An immense increase in particle density can help (figure \ref{fig:2d_grid_denser}), but the per-particle returns are minute - and this is for $2$ dimensions only. The final particles' proximity to the modes is at best dictated by how close to them the luckiest initial grid points happened to fall, and luck is undependable for high dimensional spaces.

\begin{figure}[!ht]
\captionsetup[subfigure]{width=.9\textwidth}%
\begin{subfigure}[t]{\textwidth/2}
  \centering
  \includegraphics[width=.7\textwidth]{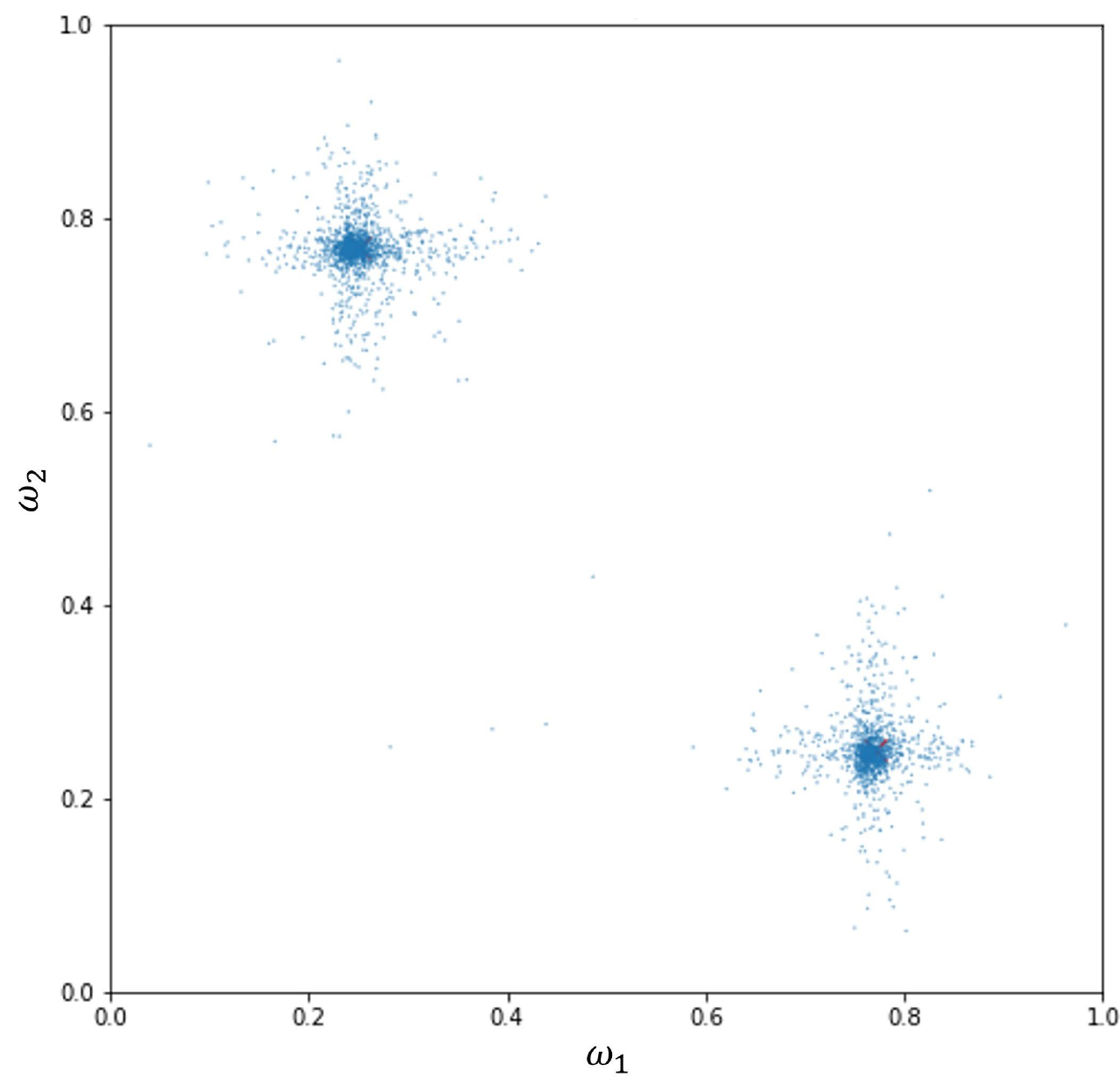}
  \caption{}
  \label{fig:sg_hmc}
\end{subfigure}%
\begin{subfigure}[t]{\textwidth/2}
  \centering
  \includegraphics[width=.7\textwidth]{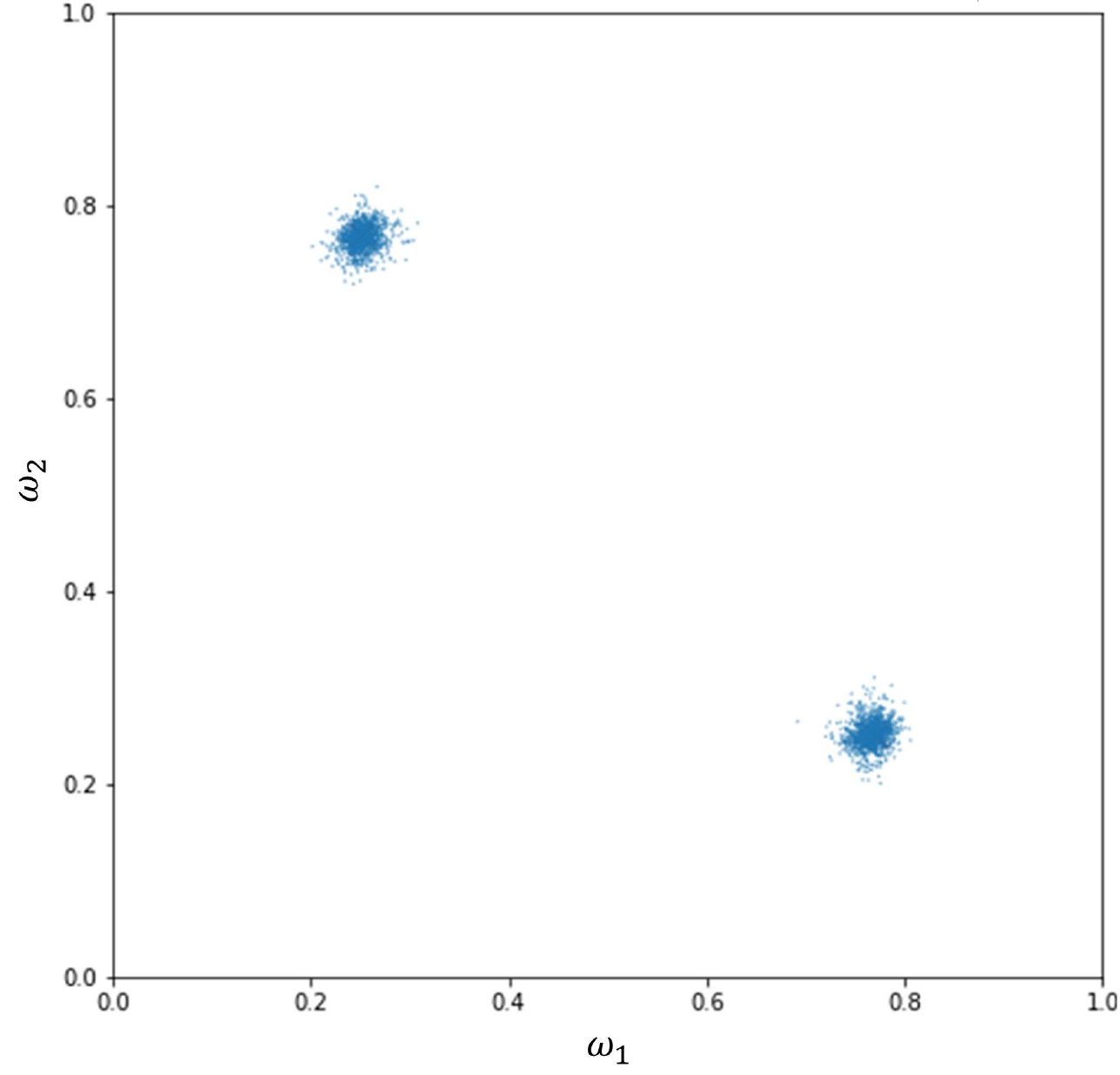}
  \caption{}
  \label{fig:sg_hmc_friction}
  \end{subfigure}%
\caption{Results of using stochastic gradients in \glsfmtshort{HMC} within \glsfmtshort{SIR} while introducing friction (\ref{fig:sg_hmc_friction}) and not (\ref{fig:sg_hmc}). The modes are marked with red 'x' markers, but are covered by the particles.}
\label{fig:sg_hmc_both}
\end{figure}

With this we can also subsample in a more general setting; namely \gls{SG}-\gls{HMC} (section \ref{sub:sg_hmc}), which hasn't been tested yet. The control variates are not applicable in this case, and so were left out. The effect of introducing friction is shown in figure \ref{fig:sg_hmc_both}. It helps control divergence, but the effect is limited.

We also test multi-dimensional generalizations of \gls{SIR}-powered adaptivity. As a reference, we use random times on $t \in [0,100]$ for a domain normalized to $1$ in all dimensions (i.e. a unit hyper-cube). In subsection \ref{sub:precession_heuristics}, an adaptive heuristic with $t_k = 1/\sigma_{k-1}$ was suggested. Unfortunately, the standard deviation is a quite less meaningful metric under multimodality. To preserve the intuition, one possibility is to cluster the particles and consider an average of the standard deviations concerning each cluster. However, clustering is a difficult task, especially in a (realistic) setting where the number of modes is not known. A more lightweight option is to consider space occupation, another measure of uncertainty. This was the selected route; occupation was computed by partitioning the space into (adaptively many) cubic cells. Finally, a third option was considered where the times are chosen offline but increase through time. Exponentially increasing times didn't do very well in this context. In short, three methods were considered:

\begin{itemize}
    \setlength\itemsep{0.2em}
    \item Random: $t^\text{(max)}=100$
    \item Adaptive: $t_k \propto \big(\texttt{occupation\_rate}_{k-1} \cdot \widehat{\text{ESS}}_{(k-1)}/\texttt{n\_particles} \big)^{-1}$
    \item Increasing random: $t^\text{(max)}_{k}= C_1 \cdot \big(\texttt{floor}(k/C_2) +1\big)$
\end{itemize}

\begin{figure}[!ht]
\captionsetup[subfigure]{width=.9\textwidth}%
\begin{subfigure}[t]{\textwidth/3}
  \centering
  \includegraphics[width=.95\textwidth]{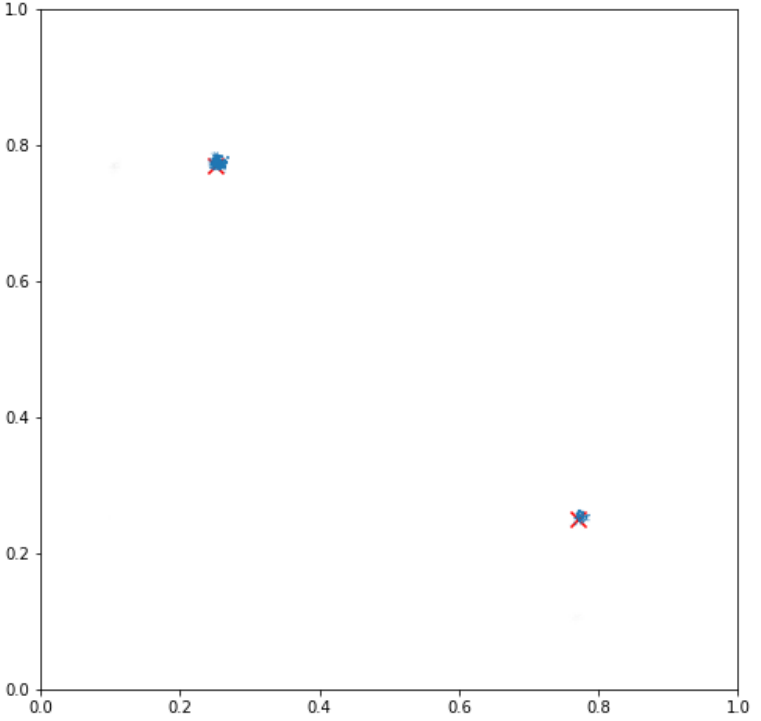}
  \caption{}
  \label{fig:2d_offline}
\end{subfigure}%
\begin{subfigure}[t]{\textwidth/3}
  \centering
  \includegraphics[width=.95\textwidth]{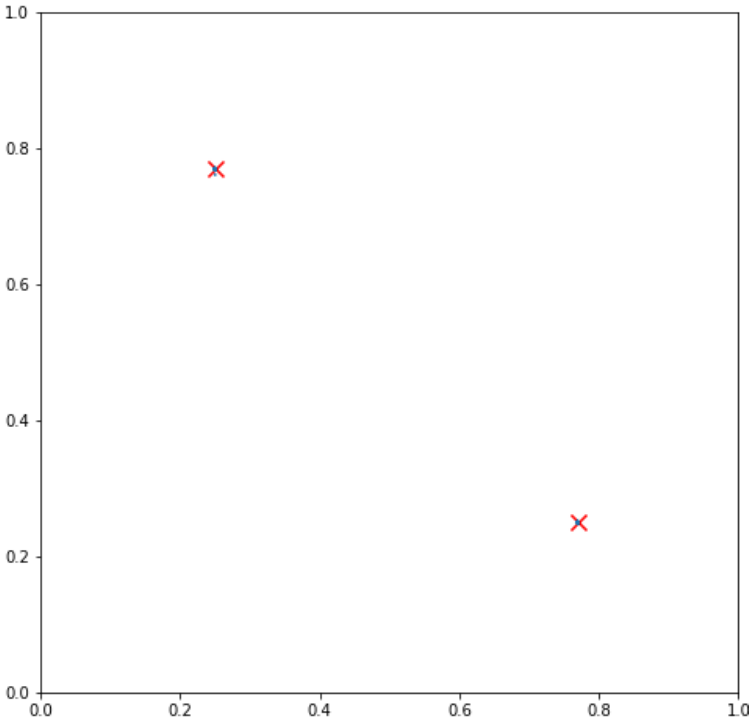}
  \caption{}
  \label{fig:2d_adaptive}
  \end{subfigure}%
\begin{subfigure}[t]{\textwidth/3}
\centering
\includegraphics[width=.95\textwidth]{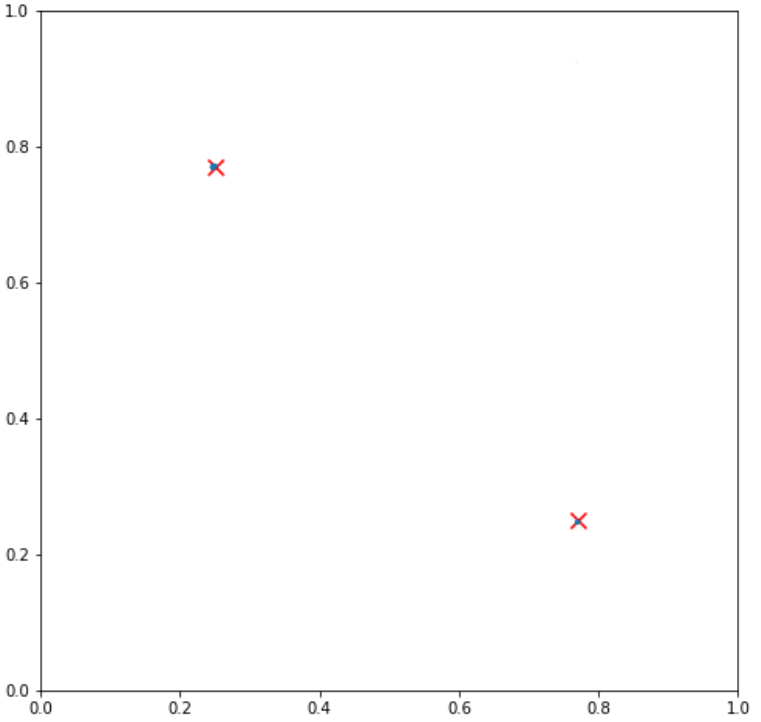}
\caption{}
\label{fig:2d_offline_inc}
\end{subfigure}%
\caption{Achieved results using random (\ref{fig:2d_offline}), adaptive (\ref{fig:2d_adaptive}) and random but increasing  (\ref{fig:2d_offline_inc}) measurement times. The modes are marked with red 'x' markers.}
\label{fig:2d_exps}
\end{figure}

The ratio of the effective sample size to the number of particles (its upper bound) is necessary to increase the times in the iterations where resampling doesn't take place. It could also be smoothed or escalated by some power. For the last case, $C_1=100$, $C_2=20$ were chosen (with a parameter sweep). Results for selected runs are shown in figure \ref{fig:2d_exps}.

\begin{figure}[!ht]
\captionsetup[subfigure]{width=.9\textwidth}%
\begin{subfigure}[t]{\textwidth/3}
  \centering
  \includegraphics[width=.95\textwidth]{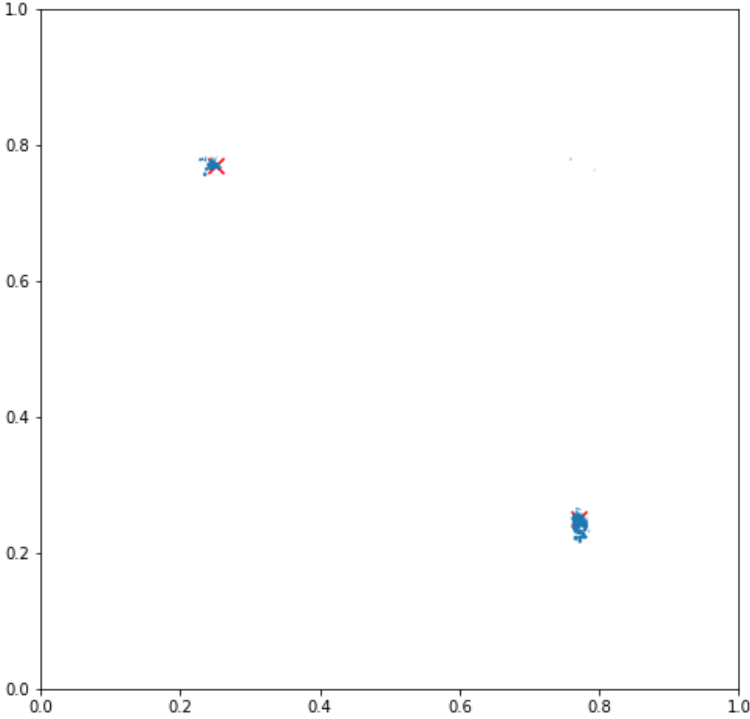}
  \caption{}
  \label{fig:2d_offline_med}
\end{subfigure}%
\begin{subfigure}[t]{\textwidth/3}
  \centering
  \includegraphics[width=.95\textwidth]{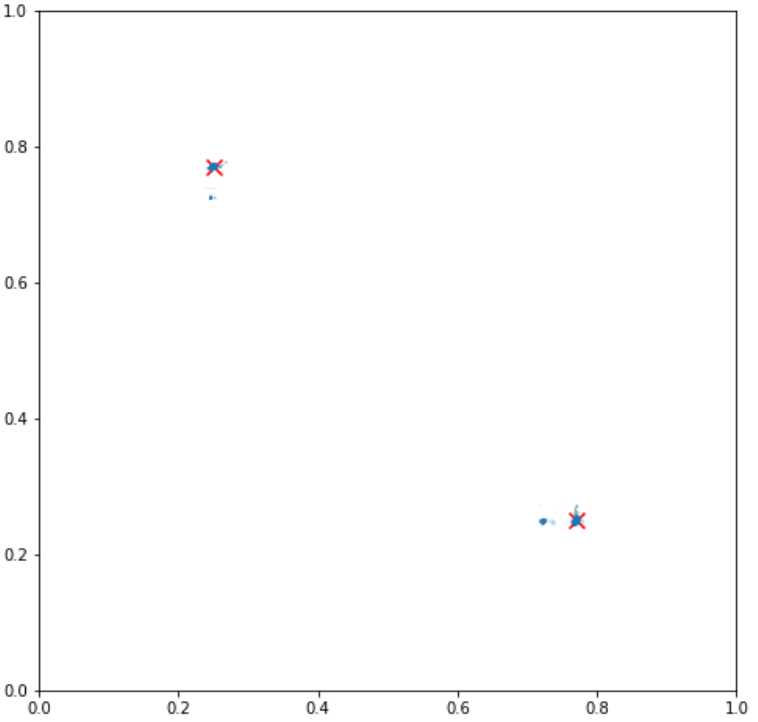}
  \caption{}
  \label{fig:2d_adaptive_med}
  \end{subfigure}%
\begin{subfigure}[t]{\textwidth/3}
\centering
\includegraphics[width=.95\textwidth]{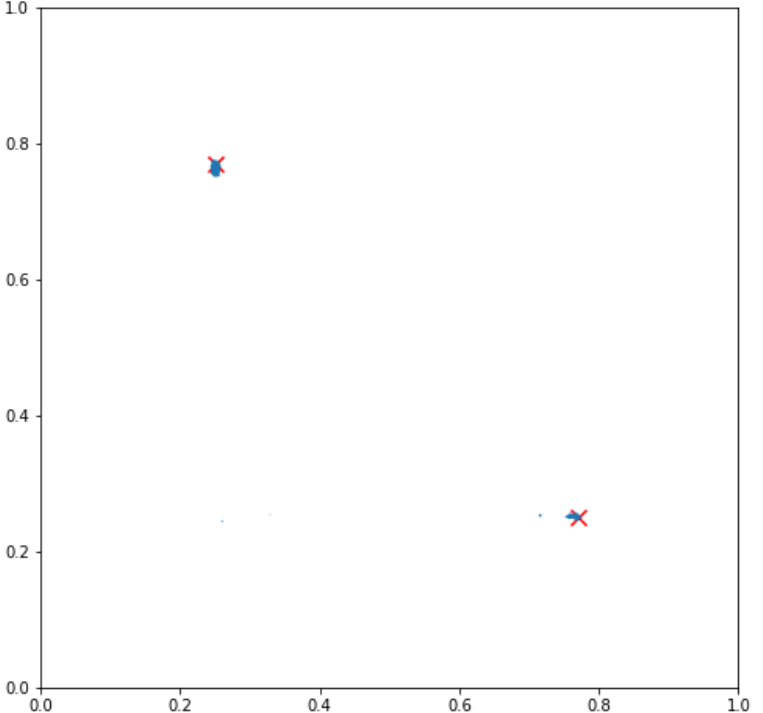}
\caption{}
\label{fig:2d_offline_inc_med}
\end{subfigure}%
\caption{Median successful results using random (\ref{fig:2d_offline}), adaptive (\ref{fig:2d_adaptive}) and random but increasing  (\ref{fig:2d_offline_inc}) measurement times. The modes are marked with red 'x' markers.}
\label{fig:2d_exps_med}
\end{figure}

The inter-run results are more variable, and taking medians shows a different picture. Figure \ref{fig:2d_exps_med} shows the final particle clouds for the successful runs with median final standard deviation.

For defining \textit{success}, assessing performance, and averaging standard deviations, the mode locations were used. Success was considered as the conjunction of accuracy, precision, correctness, and mode coverage. For evaluating these points, thresholds were considered for the average distance to a mode, the variance around each mode, the difference between real and estimated error, and the discrepancy between the number of particles attributed to each mode. Table \ref{tb:2d_exps} shows the quantitative results, including success rate. The considered metric was the weighted average of standard deviations associated with each mode, each particle being assigned to its closest one. This is what is meant by \textit{standard deviation}.

\begin{table}[ht!]
\centering
\begin{tabular}{ |g|c|c|c| } 
     \hline
     \rowcolor{gray!15}
      & \textbf{Median std. deviation} & \textbf{Mean std. deviation} & \textbf{Success rate}\\
     \hline
     \textbf{Random} & $2.0 \times 10^{-4}$ & $2.1 \times 10^{-3}$ & $65\%$ \\
     \hline
     \textbf{Adaptive} & $7.3 \times 10^{-5}$ & $1.6 \times 10^{-3}$ & $85\%$ \\
     \hline
     \textbf{Inc. random} & $1.8 \times 10^{-4}$ & $1.7 \times 10^{-3}$ & $53\%$ \\
     \hline
\end{tabular}
\caption{Results obtained for 2-dimensional estimation, using different choices of measurement times. }
\label{tb:2d_exps}
\end{table}

The adaptive method shows significant improvement in the median standard deviation, but this advantage shrinks if the mean is considered instead. This is a clear sign of instability in the inter-run performance. Another disadvantage is that the complexity of tuning and cost of computing occupation increase for higher dimensions, due to a more complicated interplay between volume and probability density. 

\begin{figure}[!ht]
\captionsetup[subfigure]{width=.9\textwidth}%
\begin{subfigure}[t]{\textwidth/4}
  \centering
  \includegraphics[width=.95\textwidth]{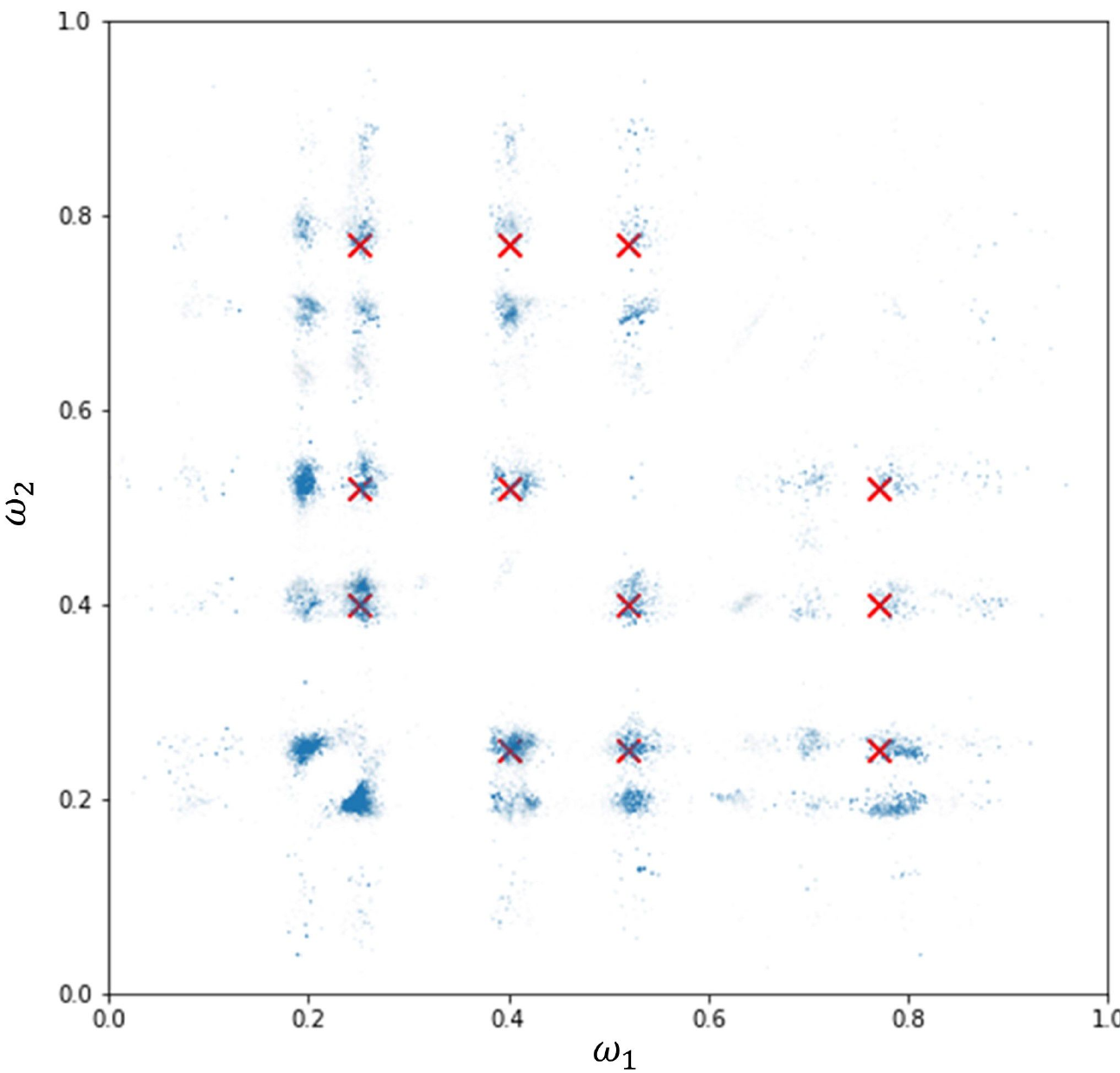}
  \caption{}
  \label{fig:4d_sir_a}
\end{subfigure}%
\begin{subfigure}[t]{\textwidth/4}
  \centering
  \includegraphics[width=.95\textwidth]{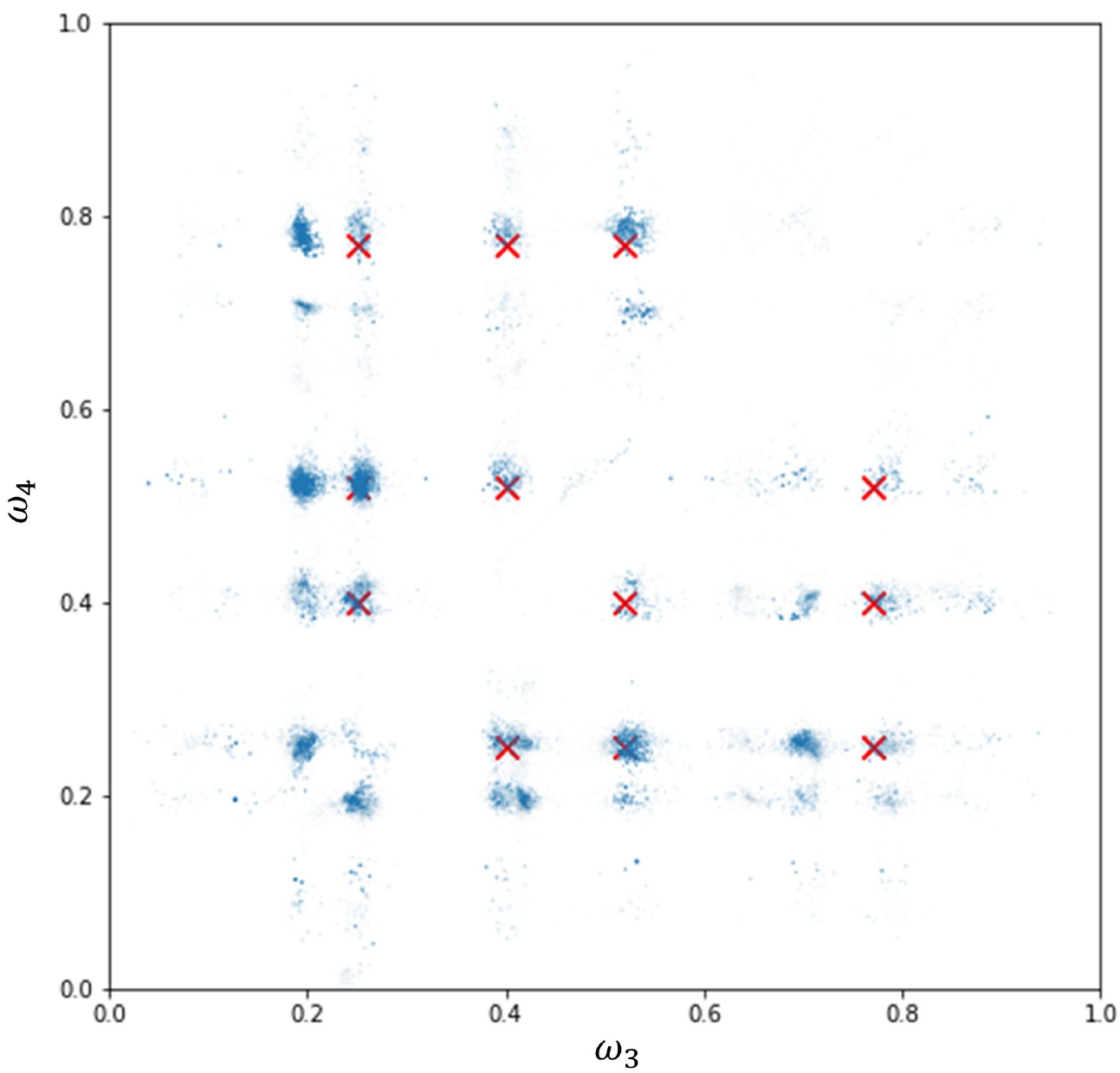}
  \caption{}
  \label{fig:4d_sir_b}
  \end{subfigure}%
\begin{subfigure}[t]{\textwidth/4}
\centering
\includegraphics[width=.95\textwidth]{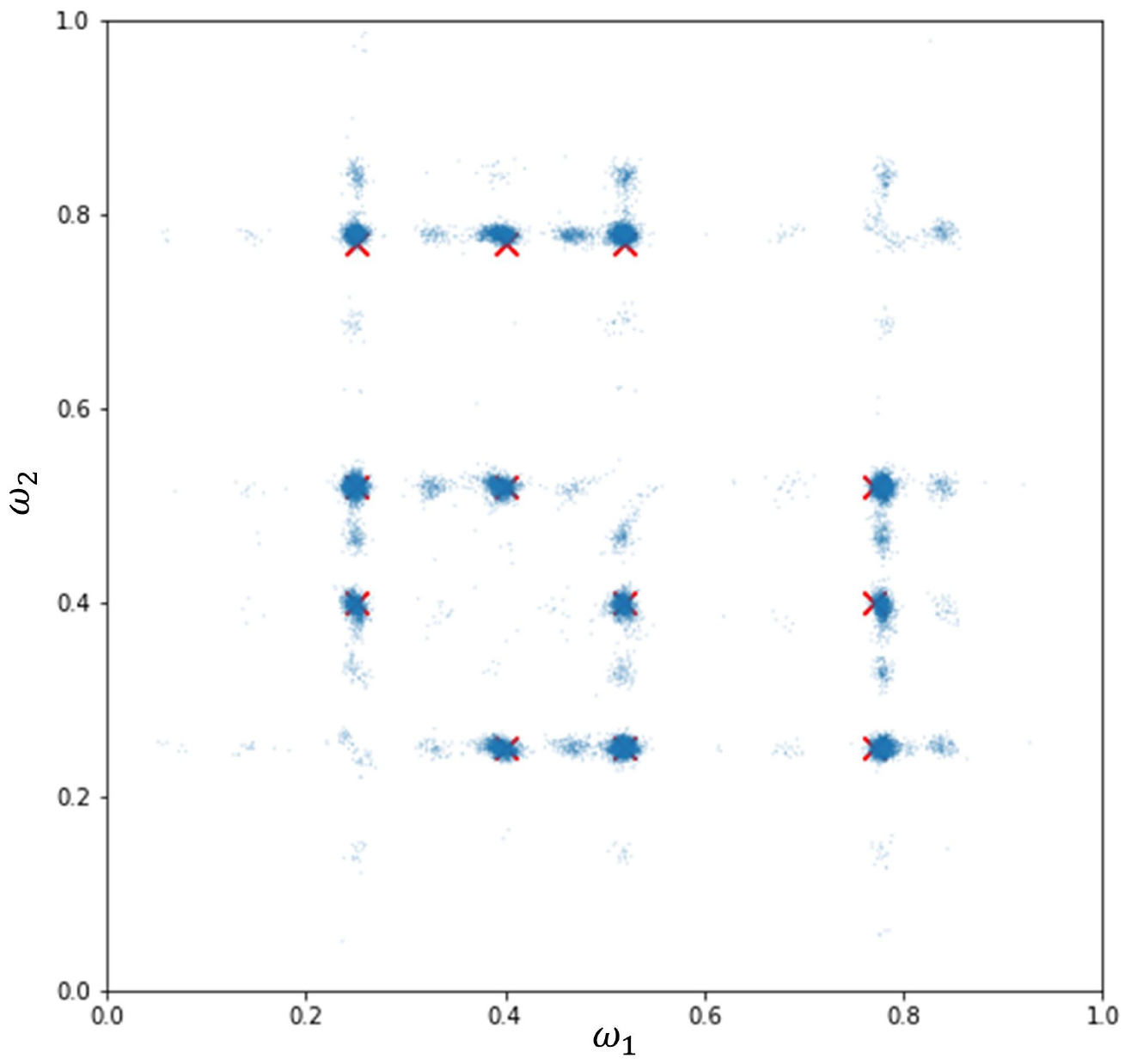}
\caption{}
\label{fig:4d_tle_a}
\end{subfigure}%
\begin{subfigure}[t]{\textwidth/4}
\centering
\includegraphics[width=.95\textwidth]{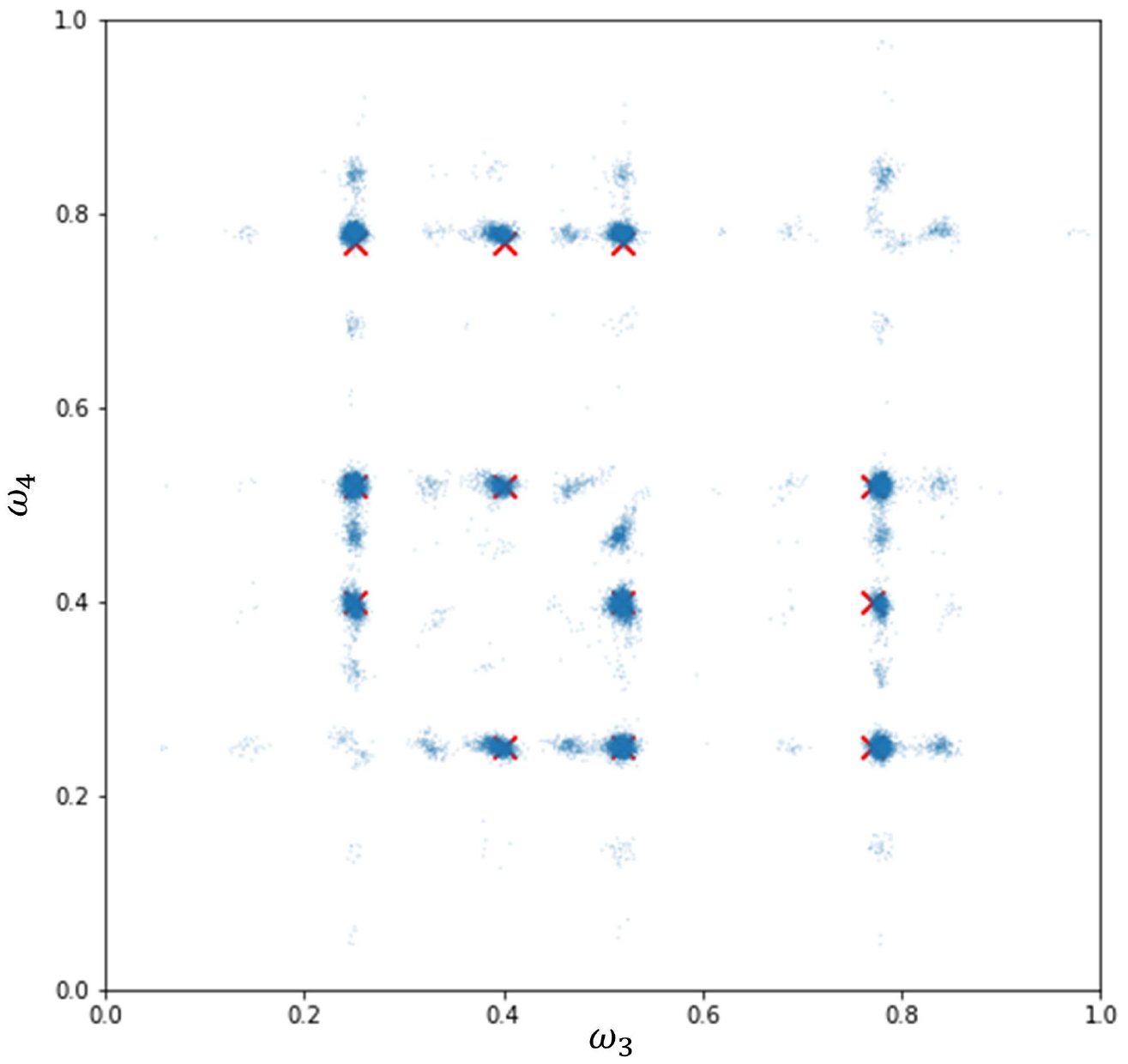}
\caption{}
\label{fig:4d_tle_b}
\end{subfigure}%
\caption{Results of 4-parameter estimation (with dimensions plotted pairwise) for two variations of \glsfmtshort{SMC}: \glsfmtshort{SIR} with the prior as importance function (\ref{fig:4d_sir_a} and \ref{fig:4d_sir_b}), and \glsfmtshort{TLE} (\ref{fig:4d_tle_a} and \ref{fig:4d_tle_b}). The modes are marked with red 'x' markers.}
\label{fig:4d_est}
\end{figure}

Finally, figure \ref{fig:4d_est} shows the results of 4-dimensional estimation ($24$ modes) using the the \gls{SMC} of algorithm \ref{alg:sir} and \gls{TLE}-\gls{SMC}, both with nearly $100\%$ \gls{HMC} moves. Both used $12^4$ particles, $250$ data, and random times on $]0,100]$. For \gls{TLE}, 5 coefficients were used. The added robustness of \gls{TLE} can be seen clearly, as can the increased complexity coming from the extra dimensions.

\chapter{Experiments on quantum hardware}
\label{cha:experiments_quantum_hardware}

This chapter presents a few quantum characterization experiments. Section \ref{sec:models_open_quantum_systems} contains a short formal description of their theory. Section \ref{sec:characterization_quantum} shows results obtained by applying the parameter estimation techniques described in the previous chapters - namely Bayesian inference, adaptive experimental design, \gls{SMC}, \gls{MCMC}, and subsampling strategies - to the open system dynamics of quantum computers.

\section{Models for open quantum systems}
\label{sec:models_open_quantum_systems}

This section will introduce the elementary theory behind the experiments whose results are to be presented in the ensuing sections. It is mostly based on \cite{Lidar_2020,Preskill_2015,Preskill_2018, qiskit_textbook,nielsen_chuang,qiskit_school_2020,qiskit_school_2021,Manzano_2020} (by order of relevance). Additional sources will be cited throughout when applicable, namely for subsection \ref{sub:spin_precession}.

Subsections \ref{sub:bloch} and \ref{sub:kraus} present methods for the visualization of mixed states and formal treatment of open quantum systems respectively. The former point is useful for the latter, as will become clear in sections \ref{sub:amplitude_damping} and \ref{sub:phase_damping} when analysing two paradigmatic noise models for open quantum systems.

At last, section \ref{sub:spin_precession} elaborates upon the precession dynamics of subsection \ref{sub:quantum_characterization_examples} to create applied examples which are at once more realizable than the original example (given the available quantum machines) and more interesting than simple open quantum system effects. When observed in quantum devices, these phenomena too are inevitably joined by environment-induced effects (such as the ones to be explored in the two preceding subsections \ref{sub:amplitude_damping} and \ref{sub:phase_damping}), bringing added complexity to characterization problems.

\subsection{Inside the Bloch sphere: visualizing mixed states}
\label{sub:bloch}

While \textit{prototypical} quantum states can be written as state vectors, which consist of a linear combination of basis states, this often fails to present a realistic description. It assumes knowledge to be limited only by fundamental quantum uncertainty, and otherwise perfect. Such a representation fails to deliver when dealing with real, noisy quantum devices. In particular, these devices are ultimately \textit{open} quantum systems, whereas this restrictive view treats but closed ones. As eloquently put in \cite{Peres_2002}, quantum phenomena happen not in a Hilbert space, but rather in a lab.

This prompts an extension of the formalism to encompass \textit{mixed} states. These are comprised of an ensemble of pure quantum states $\ket{\psi_i}$, each of which occurs with some probability $p_i$. A convenient description is brought by density operators $\rho$, which generalize the state vector description to all states, pure or not:
\begin{equation}
    \rho = \sum_i p_i \ketbra{\psi_i}{\psi_i} 
\end{equation}

\noindent, where each $\ket{\psi_i}$ is a state vector corresponding to a pure state.

Properties can then be obtained by taking statistics over the weighted sum. These are \textit{classical} statistics, an unwelcome feature because interference is limited. The amplitudes corresponding to different $\ket{\psi_i}$ don't interact with one another as they do within superposition states; instead, the ensemble behaves as a mixture of isolated states. In other words, each term in the sum is closed off, and quantum effects can be observed only within its own scope. Mixed states describe uncertainty, but not in the quantum sense; they represent flawed knowledge, these flaws extending further than the fundamental limits imposed by quantum nature.

Single qubit pure states find a nice depiction on the Bloch sphere. For this representation, we write the state vector in the form:
\begin{equation}
    \label{eq:bloch_pure}
    \ket{\psi} = \cos \left( \frac{\theta}{2} \right)\ket{0} 
    + e^{i\varphi} 
    \sin \left( \frac{\theta}{2} \right) \ket{1}
\end{equation}

\noindent, for $\{\theta,\varphi\} \in [0,\pi ] \times [0,2\pi [$.

Unlike (single qubit) pure states, mixed states can't be represented on the Bloch sphere; they can however be represented \textit{in} it. This is clear by vector algebra: the vectors corresponding to a mixed state can be obtained as a convex combination of pure state vectors, a combination which will have at most unit norm (this edge case occurring precisely in the single $i$ case, i.e. for a pure state).

A handy generalization is given by writing:
\begin{equation}
    \label{eq:bloch_mixed}
    \rho = \frac{1}{2} (\op{I} + \Vec{v} \cdot \Vec{\sigma})
    = 
    \begin{bmatrix}
    1 + v_z & v_x - iv_y\\
    v_x + iv_y & 1 - v_z
    \end{bmatrix}
\end{equation}

\noindent, for  $\Vec{\sigma} = (\opsub{\sigma}{x},\opsub{\sigma}{y},\opsub{\sigma}{z})$, $\Vec{v}=(v_x,v_y,v_z)$, and $v_x,v_y,v_z \in [0,1]$.

\begin{figure}
\captionsetup[subfigure]{width=.9\textwidth}%
\begin{subfigure}[t]{.45\textwidth}
  \centering
  \includegraphics[width=\linewidth]{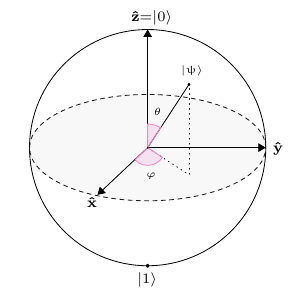}
  \caption{Bloch sphere representation of a pure state. The Bloch coordinates are the angles $(\theta,\varphi)$.}
  \label{fig:bloch_pure}
\end{subfigure}
\begin{subfigure}[t]{.45\textwidth}
  \centering
  \includegraphics[width=\linewidth]{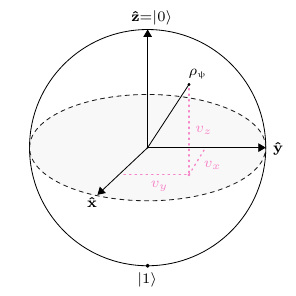}
  \caption{Bloch ball representation of an arbitrary state, pure or mixed. The Bloch coordinates are the 3-dimensional cartesian coordinates $\Vec{v}=(v_x,v_y,v_z)$, which correspond to the vector of expectations of the Pauli operators.}
  \label{fig:bloch_mixed}
\end{subfigure}
\caption{Bloch representations for pure and general states.}
\label{fig:bloch_representations}
\end{figure}

This switch from two angles (fixed-radius polar coordinates) to three lengths (cartesian coordinates) provides a convenient way to picture general states. Both representations are depicted in figure \ref{fig:bloch_representations}.

It follows easily from \ref{eq:bloch_mixed} that each element $v_i$ of the vector $\Vec{v}$ corresponds to the expectation value of the corresponding Pauli operator $\sigma_i$, given by $\langle \sigma_i \rangle = \text{Tr} [\sigma_i \rho]$. This is the reason for the convenience. As an example, the Bloch vectors pointing to the $x$/$y$/$z$ positive and negative poles correspond to the positive and negative eigenvalue eigenstates of $\opsub{\sigma}{x}$/$\opsub{\sigma}{y}$/$\opsub{\sigma}{z}$, with e.g. $\Vec{v} = (\pm 1,0,0)$ representing the $\ket{\pm}$ state. These intuitive implications are just consequences of the fact that the projection onto each axis $\op{i}$ represents the expected value of its associated observable $\opsub{\sigma}{i}$.

For any valid density matrix, the Bloch vector length satisfies $\norm{\Vec{v}} \leq 1$. It can be shown that the purity is given as a function of this norm:
\begin{equation}
    \text{Tr}\left[\rho^2 \right] = \frac{1}{2}(1 + \norm{\Vec{v}}^2)
\end{equation}

That is, the center of the sphere corresponds to the maximally mixed state, and unit length vectors correspond to pure states. Other mixed states will lie somewhere in between, inside the Bloch ball. 

Ideally, all computations would unfold on the surface of the Bloch sphere, but environment effects and experimental error pull them inside - signifying information loss. In general, error channels result in deformations of the Bloch sphere, namely compressions. These error channels often depict unwanted interactions with the environment, which may result in energy and coherence loss. The following section introduces fundamentals concerning the symbolic treatment of open quantum systems, whereas the two subsequent ones present representative cases that illustrate artifacts commonly found in real-world practice due to imperfect isolation.

\subsection{The Kraus operator sum representation}
\label{sub:kraus}

The time evolution of a quantum wavefunction is shaped by the Schrödinger equation, one of the most emblematic cornerstones of quantum mechanics. Often, we are forced to make room for incomplete information by generalizing it to density operators, which can describe mixed states. But even still, the usual form still has a propensity for being overwhelmingly broad - in that case, we must make room for incomplete \textit{observations}.

Schrödinger's equation describes the unitary time evolution of a system; this is \textit{any} system's time evolution; but we're not exactly talking about systems that fit inside a glass vial. More specifically, we speak of closed quantum systems, whose time evolution is given by:
\begin{equation}
    \label{eq:unitary_evolution}
    \rho(t) = \op{U} \rho(0) \op{U}^\dagger 
\end{equation}

\noindent, with time-dependent $\op{U}$.

Unless the system under study is perfectly isolated, it will enjoy unitary evolution \textit{jointly with its surroundings}. If we can't observe these surroundings as well, unitarity is no longer assured, and equation \ref{eq:unitary_evolution} is not to be applied directly.

We can think of any subject of study, such as a qubit, as a system coupled to an environment, or bath - such as the lab, or the entire outside world. Naturally, external effects vary in degree, and some may be negligible. Nonetheless, rarely does it ever occur that they are \textit{all} negligible, which means that realistic analyses must in general contemplate them. The result of the coupling is that our subject, strictly speaking, ceases to function as a system on its own right; in particular, information can be \textit{leaked} to the external environment. 

To partial observations on a system corresponds partial information. As such, it is not very surprising that focusing on a subsystem may cause it to behave like a mixed state. To enable the discussion of some simple instances where this happens, we will now go over the basic formalism required for describing them.

We suppose we have a system of interest $s$, which we will just call system (despite the fact that it is just a subsystem of the bigger bipartite system that includes the environment too), living on some Hilbert space $\mathcal{H}_s$. The environment $e$ has a space of its own, $\mathcal{H}_e$, and together they exist in $\mathcal{H}_s \otimes \mathcal{H}_e$.

The \textit{partial trace} over the environment is a transformation $(\mathcal{H}_s \otimes \mathcal{H}_e) \mapsto \mathcal{H}_s$ mapping density operators describing the composite system to reduced density operators describing only system $s$. It is obtained by sandwiching the composite operator between the orthonormal vectors belonging to the environment's eigenbasis:
\begin{equation}
    \rho_s \equiv \text{Tr}_e
    \left[\rho^2 \right]
    = \sum_\alpha 
    \pbra{\alpha}{e}  \ \rho \ket{\alpha}_e
\end{equation}

The $\{\ \ket{\alpha}\}$ span $\mathcal{H}_e$ and act on it alone. There are as many as the environment's Hilbert space dimension.

We assume a separable initial state, and further that the environment is initially in a pure state $\ket{\mu_0}$\footnote{This is only for the sake of simplicity, and the generalization is straightforward \cite{Lidar_2020}.}. From the joint unitary evolution governed by $U(t)$, we can find the map acting exclusively on $s$ as:
\begin{equation}
    \label{eq:osr}
    \rho_s(t) = \sum_{\alpha} \pbra{\alpha}{e} \ 
    \op{U} \Big( \rho_s(0) \otimes \ketbra{\mu_0}{\mu_0} \Big) \op{U}^\dagger 
    \ \ket{\alpha}
    =
    \sum_{\alpha} \op{K}_{\alpha} \rho_s(0)
     \op{K}_{\alpha}^\dagger
\end{equation}

Where we have defined:
\begin{equation}
    \label{eq:kraus}
    \op{K}_\alpha \equiv \bra{\alpha} \ \op{U} \ket{\mu_0}
\end{equation}

In the context of quantum information science, quantum maps are often called channels. Time evolution is itself a map, so a channel can represent the evolution $ \rho_s(0) \mapsto \rho_s(t)$. In this context, equation \ref{eq:osr} is often written alternatively as:
\begin{equation}
    \label{eq:channel_osr}
    \bm{\mathcal{E}} \big[ \rho_s \big] = 
    \sum_{\alpha} \op{K}_{\alpha} \rho_s
     \op{K}_{\alpha}^\dagger
\end{equation}

The sum form of equations \ref{eq:osr} and \ref{eq:channel_osr} is generally termed an \glsxtrfull{OSR}, and the $\op{K}_\alpha$ are  called Kraus operators. They may be time-dependent, just like $\op{U}$. 

These operators must still respect a normalization condition, which guarantees that they preserve any density operators' own normalization (unit trace) - that is, that they map valid states to still valid states. The appropriate condition can be shown to be:
\begin{equation}
    \sum_\alpha \op{K}_\alpha^\dagger \op{K}_\alpha = I
\end{equation}

If there exists a single Kraus operator, this normalization condition imposes that it be unitary, in which case we recover Schrödinger dynamics as a special case. Otherwise, the system is not closed, but thanks to the partial trace over the environment we single out its evolution. That is, tracing out the environment leaves the Kraus operators to act on it alone, describing its specific dynamics. Of course, this doesn't mean that we isolate the system, but rather that we factor in the uncertainty contributed by the fact that it is an \textit{open} system.

Considering the unitary evolution of the composite system (subsystem of interest + environment), a transformation $\op{U}$ given by $\ket{\psi_0} \mapsto \ket{\psi_t}$ can be seen as $\ketbra{\psi_t}{\psi_0}$, meaning that it \textit{eats} a state $\ket{\psi_0}$ and \textit{produces} a state $\ket{\psi_t}$. This simplifies the interpretation of \ref{eq:kraus}. Tracing out a constant pure initial state $\ket{\mu_0}$ removes the initial environment state from the side to the left of $\mapsto$, leaving still its basis states to be traced out on the right-hand side. Then, the Kraus operators can be found by ignoring the fixed environment initial state and applying to the right-hand-side \cite{Preskill_2018}:
\begin{equation}
    \op{K}_\alpha = \pbra{\alpha}{e} \ \op{U}
\end{equation}

All of these ideas will become clearer with the ensuing basic examples. In the following section, referring to the system is assumed to be the default, so the $s$ subscript will be sometimes dropped.

\subsection{Energy relaxation: the amplitude damping channel}
\label{sub:amplitude_damping}

Energy loss is a problem that plagues many applications across the fields of physics, and quantum computation is no exception. It is unavoidable that a qubit which is not in its ground state will with some probability release energy to its environment, signifying amplitude damping - and thus a loss of information. 

Here we will analyse the effect energy relaxation exerts in a single qubit, which can be visualized in the Bloch ball. We consider a perfect two-level system and assume the energy flow to be unidirectional; that is, the environment does not supply energy to the system. Under these conditions, the possible transitions for the qubit are from the excited state $\ket{1}$ to the ground state $\ket{0}$. We suppose they occur with a time-independent probability $p_1$ per unit time. The index is in chosen in anticipation of the fact that a second channel will be considered later. 

The effect of the channel is to map:
\begin{gather}
        \ket{1} \mapsto \ket{0} \text{ , with probability $p_1$}
\end{gather}

The state $\ket{0}$ is left untouched with probability $1$, whereas $\ket{1}$ is preserved only with probability $1-p_1$.

Under these conditions, one of the Kraus operators will be a lowering operator:
\begin{equation}
    \op{K}_1 = \sqrt{p_1} \ketbra{0}{1} = 
    \begin{bmatrix}
    0 & \sqrt{p_1}\\
    0 & 0
    \end{bmatrix}
\end{equation}

Another Kraus operator should map the ground state component to itself, i.e. it should represent the component $\ketbra{0}{0}$. Along with the normalization condition, this produces:
\begin{equation}
    \op{K}_0 = 
    \begin{bmatrix}
    1 & 0\\
    0 & \sqrt{1-p_1}
    \end{bmatrix}
    = \ketbra{0}{0} + \sqrt{1-p_1} \ketbra{1}{1} 
\end{equation}

The meaning of the lower-right element may become clearer by noting the extreme cases. For $p_1=0$, this operator is just the identity, and the only Kraus operator. There is no energy loss, and unitary evolution is recovered. If $p_1=1$, then it is $\ketbra{0}{0}$. The result is that $\ket{1}$ is always mapped to $\ket{0}$, because there is unit probability of decay. In the intermediate cases, the density operator element $\rho_{11}$ is reduced by the multiplying factor $\sqrt{1-p_1}$. The higher the emission probability, the larger this decrease: this compensates for the excited state population removed (brought to the ground state) by $\op{K}_1$.

The normalization condition may seem to remove intuition from the process, but it can be recovered by a different view. Assuming the system starts out in a pure state, there are two possible branches for the unknown system evolution $\psi (0) \mapsto \psi (\dd t)$:
\begin{align}
    \ket{0}_s \otimes \ket{0}_e &\mapsto 
    \ket{0}_s \otimes \ket{0}_e\\
    \ket{1}_s \otimes \ket{0}_e &\mapsto 
    \sqrt{1-p_1} \ket{1}_s \otimes \ket{0}_e
    + \sqrt{p_1} \ket{0}_s \otimes \ket{1}_e
\end{align}

\noindent, where we assume two possible environment states: having absorbed energy emitted by the system and not, respectively $\ket{1}_e$ and $\ket{0}_e$.

The operator $\op{K}_1$ is obtained by tracing over the environment state $\ket{1}_e$ - it leaves to the system a single possible evolution path, from $\ket{1}_s$ to $\ket{0}_s$. The operator $\op{K}_0$ comes from tracing over $\ket{0}_e$. In this case, two paths are possible for the system: either it was always in state $\ket{0}_s$, or it was and still is in state $\ket{1}_s$. However, not having observed the environment in state $\ket{1}_e$ reduces the likelihood that the system was originally in state $\ket{1}_s$, since this fact would have yielded non-null probability of observing a different outcome whereas $\ket{0}_s$ would not.

Now we would like a more convenient expression - one that would give us the time evolution at once. The first step is getting the explicit expression for the density matrix the channel maps to. Let us write the system's state at any time $t$ as:
\begin{equation}
    \rho(t) = \begin{bmatrix}
    \rho_{00}(t) & \rho_{01}(t)\\
    \rho_{10}(t) & \rho_{11}(t)
    \end{bmatrix}
\end{equation}

We then apply the \gls{OSR} to an initial state $\rho(0)$ to get:
\begin{equation}
    \rho(\dd t) = \op{K}_{0} \rho_s(0) \op{K}_{0}^\dagger
    + \op{K}_{1} \rho_s(0) \op{K}_{1}^\dagger
    = \begin{bmatrix}
    \rho_{00}(0)+ \rho_{11}(0) \cdot p_1 & 
    \rho_{01}(0) \cdot (1-p_1)^{1/2}\\
    \rho_{10}(0)\cdot (1-p_1)^{1/2} 
    & \rho_{11}(0) \cdot (1-p_1)
    \end{bmatrix}
\end{equation}

This is an interesting result: energy loss doesn't only reduce the excited state population, it also damps the off-diagonal elements - effectively washing off quantum coherence (though more slowly than it reduces $\rho_{11}$). Clear off-diagonals in a density matrix mean an absence of quantum effects, or quantum coherence. 

To see how the probability of finding the system in state $\ket{1}_s$ changes with time, we can consider a relaxation rate $\gamma_1 = p_1/\dd t$ and the map to be applied $n$ times until a time $t=n \dd t$. Then, $p_1 = \gamma_1 \dd t = \gamma_1 t/n$, and:
\begin{equation}
    \rho_{11}(t) = \rho_{11} (0)
    \left(1- \frac{\gamma_1 t}{n} \right)^n
\end{equation}

Taking the continuous time limit $n \rightarrow \infty $ and recognizing $\lim_{n \rightarrow \infty} \big( 1 + x/n\big)^n=e^x$ as the Euler number limit:
\begin{equation}
    \rho_{11}(t) = e^{-\gamma_1t}\rho_{11} (0) = e^{-t/T_1}\rho_{11} (0)
\end{equation}

\noindent, where $T_1 \equiv 1/\gamma_1$ is known as the characteristic energy relaxation time.

As for the other elements of the density operator, the off-diagonal terms can readily be seen to decay at a rate $\gamma_1/2$, or equivalently to enjoy a characteristic time twice as large. Finally, it follows easily from the trace-preservation property of the Kraus operators that the ground state population $\rho_{00}$ increases by the summand $\rho_{11}(t)=(1-e^{-\gamma_1t})\rho_{11}(0)$.

This can be visualized as a compression of the Bloch sphere, where all points move closer to the north pole except for the north pole itself (as long as $p_1>0$). Its shape is compressed along all directions, but twice as strongly in the $\op{z}$ direction, and $\ket{0}$ suffers no displacement. The magnitudes of both the shift and the compression are proportional to $\gamma_1$. Details on the geometric interpretation can be found in \cite{Lidar_2020}.

The key thing to note is that because we don't actually measure the environment, we have only a probabilistic understanding of what transpired, which decreases the purity of the qubit's state.

\begin{figure}[!ht]
    \centering
    \includegraphics[height=1.75cm]{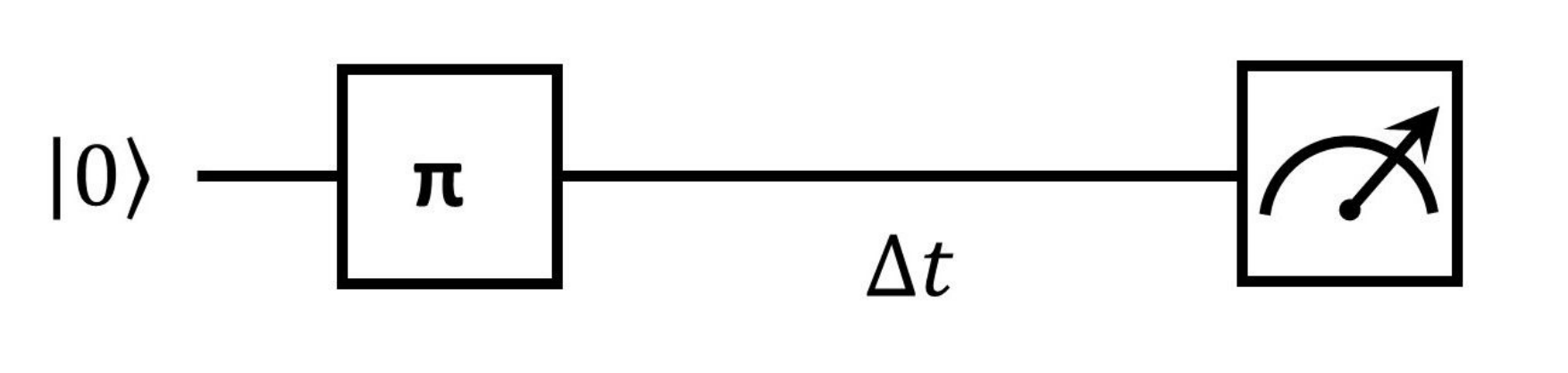}
    \caption{Pulse schedule for measuring $T_1$.}
    \label{fig:t1_circ}
\end{figure}

The characteristic time $T_1$ can be estimated using the circuit of figure \ref{fig:t1_circ} for several waiting times $\Delta t$ since the initialization at $\ket{1}$. The pi pulse realizes a rotation of $\pi$ radians in the Bloch sphere, taking the $\ket{0}$ state to the $\ket{1}$ state. It can be a $\opsub{\sigma}{x}$ gate, in which case the rotation is around the $x$ axis.

After the qubit is brought to the excited state $\ket{1}$ (at the time reference $\Delta t = 0$), the probability of measuring $\ket{1}$ decays as:
\begin{equation}
    \label{eq:t2}
    \mathbf{P}(1 \mid \Delta t) = e^{- \Delta t/T_2}
\end{equation}

This model can be used to perform inference, or to fit a curve by regression.

\subsection{Dephasing: the phase damping channel}
\label{sub:phase_damping}

Another common source of noise is \textit{phase} damping. This map is characterized by a phase flip which occurs with some probability $p_\varphi$:
\begin{gather}
    \label{eq:dephasing}
        \rho(dt) = (1-p) \cdot \rho(0) 
        + p \cdot \opsub{\sigma}{z}\rho(0) \opsub{\sigma}{z}
\end{gather}

Clearly, a possible set of Kraus operators is now:
\begin{gather}
    \op{K}_0 = \sqrt{1-p_\varphi} \cdot \op{I}\\
    \op{K}_1 = \sqrt{p_\varphi} \cdot \opsub{\sigma}{z}
\end{gather}

The second operator flips the sign of the off-diagonals, which slowly erases them since they partially cancel out their own original versions (left unaltered with probability $1-p$). It is easy to see that acting on the Bloch representation of pure states with $\opsub{\sigma}{z}$ flips the sign of the Bloch coordinate $\varphi$, the phase, hence the term \textit{dephasing}. For general states, it flips the $x$ and $y$ Bloch vector components. The evolution of the density operator is given by:
\begin{equation}
    \label{eq:dephasing_rho}
    \rho(\dd t) = \op{K}_{0} \rho_s(0) \op{K}_{0}^\dagger
    + \op{K}_{1} \rho_s(0) \op{K}_{1}^\dagger
    = \begin{bmatrix}
    \rho_{00}(0) & \rho_{01} (0)\cdot (1-2p_\varphi)\\
    \rho_{10}(0)\cdot (1-2p_\varphi) & \rho_{11}(0)
    \end{bmatrix}
\end{equation}

Taking the continuous time limit as before, now for $p_\varphi=\gamma_\varphi t/n$, yields:
\begin{equation}
    \rho(t)
    = \begin{bmatrix}
    \rho_{00}(0) & e^{-t/T_\varphi} \rho_{01} (0)\\
    e^{-t/T_\varphi} \rho_{10}(0) & \rho_{11}(0)
    \end{bmatrix}
\end{equation}  

\noindent, where $T_\varphi \equiv 1/(2\gamma_\varphi)$ is defined as the characteristic time for this noise channel.

The effect on the off-diagonal terms is similar to what we observed in the previous section, meaning that the $x$ and $y$ Bloch components shrink, but now the diagonal is left intact. In other words, this channel has no effect on the $z$ eigenbasis.

The Bloch vector can easily be found to evolve as:
\begin{equation}
    \Vec{v}(t) = (e^{-t/T_\varphi}, e^{-t/T_\varphi}, 1) \cdot \Vec{v}(0)
\end{equation}

That is, it recedes in the horizontal plane towards the vertical ($z$) axis, as its $x$ and $y$ components suffer from damping until they reach $0$ (at which point the state is maximally mixed along those directions, the pairs of eigenstates being equally probable so as to yield null expectation).

If both dephasing and energy loss are present, they combine to produce an effective coherence time $T_2^*$ given by:
\begin{equation}
    \frac{1}{T_2^*} = \frac{1}{2T_1} + \frac{1}{T_\varphi}
\end{equation}

\begin{figure}[!ht]
    \centering
    \includegraphics[height=1.75cm]{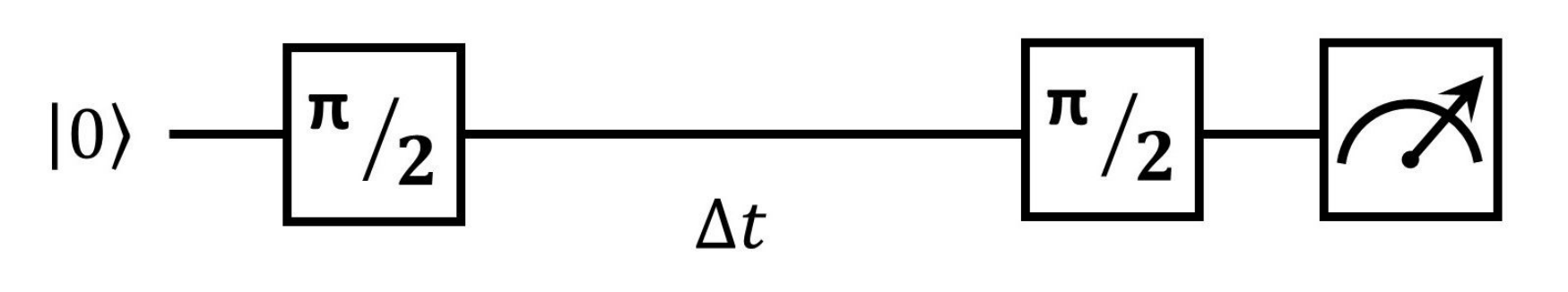}
    \caption{Pulse schedule for measuring $T_2^*$. }
    \label{fig:t2_circ}
\end{figure}

This coherence time $T_2^*$ is known as $T_2$ \textit{star} to distinguish it from another $T_2$ that will be discussed shortly. It can be estimated using the circuit \ref{fig:t2_circ} for several waiting times $\Delta t$ between half-pi pulses. 

The purpose of the half-pi pulses is to take the Bloch vector to the equatorial plane, where the $T_2^*$ is most relevant, and then back for the computational basis measurement. In the $x$ basis, the function is simply: 
\begin{equation}
    \label{eq:t2star_decay}
    \mathbf{P}(+ \mid \Delta t) = \frac{1}{2} + \frac{e^{- \Delta t/T_2^*}}{2}
\end{equation}

A Hadamard gate directly translates this to $\mathbf{P}(0 \mid \Delta t)$, but any rotation that takes the Bloch vector to the $x$-$y$ plane will do, such as rotations of $\pi/2$ radians around the $x$ or $y$ axis (though then if the angle is the same for both pulses the probability $\mathbf{P}(0 \mid \Delta t)$ is complementary to \ref{eq:t2star_decay}).

The sources of this decay can be many, but some of them are actually reversible, unlike the energy loss of the previous section. In particular, a common reason for this effect is field inhomogeneities  that lead to low frequency noise. This can occur for an ensemble of physical qubits due to local field variations that change their individual Larmor frequencies, or similarly if there are small temporal fluctuations in the qubit's resonance frequency. 

Looking at figure \ref{fig:t2_circ}, this will cause the $\pi/2$ pulses of  to be slightly off-resonance by a varying amount, causing a series of \textit{accidental} Ramsey experiments as will be described in the following section. We can think of a set of Bloch vectors that start out at state $\ket{+}$, but each of which rotates around the $z$ axis at a slightly different frequency. If we measure such an ensemble, the final states will be inconsistent across several measurements, signifying a loss of purity.

Each such vector is pulsed into the equatorial plane state and slowly precesses within it at some individual frequency $\delta$, instead of staying in place as would be desirable. After a time $\Delta t$, it will have covered an angular distance of $\delta \cdot \Delta t$. But if halfway through the evolution we apply a $\pi$ pulse, its position will be mirrored relative to the vertical plane, so that it is $\delta \cdot \Delta t$ away from $\ket{-}$.

Allowing it to evolve for the other half of the free evolution time will cancel out its former motion, leaving it deterministically at $\ket{-}$ (assuming instantaneous pulses and that $\delta$ is constant). If we have a temporal or spatial ensemble of spins, all of their shifts will be erased by this inversion pulse, which effectively reunites them - so that they can jointly describe a pure state.

\begin{figure}[!ht]
    \centering
    \includegraphics[height=1.5cm]{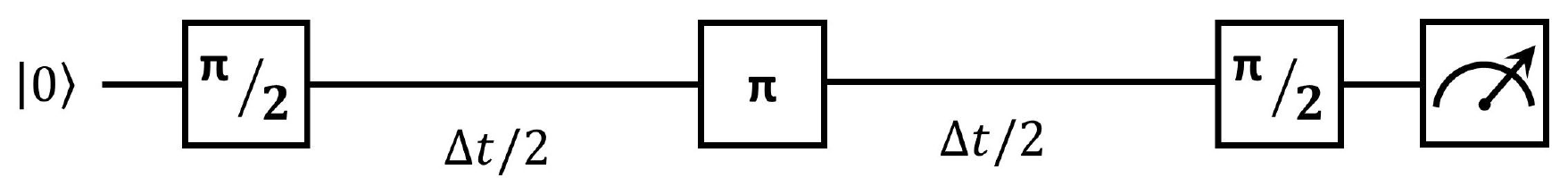}
    \caption{Pulse schedule for measuring $T_2$. This is known as a \textit{Hahn echo} or \textit{spin echo} sequence.}
    \label{fig:hahn_echo_circ}
\end{figure}

This is generally called a spin echo or Hahn echo experiment \cite{hahn_1950}, the term \textit{echo} alluding to the way the noise retraces its steps to repeat its own pattern until it cancels itself out for a measurement at $\Delta t$. The result is a longer characteristic time, called $T_2$: by refocusing the spins, it eliminates a significant source of noise, improving $T_2^*$. It essentially works as a high-pass filter. $T_2$ is often called the dephasing time constant. The circuit for measuring this time is shown on figure \ref{fig:hahn_echo_circ}. 

To prolong the coherence time still further, higher order methods can be used. Here we will use the simple circuit from \ref{fig:hahn_echo_circ}, the purpose being mostly to make the coherence time more stable. The crude $T_2^*$ is typically quite short, and suffers from strong fluctuations. This interferes with the comparison of results across multiple methods, which can hardly be expected to be consistent.

With this we conclude the open system effects to be characterized in section \ref{sec:characterization_quantum}. Despite being very simple, these two examples actually capture the essence of two important types of noise present in real quantum devices. In practice, they can stem from different and complex physical mechanisms - but these straightforward descriptions present a simple view of their net result, providing useful and widely used noise models.

\subsection{Spin precession}
\label{sub:spin_precession}

We will now consider unitary dynamics that combine with the open system effects of subsections \ref{sub:amplitude_damping} and \ref{sub:phase_damping} to produce likelihood functions whose characterization is more challenging than that of simple exponential models. They take the form of sinusoids. The main references are \cite{Zwiebach_2013,Gross_2013,Frimmer_2014,Vitanov_2015,Deutsch_2005}.

In section \ref{sub:quantum_characterization_examples}, we considered the evolution induced by the Hamiltonian:
\begin{equation}
    \label{eq:larmor_harmiltonian}
    \opsub{H}{0} =  \frac{\Omega_0}{2}\opsub{\sigma}{z}
    = \frac{1}{2}\begin{bmatrix}
    \Omega_0 & 0 \\
    0 & -\Omega_0
    \end{bmatrix}
\end{equation}

Here we change the $\omega$ of before into $\Omega_0$ for reasons that will shortly become clear. This Hamiltonian describes \textbf{Larmor precession}, and it can be realized by applying a magnetic field in the $\op{z}$ direction \cite{Zwiebach_2013}. We saw that the resulting wavefunction evolution was given for some initial Bloch coordinates $(\theta_0,\varphi_0)$ by:
\begin{equation}
     \ket{\psi (t)} =
     \cos(\theta_0 /2)\ket{0} + e^{i\varphi_0 + \Omega_0 t} \sin(\theta_0 /2)\ket{1}
\end{equation}

The angular frequency $\Omega_0$, called the Larmor frequency, is proportional to the magnitude of the applied field, $\Omega_0 = \gamma B$ (using Planck units so that $\hbar=1$).

Similar phenomena can be observed by building on the Hamiltonian of \ref{eq:larmor_harmiltonian}, some of which are useful for control or characterization of two level systems. Considering a qubit subject to \ref{eq:larmor_harmiltonian}, we can add a $\opsub{\sigma}{x}$ contribution which rotates around the $z$ axis:
\begin{equation}
    \label{eq:rabi_harmiltonian}
    \op{H} = \opsub{H}{0} + \opsub{H}{d} =
    \frac{\Omega_0}{2}\opsub{\sigma}{z}
    + \frac{\Omega_d}{2} 
    ( e^{i\omega t}\opsub{\sigma}{+} 
    + e^{-i\omega t}\opsub{\sigma}{-} )
    = \begin{bmatrix}
    \Omega_0 & \Omega_d e^{-i\omega t} \\
    \Omega_d e^{i\omega t} & -\Omega_0
    \end{bmatrix}
\end{equation}

\noindent, where $\opsub{\sigma}{+}=\ketbra{1}{0}$ and $\opsub{\sigma}{-}=\ketbra{0}{1}$ are the raising and lowering operators for the $z$ basis. This addition corresponds to a periodic driving field that rotates around the $z$ axis (i.e. perpendicularly to the polarization).

Applying the time-dependent Schrödinger equation to a state vector in the $z$ basis, written in the form $\ket{\psi (t)} = a(0 \mid t)\ket{0}+a(1 \mid t)\ket{1}$, produces a system of two equations - one for each amplitude's derivative. The explicit time dependence on $e^{i\omega t}$ can be removed by transforming to a rotating frame of reference (e.g. by applying the unitary transformation $\op{U}=e^{-i\omega t/2}\ketbra{0}{0} + e^{i\omega t/2}\ketbra{1}{1}$), which yields a simplified pair of equations. These correspond to an \textit{effective} Hamiltonian that is time-independent:
\begin{equation}
    \label{eq:rabi_effective_h}
    \op{H}_\textit{eff} = \frac{1}{2}
    \begin{bmatrix}
    \delta & \Omega_d \\
    \Omega_d & -\delta
    \end{bmatrix}
    \quad , \ \delta \equiv \Omega_0 - \omega
\end{equation}

This Hamiltonian describes the same dynamics as the previous one, but now in a rotating referential. Comparing with \ref{eq:rabi_harmiltonian}, we can see that in the rotating frame the driving field appears at rest ($\omega=0$), and the Larmor frequency has been replaced by the difference $\delta$ between itself and the normal field's rotation frequency, this difference being called the \textit{detuning}.

Away from resonance, $\abs{\delta} \gg 0$, the detuning $\delta$ induces dynamics similar to those of \ref{eq:larmor_harmiltonian}. However, the driving field leads to coherent switching between basis states, and in resonance ($\omega \approx \Omega_0 \leftrightarrow \delta \approx 0$) the most relevant phenomenon are the $\ket{0} \leftrightarrow \ket{1}$ transitions.

The time evolution of the state vector can most easily be described by diagonalizing $\op{H}_\textit{eff}$, i.e. solving the characteristic polynomial in the rotating frame and then the eigenequation (which is just the time-independent Schrödinger equation) for each eigenvalue $E_\pm$. Working in the energy eigenbasis of $\op{H}_\textit{eff}$ and assuming a $\ket{0}$ initial state, we get:
\begin{equation}
    \label{eq:rabi_psi}
    \ket{\psi (t)} = \cos(\theta/2)e^{-itE_+}
    \ket{E_+}
    + \sin(\theta/2)e^{-itE_-}\ket{E_-}
\end{equation}

These energy eigenstates $\ket{E_\pm}$, around which the Bloch vector precesses, belong to the $x$-$z$ plane, making an angle $\alpha= \tan^{-1}(\Omega_d/\delta)$ with the $z$ axis. This amounts to a rotation of the quantization axis. That is, the basis of the Bloch sphere is altered according to the ratio between the driving amplitude $\Omega_d$ and the detuning $\delta$. The larger the first one is compared to the second, the further the axes move away from the $z$ axis, and the closest to the $x$ axis.

If there is no driving ($\Omega_d=0,\omega=0$), then $\alpha = 0$ and the eigenvectors overlap with those of the $z$ basis. In that case, the Bloch vector precesses around the $\op{z}$ direction and the frequency is $\Omega_0$; we recover the original Larmor oscillations. For no detuning (in resonance, $\delta=0$), they instead coincide with those of the $x$ basis, and the frequency is $\Omega_d$. All of this can be seen intuitively from the Hamiltonian \ref{eq:rabi_effective_h}.

From \ref{eq:rabi_psi}, we can get the probability of state $\ket{1}$ at a time $t$:
\begin{equation}
    \label{eq:rabi_p1}
    \mathbf{P}(1 \mid t) = \frac{\Omega_d^2}{\Omega_d^2+ \delta^2}
    \sin^2 \left( \frac{\sqrt{\Omega_d^2+ \delta^2}}{2}
    \cdot t \right)
\end{equation}

This formula describes what is called \textbf{Rabi flopping}, the cyclic evolution of the amplitudes of the two energy levels of a qubit under an oscillatory driving field \cite{Gross_2013}. For finite detuning, the oscillation frequency is increased and the amplitude is reduced.

This can be used to implement quantum gates, since equation \ref{eq:rabi_p1} can be solved for the time given a desired Bloch sphere position. One can initialize the qubit at $\ket{0}$ and apply resonant microwave pulses for some duration $\Omega_d \cdot t = 2\theta_X$ to move the Bloch vector by an angle $\theta_X$ (this $\theta$ being the Bloch sphere angle with the $z$ axis, and the same of figure \ref{fig:bloch_pure} and equation \ref{eq:bloch_pure}). The direction too can be chosen according to the intended results. 

Other than control, a possible application for this type of setup is characterization. Equation \ref{eq:rabi_p1} can be used to determine the qubit frequency $\Omega_0$ to some precision, by applying transverse radiation of known parameters and observing the system's behaviour. For this purpose, better results can be achieved by using another method, called Ramsey's method \cite{Vitanov_2015,Frimmer_2014}. Norman Ramsey was originally a student of the physicist responsible for the original technique, Isidor Rabi, and improved upon it.

The idea is to replace the application of a field for some evolution time $\Delta t$ by two shorter duration pulses separated in space/time. Shortening the interaction length is positive, because it reduces the degree to which the oscillations are affected by experimental error (e.g. field inhomogeneities). 

The experimental scheme is based on two $\pi/2$ pulses of frequency $\omega$ detuned by $\delta$ with respect to an estimate of the qubit's resonant frequency $\widetilde{\Omega}_0$ (e.g. a difference of the order of $MHz$ vs. $GHz$). Their application is separated by free evolution for a time $\Delta t$ during which the only action on the system is the parallel field $\opsub{H}{0}$. During this period, with no driving, the Bloch vector oscillates at a frequency $\delta = \Omega_0-\omega$ around the $z$ axis, $\Omega_0$ being the \textit{real} resonant frequency and the quantity of interest.

The last applied pulse's frequency determines the free evolution, during which the effective Hamiltonian is the Larmor-type Hamiltonian given by equation \ref{eq:rabi_effective_h} when $\Omega_d=0$. During this period, the driving is off, so its amplitude is indeed null. In the end, we apply a second identical pulse, after which measuring in the $z$ basis gives (again assuming initialization at $\ket{0}$):
\begin{equation}
    \label{eq:ramsey_p1}
    \mathbf{P}(1 \mid t) = \cos^2
    \left( \frac{\delta}{2}\cdot \Delta t \right)
\end{equation}

If the pulse frequency $\widetilde{\Omega}_0$ exactly matched $\Omega_0$, no oscillations would be observed, because under those circumstances $\delta=0$. In the likely case that it doesn't, the difference between these values can be learned: it corresponds to the observed oscillation frequency $\delta_\text{meas}$, which can be estimated by performing measurements for several $\Delta t$ and performing inference of fitting a curve. Its value can then be used to correct the estimate.

\begin{figure}[!ht]
    \centering
    \includegraphics[height=4cm]{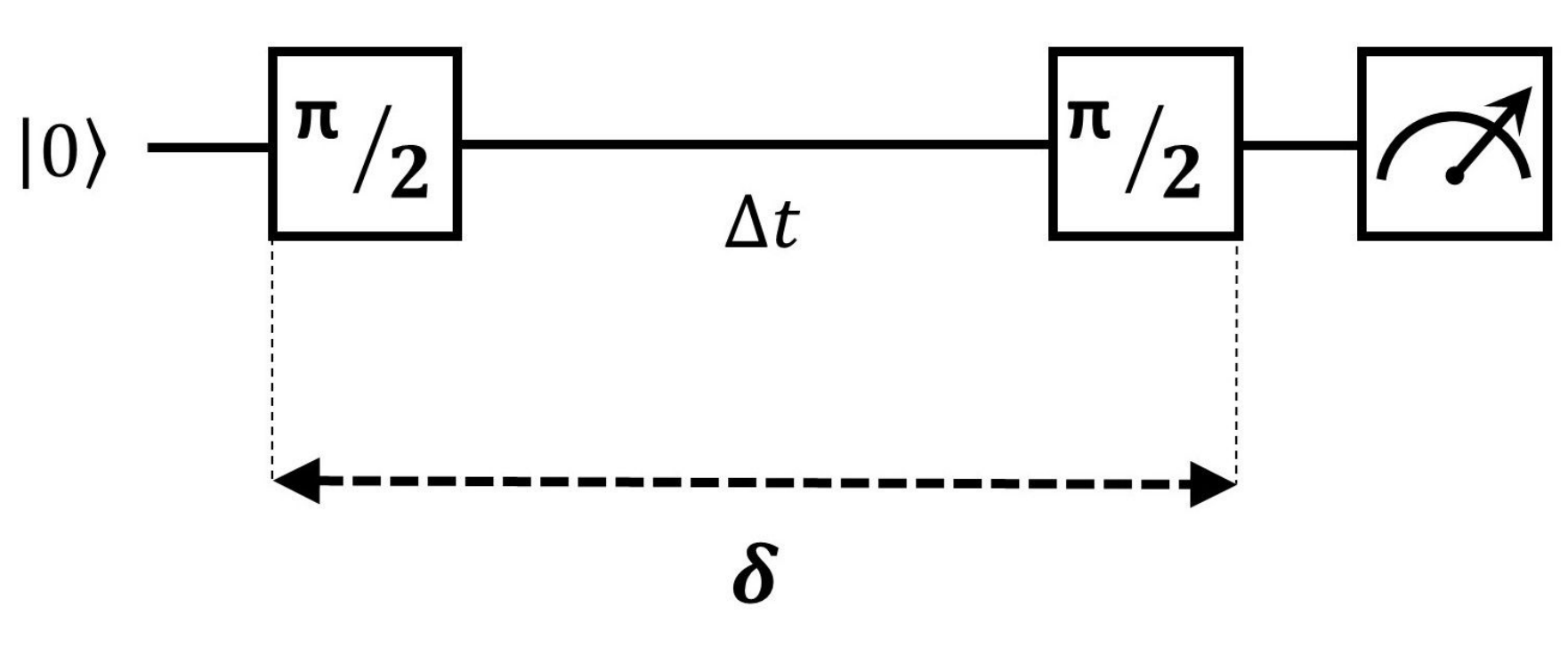}
    \caption{Pulse schedule for estimating the detuning $\delta_\text{meas}$ in a Ramsey experiment. This is known as a \textit{Ramsey sequence}. Note that $\boldsymbol{\delta}$ is an experimental control determining the pulse frequency. The frequency estimate can then be corrected to higher precision as $\Omega_0 = \widetilde{\Omega}_0+\delta_\text{meas}$}.
    \label{fig:ramsey_circ}
\end{figure}

This technique is most commonly called \textbf{Ramsey interferometry}, and its formulation has been central for high precision metrology (including applications as sensitive as atomic clocks \cite{Vitanov_2015}). The basic sequence to be used for the Ramsey experiments is depicted in figure \ref{fig:ramsey_circ}. 

In practice, the oscillation pattern is unlikely to exactly recreate equation \ref{eq:ramsey_p1}, due to experimental error and environment effects. The protocol suffers from decoherence, similar to what has been described in section \ref{sub:phase_damping}, as well as other sources of error such as fluctuations in the detuning. The result is that the sinusoidal wave suffers from damping, which can be modeled by an exponential decay envelope:
\begin{equation}
    \label{eq:ramsey_p1_decay}
    \mathbf{P}(1 \mid t) = e^{- t/T_2^*}\cos^2
    \left( \frac{\delta}{2}\cdot \Delta t \right)
    + \frac{1-e^{-t/T_2^*}}{2}
\end{equation}

This imposes a symmetric effect on the basis states, since decoherence is expected to affect $\ket{+}$ and $\ket{-}$ equally (the $\pi/2$ pulses translate these results into the computational basis). This expression can also be derived by considering Lorentzian noise in the real parameter, and replacing it with a pair of hyperparameters \cite{Granade_2012}.

\begin{figure}[!ht]
    \centering
    \includegraphics[height=3.5cm]{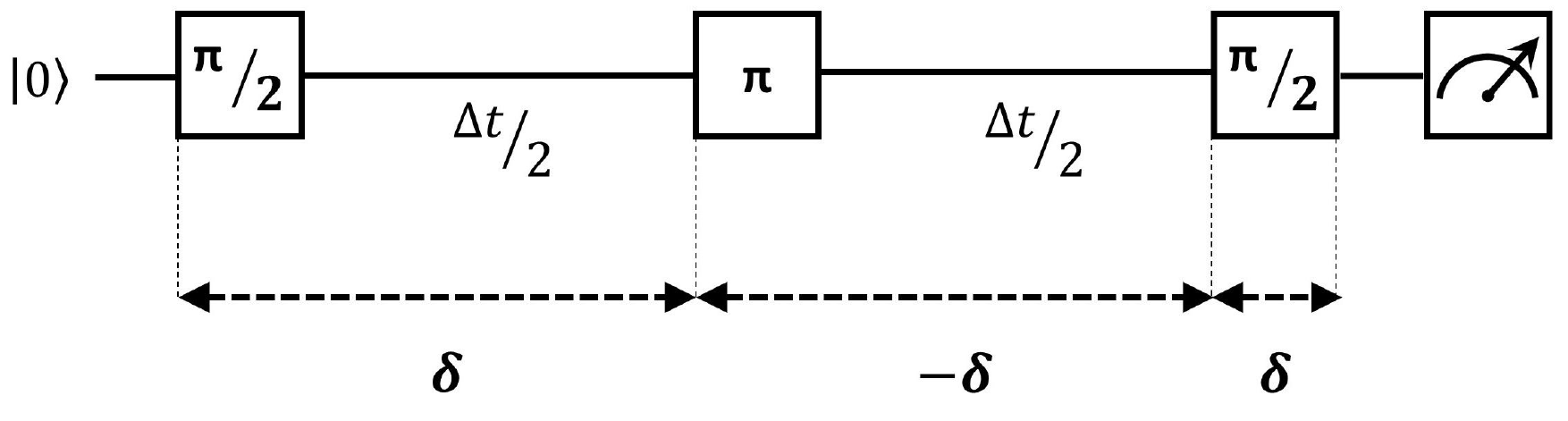}
    \caption{Pulse schedule for measuring $\Omega_0 = \widetilde{\Omega}_0+\delta_\text{meas}$ in an echoed Ramsey experiment. Note that $\boldsymbol{\delta}$ is an experimental control determining the pulse frequency. This is a simplification of the ideas in \cite{Vitanov_2015}, who called this type of setup a \textit{Hahn-Ramsey sequence}.}
    \label{fig:echoed_ramsey_circ}
\end{figure}

Just as some of the damping effects of section \ref{sub:phase_damping} were remediable, Ramsey sequences can benefit from error mitigation techniques for improved robustness. We refer to \cite{Vitanov_2015} for discussion on how to realize these refinements. Based on the ideas exposed therein, we construct a very simple improved version of the circuit from figure \ref{fig:ramsey_circ}, shown in figure \ref{fig:echoed_ramsey_circ}. More sophisticated enhancements can be found in \cite{Vitanov_2015}. 

This is called by the authors a Hahn-Ramsey scheme, owing to the way it borrows from the Ramsey and Hahn echo experiments. Due to cancelling out some random errors, it improves the contrast of the Ramsey fringes, and eases their decay. Under the presence of noise, the achievable precision in determining $\delta$ greatly benefits from this.

Within the scope of this work, the purpose of such an arrangement is to make the coherence time large relative to the duration of at least a few oscillation cycles. One may then assume it to be negligible in the scale of the measurement times, which takes us from equation \ref{eq:ramsey_p1_decay} back to \ref{eq:ramsey_p1} - effectively halving the number of parameters. This allows for testing methods that would otherwise not be feasible due to processing constraints (namely \gls{TLE} variations).

Unlike the ones proposed in subsections \ref{sub:amplitude_damping} and \ref{sub:phase_damping}, Ramsey-type experiments necessarily require pulse-level control; to enforce the chosen detuning $\delta$, the frequency of the pulses should be customizable, whereas pre-defined unitaries are tuned to the resonant frequency. 

The $\pi/2$ pulses may be constructed from scratch by running a Rabi fit to determine the amplitude given the intended duration, or vice-versa. Calibrating these pulses is the type of task the Ramsey experiment should realize, but naturally a starting point is required. An alternative is to adapt standard operations, namely rotations around the $x$ or $y$ axis. Since gates are calibrated to be in resonance, the frequency should be altered by a small detuning $\delta$.

Of course, changes in frequency necessitate a correction in the pulse duration and/or amplitude (for a fixed angle, e.g. $\pi/2$), which should be adapted by solving \ref{eq:rabi_p1} for $\Delta t$. Depending on the intended precision, bringing the pulses slightly off-resonance while maintaining the amplitude $\Omega_d$ and the pulse duration may be sufficiently accurate. For a $\delta$ small enough as compared to the resonant frequency $\Omega_0$, the effect in the pulse duration should be negligible.

\section{Characterization of quantum devices}
\label{sec:characterization_quantum}

This section presents the results of applying Bayesian inference to the problem of learning the dynamical parameters of quantum devices (open quantum systems), using various sampling strategies. The subjects of characterization were a few IBM quantum machines. 

Subsection \ref{sub:exp_setup} details considerations relevant to the experiments, namely how they were executed and their performance assessed. Subsections \ref{sub:exp_t2} and \ref{sub:exp_t1} present the results of coherence time estimation ($T_2$ from subsection \ref{sub:phase_damping} and $T_1$ from \ref{sub:amplitude_damping} respectively). Subsections \ref{sub:exp_ramsey_2d} and \ref{sub:exp_ramsey_1d} concern Ramsey estimation, as per subsection \ref{sub:spin_precession}. In the former, the basic pulse sequence is taken, so the induced oscillation frequency $\delta$ is targeted along with $T_2^*$. In the latter, the results of applying a refocusing pulse are demonstrated, and more resource-intensive techniques are applied to the characterization of the frequency, which becomes one-dimensional (assuming $T_2^* = \infty$ for short enough evolution times).

\subsection{Experimental setup}
\label{sub:exp_setup}

The experiments were performed using IBM quantum services, Qiskit \cite{Qiskit} and Qiskit-Pulse \cite{Alexander_2020}, a pulse-level programming kit. 

 Except where indicated differently, measurement times were chosen non-adaptively on a grid, and the posterior distribution was discretized via \gls{SIR} with the prior as importance function (algorithm \ref{alg:sir}, subsection \ref{sub:sir}) and \gls{MCMC} move steps (algorithm \ref{alg:mcmc}, (sub)sections \ref{sec:mcmc} and \ref{sub:mcmc_smc}). Additionally, the chosen Markov kernel was unless otherwise stated \gls{RWM} (subsection \ref{sub:rwm}) with a proposal variance proportional to the current \gls{SMC} estimate of the variance. This proportionality was tuned to obtain a stable particle acceptance rate around $65\%$.

For each set of results, the backend estimates at the time of data collection are presented where relevant and applicable. The latest known coherence times are supplied by IBM at any given time. However, these estimates are not very reliable: variations as high as $50\%$ were observed between consecutive calibrations. Intervals of hours between calibrations are not uncommon, sometimes lasting up to a full day. Within that time span, the time constants tend to suffer from significant fluctuations.

As such, curve fits were used as a benchmark to validate and contextualize results. They additionally allow for comparing the achieved uncertainty in view of the data requirements, yielding a more complete performance assessment. The fitting was done using two methods: the SciPy \cite{Scipy} function \texttt{curve\_fit} (non-linear least squares fit) with default parameters, and Qiskit's native coherence fitters (from the \texttt{BaseCoherenceFitter} class).

The former was used for direct comparison, since it can work on the same data (measurement times and pulse schedules) as the inference protocol, whereas the default fitters take only gate sequences. Regardless, the methods were approximately matched for maximum evolution time, and for each of them the measurement times were evenly spaced from some near-zero value to this upper limit. 

The SciPy fitter is the same tool that Qiskit coherence fitters use internally, but Qiskit's error estimates are more reliable. They are adjusted according to the variance of each measurement over all shots, through the optional argument \texttt{sigma} of \texttt{curve\_fit}. It was made evident by the obtained results that the default parameters result in the errors' being grossly overestimated. For a fair comparative analysis, the quantitative results of the Qiskit fitters are the ones presented.


In the case of the inference, whose performance is more variable due to using fewer data, these quantitative results were obtained using a number of runs and taking median values for both the parameter estimates (means) and their associated uncertainties (standard deviations). In some cases, the evolution of the uncertainty through the iterations of the \gls{SMC} algorithm is also represented. Finally, the results will where relevant be represented graphically as the curve associated with the determined parameter(s).

In the last case, the same graphs present data points corresponding to the \textit{(measurement time, probability)} tuples obtained using custom pulse schedules and a curve fitted specifically to them. This is done using external (non-Qiskit) curve fitting. The data used for the points and for these curve fits are exactly the same, whereas the inference results use a lower number of shots. This serves as an assessment of predictive power. Curve fits tend to require higher resolution if reasonable results are to be obtained.

\begin{figure}[ht]
\captionsetup[subfigure]{width=.9\textwidth}%
\begin{subfigure}[t]{.45\textwidth}
  \centering
  \includegraphics[width=\linewidth]{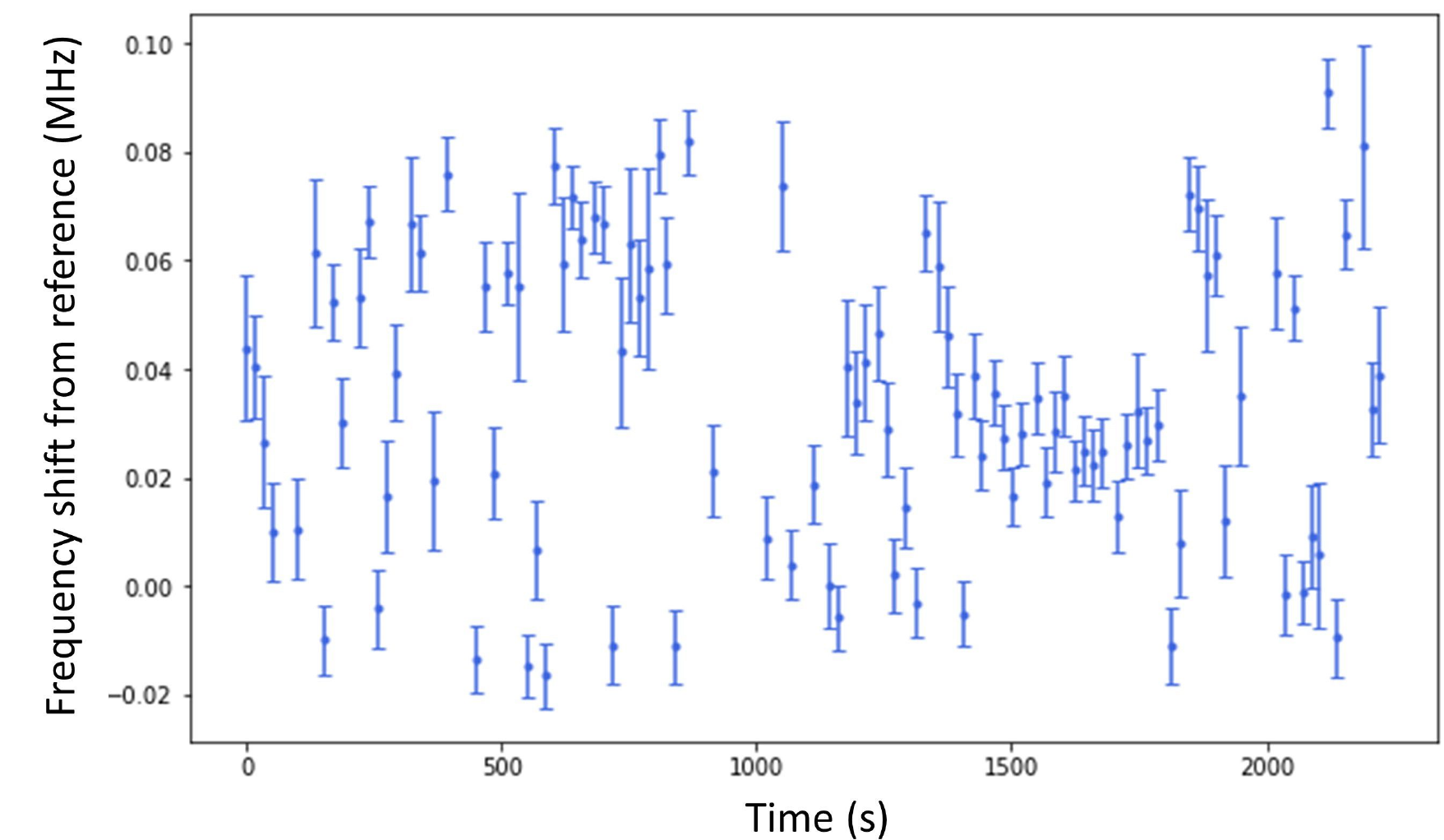}
  \caption{Estimated discrepancy between the latest available estimate of the resonance frequency - $\mathcal{O}(GHz)$ - and its current estimate.}
  \label{fig:f_fluctuation}
\end{subfigure}
\begin{subfigure}[t]{.45\textwidth}
  \centering
  \includegraphics[width=\linewidth]{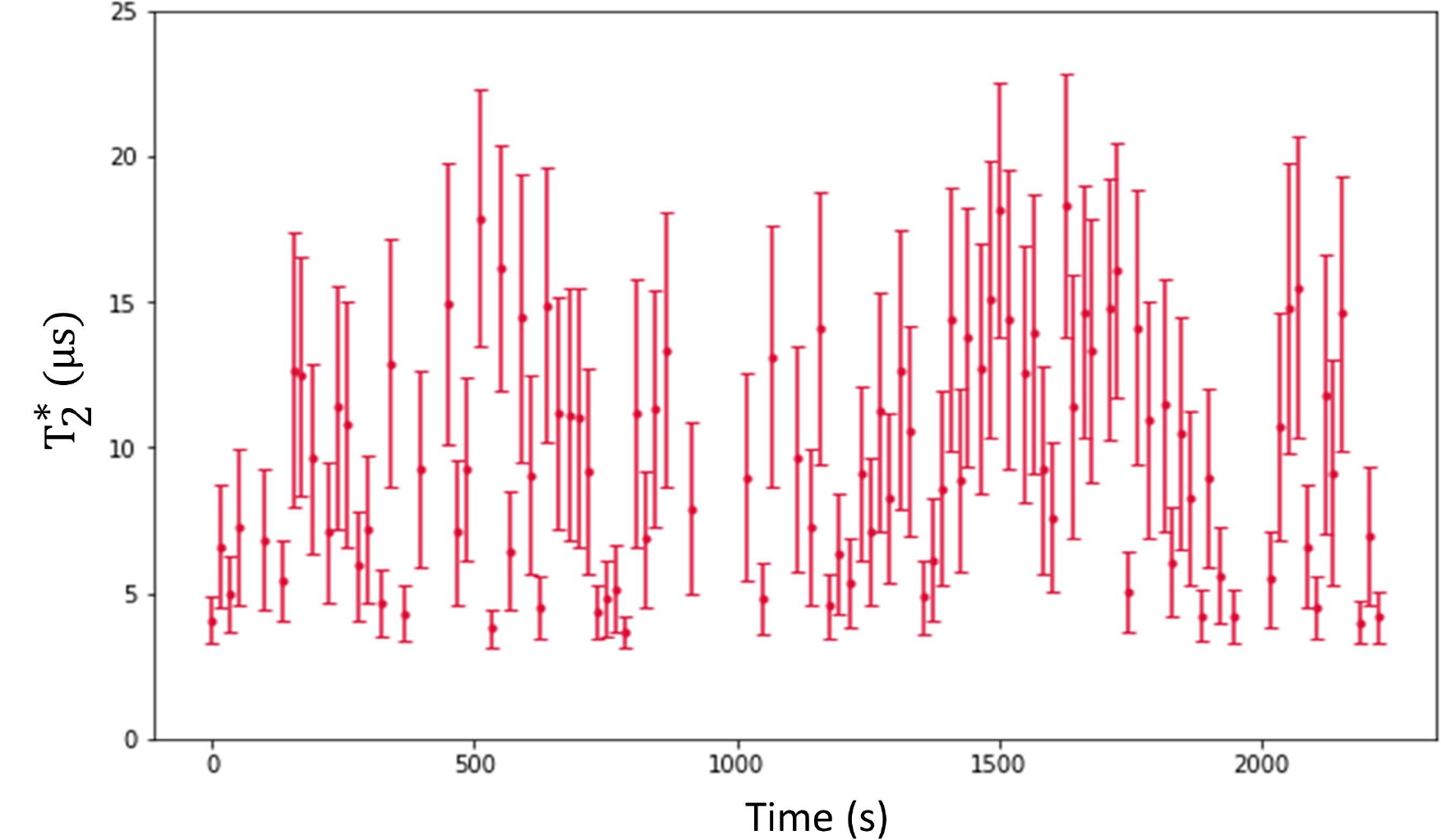}
  \caption{Estimated (unechoed) dephasing time, as a function of the starting time of data collection.}
  \label{fig:T2_fluctuation}
\end{subfigure}
\caption{Estimated time fluctuations of two quantities of interest for the IBMQ backend \texttt{ibmq\_armonk}. These are the parameters to be targeted in the Ramsey experiment of section \ref{sub:exp_ramsey_2d}. The $x$ coordinates are the starting times of data collection. The error bars represent standard deviations.}
\label{fig:fluctuation}
\end{figure}

In the cases where direct quantitative comparison is intended, the same backend was used, and maximal time proximity for the measurements across several methods was aimed for. Even still, unavoidable factors make it so that gathering datasets in succession where possible does not guarantee that each registered identical behaviour. Figure
\ref{fig:fluctuation} shows the time evolution of the estimated values using inference (whose results can be more local in time due to relying on fewer shots). The delays are due to queuing, execution times and repetition of runs to extract statistics. As such, discrepancies may in some cases be attributable to sources other than error in the estimation.

Unlike curve fits, inference can work with all-different times for a single measurement each. However, Qiskit only allows up to $75$ different pulse schedules per job submitted, each of which can be repeated for up to $8192$ shots. To streamline the data collection process, a maximum of $75$ different times is used for the data vectors, the total desired number of experiments being distributed by these times.

It should be noted that the problem of estimating time constants is fairly simple, and not necessarily a case where inference stands out; oscillations are a more interesting application. Even so, they make an interesting test case, as they elucidate the impact of intrinsic differences between likelihood models (in terms of learning rate, most favorable \gls{SMC} sequences, particle number, effective sample size, order of magnitude of the evolution times, etc.). They also help understand the difficulties estimating an oscillation frequency along with a damping constant brings, due to conflicts between what is optimal for each.

For the Ramsey experiments, the probabilities of $0$ and $1$ are switched with respect to equations \ref{eq:ramsey_p1} and \ref{eq:ramsey_p1_decay}. This is because a the half-pi pulses were adapted from a Hadamard gate - using the default instruction \texttt{u2}$(0,\pi)$ -, which is an involution. Because it represents a rotation of $\pi$ around the $x+z$ axis, when applied twice to $\ket{0}$ it brings the state back to $\ket{0}$ rather than to $\ket{1}$. This is without prejudice to the covered angle's being as intended $\pi/2$ on the $x-z$ plane.

\subsection{Hahn echo experiment (T2 estimation)}
\label{sub:exp_t2}

This subsection presents the results of estimating the dephasing time $T_2$ (subsection \ref{sub:phase_damping}) of a few IBM quantum devices.

In the case of exponential decay models such as this one, the sampler benefits from long evolution times being considered sooner, because near the origin the information is more challenging. This stands in contrast with the precession frequency example taken as an example throughout chapters \ref{cha:quantum_parameter_estimation} and \ref{cha:monte_carlo_posterior_sampling}, but follows the same line of reasoning.

\begin{figure}
\captionsetup[subfigure]{width=.9\textwidth}%
\begin{subfigure}[t]{.45\textwidth}
  \centering
  \includegraphics[width=\linewidth]{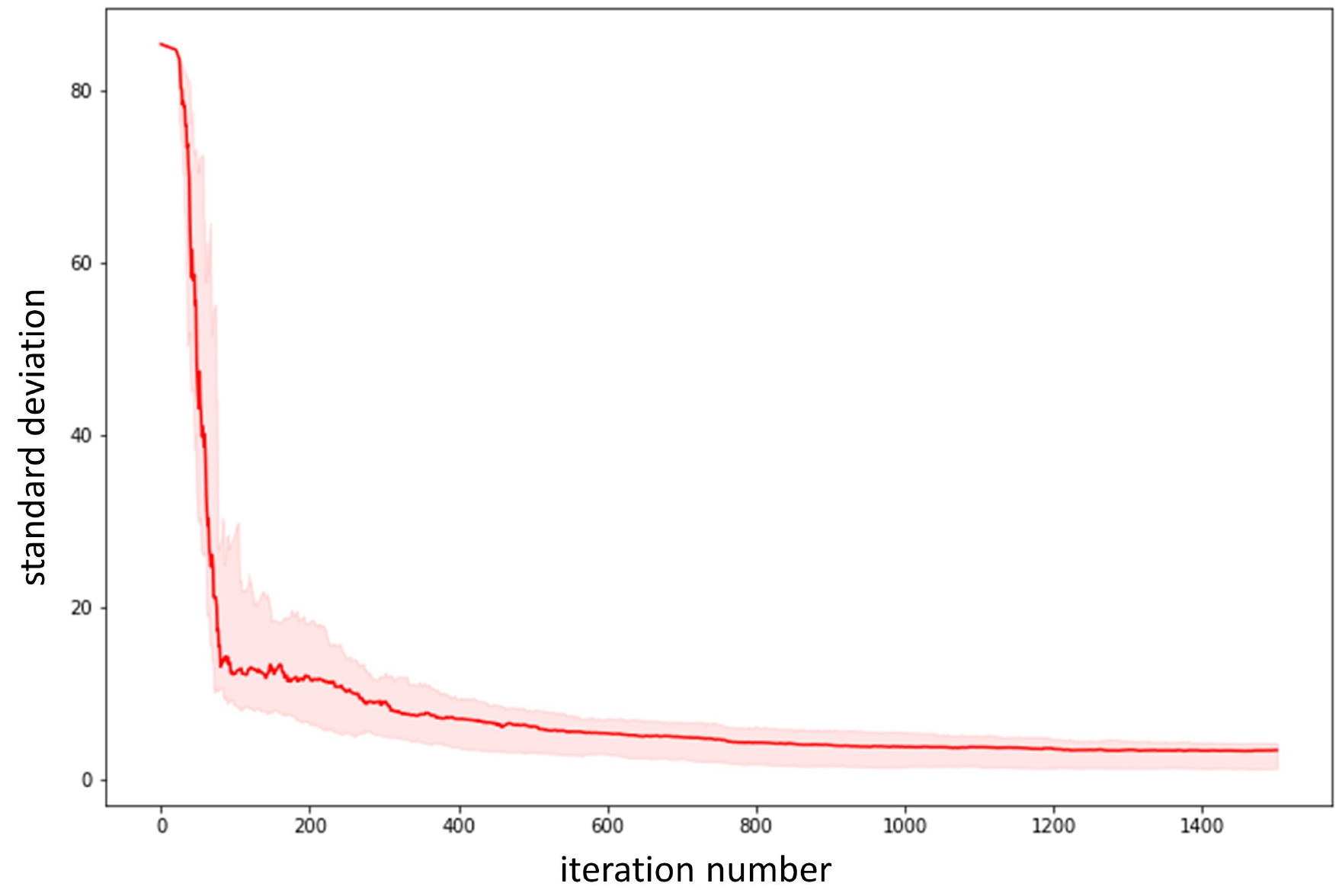}
  \caption{Data added by order of ascending evolution time.}
  \label{fig:T2_non_reversed}
\end{subfigure}
\begin{subfigure}[t]{.45\textwidth}
  \centering
  \includegraphics[width=\linewidth]{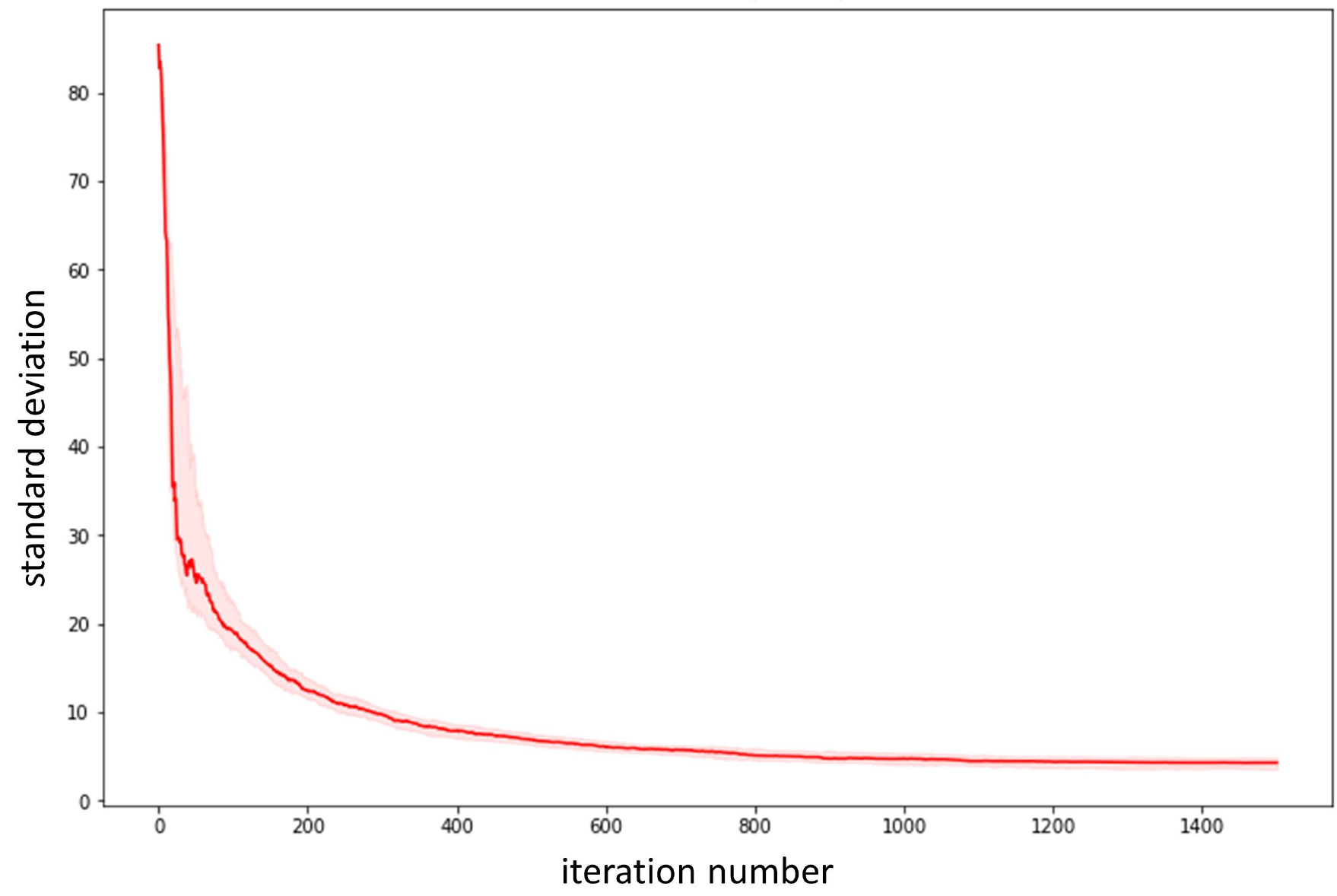}
  \caption{Data added by order of descending evolution time.}
  \label{fig:T2_reversed}
\end{subfigure}
\caption{Evolution of the standard deviation of the dephasing time $T_2$ during the inference process, for the IBMQ device \texttt{ibmq\_rome} and two different orders of data addition. Both results were based on the same dataset containing $1500$ observations. The shaded region represents the interquartile range.}
\label{fig:T2_evolution}
\end{figure}

Figure \ref{fig:T2_evolution} illustrates this fact. The only difference between graphs is the order by which the data are fed to the sampler. In the case of figure \ref{fig:T2_non_reversed} the shortest times were considered first, whereas in that of \ref{fig:T2_reversed} the opposite was done.

The sharp drop in the uncertainty followed by a plateau and the larger inter-run variability in the first case are a sign of a sharp drop in effective sample size and undue particle concentration (owing to particle collapse). Conversely, the second graph  shows a steadier learning process. Thus, the longer measurement times were included earliest in the sequence of \gls{SMC} target distributions.

\begin{figure}[!ht]
\captionsetup[subfigure]{width=.9\textwidth}%
\begin{subfigure}[t]{.45\textwidth}
  \centering
  \includegraphics[width=\linewidth]{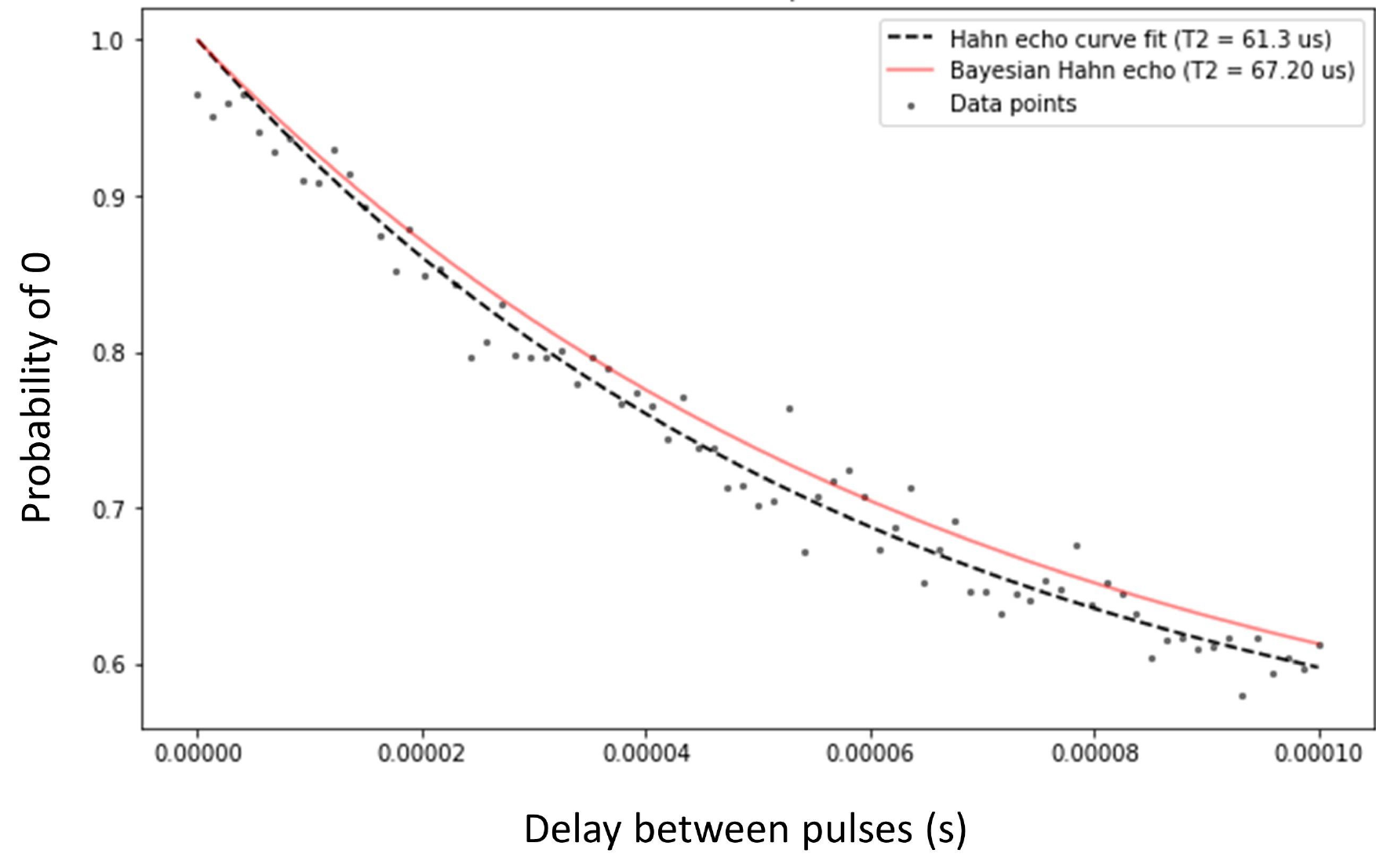}
  \caption{Results obtained using Bayesian inference (solid red line) and a curve fit (dashed black line), superposed on the data points used for the latter (gray dots), which used 512 shots each. For the inference, only $4\%$ (20) of these shots per point were used. }
  \label{fig:T2_scipy_bayes}
\end{subfigure}
\begin{subfigure}[t]{.45\textwidth}
  \centering
  \includegraphics[width=\linewidth]{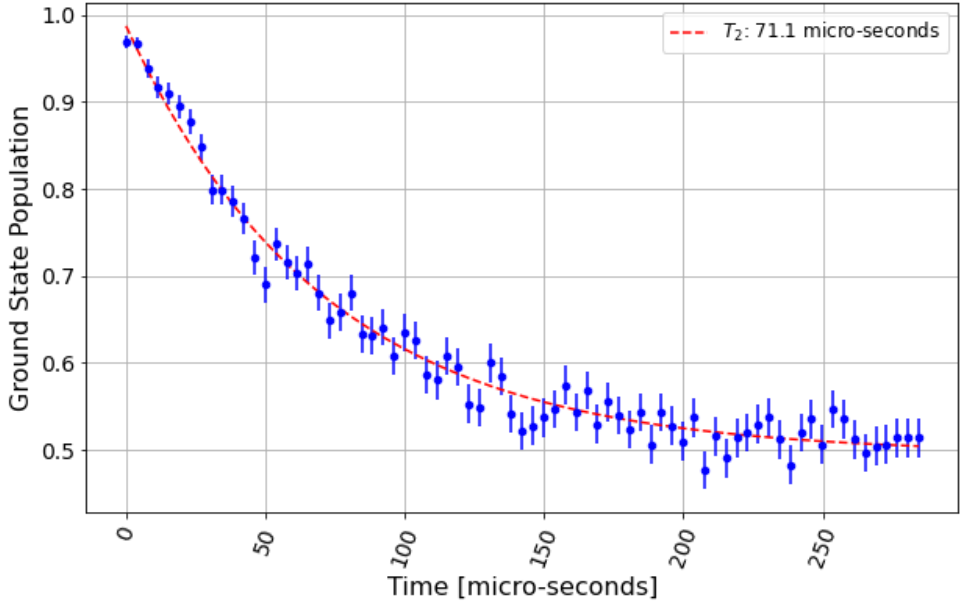}
  \caption{Results obtained using the Qiskit coherence fitter for $T_2$, with 512 shots per data point. The model of equation \ref{eq:t2} is relaxed to adapt the coordinates at $t=0$ and $t=\infty$.}
  \label{fig:T2_Qiskit}
\end{subfigure}
\caption{Curves produced by $T_2$ estimates for the IBMQ device \texttt{ibmq\_rome}, using $75$ different measurement times. In figure \ref{fig:T2_scipy_bayes} the model of equation \ref{eq:t2} was used, whereas \ref{fig:T2_Qiskit} relaxes it to include two constants $A$ and $B$.}
\label{fig:T2_fixed_coord}
\end{figure}

Due to systematic errors, the coefficient and constant of equation \ref{eq:t2} may not be exactly $1/2$. In that case, it may be beneficial to allow for more freedom in the coordinate at origin and horizontal asymptote, which is done by default in the Qiskit fits. This problem is illustrated in figure \ref{fig:T2_fixed_coord}. Equation \ref{eq:t2} should then be changed to $\mathbf{P}(1 \mid \Delta t)=A\exp(-t/T_2)+B$.

\begin{figure}[!ht]
    \centering
    \includegraphics[width=12cm]{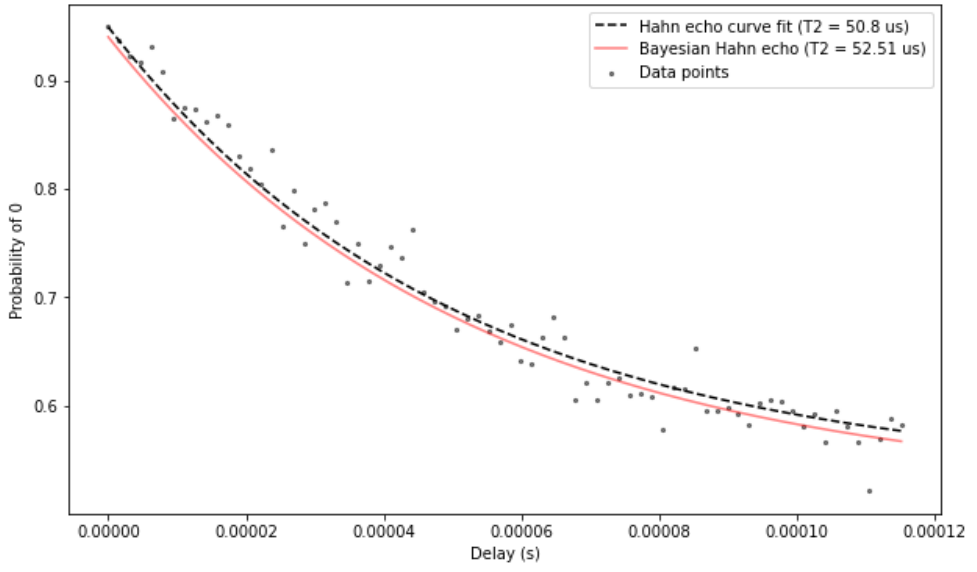}
    \caption{Curves produced by $T_2$ estimates for the IBMQ device \texttt{ibmq\_casablanca}, using also estimates for $A,B \not\equiv 1/2$. The results were obtained using Bayesian inference (solid red line) and a curve fit (dashed black line), superposed on the data points used for the latter (gray dots). The curve fit relied on $512$ shots per point, including the extremes to which it was adapted. For the inference, $20$ shots per point were used, and $100$ for each of the hyperparameter-adjusting ($A$ and $B$) measurements.}
    \label{fig:T2_bayes_adj}
\end{figure}

To remedy this, one can include a rough estimate of the hyperparameters $A$ and $B$ in the inference model. This can be achieved with a higher shot count for $t=0$ and some large $t \rightarrow \infty$ (implying a constant measurement overhead). The fraction of $0$ outcomes for each of these times can be identified with $A$ and $B$ respectively. The result of using $512$ shots for these model calibration measurements before the inference process is shown in figure \ref{fig:T2_bayes_adj}. Though they were chosen to match the fitter, rougher estimates can be used, e.g. $50$. The curves can be seen to fit the data better as compared to figure \ref{fig:T2_scipy_bayes}.

With these adjustments in mind, we can compare the quantitative results of inference and the Qiskit fitter. The experiments were in this case done on the \texttt{ibmq\_casablanca} backend.

The conditions for the inference were as follows. A $50$ particle \gls{SMC} approximation was used, with a flat prior on $]0,250]\mu s$. A single Markov (\gls{RWM}) move was used per particle per step. The threshold effective sample size for resampling was set at $\widehat{\text{ESS}}=0.8$. A total of $75$ different measurement times was used for 20 shots each, signifying 1500 data/steps, plus $2 \cdot 512$ additional shots to customize $A$ and $B$. Of the total steps, 33 (2\%) triggered a resampling stage on average. The measurement times were chosen in constant increments within $t \in ]0,115] \mu s$. The results were obtained by taking medians over $100$ runs split equally by $10$ different datasets. 

\begin{figure}[!ht]
    \centering
    \includegraphics[width=8cm]{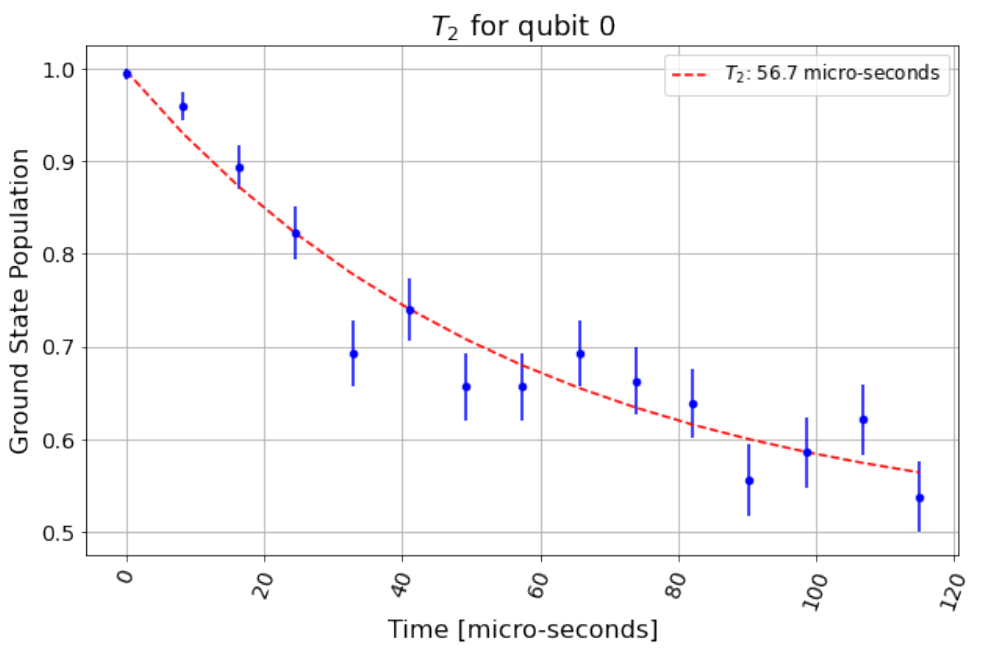}
    \caption{Curve produced by the Qiskit coherence fitter for $T_2$ of device \texttt{ibmq\_casablanca}, using $15$ different measurement times for $169$ shots each (a total of $2 \, 535$ shots). }
    \label{fig:T2_matched_shots}
\end{figure}

The results are shown in table \ref{tb:t2}, and compared to those of the Qiskit fitter. For comparison purposes, the latter was applied in two sets of conditions. First, using the same number of measurement times and range, but for $512$ shots each instead of $20$. In this case, both the obtained values and their associated uncertainties were close, in spite of the discrepancy between shot counts. Second, allowing the curve fitter roughly as many total measurements as were used for the inference; the associated graph can be seen in figure \ref{fig:T2_matched_shots}. The measurement times were uniformly spaced out in the same range as for the inference, but with fewer different times and more shots. The achieved error was over 3 times larger than that obtained by inference.

\begin{table}
\centering
\begin{tabular}{ |g|c|c|c| } 
     \hline
     \rowcolor{gray!15}
      & $\mathbf{T_2}$ ($\mu s$) & \textbf{Standard deviation} & \textbf{Shot number}\\
     \hline
     \textbf{Bayesian inference} & $52.51$ & $4.4$ & \multicolumn{1}{|r|}{$20 \cdot 75 + 512 \cdot 2 \ = \phantom{3}2 \, 524$} \\
     \hline
     \textbf{Qiskit fitter} & $52.39$ & $3.7$  & 
     \multicolumn{1}{|r|}{$512\cdot 75 \quad \ = 38 \, 400$} \\
     \hline
      \textbf{Qiskit fitter} & $56.70$ & $13.9$  & 
      \multicolumn{1}{|r|}{$169 \cdot 15 \quad \: = \phantom{3}2 \, 535$} \\
     \hline
\end{tabular}
\caption{Dephasing time estimation results for the IBMQ device \texttt{ibmq\_casablanca}. The registered backend estimate was at the time $T_2 = 57.58\mu s$ for the two first lines, and $48.31\mu s$ for the last one (as mentioned before, variations such as these are just an experimental artifact).}
\label{tb:t2}
\end{table}

The reason for using different experiments is that simply providing the fitter the exact data that was used for the inference yielded an error in the order of the thousands of microseconds. Starting from there, the variance could be reduced by reducing the number of measurement times, originally $75$, by which the shots were distributed. This was done with a granularity of $10$ times. After a point, the trend was reversed; in the table, the lowest achieved standard deviation is presented. Though the measurement schemes were not exhaustively picked to be the best, still less were those chosen for the inference: they were simply arrayed on intervals for which the signal was still sure to be passably clear. 

In conclusion, both comparative cases strongly suggest that that inference is more robust for  extracting information from a small dataset.

\subsection{Quantifying energy loss (T1 estimation)}
\label{sub:exp_t1}

\begin{figure}[!ht]
    \centering
    \includegraphics[width=8cm]{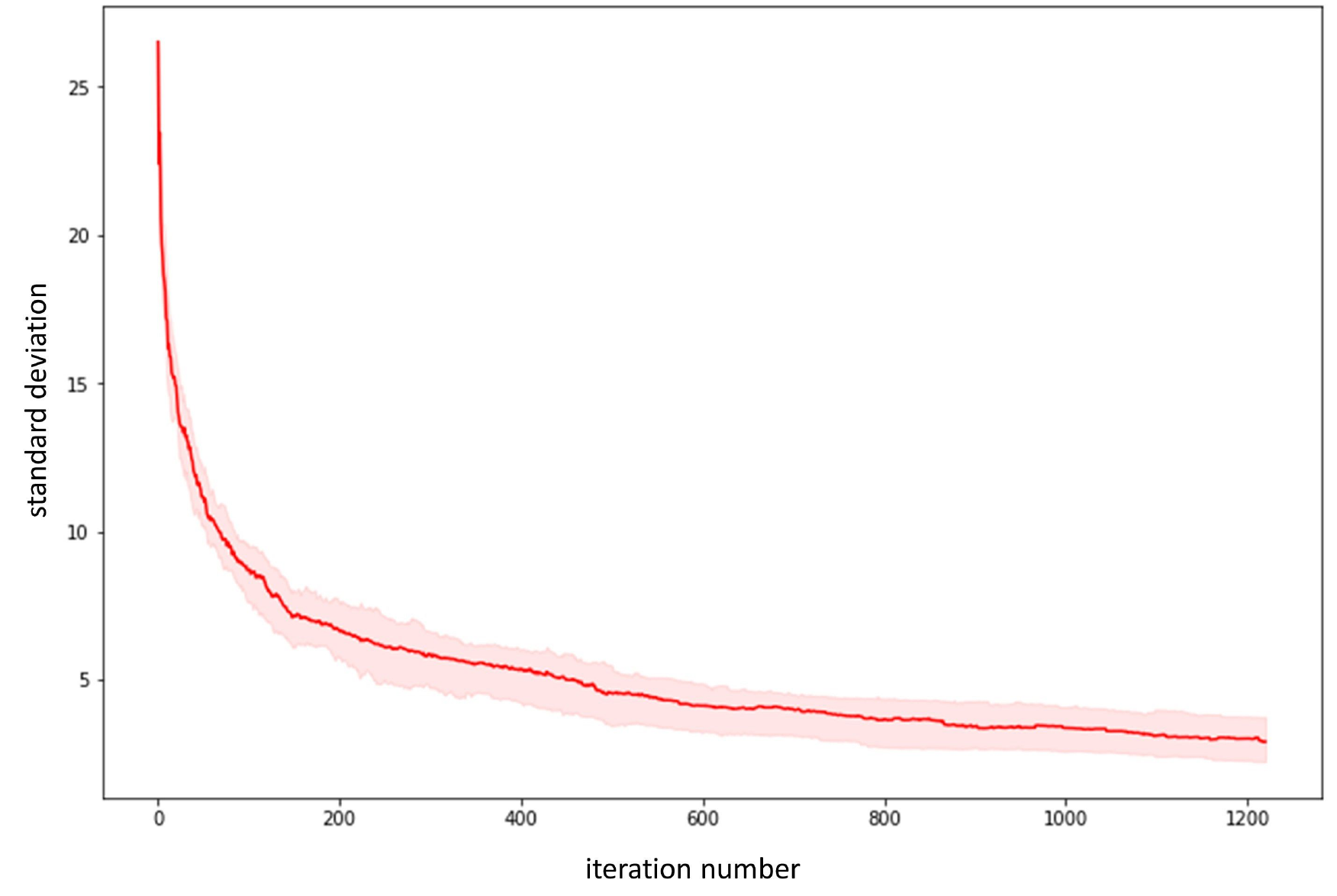}
    \caption{Evolution of the standard deviation during the estimation of the energy relaxation time constant $T_1$ of IBMQ device \texttt{ibmq\_guadalupe}. The dataset contemplated $1500$ observations. The shaded region represents the interquartile range.}
    \label{fig:t1_evolution}
\end{figure}

For this subsection, the parameter of interest is the energy relaxation time $T_1$ (subsection \ref{sub:amplitude_damping}). The experiment is identical to that of the previous section, except for the fact that the occupation of $\ket{1}$ tends to $0\%$ and not $50\%$.  Figure \ref{fig:t1_evolution} shows the evolution of the uncertainty as the inference proceeds.

The conditions were similar to section \ref{sub:exp_t2}, apart from two points. The prior was set on $]0,100]\mu s$ (this doesn't have much of an impact after the first few iterations), and the measurement times on $]0,50]\mu s$. A standard deviation of $\sigma = 2.9 \mu s$ for an estimate of $T_1=62.45\mu s$ was achieved.

\subsection{Ramsey experiment (2-dimensional estimation)}
\label{sub:exp_ramsey_2d}

This subsection presents the results of performing Ramsey experiments (subsection \ref{sub:spin_precession}) on IBM quantum devices. The aim was to use the circuit from figure \ref{fig:ramsey_circ} to estimate the detuning $\delta$ and coherence time $T_2^*$ from equation \ref{eq:ramsey_p1_decay}. Instead of targeting the latter directly, the factor $\gamma_2^*=1/T_2^*$ is considered, so that the two parameters are matched in units (though not necessarily scale) as suggested in \cite{Granade_2012}. 

As for the frequency estimation, relevant considerations have been laid out in subsection \ref{sub:quantum_characterization_examples} and the ensuing sections. Unlike for the isolated characteristic time estimation of subsections \ref{sub:exp_t2} and \ref{sub:exp_t1}, the data were considered by order of increasing evolution times, as this was found to work best in this case. Note that although this tends to streamline \textit{frequency} estimation, it has the opposite effect for the characteristic time, as demonstrated in subsection \ref{sub:exp_t2} for the non-oscillatory case. Of course, once these parameters are joined together in a likelihood function, there exists a mutual influence in their learning rates, creating a more delicate interplay. The tension between what is best for each parameter complicates the decision process, given that they play such different but inseparable roles. 

\begin{figure}[!ht]
    \centering
    \includegraphics[width=8cm]{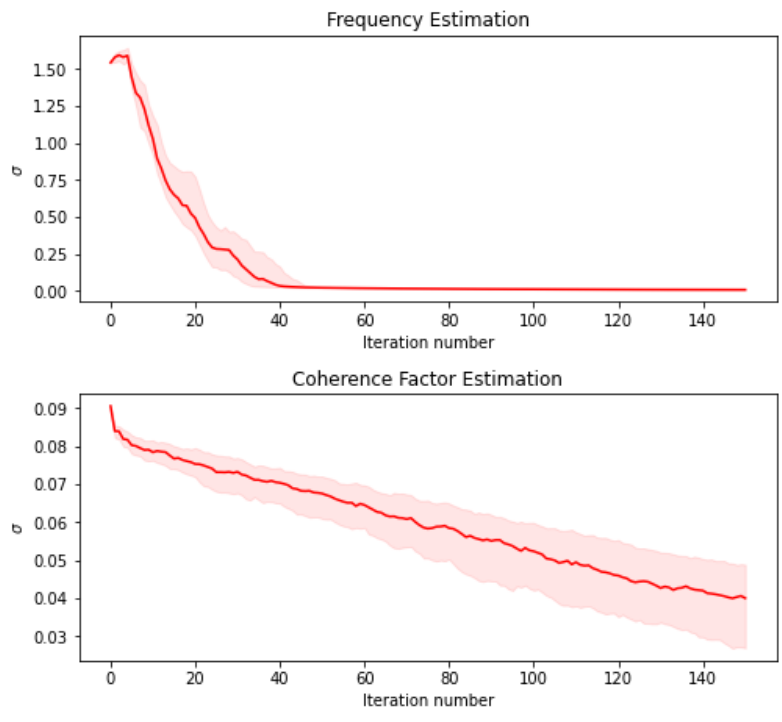}
    \caption{Evolution of the standard deviations during the estimation of the detuning frequency $\delta$ and coherence factor $\gamma_2^*$ of IBMQ device \texttt{ibmq\_armonk}. The dataset contemplated $150$ observations. The solid curves are the median results for 300 runs split by 100 datasets. The shaded regions represent interquartile ranges.}
    \label{fig:ramsey_evol}
\end{figure}

An example the evolution of standard deviations through the iterations is shown in figure \ref{fig:ramsey_evol}, where the difference between parameter learning rates is evident.

\begin{figure}[!ht]
\captionsetup[subfigure]{width=.9\textwidth}%
\begin{subfigure}[t]{.45\textwidth}
  \centering
  \includegraphics[width=\linewidth]{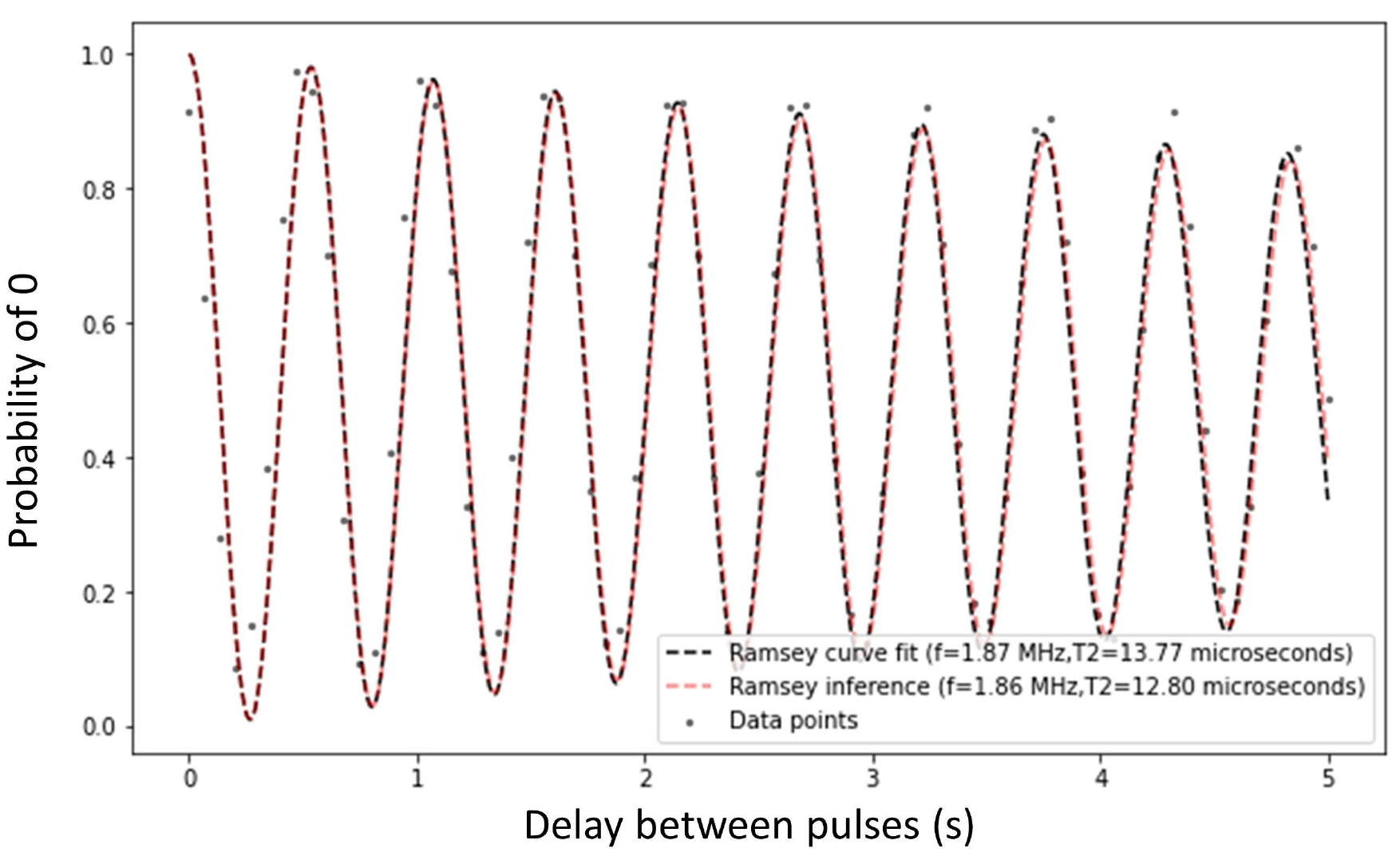}
  \caption{Results obtained using Bayesian inference (solid red line) and a curve fit (dashed black line), superposed on the data points used for the latter (gray dots), which used 512 shots each. For the inference, only $0.4\%$ (2) of the shots per point were used.}
  \label{fig:ramsey_curves}
\end{subfigure}
\begin{subfigure}[t]{.45\textwidth}
  \centering
  \includegraphics[width=\textwidth]{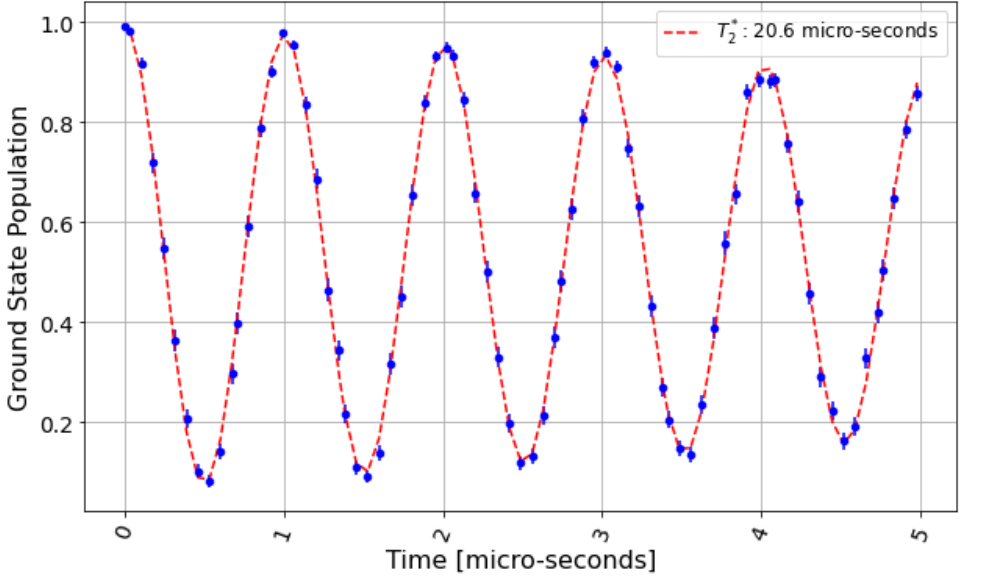}
  \caption{Results obtained using the Qiskit coherence fitter for $T_2^*$, with 512 shots per data point.}
  \label{fig:ramsey_qiskit}
\end{subfigure}
\caption{Curves produced by $\delta$ (labeled $f$) and $T_2^*$ estimates for the IBMQ device \texttt{ibmq\_guadalupe}, using $75$ different measurement times. }
\label{fig:ramsey_curves_all}
\end{figure}

The default Qiskit fitter for $T_2^*$ takes as an input parameter the number of oscillations to be induced during the time span, rather than a detuning frequency. It is geared towards coherence time estimation, and not frequency calibration. Thus, this is the quantity whose estimation is to be compared, though the frequency still brings added complexity. An example of the results obtained by inference and by curve fitting is presented in figure \ref{fig:ramsey_curves_all}. For reference, the uncertainties in the estimation of the induced oscillation frequency were $0.4\%$ for inference, nearly $0\%$ for curve fitting with $256$ times more shots, and $15\%$ for curve fitting with the same shots.

As discussed in section \ref{sub:phase_damping}, $T_2^*$ is susceptible to large variations, so any comparison should be taken with a grain of salt. Qiskit does not make estimates of this constant available (despite still supplying a built-in fitter for it), likely due to its volatility.

We will now present some quantitative final results for a specific setting. For the inference, a $225$ particle \gls{SMC} approximation was used, with a flat prior on $f \in ]0.0,5.0]MHz$, $T_2^* \in [3.00,25.00]\mu s$. Three Markov moves were used per particle per step. The threshold effective sample size for resampling was set at $\widehat{\text{ESS}}=0.8$. A total of $75$ different measurement times was used for 2 shots each, signifying 150 data/steps. Of these, 28 (19\%) triggered a resampling stage on average. The measurement times were chosen in constant increments within $t \in ]2.0,5] \mu s$. The results were obtained by taking medians over $100$ runs split equally by $10$ different datasets.

For a detuning of $\delta = 1.83MHz$, the final estimates were $\delta = 1.863 MHz$ and $T_2^*=9\mu s$, and their associated uncertainties $\sigma = 0.007 MHz$ and $3\mu s$. Table \ref{tb:ramsey} compares the $T_2^*$ results between this method and Qiskit's. We again observe that the \textit{real} values are quite fickle, which explains the disparities.

\begin{table}[ht!]
\centering
\begin{tabular}{ |g|c|c|c|c| } 
     \hline
     \rowcolor{gray!15}
      & $\mathbf{T_2^*}$ ($\mu s$) & \textbf{Standard deviation} & \textbf{Total shot count}\\
     \hline
     \textbf{Bayesian inference} & $9$ & $3$ & 
     \multicolumn{1}{|r|}{$75 \cdot 2 \ \, = \phantom{38} \, 150$}\\
     \hline
     \textbf{Qiskit fitter} & $21$ & $1$ & 
     \multicolumn{1}{|r|}{$75 \cdot 512 = 38 \, 400$}\\
     \hline
     \textbf{Qiskit fitter} & $9$ & $30$ & 
     \multicolumn{1}{|r|}{$15 \cdot 10 \; = \phantom{38} \, 150$}\\
     \hline
\end{tabular}
\caption{Unechoed dephasing time estimation results for the IBMQ device \texttt{ibmq\_armonk}.}
\label{tb:ramsey}
\end{table}

\begin{figure}[!ht]
    \centering
    \includegraphics[width=8cm]{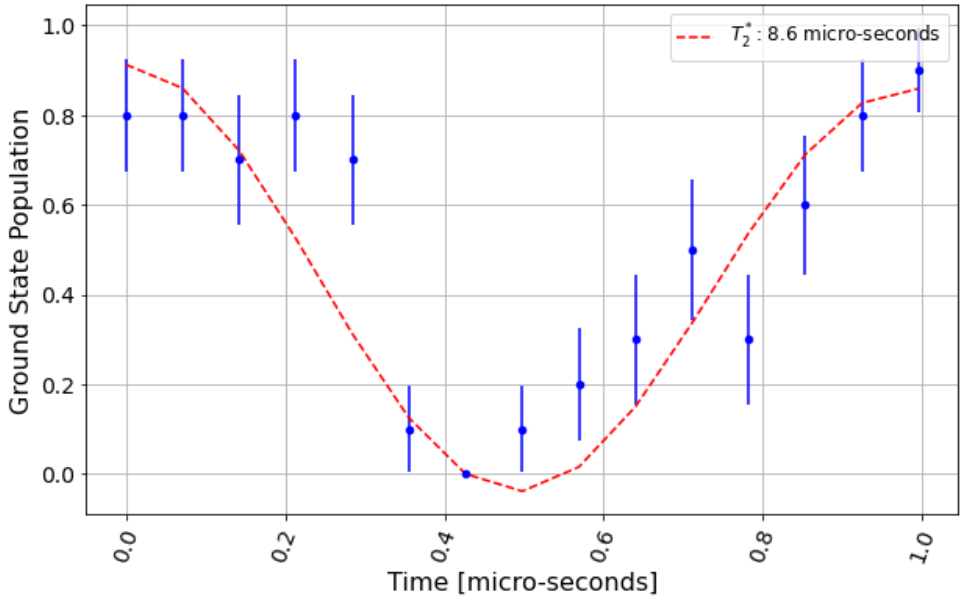}
    \caption{Curve produced by the Qiskit coherence fitter for $T_2^*$ of device \texttt{ibmq\_guadalupe}, using $10$ different measurement times for $15$ shots each (a total of $150$ shots). }
    \label{fig:ramsey_qiskit_matched}
\end{figure}

Once more, the last row presents the result of running a curve fit for as many total measurements as were included in the inference's dataset (the corresponding graph is shown in figure \ref{fig:ramsey_qiskit_matched} for visualization), which was $10$ times worse than the inference's. The registered standard deviation was the best one obtained among multiple good-faith attempts. They were chosen as for the case of table \ref{tb:t2}, only now the time range too had to be adapted (shortened) to obtain reasonable results. These adaptations were at all times in benefit of the curve fitter. Most choices of running curve fits with $150$ total shots yielded infinite variance, including when matching the scheme used for inference. Again, the latter dealt better with limited information. 

For completeness, we note that the fitter achieved uncertainties of $0$ and $15\%$ for the frequency, for the higher and equal shot counts respectively (second and third lines of table \ref{tb:ramsey}). The key point is that once the data allowance was matched, the fitter performed roughly $40$ times worse for this parameter, an even starker contrast. The induced oscillation frequencies were matched for order of magnitude. 

\begin{figure}[!ht]
\captionsetup[subfigure]{width=.9\textwidth}%
\begin{subfigure}[t]{.45\textwidth}
  \centering
  \includegraphics[width=\linewidth]{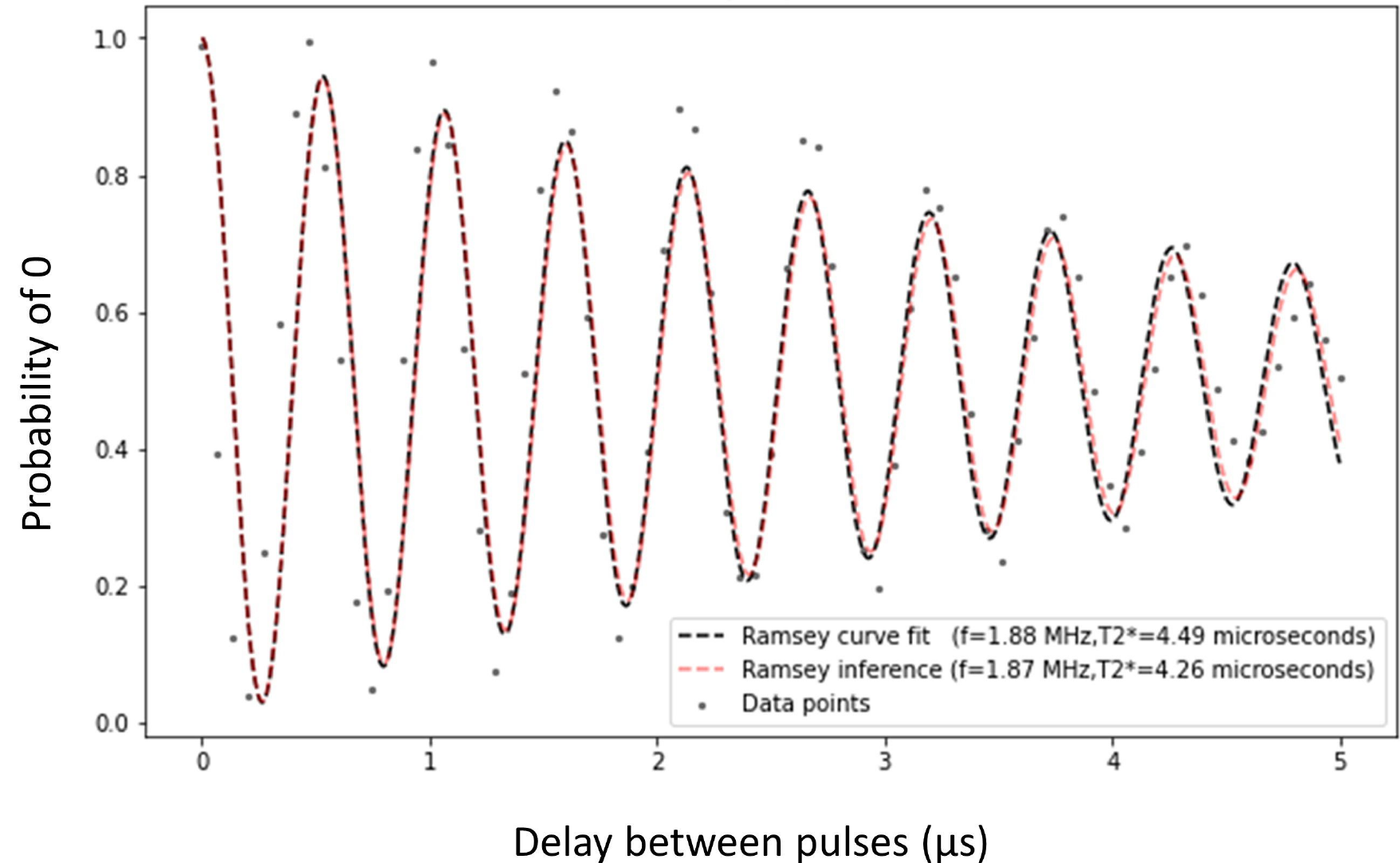}
  \caption{Bayesian inference results plotted along with the data they were based on ($4\%$ of the information contained by these data was used).}
  \label{fig:ramsey_closest}
\end{subfigure}
\begin{subfigure}[t]{.45\textwidth}
  \centering
  \includegraphics[width=\textwidth]{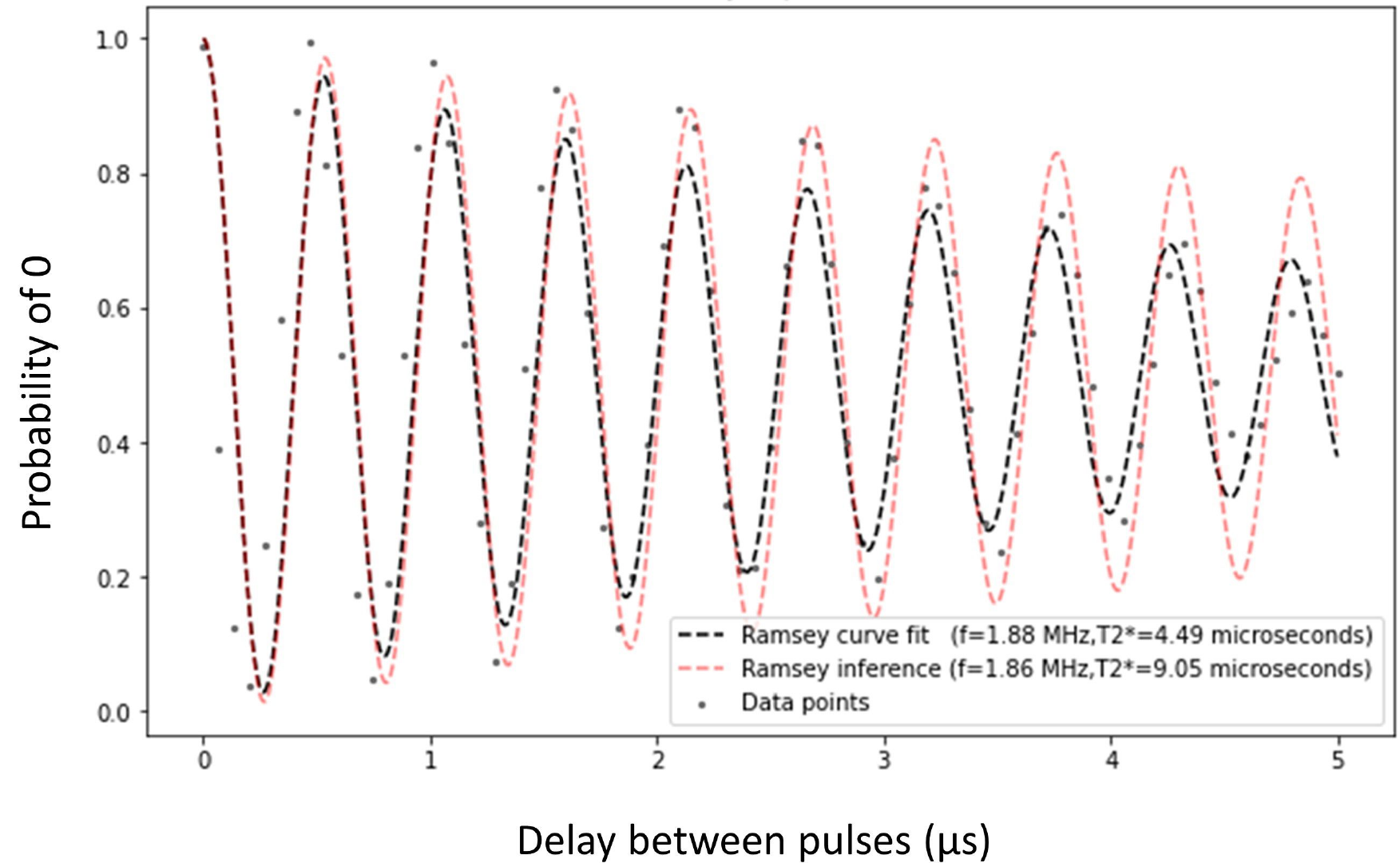}
  \caption{Bayesian inference results plotted along with a preliminary set of gathered data, which preceded most of the data the inference was based on.}
  \label{fig:ramsey_median}
\end{subfigure}
\caption{Ramsey experiment inference results obtained using 100 runs over the same dataset, or distributed by successive datasets spanning a longer period of time. The inference results (dashed red lines) are plotted against the same set of data points (gray dots), from which the data for the inference was extracted in the case of figure \ref{fig:ramsey_closest}. A curve fit (dashed black lines) tailored to these points is included for reference. Each of the data points used $512$ shots, whereas for the inference only $2$ were used for each time. The targeted device was \texttt{ibmq\_armonk}, with $75$ different measurement times.}
\label{fig:ramsey_time}
\end{figure}

Figure \ref{fig:ramsey_curves} suggests that the inference results fit the data well. Expectably, the results are more variable when the coherence times are shorter. In particular, they may be quite local in time; due to using few shots, they capture short-time pictures of the system (see also figure \ref{fig:fluctuation}). The results tend to be more consistent for evolution times well within the coherence time, where the signal is less noisy. This is illustrated in figure \ref{fig:ramsey_time}. In \ref{fig:ramsey_closest}, inference was performed based on a dataset extracted from the presented data points; it underestimated the earliest data points' oscillation amplitude. Using a more general dataset (figure \ref{fig:ramsey_median}) tends to fit the earliest points of the dataset better than inference based on the dataset itself.

\subsection{Echoed Ramsey experiment (1-dimensional estimation)}
\label{sub:exp_ramsey_1d}

This subsection presents the results of performing \textit{Hahn-Ramsey} experiments on a few IBM quantum devices. The aim was to use the circuit from figure \ref{fig:echoed_ramsey_circ} to estimate the detuning $\delta$ from equation \ref{eq:ramsey_p1}. Since learning a single parameter is less demanding than learning two, this allows for testing methods that require more computational resources while still targeting a more interesting example than those of sections \ref{sub:exp_t2} and \ref{sub:exp_t1}.

\begin{figure}[!ht]
\captionsetup[subfigure]{width=.9\textwidth}%
\begin{subfigure}[t]{.45\textwidth}
  \centering
  \includegraphics[width=\linewidth]{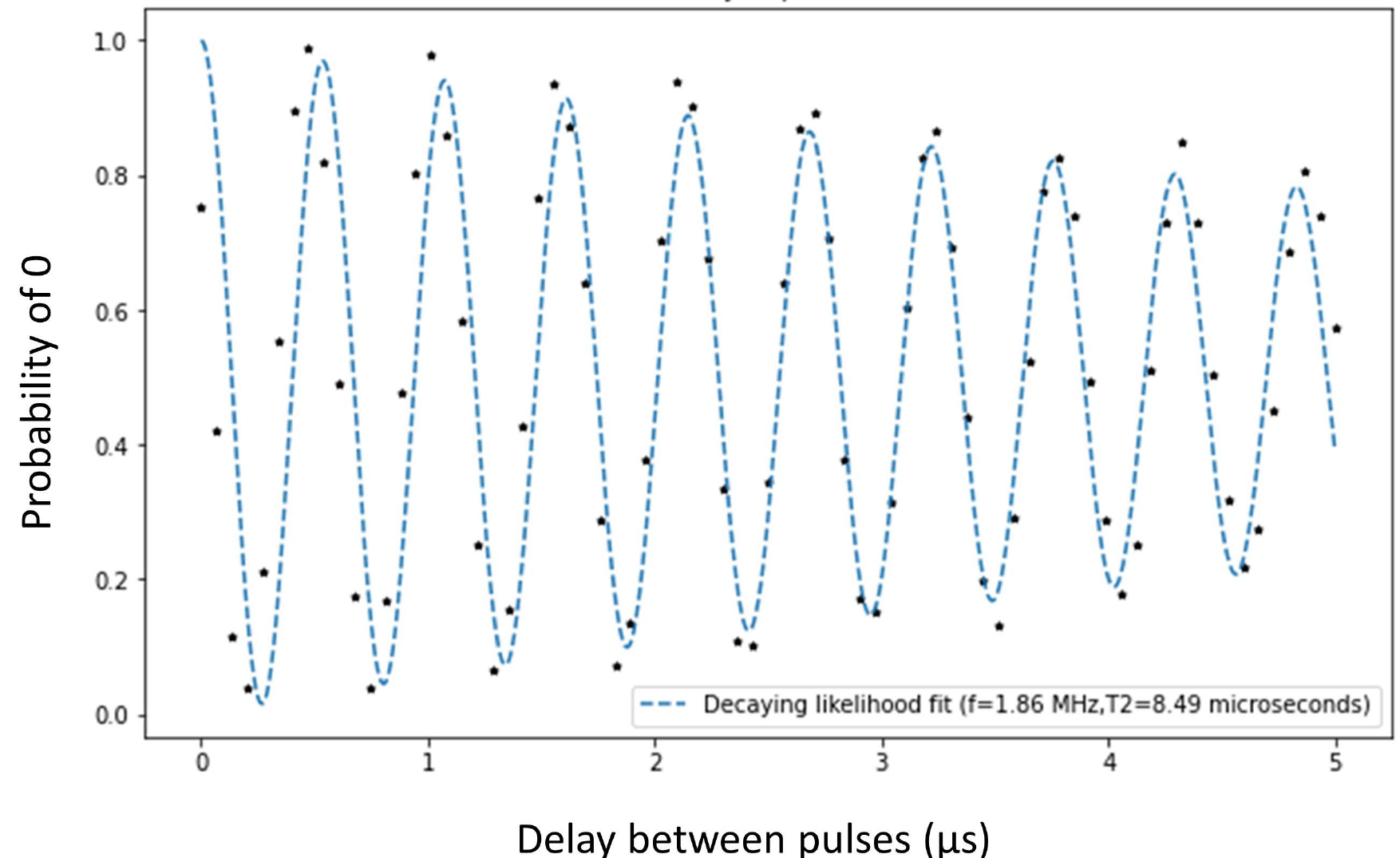}
  \caption{Data points generated using the Ramsey sequence of figure \ref{fig:ramsey_circ}.}
  \label{fig:unechoed}
\end{subfigure}
\begin{subfigure}[t]{.45\textwidth}
  \centering
  \includegraphics[width=\textwidth]{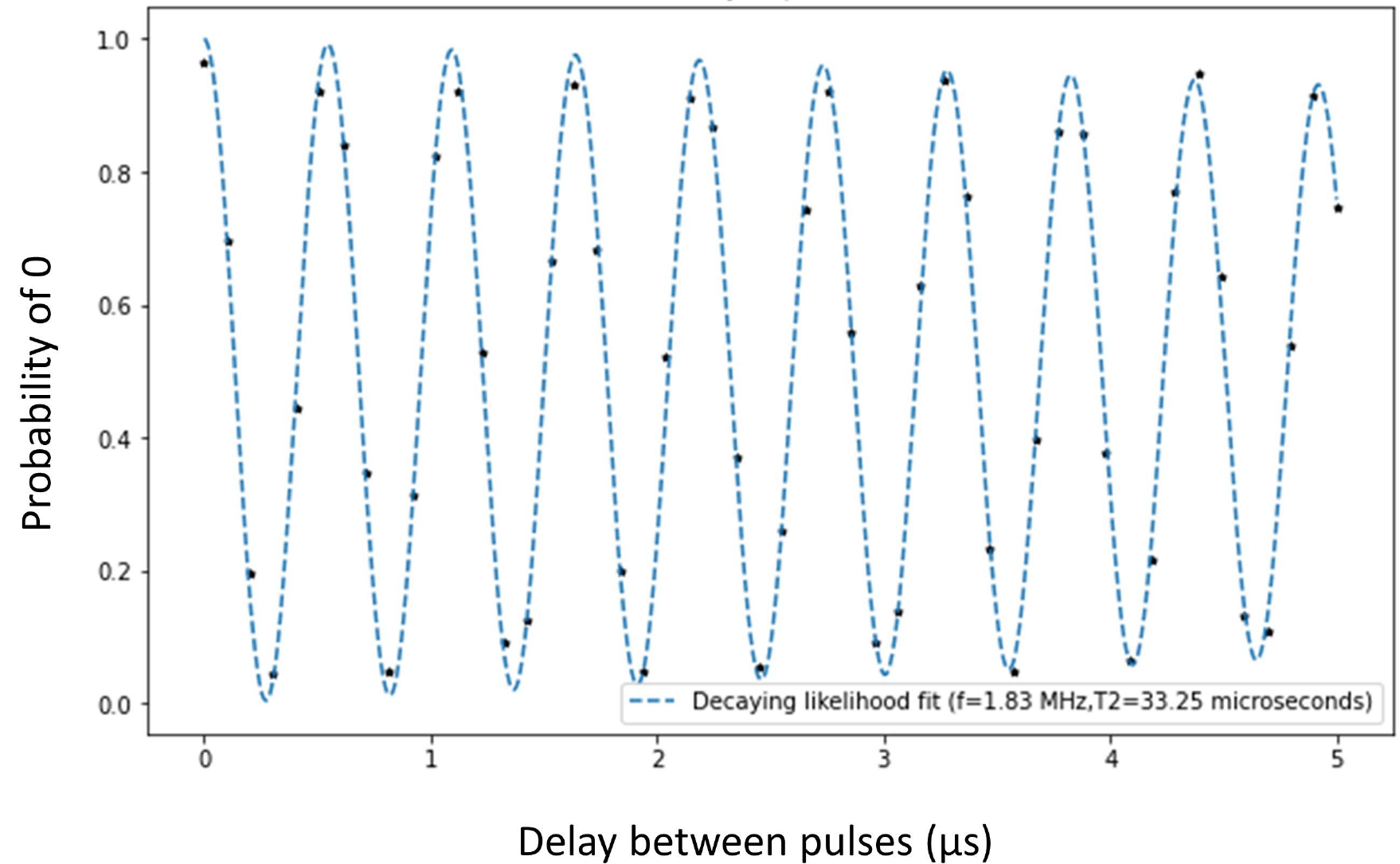}
  \caption{Data points generated using the Ramsey sequence of figure \ref{fig:echoed_ramsey_circ}.}
  \label{fig:echoed}
\end{subfigure}
\caption{Effect of inserting a refocusing pulse in a Ramsey sequence for \texttt{ibmq\_armonk}. Each of the $75$ data points (black dots) was obtained using $512$ shots. To aid visualization, a curve fit (dashed blue lines) performed using SciPy for the function of equation \ref{eq:ramsey_p1} is plotted along.}
\label{fig:echoed_vs_not}
\end{figure}

The impact of the pi-pulse is illustrated in figure  \ref{fig:echoed_vs_not}. When comparing figure \ref{fig:echoed} to the original case of figure \ref{fig:unechoed}, we can see that the decoherence in the represented interval (up to $5\mu s$) is very slight. For measurements within that interval, we can approximate the amplitude as $1$ while still obtaining reasonable results. The coherence time was improved by around $400\%$.

\begin{figure}[!ht]
    \centering
    \includegraphics[width=8cm]{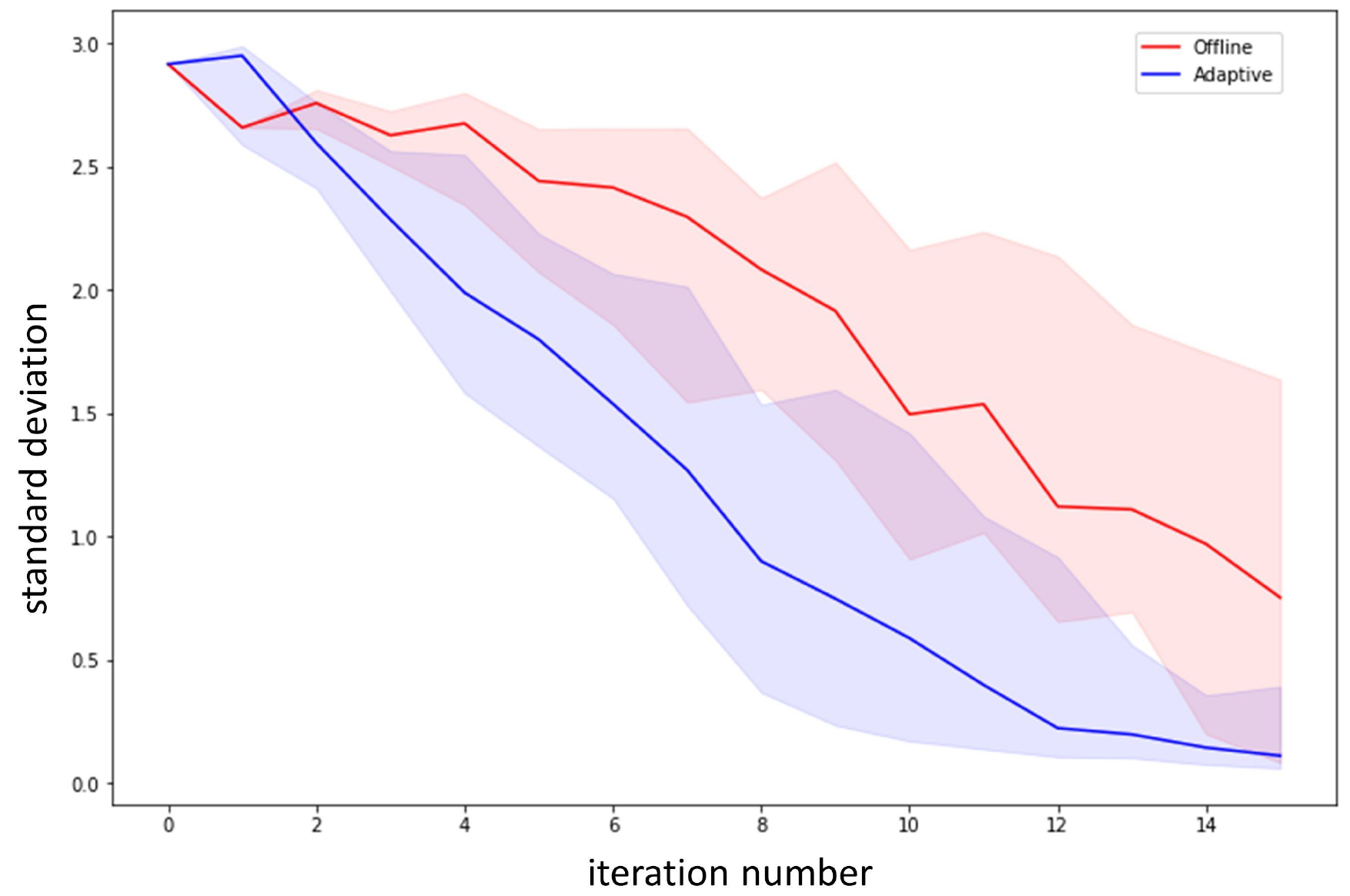}
    \caption{Evolution of the standard deviation during the inference process for adaptive (blue line) and offline (red line) estimation. The shaded regions represent interquartile ranges. }
    \label{fig:adaptive}
\end{figure}

This was used to test adaptive experimental design on \texttt{ibmq\_armonk} (section \ref{sec:bayesian_experimental_design}). The strategy
was compared to the non-adaptive (offline) estimation of before. This serializes the data collection process. Due to limited usage of the IBMQ devices, only $15$ data/steps were used for both methods, and $100$ runs. Still, the difference was clear (figure \ref{fig:adaptive}).

A $100$ particle \gls{SMC} approximation was used, with a flat prior on $\delta \in ]0.0,10.0]MHz$. One Markov move was used per particle per step. The threshold effective sample size for resampling was set at $\widehat{\text{ESS}}=0.5$. A total of $15$ different measurement times was used for a single shot each, signifying $15$ steps. Of these, $7$ ($47\%$) and $9$ ($60\%$) triggered a resampling stage on average for the adaptive and offline methods respectively. The offline measurement times were chosen in constant increments within $t \in ]0.2,2] \mu s$, and the adaptive ones were chosen from $20$ guesses around $1/\sigma_\text{curr}$ to have the lowest variance (as suggested in subsection \ref{sub:precession_heuristics}). The results were obtained by taking medians over $100$ runs split equally by $10$ different datasets. The quantitative results are presented in table \ref{tb:ramsey_adaptive}. 

\begin{table}[ht!]
\centering
\begin{tabular}{ |g|c|c|c|c|c| } 
     \hline
     \rowcolor{gray!15}
      & $\boldsymbol{\delta}$ ($MHz$) & \textbf{Standard deviation} & \textbf{Total shot count} & 
      \textbf{Precision} ($\sigma^2\Delta t_\text{acc}$)\\
     \hline
     \textbf{Adaptive} & $1.8$ & $0.1$ & $15$ & $0.38$ \\
     \hline
     \textbf{Offline} & $2.3$ & $0.8$ & $15$ & $9.3$ \\
     \hline
\end{tabular}
\caption{Results of using adaptive and offline inference to learn a detuning frequency on the IBMQ device \texttt{ibmq\_armonk}. The detuning relative to the prevailing backend estimate was $1.83 MHz$. }
\label{tb:ramsey_adaptive}
\end{table}

An additional quantity was considered: the \textit{precision}, defined as in \cite{Santagati_2019} ($\sigma^2\Delta t_\text{acc}$, where $\Delta t_\text{acc}$ is the cumulative evolution time). This is meant to penalize long measurement times, and judge whether the adaptive method unfairly benefits from lengthier evolutions. According to this metric, adaptivity still outperforms the offline method. Note that penalties of this type can be incorporated into the utility function, if curtailing simulation runtime is critical.

\begin{figure}[!ht]
    \centering
\captionsetup[subfigure]{width=.9\textwidth}%
\begin{subfigure}[t]{.35\textwidth}
  \centering
  \includegraphics[height=5cm]{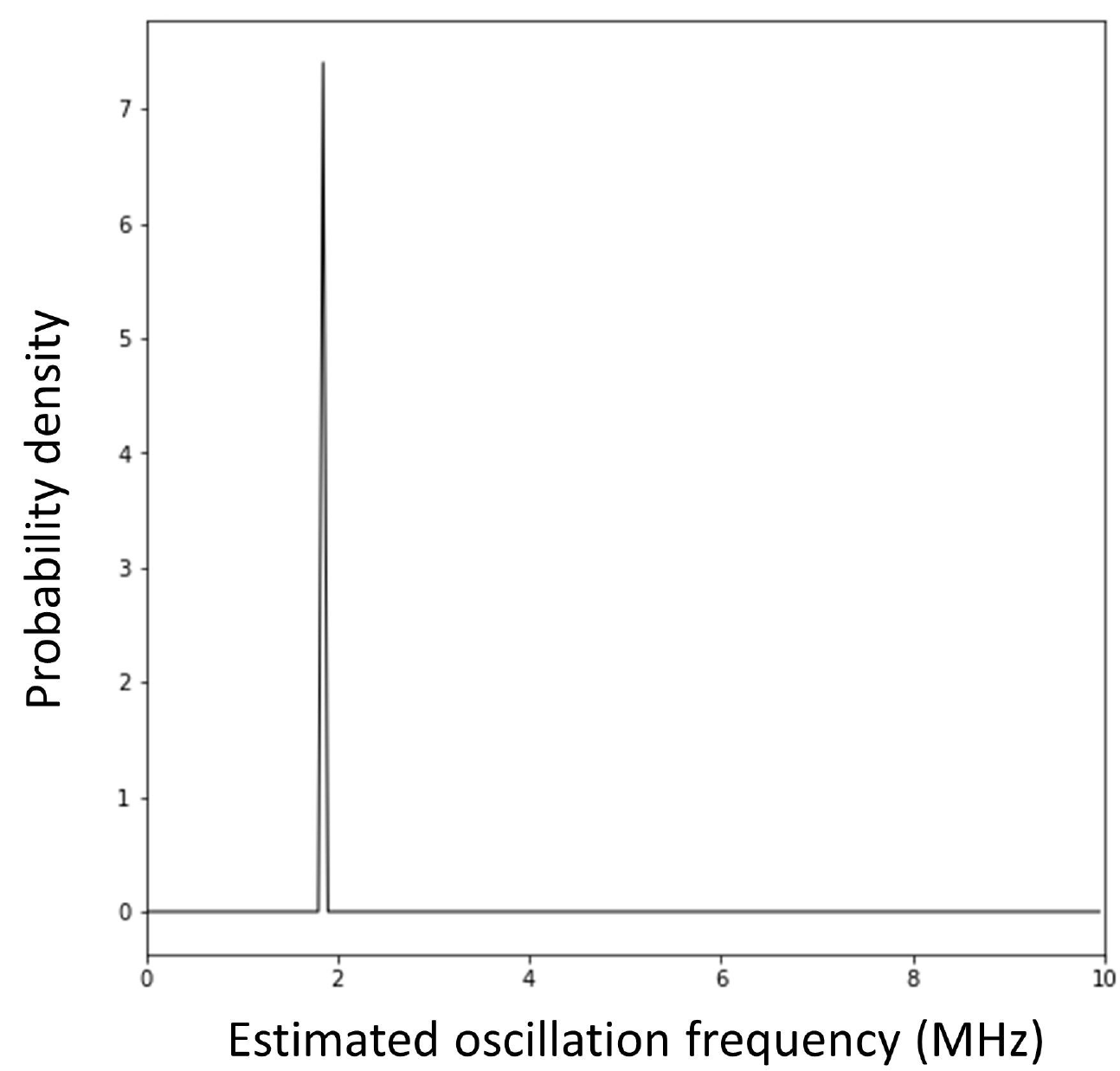}
  \caption{Kernel density estimate of the posterior distribution.}
  \label{fig:tempered_hmc}
\end{subfigure}
\begin{subfigure}[t]{.55\textwidth}
  \centering
  \includegraphics[height=5cm]{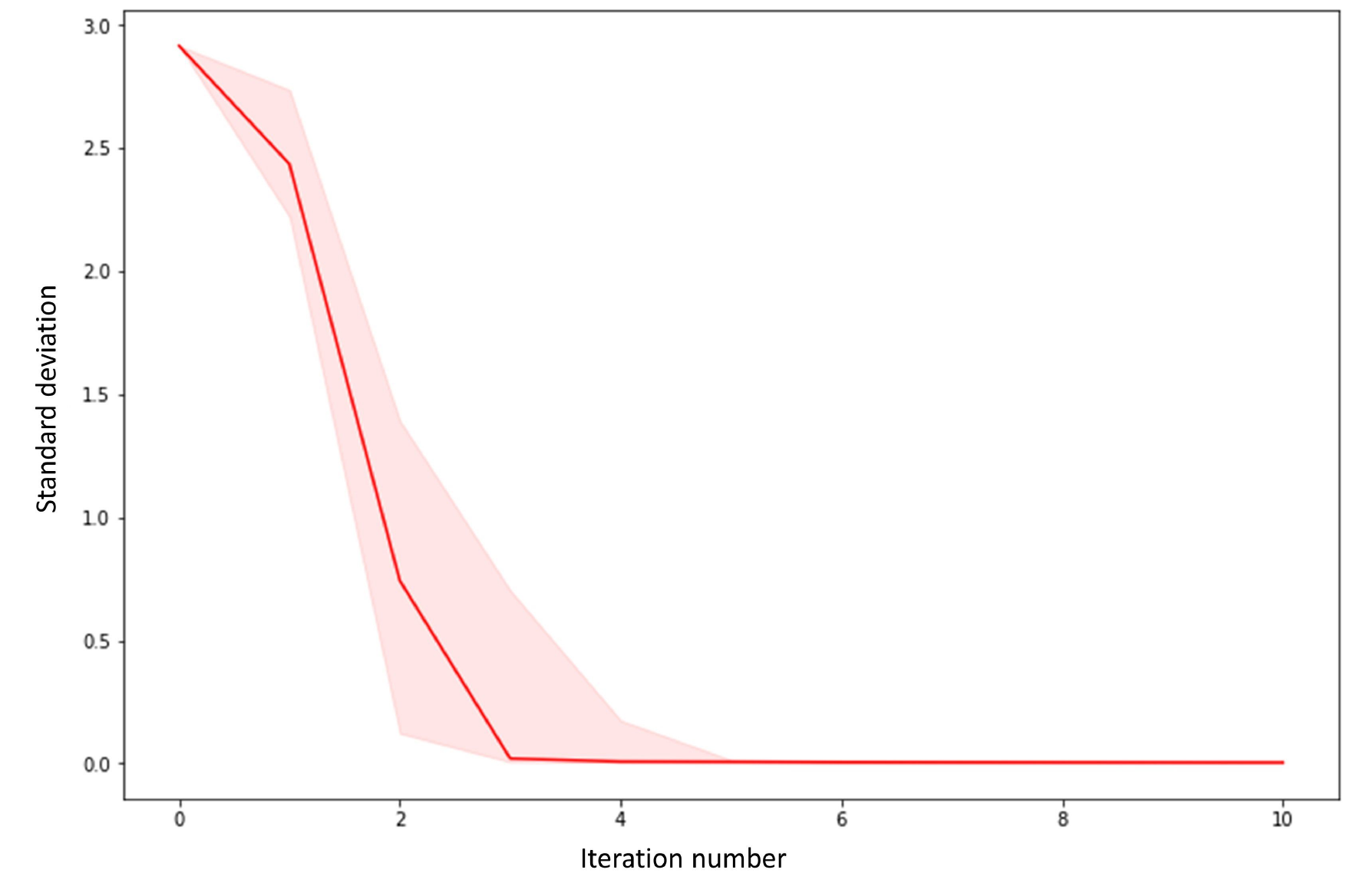}
  \caption{Evolution of the standard deviation during the inference process. The shaded region represents the interquartile range.}
  \label{fig:tle_hmc_evolution}
\end{subfigure}
\caption{Results of using \glsfmtshort{TLE} with \glsfmtshort{HMC} move steps to learn a detuning frequency on the IBMQ device \texttt{ibmq\_armonk}. The dataset contemplated $75$ observations, and ten tempering coefficients were used. }
\label{fig:tempered_hmc_all}
\end{figure}

Next, \gls{TLE} with \gls{HMC} move steps was attempted (figure \ref{fig:tempered_hmc_all}) for the backend \texttt{ibmq\_armonk}. A $100$ particle \gls{SMC} approximation was used, with a flat prior on $f \in ]0.0,10.0]MHz$. One Markov (\gls{HMC}) move was used per particle per step. The \gls{HMC} hyperparameters were set to $M=\/\sigma_\text{curr}^2$, $L=20$, $\epsilon = 0.001$, producing an acceptance rate of $68\%$. The threshold effective sample size for resampling was set at $\widehat{\text{ESS}}=0.5$. A total of $75$ different measurement times was used for a single shot each. The tempering coefficients were spaced on a grid for a total of $10$, signifying $10$ steps. Of these, $3$ ($30\%$) triggered a resampling stage on average. The measurement times were chosen in constant increments within $t \in ]0.2,10] \mu s$. The results were obtained by taking medians over $100$ runs split equally by $10$ different datasets. The quantitative results are presented in table \ref{tb:ramsey_1d}. 

\begin{table}[ht!]
\centering
\begin{tabular}{ |g|c|c|c|c| } 
     \hline
     \rowcolor{gray!15}
      & $\boldsymbol{\delta}$ ($MHz$) & \textbf{Standard deviation} & \textbf{Total shot count}\\
     \hline
     \textbf{Bayesian inference} & $1.830$ & $6 \times 10^{-3}$ & $75$ \\
     \hline
\end{tabular}
\caption{Hahn-Ramsey frequency estimation results for the IBMQ device \texttt{ibmq\_armonk}, using \glsfmtshort{TLE} with \glsfmtshort{HMC} move steps. The detuning relative to the prevailing backend estimate was $1.83 MHz$. }
\label{tb:ramsey_1d}
\end{table}

\begin{figure}[!ht]
    \centering
\captionsetup[subfigure]{width=.9\textwidth}%
\begin{subfigure}[t]{.45\textwidth}
  \centering
  \includegraphics[width=\linewidth]{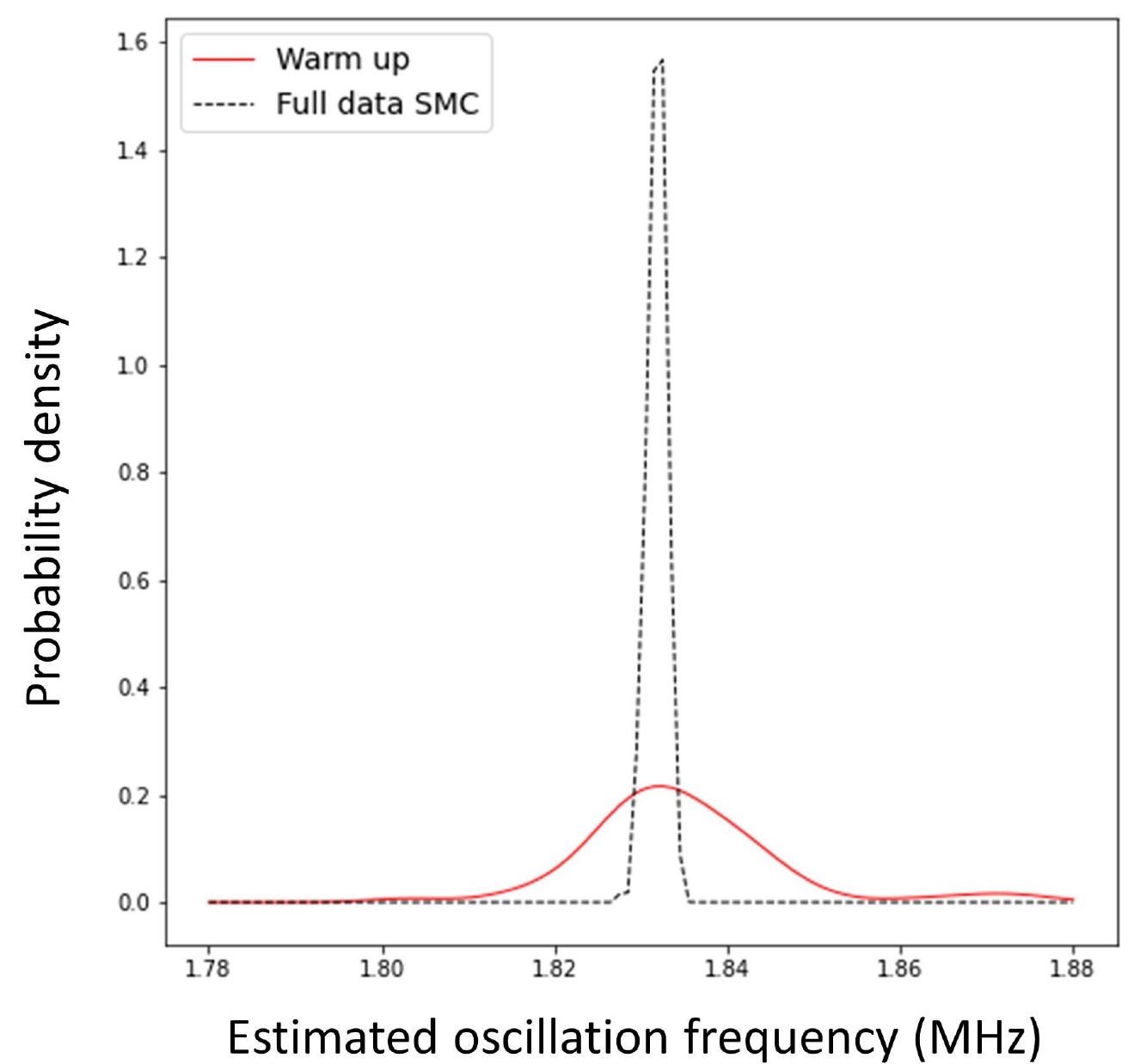}
  \caption{Kernel density estimates of the full data posterior (375 observations) and of the warmed up distribution (150 observations).}
  \label{fig:subs_warmup}
\end{subfigure}
\begin{subfigure}[t]{.45\textwidth}
  \centering
  \includegraphics[width=\linewidth]{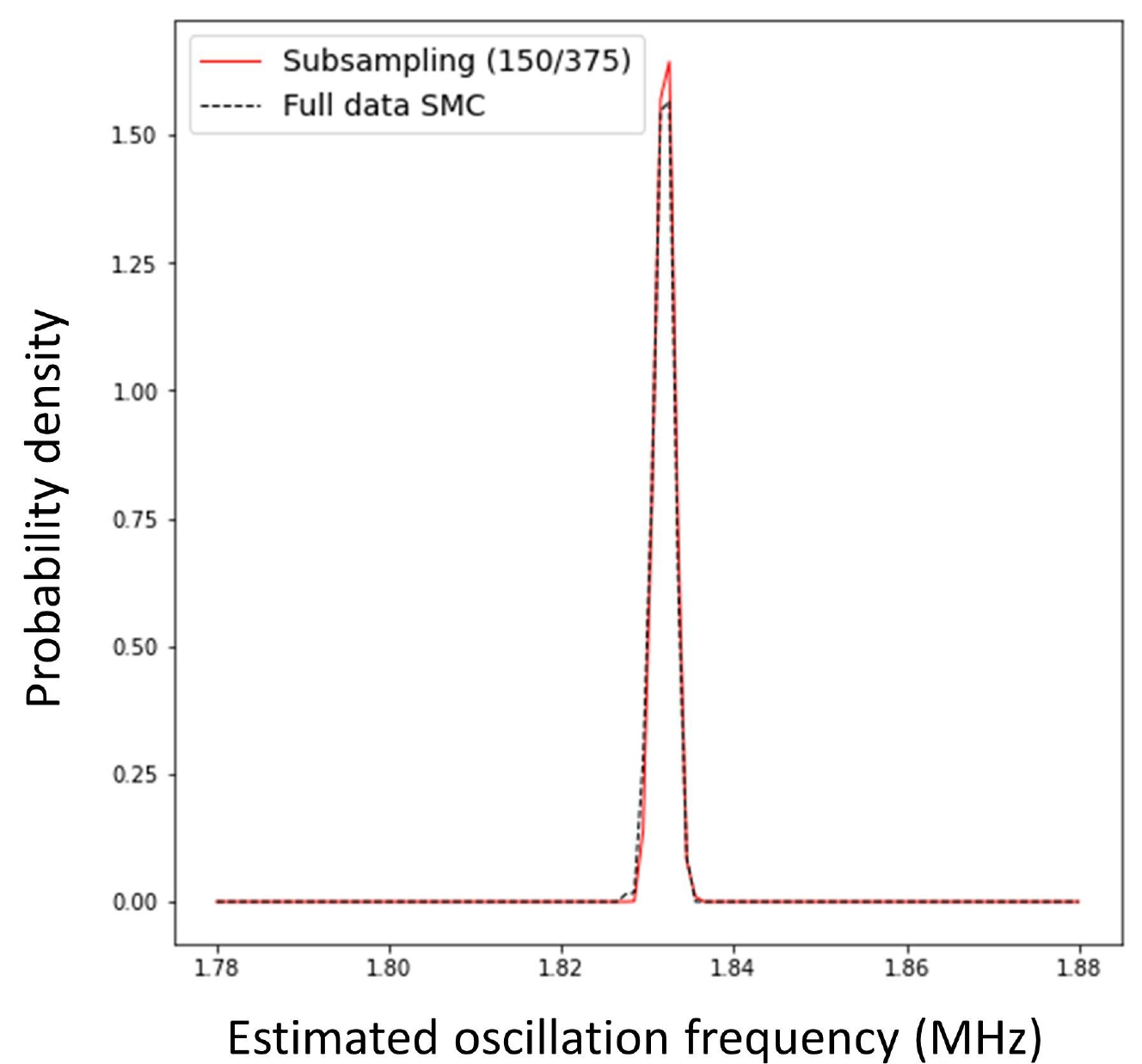}
  \caption{Kernel density estimates of the full data posterior (375 observations) and of the subsampling posterior (150 observations).}
  \label{fig:subs_final}
\end{subfigure}
\caption{Results of using \glsfmtshort{TLE} with \glsfmtshort{RWM} move steps and data subsampling with control variates to learn a detuning frequency on the IBMQ device \texttt{ibmq\_armonk}. A close up of the distribution is presented, though the prior covered $]0,10]MHz$. The dataset contemplated $375$ observations, and $10$ tempering coefficients were used, $3$ of which were employed in the warm up in the subsampling case. When warming up or subsampling, each likelihood evaluation relied on $150/375$ data, in the former case the $150$ ones with lower evolution times.}
\label{fig:subs_tle}
\end{figure}

Finally, subsampling \gls{TLE} with \gls{RWM} move steps with control variates was tested. The methods are those of subsection \ref{sec:subsampling}. As has been mentioned in chapter \ref{cha:applied_examples}, a narrowed down the prior is required for accuracy due to the control variates. This was achieved via a warm up on the lower times. The results are presented in figure \ref{fig:subs_tle}. 

The experiment was performed in similar conditions to the previous one, except for a few aspects. The number of particles was $200$, and they were resampled at each step. A total of $75$ different measurement times was used for $5$ shots each, totaling $375$ data.  The quantitative results are presented in table \ref{tb:ramsey_1d_subs} 

\begin{table}[ht!]
\centering
\begin{tabular}{ |g|c|c|c|c| } 
     \hline
     \rowcolor{gray!15}
      & $\boldsymbol{\delta}$ ($MHz$) & \textbf{Standard deviation} & \textbf{Data batch size}\\
     \hline
     \textbf{Full data \gls{TLE}} & $1.831$ & $1.2 \times 10^{-3}$ & $375$ \\
     \hline
     \textbf{Subsampling \gls{TLE}} & $1.832$ & $1.0 \times 10^{-3}$ & $150$ \\
     \hline
     \color{black!60}
     \textbf{Warm-up} & $\color{black!60} 
     1.83$ & $\color{black!60} 1.3 \times 10^{-2}$
     & $\color{black!60} 150$ \\
     \hline
\end{tabular}
\caption{Hahn-Ramsey frequency estimation results for the IBMQ device \texttt{ibmq\_armonk}, using \glsfmtshort{TLE} with \glsfmtshort{HMC} move steps based on the full data or subsampling. The detuning relative to the prevailing backend estimate was $1.83 MHz$. }
\label{tb:ramsey_1d_subs}
\end{table}

The difference in the estimator between the full data and subsampling cases was negligible. The standard deviation, albeit slightly more variable, was also close between methods.

\typeout{NT FILE chapter8.tex}

\chapter{Discussion and conclusions}
\label{cha:conclusions}

The alliance between Bayesian inference and Monte Carlo methods is a powerful tool for quantum metrology, and one which can negotiate costs on all fronts. It offers a springboard for achieving optimal balance between accuracy and each of the key resources: classical processing and memory, number of quantum measurements, number of quantum simulation runs, and simulation runtime.

Moreover, these strategies enjoy remarkable generality, barely making assumptions about the problem at hand. This is a large part of what popularized them as versatile instruments for data science, but applies even within the realm of quantum mechanics. They supply a way of characterizing quantum systems via experimental outcomes, in the broadest possible sense. These \textit{quantum systems} can be digital computers, sensors, analog simulators, circuits, or others; and the experiments can be any, as long as their outcomes provide an insight into the system's inner workings. This makes them a suitable solution for countless quantum problems, including archetypal characterization tasks (tomography, parameter estimation) but also routines as fundamental as phase estimation. Thus, the methodologies exposed herein are more than means for the validation of quantum systems: they are also one more example of how machine learning and other classical processing strategies can serve pragmatic quantum algorithms, and assist in the extraction of maximal utility from state-of-the-art quantum devices.

One of the scenarios where the Bayesian approach shines the most is when the available information is scarce, or its obtention costly. For small datasets, its performance was shown to be vastly superior to that of standard regression-based curve fitting. Other notable contexts for application are those where processing data as it arrives is an advantageous. In particular, this capability paves the way for achieving fundamental limits of metrology in the quantum regime. Adaptive quantum measurements informed by Bayesian experimental design, another perk afforded by the paradigm's far-reaching nature, present an interesting path towards Heisenberg-limited estimation. Importantly, this approach relinquishes onerous state preparation and prohibitively long coherence times in favor of classical resources. As with other \glsxtrshort{NISQ}-friendly protocols, this grants current devices the opportunity to observe quantum advantage with greatly eased requirements. Even with the simple experiments and sparing use of optimization considered here, their potential was clear, as empirical strategies for roughly locally optimal designs greatly improved the achievable precision. Equivalently, adaptivity promises to bring a sharp drop in the number of measurements required for attaining some precision threshold.

Another aspect that can have a drastic impact in performance is the representation of the posterior distribution. In Bayesian inference, given a generative model for the data, this is in principle the only source of inexactitude, as the results aren't otherwise assumed asymptotic. All sensible inquiries (in a statistical/measure-theoretic sense) to this distribution, and thus all results, are entrusted to this representation through expectation integrals. Such an ability to forsake optimization is crucial for dependability given complex models; and the lack thereof is the pitfall of many machine learning algorithms, whose naiveté becomes fatal in high-dimensional spaces.

The literature on Bayesian algorithms for quantum phenomena has favored uncomplicated strategies for relaying the results, but these are bound to become ineffective as more complex systems are targeted. And who better than modern day inference to foretell of their prospective future? Simple methods are fitting for simple problems, and/or when processing speed or memory requirements are the main concerns. However, they can quickly become unworkable, forfeiting the richness - and even the correctness - of Bayesian statistics. Not only does this compromise the results, it may also misinform experimental decisions. Ultimately, problems such as these will frustrate attempts at learning Hamiltonians with many parameters; and naturally, generalizing the description to open systems only adds to the difficulty.

In that case, high performance \gls{MCMC} methods are the foremost solution, as anticipated by their widespread use in the statistical sphere (and for Bayesian inference in particular). These Markov chains can be run independently, or coupled within a \gls{SMC} scheme. Due to benefitting from informative inter-particle interactions, the latter option can create especially robust and fast algorithms; as a bonus, the ensemble offers complimentary estimators of the the model evidence. Keeping \gls{SIR} but substituting \gls{LWF} with Markov kernels is particularly interesting, due to the fact that many advantages are kept, such as the compatibility with online processing. Markov moves inside \gls{SMC} are a widely used strategy in statistics and across the multitude of fields that rely on it, especially when correctness and generality are a concern. For offline processing, \gls{TLE} may offer still more correctness, namely for multimodal and high-dimensional distributions. It also offers a more stable understructure for subsampling, which may overcome its typically higher resource count to bring quite favorable scaling.

As for the construction of the Markov transitions themselves, the best practice is to tailor them to the problem. For continuous probability densities, \gls{HMC} is doubtlessly a strong contender, owing to the solid geometric foundations that underpin its many empirical successes. In some cases, its implementation may require adjustments; in others, such an elaborate method may be unwarranted, as others can fill in for a fraction of the cost. Regardless, it is to be expected that scaling up the complexity will eventually require upskilling the walker, and \gls{HMC}'s is arguably the most skillful of them all - the more so as the original framework is built upon.

From scrupulous experimental design to advanced sampling strategies, the tactics explored throughout this dissertation are interesting directions to be expanded further in future research, as attested by their aptness for learning the dynamical parameters of noisy quantum systems based on meager datasets. Resource intensive as they may seem, they are ultimately more cost-effective once the problem becomes complex enough - which this problem is bound to. For any progress that is made, the means for its validation should trail after; this is a necessary - if challenging - imposition for reaching new regimen of physics in applied practice. As science pushes the frontiers of what is realisable, efficient characterization will play a fundamental role in guarding the path to quantum advantage.

First of all I would like to express my gratitude to the people who made this learning experience possible. In particular, to Raffaele Santagati, for the opportunity to undertake this project and for his guidance throughout the year; to Ernesto Galvão, for his invaluable help; to Nathan Wiebe, for his insightful comments; and to professor Luís Barbosa, for being a great teacher.  

I would also like to thank my family and friends, who supported me in other ways - especially my sister Mafalda, for being my biggest companion throughout this journey.

Finally, I acknowledge the use of the use of IBM Quantum services.

\printunsrtglossaries

\bibliographystyle{plain} 
\bibliography{bibliography.bib}

\end{document}